\RequirePackage{lineno}
\documentclass[final,3p,times]{elsarticle}

\biboptions{sort&compress}
\usepackage{verbatim}
\usepackage{overpic}
\usepackage{graphicx}
\usepackage{bm} 
\usepackage{amsmath,amssymb}
\usepackage{enumerate}
\usepackage{lineno}
\usepackage[section]{placeins}
\usepackage{color}
\usepackage[figuresright]{rotating}
\usepackage{setspace}
\usepackage{booktabs} 
\usepackage{threeparttable} 
\usepackage{braket}
\graphicspath{{figures/}}
\usepackage{float}
\usepackage{epsfig}
\usepackage{epstopdf}
\usepackage{indentfirst}
\usepackage{array}
\usepackage{dcolumn}
\usepackage{multirow}
\usepackage{pifont} 
\usepackage{comment}
\usepackage{threeparttable} 
\usepackage[colorlinks,linkcolor=red,anchorcolor=blue,citecolor=blue]{hyperref}
\usepackage{cleveref}

\usepackage{hyperref}
\usepackage{doi}

\usepackage{ulem}
\usepackage{orcidlink}

\newcommand{\be}{\begin{equation}}
\newcommand{\ee}{\end{equation}}
\newcommand{\bea}{\begin{eqnarray*}}
\newcommand{\eea}{\end{eqnarray*}}
\newcommand{\beas}{\begin{eqnarray*}}
\newcommand{\eeas}{\end{eqnarray*}}
\newcommand{\X}{X(3872)}
\newcommand{\Xb}{X_b}
\newcommand{\Yb}{\Upsilon(10753)}
\newcommand{\Yfour}{\Upsilon(10580)}
\newcommand{\Yfive}{\Upsilon(10860)}
\newcommand{\Ysix}{\Upsilon(11020)}
\newcommand{\Dszero}{D^*_{s0}(2317)}

\newcommand{\ves}{{\bm s}}

\newcommand{\ver}{{\bm r}}
\newcommand{\vej}{{\bm j}}
\newcommand{\vel}{{\bm l}}

\newcommand{\veE}{{\bm E}}

\newcommand{\veS}{{\bm S}}
\newcommand{\veL}{{\bm L}}
\newcommand{\veJ}{{\bm J}}
\newcommand{\gev}{GeV/$c^2$}
\newcommand{\mev}{MeV/$c^2$}
\newcommand{\BR}{{\cal B}}

\newcommand{\dash}{--}

\journal{Physics Reports}

\begin{document}

\title{Exotic hadrons associated with $b$-quark}

\author[Tsinghua]{Xinchen Dai\orcidlink{0000-0003-3395-7151}\footnotemark[1]}
\ead{xdai@cern.ch}

\author[SEU]{Sen Jia\orcidlink{0000-0001-8176-8545}\footnotemark[1]}
\ead{jiasen@seu.edu.cn}

\author[Bonn]{Alexey Nefediev\orcidlink{0000-0002-9988-9430}\footnotemark[1]}
\ead{nefediev@hiskp.uni-bonn.de}

\author[Valencia]{Juan Nieves\orcidlink{0000-0002-2518-4606}\footnotemark[1]}
\ead{Juan.M.Nieves@ific.uv.es}

\author[FUDAN,Zhengzhou]{Chengping Shen\orcidlink{0000-0002-9012-4618}\corref{cor1}}
\cortext[cor1]{Corresponding author}
\ead{shencp@fudan.edu.cn}

\author[Tsinghua]{Liming Zhang\orcidlink{0000-0003-2279-8837}\corref{cor2}}
\cortext[cor2]{Corresponding author}
\ead{liming_zhang@mail.tsinghua.edu.cn}

\address[Tsinghua]{Department of Engineering Physics, Tsinghua University, Beijing 100084, China} 

\address[SEU]{School of Physics, Southeast University, Nanjing 211189, China} 

\address[Bonn]{Universit\"at Bonn, Helmholtz-Institut f\"ur Strahlen- und Kernphysik, D-53115 Bonn, Germany} 

\address[Valencia]{Instituto de F\'isica Corpuscular (centro mixto CSIC-UV), Institutos de Investigación de Paterna, C/Catedr\'atico Jos\'e Beltr\'an 2, E-46980 Paterna, Valencia, Spain} 

\address[FUDAN]{Key Laboratory of Nuclear Physics and Ion-beam Application (MOE) and Institute of Modern Physics, Fudan University, Shanghai 200443, China} 

\address[Zhengzhou]{School of Physics, Zhengzhou University, Zhengzhou 450001, China}

\footnotetext[1]{These authors equally contribute to this work.}

\begin{abstract}

Compared to charmonium-like states, exotic hadrons associated with $b$-quark offer distinct advantages for exploring the nature of multiquark phenomena and the dynamics of the strong interaction. Due to the heavier bottom quark mass, theoretical calculations, particularly those based on effective field theories and potential models, tend to be more reliable and under better control in the bottomonium sector. With its clean $e^+e^-$ collision environment and high luminosity, the Belle and Belle II experiments are ideally suited to explore these exotic hadrons associated with $b$-quark, including $Z_b$, $\Xb$, and $Y_b$ states, and charmonium-like states in $B$ decays. Utilizing the large proton--proton collision dataset, the LHCb experiment has conducted extensive investigations of heavy-flavor multiquark states through $B$ and $\Lambda_b$ decay channels‌. The relevant phenomenological interpretations are also reviewed.

\end{abstract}


\maketitle

\tableofcontents


\section{Introduction}

The spectroscopy of exotic hadrons has become a central topic in heavy-flavor physics that constantly brings many surprises and new experimental discoveries, and calls for considerable theoretical efforts to describe the newly observed exotic states, disclose their nature, and understand their properties. Decays of hadrons containing the bottom quark --- hereinafter referred to as $b$-hadrons for brevity --- provide a particularly suitable environment for such studies. Indeed, the large mass of the $b$ quark and a rich family of the decay topologies of $b$-hadrons allow one to access a wide variety of multi-body final states, in which non-conventional multi-quark configurations may naturally manifest themselves. 

%
Our contemporary understanding of the hadron spectrum, its classification, and the associated naming scheme is rooted in the flavor classification framework, obtained by extending the isospin symmetry to the unitary group SU(3), which was independently proposed by Gell-Mann \cite{Gell-Mann:1961omu} and Ne'eman \cite{Neeman:1961jhl} in 1961. In 1964, this framework received a microscopic interpretation through the quark model proposed by Gell-Mann \cite{Gell-Mann:1964ewy} and Zweig \cite{Zweig:1964ruk}. The quark--antiquark mesons and three-quark baryons, since then regarded as ordinary hadrons, were argued to fill the corresponding SU(3) multiplets according to the reductions 
\be
3\otimes \bar{3}=1\oplus 8,\quad 3\otimes 3\otimes 3=1\oplus 8 \oplus 8\oplus 10
\ee
of the flavor wave functions into irreducible representations for mesons and baryons, respectively. Within this picture, all hadrons known at the time could be understood in terms of the three light quark flavors, namely $u$, $d$, and $s$. One of the natural outcomes of the proposed scheme --- the relations between the masses of such hadrons (the so-called Gell-Mann--Okubo mass formulas) \cite{Gell-Mann:1961omu,Okubo:1961jc} --- is fulfilled to a high precision, and missing states came as predictions of the model that could be verified experimentally. 
Since then, the quark model developed from a symmetry-based classification scheme for hadronic states to a powerful theoretical method that allows one not only to predict the existence of a new meson or baryon but also to calculate its mass and partial decay probabilities, estimate its total width, and so on. 
Today, not only is the number of quark models used in the literature on hadronic physics enormous, but there is also a wide variety of these models, ranging from the simplest constituent potential models to more sophisticated ones based on quantum field theory and using nonlocal potentials and quasi-potentials, among others.
Since any quark model operates with quarks and describes their motion inside of a hadron, the relevant quantum numbers to be used as building blocks are the spins of the quarks and antiquarks as well as their relative angular momenta that finally sum to the total spin of the hadron. 
Thus, the classification scheme for hadrons, inspired by the quark model and used in high-energy physics for more than half a century, was based on these quantum numbers and consisted in assigning each meson or baryon a specific letter (Greek or Latin) according to its quantum numbers $J^{PC}$. For instance, light unflavored isospin-1 quark--antiquark mesons composed of the $u$ and $d$ quarks are conventionally tagged as $\pi$'s (pseudoscalars), $a_0$'s (scalars), $\rho$'s (vectors), and so on. As soon as the scheme is extended to the $s$ quark, unflavored isospin-0 $\eta$'s (pseudoscalars), $f_0$'s (scalars), $\phi$'s (vectors), and so on are included.

The existence of the $c$ quark in nature was anticipated already in 1964 by Bjorken and Glashow \cite{Bjorken:1964gz} and basically predicted by Glashow, Iliopoulos, and Maiani in 1970 \cite{Glashow:1970gm}. The first hadron composed of the $c\bar{c}$ pair was observed shortly after that, in 1974 \cite{E598:1974sol,SLAC-SP-017:1974ind}. Although the $c$ quark is heavy (unlike the $u$, $d$, and $s$ quarks, its mass exceeds several times the typical scale of QCD $\Lambda_{\rm QCD}\simeq 300~\mbox{MeV}$) and, consequently, the flavor group of the light quarks can not be extended to hadrons with charm, the same basic naming principles naturally apply to such hadrons. In particular, the $c\bar{c}$ mesons (charmonia) are conventionally tagged as $\eta_c$'s for pseudoscalars, $\psi$'s for vectors (with the only exception for the lightest one known as $J/\psi$ for historical reasons), $\chi_{cJ}$'s for the positive-parity states with $J=0,~1,~2$, and so on.
This naming scheme naturally extends to the $b$ quark --- another heavy quark in the Standard model set of elementary particles. 
This way the names $\eta_b$'s, $\Upsilon$'s, $\chi_{bJ}$'s, and so on appear
for the $b\bar{b}$ states (bottomonia) with different quantum numbers. 
The existence of more complicated multiquark configurations than the generic $q\bar{q}$ mesons and $qqq$ baryons was anticipated by many physicists already in the first years of the quark model. However, in the absence of clear experimental signatures for such states in nature, no standard names were invented for them at that time.

The naming scheme for hadrons outlined above was challenged in 2003 when Belle discovered the first unconventional hadronic state in the spectrum of charmonium \cite{Belle:2003nnu}. Very unusual and unexpected properties of this state, if viewed from the position of the quark model for a generic $c\bar{c}$ meson, were reflected in regarding it as exotic and further ``encoded'' in the name assigned to it --- to emphasize its unknown quantum numbers and internal structure, 
the discovered state was tagged $\X$, where the letter was chosen from the end of the Latin alphabet and the number in parentheses corresponded to its measured mass. 
The quantum numbers of this state were not firmly established for over a decade after its discovery, so it became a tradition to name new candidates for exotic states with unassigned quantum numbers as $X$ states.
Another, this time vector, state with unusual properties, discovered by BaBar in the charmonium spectrum in 2005 \cite{BaBar:2005hhc} and originally named $Y(4260)$, gave rise to yet another family of exotic states, which were designated as $Y$ hadrons or simply $Y$ states.
Finally, a charged (isovector) exotic state with the quantum numbers\footnote{Hereinafter the $C$-parity of the charged isovector states is defined for their neutral components.} $J^{PC}=1^{+-}$ found in the spectrum of charmonium in 2007
\cite{Belle:2007hrb} was named $Z(4430)$ while the further resonances belonging to the same class observed in 
2013 were named $Z_c(3900)$ \cite{BESIII:2013ris,BESIII:2015cld,Belle:2013yex,BESIII:2013qmu} and $Z_c(4020)$ \cite{BESIII:2013ouc,BESIII:2014gnk,BESIII:2013mhi,BESIII:2015tix}, with the subscript indicating their association with a particular heavy-quark sector. This way a new naming scheme, often referred to as the $XYZ$ scheme, was originated by the above discoveries. 
This scheme is straightforwardly extended to the spectrum of bottomonium, with the natural substitution of names $X\to X_b$, $Y\to Y_b$, and $Z_c\to Z_b$. For a comprehensive overview of the experimental and theoretical status of the 
$XYZ$ states up to 2019, see Ref.~\cite{Brambilla:2019esw}.
Notice that, while two representatives of the $Z_b$-family --- the $Z_b(10610)$ and $Z_b(10650)$ --- were already unambiguously established by Belle in 2011 \cite{Belle:2011aa} and a good candidate to a $Y_b$ state --- the $\Yb$ --- was reported by Belle in 2019 \cite{Belle:2019cbt}, experimental searches for the $X_b$ --- a heavy flavor partner of the $\X$ --- have not revealed any candidate yet. Below we give a comprehensive introduction to various aspects of the physics related to the two discovered $Z_b$'s as well as their predicted spin partners and then discuss the current experimental and theoretical situation with the studies of the $X_b$ and $Y_b$ bottomonium-like states. 

It is noteworthy that in recent years, motivated by the rapid increase in the number of exotic states observed experimentally and predicted theoretically, the Particle Data Group (PDG) extended the naming scheme for hadrons containing heavy quarks, so now it both incorporates the conventions inherited from the quark model and naturally applies to exotic states with different quark content.
Indeed, once the collider experiments started their physics work programs at the energies above the open-flavor thresholds, they inevitably had to deal with hadrons that could be observed not only as compact quark states but, with a large probability, as hadron-hadron systems as well. Thus, the wave functions of such hadrons are expected to have multiple components, including both quark and hadronic ones simultaneously. 
Although such hadronic states are still conventionally regarded as exotic, the conjecture that their wave functions go beyond the simple $q\bar{q}$ or $qqq$ assignment appears not only plausible but also most natural at energies where the creation of open-flavor hadron pairs is energetically allowed. Within the updated hadronic nomenclature, the particle name is defined so as to reflect both its quantum numbers and its minimal admissible quark content, now also for exotic hadrons. All generic quark model states preserve their old famous names in the proposed scheme. In the meantime, the progenitor of the entire family of exotic states --- the $\X$ --- should now be named $\chi_{c1}(3872)$ pointing to its established quantum numbers $1^{++}$ and at the same time indicating that this state may not be a generic radially excited $c\bar{c}$ charmonium --- the latter would be denoted as $\chi_{c1}(2P)$ instead. Similarly, the charged $Z_c$ states in the spectrum of charmonium and $Z_b$ states in the spectrum of bottomonium should be named $T_{c\bar{c}1}$ and $T_{b\bar{b}1}$, respectively. These new names can be decoded as tetraquarks (the minimal possible quark content of the charged $Z$'s)
with a $c\bar{c}$ or $b\bar{b}$ pair on board and with the total spin $J=1$. The light-quark ($u$ and $d$) content is not stated explicitly and can be unambiguously established from the quantum numbers. 
Similarly, the states $T_{cs0}$, $T_{c\bar{s}0}$, $T_{c\bar{c}\bar{s}1}$ discussed in Sec.~\ref{sec:Bmeson} below correspond to tetraquarks with the strange/heavy quark content being $cs$, $c\bar{s}$, and $c\bar{c}\bar{s}$, respectively (plus one or two light quarks/antiquarks) and the total spin $J=0$, for the two former states, and $J=1$, for the latter. Similar rules apply to other multiquark configurations like pentaquarks --- the designation of the $P_{c\bar{c}s}$ family discussed in Subsec.~\ref{sec:penta} is interpreted as referring to an exotic hadron with a minimal five-quark content, including a $c\bar{c}$ pair and one $s$ quark. Further details of the contemporary PDG naming scheme can be found in Review of Particle Physics (RPP) \cite{ParticleDataGroup:2024cfk}.

Finally, the discovery of exotic hadrons led to yet another paradigm change. Indeed, for a generic isolated quark model state, with its line shapes well approximated by a Breit--Wigner (BW) distribution, the mass as a parameter of this distribution and the mass as the real part of the corresponding pole on the unphysical Riemann sheet on the complex energy plane could be used interchangeably. However, line shapes for exotic near-threshold states have the structure that typically departs far away from the BW one, so the masses of such exotic states should be understood as the real parts of the corresponding poles of the amplitude. Furthermore, in some cases, the amplitude may possess a nontrivial multipole structure like, for example, a two-pole structure now commonly adopted for the $\Lambda(1405)$ baryon \cite{ParticleDataGroup:2024cfk}; see also Ref.~\cite{Meissner:2020khl} for a detailed and instructive discussion of the case. Similar nontrivial structures may be naturally expected to exist in the spectrum of hadrons containing $b$-quark, too \cite{Kolomeitsev:2003ac,Guo:2006fu,Guo:2006rp}.

This review aims to summarise the progress achieved over the past several years in the experimental observation and theoretical description of exotic states that either contain $b$ quarks or are produced in decays of $B$ mesons. Such a summary is particularly timely in view of the ongoing data taking and the numerous related discoveries recently reported by two major international experiments: Belle~II at the KEK laboratory in Japan and LHCb at CERN in Europe. Accordingly, our decision on whether to discuss a given exotic state is primarily guided by the availability of recent experimental information.
As an illustration, we do not consider the well-known double-charm tetraquark $T_{cc}(3875)^+$, recently observed by LHCb~\cite{LHCb:2021vvq}. Although the discovery of this state has triggered an active discussion in the hadronic community, including the prediction of its possible heavy-quark flavor partners $T_{bb}$ and $T_{bc}$, $T_{cc}$, it has not yet been observed at Belle~II, while its partner states can not be accessed at this experiment at all. In principle, LHCb could search for the particularly intriguing $T_{bc}$ state, which may exhibit a non-trivial interplay between compact multiquark and molecular degrees of freedom; however, no experimental information on this state is currently available.

Another disclaimer regarding the presentation and interpretation of the experimental material in this review is also in order here. Typical event distributions reported by experimental collaborations in energy regions where multiple strong thresholds are present often exhibit nontrivial shapes, with numerous structures such as peaks and dips. In such cases, strong coupled-channel effects and final-state interactions may play a more significant role in generating these features than additional poles of the amplitude in the complex energy plane, which are conventionally interpreted as physical states. However, currently available reliable, model-independent theoretical approaches can provide a controlled description of the line shapes only within a relatively narrow near-threshold region and can not account for the entire energy range covered by the experimental signal. Conversely, theoretical frameworks designed to fit the data over a sufficiently broad energy interval inevitably sacrifice model independence and often involve a large number of parameters that are not well constrained by the available data. Under such circumstances, experimental data analyses frequently rely on relatively simple fits that incorporate multiple assumed hadronic resonances, with their parameters determined directly from the data. It therefore remains an open question whether the inclusion of such multiple seed poles in the amplitude is indeed justified, or whether the observed structures could instead be explained by various multichannel re-scattering effects. In the absence of reliable and universal theoretical tools for analyzing these data, this review adopts a pragmatic approach: we report the details of the analyses and their results as presented in the corresponding experimental publications, without attempting a critical assessment of the physical interpretation underlying those analyses.

The review has the following structure. 
In Sec.~\ref{sec:exp}, we give a pedagogical introduction to different experiments employed in studies of exotic states associated with the $b$ quark. In particular, we discuss the former Belle and current Belle II experiments at the $e^+e^-$ collider in Japan and the LHCb experiment at the proton--proton collider at CERN. In Sec.~\ref{sec:bquark}, we concentrate on the exotic states that contain $b$ quark in their content. In particular, we (i) introduce the $Z_b$ states discovered and studied by Belle, (ii) discuss their theoretically predicted spin partners $W_{bJ}$, (iii) overview the current experimental and theoretical situation with the candidates to vector bottomonium-like states $Y_b$ (with a focus on the $\Yb$), and (iv) comment on the current searches for the $\Xb$ --- a gedanken heavy flavor partner of the $\X$. In Sec.~\ref{sec:Bmeson} we proceed to the exotic states produced in $B$ decays. Finally, we discuss future prospects of the field in Sec.~\ref{sec:future} and summarize in Sec.~\ref{sec:summary}.

As a general rule throughout this review, if an experimental measurement is given with two uncertainties quoted one after another, they are the statistical and systematic one, respectively.

\section{Introduction to different experiments}
\label{sec:exp}

\subsection{The Belle and Belle II experiments}

\begin{figure*}[t]
\centering
\includegraphics*[width=0.7\textwidth]{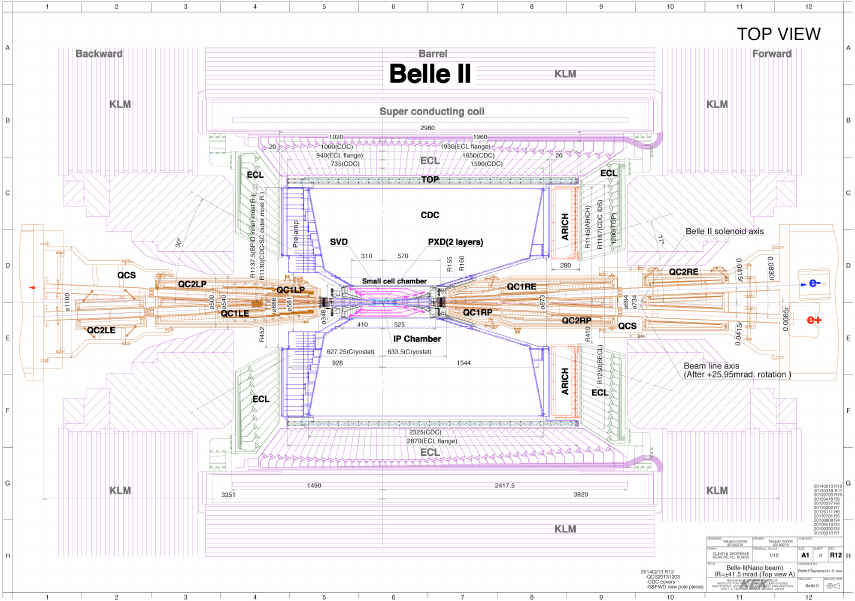}
\caption{Belle detector from top review~\cite{Belle-II:2010dht,Adachi:2018qme}.
}\label{fig:Belle2_detector}
\end{figure*}

The Belle detector~\cite{Belle:2000cnh,Belle:2012iwr} collected data from collisions of 8-GeV electrons on 3.5-GeV positrons produced by the KEKB collider~\cite{Kurokawa:2001nw,Abe:2013kxa} at KEK. An upgraded detector, Belle II~\cite{Belle-II:2010dht}, collects collisions of 7-GeV electrons on 4-GeV positrons produced by the SuperKEKB successor collider~\cite{Akai:2018mbz}. 
A detailed description of the Belle detector can be found in Refs.~\cite{Belle:2000cnh,Belle:2012iwr}. 
The Belle II detector is an upgraded version of the Belle detector that contains several completely new subdetectors, as well as substantial upgrades to other subsystems. The innermost subdetector is the vertex detector (VXD), which uses position-sensitive silicon sensors to precisely sample the trajectories of charged particles (tracks) in the vicinity of the interaction point. The VXD includes two inner layers of pixel sensors and four outer layers of double-sided silicon microstrip sensors.
Charged-particle momenta and charges are measured by a new large-radius, helium-ethane, small-cell central drift chamber (CDC), which also gives charged-particle-identification information through a measurement of particles’ specific ionization. The Belle particle identification system has been replaced. A Cherenkov-light angle and time-of-propagation detector surrounding the CDC provides charged-particle identification in the central detector volume, supplemented by a proximity-focusing aerogel ring-imaging Cherenkov detector in the forward region. 
The readout electronics of the electromagnetic calorimeter comprised of CsI(Tl) crystals has been upgraded.
The resistive plate chambers in the endcaps and the inner two layers of the barrel have been replaced by scintillator strips to detect $K^0_L$ mesons and identify muons. 
Figure~\ref{fig:Belle2_detector} shows the structure of the Belle II detector~\cite{Belle-II:2010dht,Adachi:2018qme}.
A detailed description of the Belle II detector can be found in Ref.~\cite{Belle-II:2010dht}. 

The Belle detector took data from 1999 to 2010. Since the beginning of 2019, its successor experiment Belle II has been in operation. 
Integrated luminosities of data samples accumulated by Belle and Belle II experiments are listed in Table~\ref{data}~\cite{Belle:2012iwr,Belle-II:2024vuc}.
The $\Yfour$ resonance\footnote{As explained in the Introduction, hadrons that can be identified as conventional quark-model states are denoted by their spectroscopic quantum numbers in parentheses. Accordingly, following the RPP, we refer to the three lightest vector bottomonia as $\Upsilon(1S)$, $\Upsilon(2S)$, and $\Upsilon(3S)$ \cite{ParticleDataGroup:2024cfk}. However, when the internal structure of a hadron is under discussion, we refer to it by its mass instead. Thus, although for historical reasons the higher vector states in the bottomonium spectrum are often denoted as $\Upsilon(4S)$, $\Upsilon(5S)$, and $\Upsilon(6S)$, hereinafter we refer to them as $\Yfour$, $\Yfive$, and $\Ysix$, respectively, in order to emphasize that they reside above the open-bottom threshold and, as such, are expected to possess a more intricate structure than conventional $b\bar b$ mesons.\label{foot:Ups}} has the highest integrated luminosity, as it is the optimal center-of-mass (C.M.) energy for the production of $B\bar B$ pairs used in $B$ physics analysis.
The continuum data was collected at $\sqrt{s}$ = 10.52 GeV to study the $e^+e^-\to q\bar q$ ($q=u,~d,~s,~c$) events.
Belle and Belle II recorded a series of unique data sets at the $\Upsilon(1S)$, $\Upsilon(2S)$, $\Upsilon(3S)$, $\Yb$, and $\Yfive$ resonances.
The data sets at the $\Yb$ and $\Yfive$ resonances are of special interest in exotic hadrons associated with $b$-quark --- see Sec.~\ref{sec:bquark} below.
There are some scan data points between $\Upsilon(10580)$ and $\Upsilon(11020)$ with low statistics for a measurement of the inclusive hadronic cross sections.
Until 2025, the total integrated luminosity at Belle and Belle II exceeds 1.5 ab$^{-1}$.
The integrated luminosity at Belle II is expected to achieve 4 ab$^{-1}$ and 10 ab$^{-1}$ in 2029 and 2034, respectively~\cite{luminosity}.

To summarize, the Belle~II experiment, operating at a high-luminosity electron--positron collider, 
offers a particularly clean experimental tool with low background 
level and well-identification final states. By collecting data at the C.M. energies 
in the vicinity of the $\Upsilon$ resonances, Belle~II provides a promising platform 
for the systematic study of exotic bottomonium-like hadrons, including the $Z_b$, $X_b$, 
and $Y_b$ states. The established collision energy and clean environment
allow for a partial reconstruction technique to increase the efficiency and measure the absolute branching fractions.
The production cross sections of these exotic hadronic states at electron--positron colliders are relatively small, so accumulation of large data samples is essential to achieve sufficient precision for detailed 
studies of their production mechanisms, decay patterns, and internal structures. See also the review \cite{Eidelman:2020iul}.

\begin{table*}[t!]
\renewcommand\arraystretch{1.2}
\setlength{\tabcolsep}{2pt}
\centering
\caption{Until 2025, the integrated luminosities (fb$^{-1}$) of data samples at different energies accumulated by Belle and Belle II experiments~\cite{Belle:2012iwr,Belle-II:2024vuc}.}\label{data}
\vspace{0.2cm}
\begin{tabular}{cccccccccc}
\hline\hline
Experiment & $\Upsilon(1S)$ & $\Upsilon(2S)$ & $\Upsilon(3S)$ & continuum & $\Yfour$ & $\Yb$ & $\Yfive$ & Scan & SUM \\\hline
Belle & 6 & 25 & 3 & 89 & 711 & - & 121 & 28 & 983 \\
Belle II & - & - & - & 61 & 498 & 20 & - & - & 579 \\
\hline\hline
\end{tabular}
\end{table*}

\subsection{The LHCb experiment}

\begin{figure}[t!]
\centering
\includegraphics[width=0.45\linewidth]{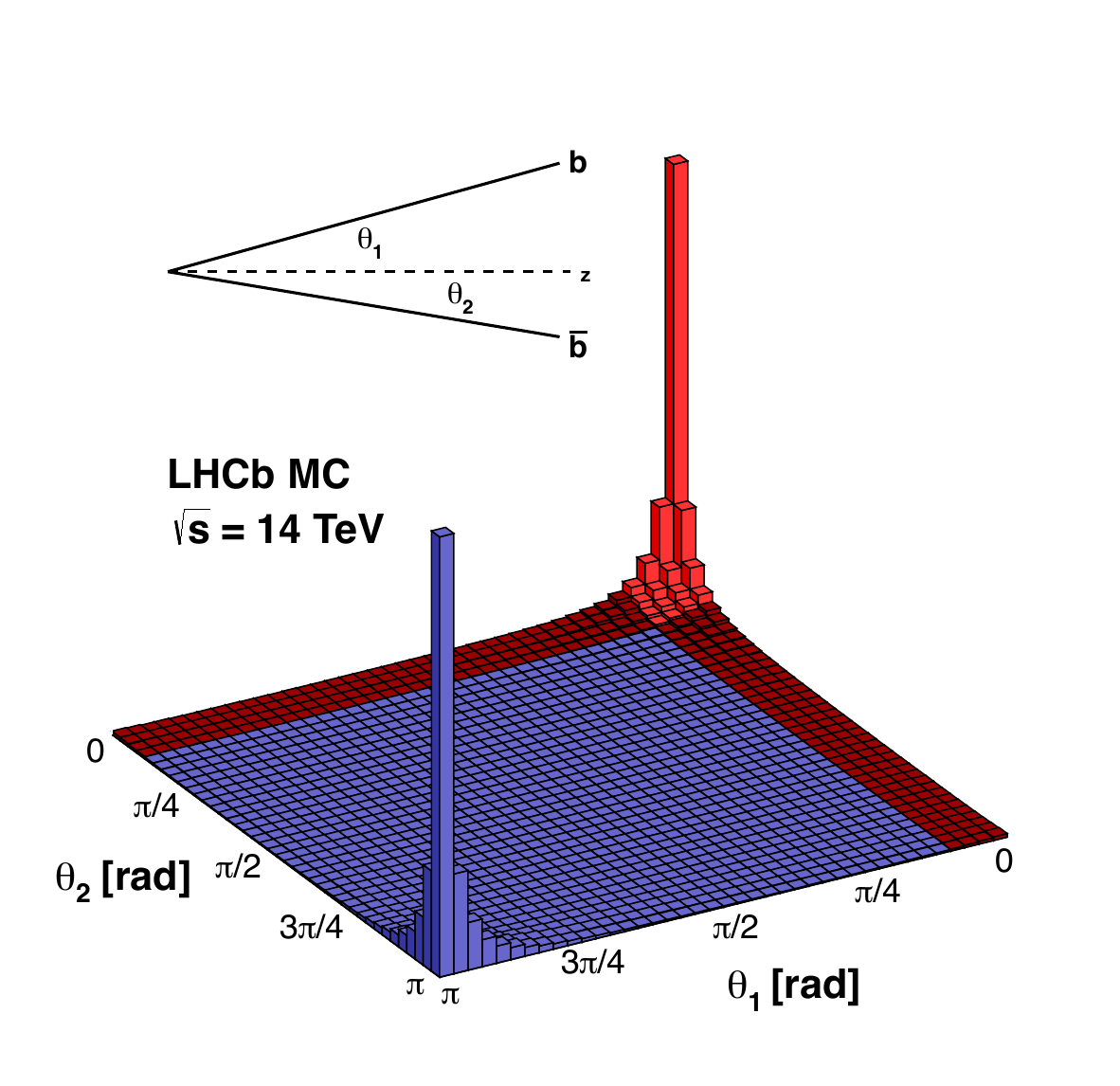}
\qquad
\includegraphics[width=0.5\linewidth]{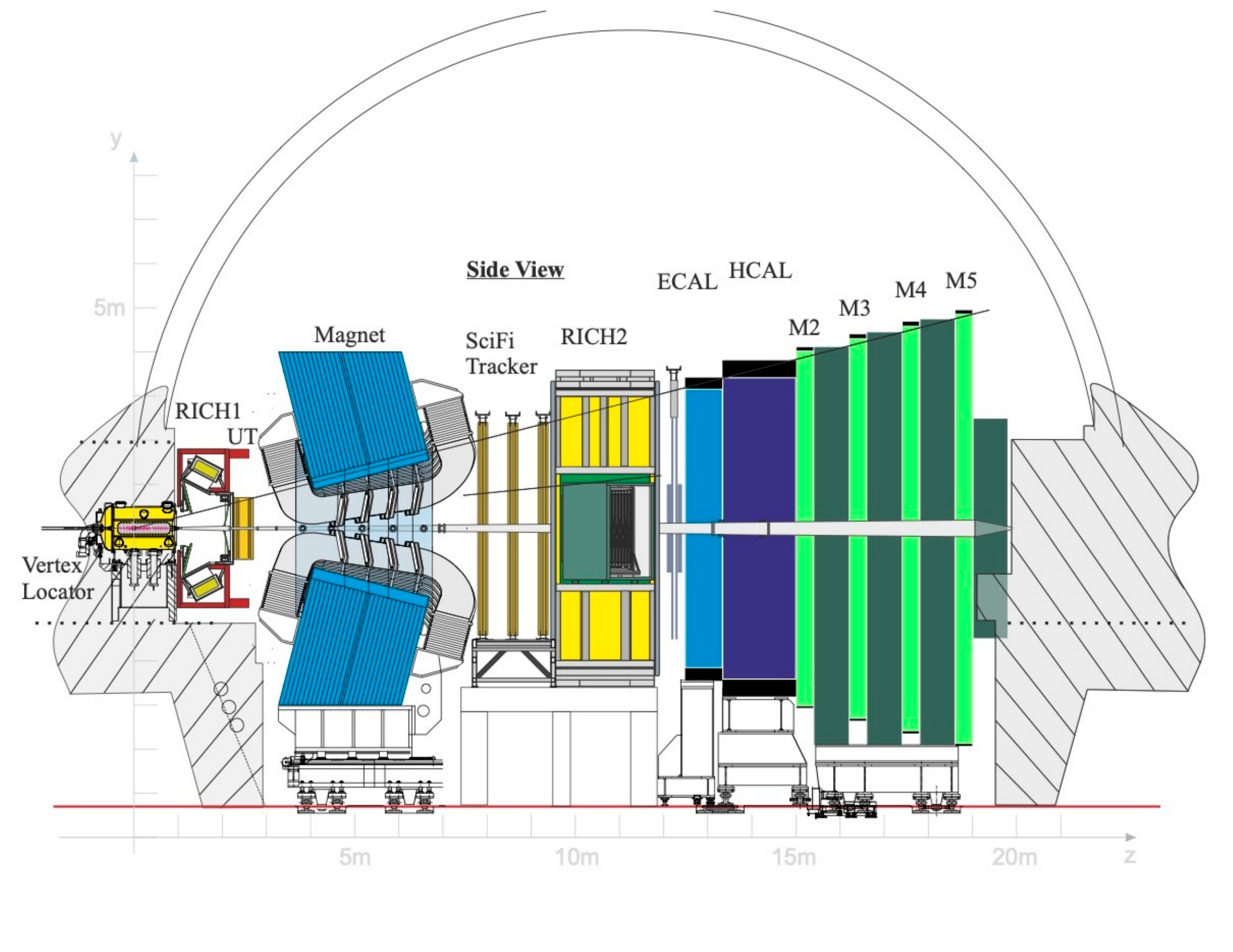}
\caption{At the LHC, heavy-flavor quark pairs are produced in the forward or backward direction~(left). Layout of the upgraded LHCb detector (right)~\cite{LHCb:2023hlw}. }
\label{fig:boost_angle_detector}
\end{figure}

The LHCb detector is specifically designed 
to the study of heavy-quark hadron production and decay processes~\cite{LHCb:2008vvz}. Owing to the kinematics of $b\bar b$ and $c\bar c$ production at the LHC, heavy-flavor hadrons are predominantly produced in the forward direction. This motivates the forward geometry of the LHCb detector, which instruments the pseudorapidity region $2 < \eta < 5$, where heavy-flavor hadrons are produced abundantly and can be efficiently reconstructed within a relatively small solid angle, as shown in Fig.~\ref{fig:boost_angle_detector}.
The tracking system consists of several subsystems arranged along the beam direction. Closest to the interaction point is the vertex locator (VELO), composed of silicon pixel detectors surrounding the beam line. It is followed by the silicon-strip upstream tracker located upstream of a dipole magnet, and by scintillating fibre detectors (SciFi Tracker) downstream of the magnet. The VELO provides precise measurements of proton--proton collision points, known as the primary vertices, displaced decay vertices of $b$ and $c$ hadrons, and the distances between them, enabling an efficient separation of heavy-flavor decays from promptly produced background.
The tracking system measures the momentum $p$ of charged particles with a relative uncertainty ranging from about 0.5\% at low momentum to approximately 1.0\% at $p \simeq 200~\mathrm{GeV}/c$. This performance translates into an invariant-mass resolution of typically 10\dash20~\mev~for fully reconstructed $b$-hadron decays, allowing different heavy-flavor hadron species to be clearly separated. Such mass resolution is a key ingredient for spectroscopy studies, including searches for narrow exotic structures.
Particle identification at LHCb is provided by two ring-imaging Cherenkov detectors, a shashlik-type electromagnetic calorimeter, an iron--scintillator tile hadronic calorimeter, and four muon stations. These subsystems deliver excellent discrimination between pions, kaons, protons, electrons, photons, and muons over a wide momentum range, which is essential for suppressing background and for reconstructing complex multibody final states relevant to exotic-hadron searches. 
Combined with a trigger system optimized for heavy-flavor signatures, the LHCb experiment achieves high trigger and reconstruction efficiencies, thereby maximising the available signal yields.

During LHCb Run~I, data were collected at the C.M. energies of 7~TeV (2011) and 8~TeV (2012), while Run~II corresponds to data-taking at 13~TeV from 2015 to 2018. In both Run~I and Run~II, the instantaneous luminosity delivered to LHCb was levelled to a maximum of about $4 \times 10^{32}~\mathrm{cm}^{-2}\mathrm{s}^{-1}$. The ongoing Run~III, which started in 2022, operates at a higher levelled luminosity of approximately $2 \times 10^{33}~\mathrm{cm}^{-2}\mathrm{s}^{-1}$. With the expected data-taking through Run~IV until around 2032, the total integrated luminosity collected by LHCb is projected to reach about $50~\mathrm{fb}^{-1}$. This large data sample will enable significantly more precise measurements and substantially extend the sensitivity of exotic-hadron spectroscopy studies.

To summarize, at hadron colliders including LHC, the production of heavy quarks is dominated by strong-interaction processes, in particular gluon--gluon fusion, resulting in heavy-flavor production rates that exceed those available at electron--positron colliders by orders of magnitude~\cite{LHCb:2010wqx,LHCb:2013xam}. This abundance of heavy-flavor hadrons enables detailed spectroscopic studies with high statistical precision. At the same time, the hadronic collision environment gives rise to large combinatorial and physical backgrounds, making precise vertexing, momentum measurement, particle identification, and highly selective trigger strategies essential prerequisites for the successful study of exotic hadron states.

\section{Exotic states with $b$ quark}
\label{sec:bquark}

In this section we mainly resort to the $XYZ$ naming scheme for the exotic states to better comply with the conventions adopted in the works by Belle and Belle II.

\subsection{$Z_b$ states (also known as $T_{b\bar{b}1}$'s)}
\label{sec:Zb}

\subsubsection{Experimental observation and studies}

\begin{figure*}[t!]
\centering
\includegraphics*[width=0.95\textwidth]{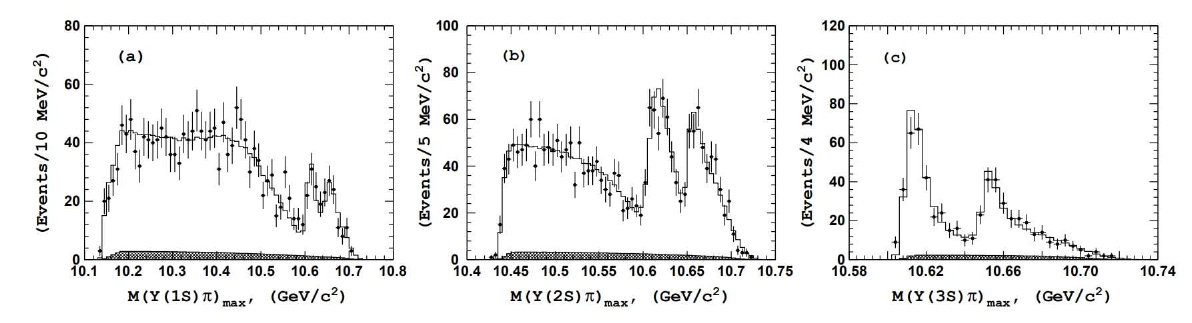}
\caption{Belle Collaboration data for the mass of the $\Upsilon\pi$ subsystem in the transitions from the $\Yfive$ state to $\pi\pi\Upsilon(1S)$ (first plot),
$\pi\pi\Upsilon(2S)$ (second plot), and $\pi\pi\Upsilon(3S)$ (third plot)~\cite{Belle:2011aa}.}\label{fig:5SpipiUps}
\end{figure*}

Using data of 21.7 fb$^{-1}$ near the peak of the $\Yfive$ resonance, Belle reported the first observation of the reactions $e^+e^-\to\pi^+\pi^-\Upsilon(1S)$ and $e^+e^-\to\pi^+\pi^-\Upsilon(2S)$, and first evidence for $e^+e^-\to\pi^+\pi^-\Upsilon(3S)$ at $\sqrt{s}$ = 10.87 GeV~\cite{Belle:2007xek}.
Assuming all signals are from the $\Yfive$ decays, the partial widths are 
\be
\begin{split}
&\Gamma[\Yfive\to\pi^+\pi^-\Upsilon(1S)]=(0.59\pm 0.04\pm 0.09)~\mbox{MeV},\\
&\Gamma[\Yfive\to\pi^+\pi^-\Upsilon(2S)]=(0.85\pm 0.07\pm 0.16)~\mbox{MeV},\\
&\Gamma[\Yfive\to\pi^+\pi^-\Upsilon(3S)]=(0.52^{+0.20}_{-0.17}\pm 0.10)~\mbox{MeV}.
\label{5spipi}
\end{split}
\ee
These decay widths are several orders of magnitude larger than those for the transitions from $\Upsilon(2S)$ and $\Upsilon(3S)$.
Using a larger data sample of 121.4 fb$^{-1}$ at $\sqrt{s}$ = 10.87 GeV, Belle updated the measurements of $e^+e^-\to\pi^+\pi^-\Upsilon(nS)$ $(n=1,~2,~3)$~\cite{Belle:2011aa}.
The large partial widths of $\Yfive$ were confirmed.
Furthermore, two charged bottomonium-like resonances were observed in the $\pi^{\pm}\Upsilon(nS)$ mass spectra, as shown in Fig.~\ref{fig:5SpipiUps}.
Their masses are $(10607.2\pm2.0)$~\mev~and $(10652.2\pm1.5)$~\mev, and widths are $(18.4\pm2.4)$ MeV and $(11.5\pm2.2)$ MeV. Following Belle, we call these two states as $Z_b(10610)$ and $Z_b(10650)$.
Further experimental details of the $Z_b(10610)$ and $Z_b(10650)$ production in the decays of $\Yfive$ and $\Upsilon(11020)$ can be found in the review \cite{Brambilla:2019esw}.

\begin{figure}[t!]
\centering
\begin{tabular}{cc}
\includegraphics[width=0.45\textwidth]{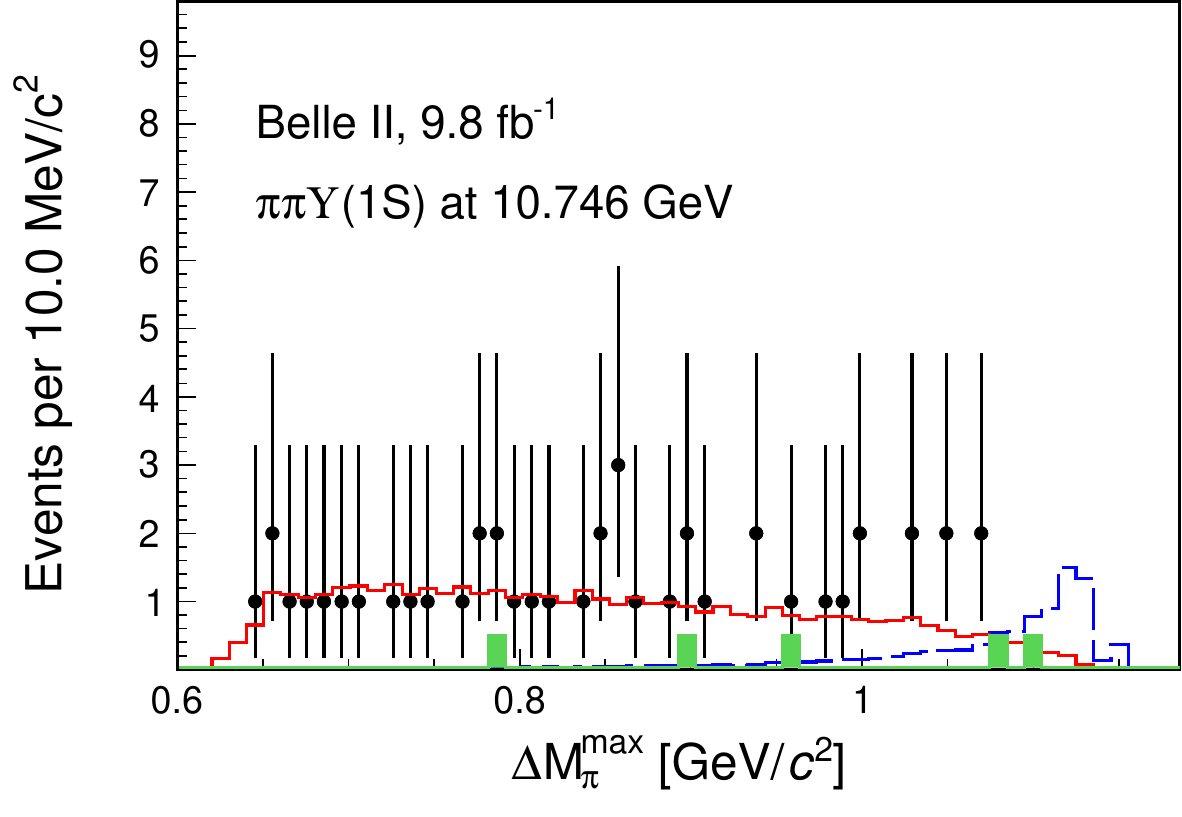}&
\includegraphics[width=0.45\textwidth]{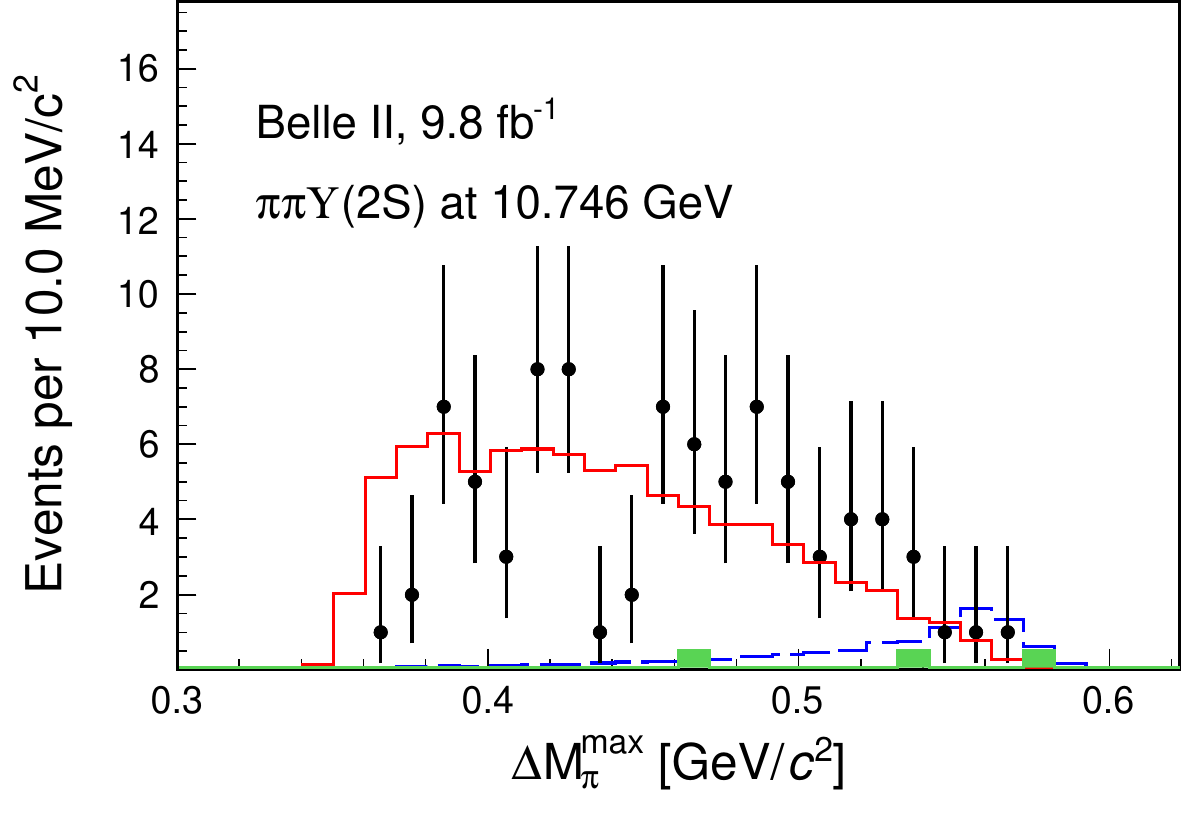}\\
\includegraphics[width=0.45\textwidth]{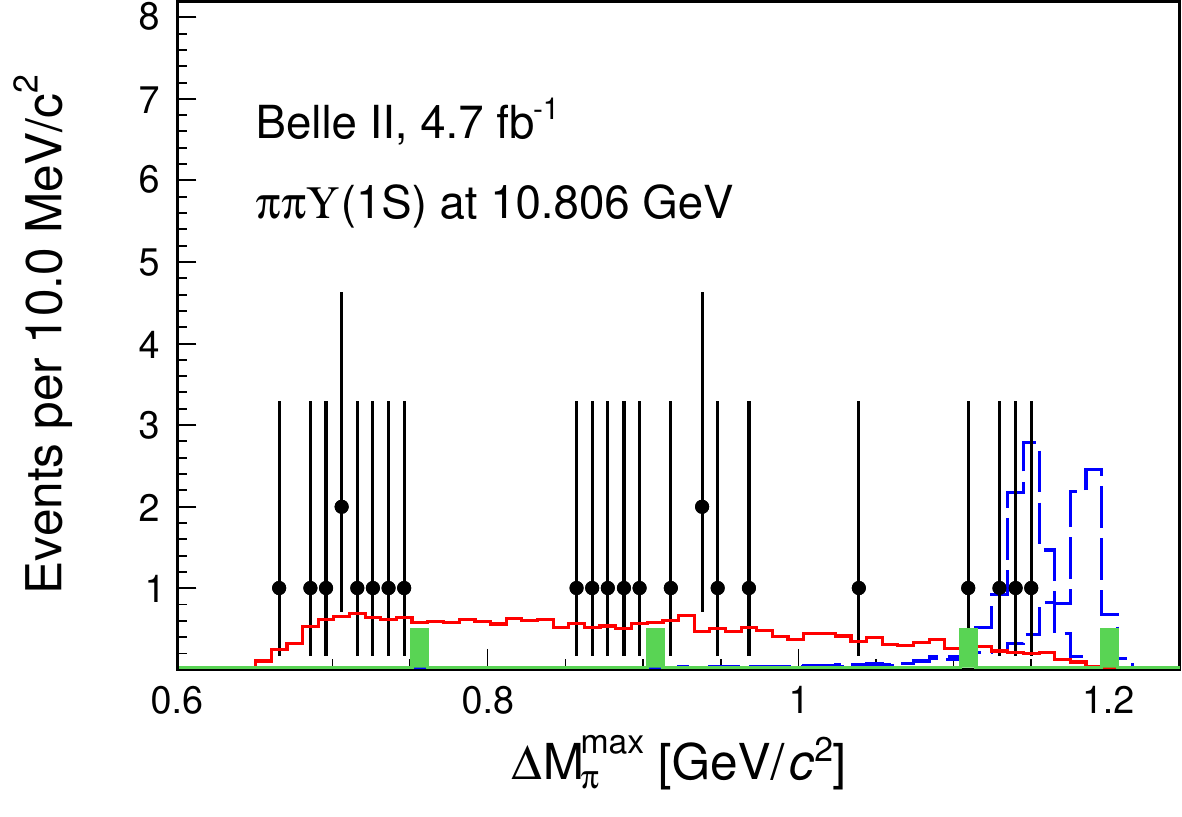}&
\includegraphics[width=0.45\textwidth]{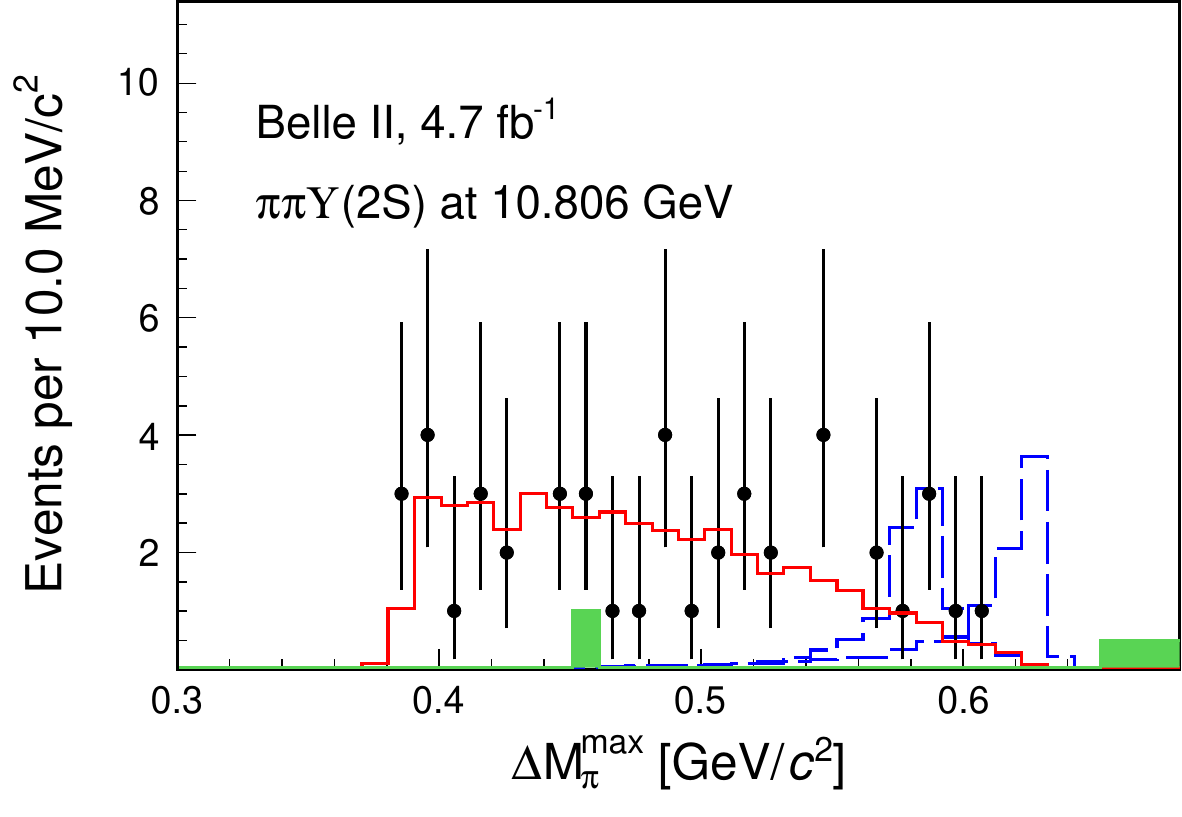}
\end{tabular}
\caption{The distributions of $\Delta M^{\rm max}_{\pi}$ in $e^+e^-\to\pi^+\pi^-\Upsilon(1S,2S)$ at $\sqrt{s}$ = 10.746 GeV and 10.805 GeV~\cite{Belle-II:2024mjm}.
Dots with error bars are from data, the green shaded histograms show the events in the $\Upsilon(10753)$ sideband region, red histograms show the phase space simulated events, and blue dashed histograms show the $Z_b(10610/10650)$ from simulations. 
}\label{B2_fig_Zb}
\end{figure}

Using data of 9.8 fb$^{-1}$ at $\sqrt{s}$ = 10.746 GeV and 4.7 fb$^{-1}$ at $\sqrt{s}$ = 10.805 GeV, Belle II searched for $Z_b$ states in the $\pi\Upsilon$ final states in $e^+e^-\to \pi\pi\Upsilon(1S,2S)$~\cite{Belle-II:2024mjm}.
Figure~\ref{B2_fig_Zb} shows the distributions of maximal difference between the $\pi^{\pm}\mu^{+}\mu^{-}$ mass and the $\mu^{+}\mu^{-}$ mass ($\Delta M^{\rm max}_{\pi}$).
The distributions in data can be described by the phase space simulations. No excesses were found. 
The blue dashed histograms show the $Z_b(10610/10650)^{\pm}$ signals from simulations, where the events are normalized arbitrarily.
The upper limits at 90\% confidence level (C.L.) on the products of the Born cross sections and branching fractions are listed in Table~\ref{B2_tab_Zb}.

\begin{table*}[t!]
\renewcommand\arraystretch{1.2}
\setlength{\tabcolsep}{2pt}
\centering
\caption{The upper limits at 90\% C.L.\ on the products of the Born cross sections and branching fractions~\cite{Belle-II:2024mjm}. The $\sqrt{s}$ is $e^+e^-$ C.M.\ energy. The $\sigma$ denotes the Born cross section for $e^+e^-\to \pi Z_b(10610,10650)$. The $\BR$ denotes the branching fraction for $Z_b(10610,10650)\to\pi\Upsilon(1S,2S)$.}\label{B2_tab_Zb}
\vspace{0.2cm}
\begin{tabular}{ccc}
\hline\hline
Mode & $\sqrt{s}$ (GeV) & $\sigma\times\BR$ (pb) \\\hline
$Z_b(10610)\to\pi\Upsilon(1S)$ & ~~~10.746, 10.805~~~ & ~~~$<$ 0.13, $<$ 0.43~~~ \\\hline
$Z_b(10610)\to\pi\Upsilon(2S)$ & 10.746, 10.805 & $<$ 0.14, $<$ 0.35 \\\hline
$Z_b(10650)\to\pi\Upsilon(1S)$ & 10.805 & $<$ 0.28 \\\hline
$Z_b(10650)\to\pi\Upsilon(2S)$ & 10.805 & $<$ 0.30 \\\hline
\hline
\end{tabular}
\end{table*}

\subsubsection{Theoretical description and spin partners}

Because the $b$-quark mass is significantly greater than $\Lambda_{\rm QCD}$, the typical scale of strong interactions, bottomonium states ($b\bar{b}$ bound states) can often be approximated as nonrelativistic quantum-mechanical systems.
The latter can be characterized by the total spin of the quark pair $\veS=\veS_b+\veS_{\bar{b}}$ and the total angular momentum $\veJ=\veL+\veS$, with $\veL$ for the orbital angular momentum in the quark--antiquark system. The spatial and charge parities of a generic $b\bar{b}$ system are straightforwardly related to its total spin and orbital angular momentum,
\be
P=(-1)^{L+1},\quad C=(-1)^{L+S},
\label{PCbb}
\ee
while the physical states can mix terms that correspond to different partial waves of the same parity.
Consequently, the spectrum of the ordinary bottomonium consists of the towers of radially excited states with $n=1,~2,~\ldots$ and different quantum numbers $J^{PC}$, namely,
\begin{itemize}
\item $\eta_b$'s --- spin-singlet pseudoscalar states with $J^{PC}=0^{-+}$,
\item $\Upsilon$'s --- spin-triplet vector states with $J^{PC}=1^{--}$,
\item $\chi_{bJ}$'s --- three spin-triplet states with the quantum numbers $J^{++}$ ($J=0,~1,~2$),
\item $h_b$'s --- spin-singlet axial-vector states with the quantum numbers $1^{+-}$,
\end{itemize}
and so on. Since, as was mentioned above, there is no one-to-one correspondence between the nonrelativistic $n{}^{2S+1}L_J$ scheme and the relativistic $nJ^{PC}$ one, some generic bottomonia may come as mixtures of the $b\bar{b}$ states with different angular momenta. In particular, vector bottomonia are expected to mix both $S$- and $D$-wave components. In the meantime, the admixture of the $D$ wave in low lying $\Upsilon$-states is expected to be small, so they are conventionally identified as $\Upsilon(nS)$, with $n=1,~2,~3$. However, as was already mentioned above (see footnote~\ref{foot:Ups}), identification of the observed high-lying vector bottomonia $\Yfour$, $\Yfive$, and $\Ysix$ with the quark model states $\Upsilon(4S)$, $\Upsilon(5S)$, and $\Upsilon(6S)$, respectively, is questionable given that they reside above various open-bottom thresholds, so their wave functions are expected to have more components than just a plain quark--antiquark one. The nature of the vector state $\Yb$ is under active debates and will be reviewed in Sec.~\ref{sec:Yb} below.

Transitions between spin-singlet and spin-triplet states of a genuine bottomonium imply a heavy-quark spin flip. Meanwhile, since the spin-dependent interaction responsible for such transitions \cite{Eichten:1980mw,Gromes:1984ma} in $b\bar{b}$ systems scales as $\Lambda_{\rm QCD}^2/m_b^2\ll 1$, such spin flips are strongly suppressed. This observation results in the emergence of an approximate but rather accurate symmetry of strong interactions --- heavy quark symmetry \cite{Isgur:1991wq}. The latter is the cornerstone of the heavy quark effective theory (HQET) that provides a systematic expansion of the mass and other characteristics of a heavy-light quark-antiquark system in terms of the inverse heavy quark mass --- see, for example, a classical book \cite{Manohar:2000dt} and references therein. More generally, heavy quark spin symmetry (HQSS) allows one to interrelate the properties of the hadronic systems with different orientations of the heavy quark spin conventionally referred to as spin partners. Importantly, while the possibility to employ HQSS to deduce such predictions is universal, the actual predictions depend on the particular nature of the hadronic system at hand. For instance, an immediate consequence of HQSS for generic quark model states with heavy quarks would be a strong suppression of the kinematically allowed transitions between $\Upsilon$ and $h_b$ bottomonia in comparison with similar transitions between $\Upsilon$'s both in the initial and final state. In the meantime, as will be reviewed below, this conclusion may not hold if hadronic states with the nature different from a plain bottomonium are involved in the transition.

The di-pion transitions from vector bottomonia to lower lying vector and axial-vector ones are amongst the most studied decays of heavy quarkonia both theoretically and experimentally. Remarkably, the di-pion transition $\Upsilon(2S)\to\pi\pi\Upsilon(1S)$ admits a purely theoretical treatment based on matching different low-energy degrees of freedom in QCD. Indeed, since the pions involved in the decay are soft, they must come as a result of hadronisation of soft-gluon exchanges between the two bottomonia. theoretical foundations of the soft-gluon-driven interactions in heavy-quark systems were explored in Refs.~\cite{Gottfried:1977gp,Voloshin:1978hc,Peskin:1979va,Bhanot:1979vb} and then developed in many subsequent publications; see also more recent publications where the same picture is obtained by matching the effective field theory (EFT) Lagrangian for an $S$-wave heavy field in the external chromoelectric field \cite{Luke:1992tm,Brambilla:2015rqa}
with the effective Lagrangian of weakly coupled potential nonrelativistic QCD ~\cite{Pineda:1997bj,Brambilla:1999xf,Brambilla:2005yk}. In this picture, the long-range interaction between color-neutral objects resembles the electromagnetic interaction between electrically neutral composite particles known as the van der Waals force and can be described employing the multipole expansion of the QCD Hamiltonian. Then the leading dipole term arises in the form
\be
\delta H_{\rm QCD}=-\frac12\xi^a\;\ver\cdot\veE^a,
\ee
where $\xi^a=t_1^a-t_2^a$, with $t^a$'s for the color generators of the heavy quark and antiquark, $\ver$ is the vector of their relative position, and $\veE^a$ is the chromoelectric field. Then, the matrix element for the di-pion transition from $\Upsilon(2S)$ to $\Upsilon(1S)$ can be written in the form
\be
{\cal M}(\Upsilon(2S)\to\pi\pi\Upsilon(1S))=\beta_{21}\braket{\pi\pi|\veE^a\cdot\veE^a|0}.
\label{M21}
\ee
Here the overall strength factor, known as the off-diagonal chromopolarisability,
\be
\beta_{21}=\frac{1}{48}\braket{\Upsilon(1S)|\xi^a\ver\; G_O\;\ver\; \xi^a|\Upsilon(2S)},
\ee
with $G_O$ for the propagator of the $b\bar{b}$ pair in the octet state, can be calculated theoretically employing a suitable quark model or extracted from the data. In any case the constant $\beta_{21}$ does not affect the energy dependence of the amplitude in Eq.~\eqref{M21} that comes entirely from the matrix element $\braket{\pi\pi|\veE^a\cdot\veE^a|0}$. The latter can be evaluated employing the relation between the gluonic operator $\veE^a\cdot\veE^a$ and the trace anomaly \cite{Voloshin:1980zf,Novikov:1980fa,Pineda:2019mhw}, on the one hand, and the chiral algebra \cite{Gasser:1983yg} (see, for example, pedagogical reviews on chiral perturbation theory in Refs.~\cite{Bernard:2006gx,Scherer:2005ri}), on the other hand. Then theoretically predicted di-pion mass spectrum (agnostic to the heavy quark flavor and, therefore, the same for both $2S\to 1S$ transitions in charmonia and bottomonia) has a specific shape with a slow growth at small $M_{\pi\pi}$'s and a single hump near the upper edge of the spectrum \cite{Brown:1975dz,Voloshin:2007dx}. This form of the $\pi\pi$ mass spectrum
in the decay $\Upsilon(2S)\to\pi\pi\Upsilon(1S)$ was confirmed in the experiment \cite{CLEO:1993fsd,CLEO:2007rbi}. Since applying the theoretical framework described above to higher bottomonium states is expected to yield limited predictive accuracy, the corresponding chromopolarizabilities are more reliably extracted through fits to experimental data \cite{Chen:2019gty}. However, in contrast to the expectations based on the QCD multipole expansion and the chromopolarizability approach, measurements of dipion transitions from higher vector bottomonia have revealed an unexpected double-peaked structure in the dipion invariant-mass distribution. Specifically, one peak appears at low values of $M_{\pi\pi}$, while the second is located close to the upper kinematic limit. This phenomenon was first observed by the CLEO Collaboration in the transition $\Upsilon(3S)\to\pi\pi\Upsilon(1S)$ \cite{CLEO:1993fsd,CLEO:2007rbi}. Subsequently, similar structures were reported in the decays $\Yfour\to\pi\pi\Upsilon(1S)$, first by the Belle Collaboration \cite{Belle:2005vil,Belle:2006wip} and later confirmed by the BaBar Collaboration, which also measured the transition $\Yfour\to\pi\pi\Upsilon(2S)$ \cite{BaBar:2006udk}. These characteristic double-peaked distributions have attracted considerable attention and have been studied extensively in the literature; see Refs.~\cite{Voloshin:1987rp,Kuang:2006me,Guo:2004dt} for early reviews and, for example, Refs.~\cite{Simonov:2008ci,Simonov:2008qy,Voloshin:2006ce} for more recent developments. An essential difference is that, while the dipion invariant mass in the transition $\Upsilon(2S)\to\pi\pi\Upsilon(1S)$ is limited to about 560~\mev, dipion transitions from higher-lying $\Upsilon$ states typically probe a much broader kinematic region. This provides greater flexibility in describing the pion--pion final-state interaction and allows for additional mechanisms that may generate nontrivial structures in the observed line shapes. In particular, the importance of coupled-channel effects involving open-bottom intermediate states was discussed in Refs.~\cite{Lipkin:1988tg,Zhou:1990ik,Simonov:2008qy}; the possible contribution of an isovector hidden-bottom tetraquark state coupled to the $\Upsilon\pi$ channel was proposed in Refs.~\cite{Anisovich:1995zu,Voloshin:1982ij,Guo:2004dt,Guo:2006ai}; strong final-state $\pi\pi$ interactions were emphasized in Refs.~\cite{Komada:2001pxu,Ishida:2001pt,Gardner:2001gc,Uehara:2002wh,Belanger:1988hs,Chakravarty:1993er,Guo:2004dt,Surovtsev:2015mba}; and relativistic corrections were investigated in Ref.~\cite{Voloshin:2006ce}, among other effects. In the meantime, a broad consensus emerged that the widths of such dipion transitions should be of the order of a few keV, as predicted by theoretical studies \cite{Yan:1980uh,Kuang:1981se,Simonov:2008ci} and subsequently confirmed experimentally for the transitions $\Upsilon(nS)\to\pi\pi\Upsilon(n'S)$ with $n'<n<4$; see, for example, Refs.~\cite{BaBar:2006udk,Belle:2006wip}.

However, as was already mentioned above, in 2007, Belle reported \cite{Belle:2007xek} the first measurement of the transitions from $\Yfive$ to the final states $\pi\pi\Upsilon(nS)$ ($n=1,~2,~3$) with the probabilities about two orders of magnitude bigger than those for the transitions from $\Upsilon(nS)$, with $n\leqslant 4$ --- see Eq.~\eqref{5spipi}. Although certain attempts were undertaken in the literature to theoretically describe these striking experimental results by re-thinking the methods previously applied to the di-pion transitions from $\Upsilon(nS)$, with $n<5$, \cite{Simonov:2008ci}, a general consensus adopted in the hadronic community was that new phenomena had to be invoked for a successful explanation of the experimental facts.
The latter expectation became particularly relevant after the above experimental results were re-confirmed by Belle in 2011 \cite{Belle:2011aa} and, in addition, di-pion transitions $\Yfive\to\pi\pi h_b(mP)$ ($m=1,~2$) were measured with yet another striking result of the corresponding probabilities taking similar values to those for the vector bottomonia $\Upsilon(nS)$ ($n=1,~2,~3$) in the final state as given in Eq.~\eqref{5spipi}. This experimental observation is strongly at odds with the expectations from the quark model outlined above that the hadronic transitions involving a heavy quark spin flip must be strongly suppressed in comparison with similar transitions that do not require a spin flip. Furthermore, two distinct peaking structures located near 10.61~\gev~and 10.65~\gev~were reliably identified in the $\pi\Upsilon(nS)$ and $\pi h_b(mP)$ subsystems across all five hidden-bottom final states studied by Belle. In other words, two charged bottomonium-like resonances of an unknown nature were observed in the di-pion transitions from $\Yfive$. Given the presence of a $b\bar{b}$ pair in these resonances, their minimal quark content could only be four-quark, so their exotic nature was undoubted.
These two exotic bottomonium-like states were tagged $Z_b(10610)$ and $Z_b(10650)$ (sometimes $Z_b$ and $Z_b'$ as a shorthand notation), and disclosing their nature and further properties immediately became one of hot topics of hadronic spectroscopy --- see the review \cite{Brambilla:2019esw} and references therein.

\begin{figure*}[t]
\centering
\includegraphics[width=0.99\textwidth]{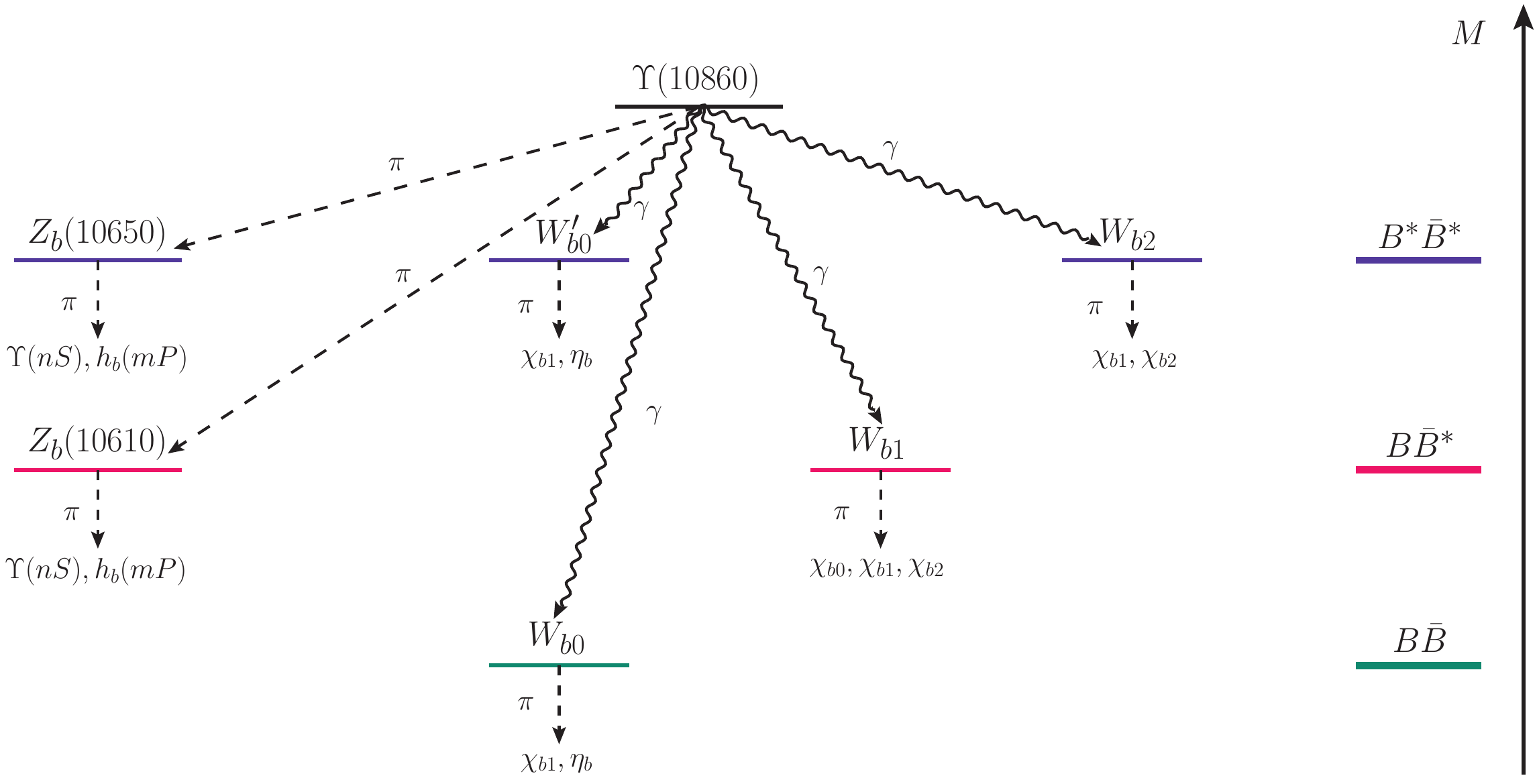}
\caption{Allowed transitions from the vector bottomonium $\Yfive$ to lower lying states of bottomonium through the formation of the $Z_b$'s or their spin partners $W_{bJ}$ ($J=0,~1,~2$). To guide the eye, the relevant open-bottom $B^{(*)}\bar{B}^{(*)}$ thresholds are schematically shown on the right-hand side of the plot. See Ref.~\cite{Voloshin:2011qa} for further details.}\label{fig:Ups5Sdecays}
\end{figure*}

\begin{figure*}[t!]
\centering
\includegraphics*[width=0.35\textwidth]{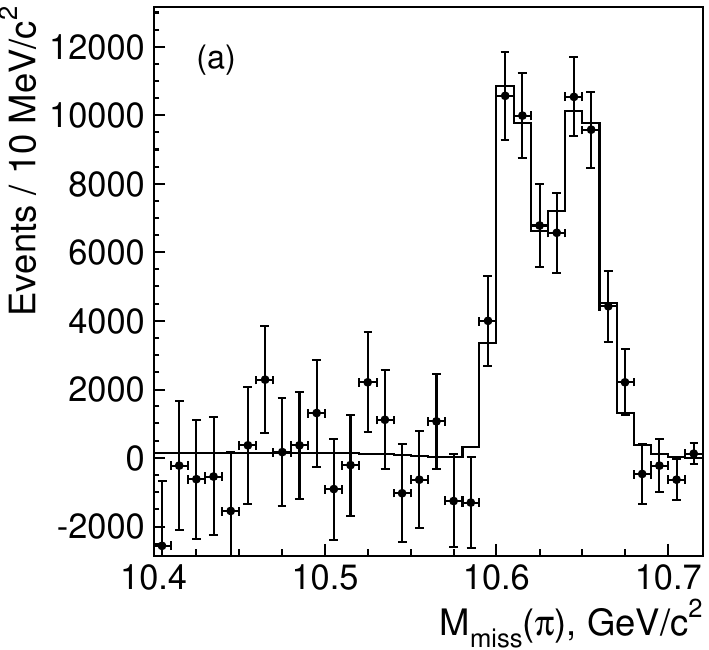}
\hspace*{0.1\textwidth}
\includegraphics*[width=0.35\textwidth]{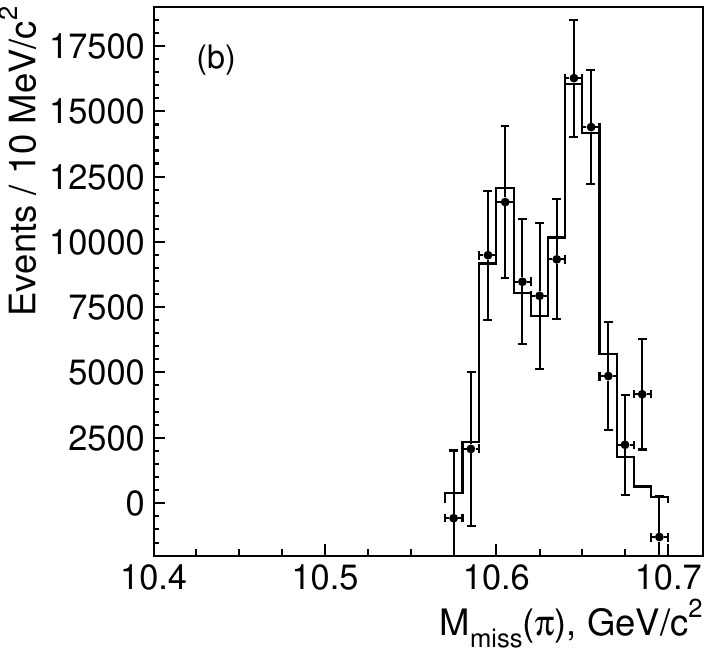}
\caption{Belle Collaboration data for di-pion transitions from $\Yfive$ state to $\pi\pi h_b(1P)$ (first plot) and $\pi\pi h_b(2P)$ (second plot)
\cite{Belle:2011aa}.}\label{fig:5Spipihb}
\end{figure*}

\begin{figure*}[t!]
\centering
\includegraphics[width=0.47\textwidth]{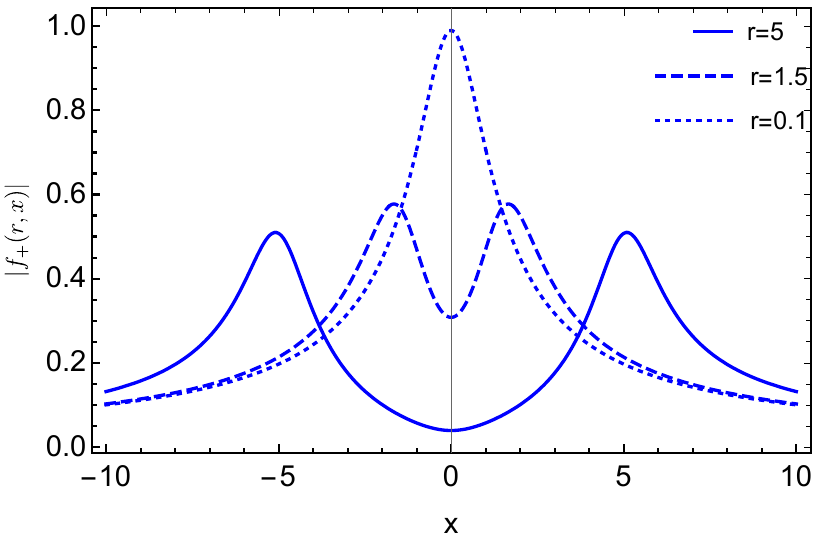}\hspace*{0.05\textwidth}\includegraphics[width=0.47\textwidth]{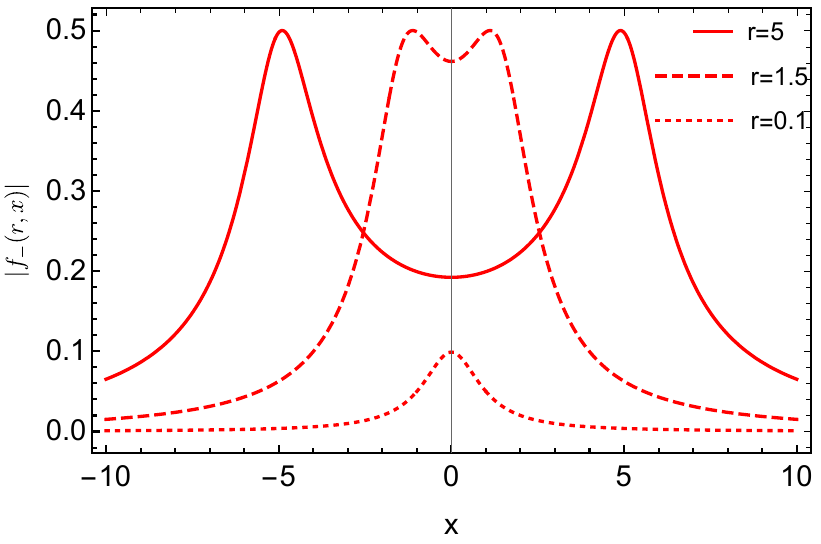}
\caption{The ``line shapes'' (absolute values of the amplitudes in some dimensionless units) $|f_+(r,x)|$ (left panel) and $|f_-(r,x)|$ (right panel) as functions of dimensionless energy $x$ (with the origin chosen in the middle of the energy interval between the positions of the resonances $Z_b$ and $Z_b'$). The function $f_+(r,x)$ ($f_-(r,x)$) shows a constructive (destructive) interference of the two $Z_b$'s for the $\pi\pi\Upsilon(nS)$ ($\pi\pi h_b(mP)$) final states outside the peaks region as described in the text~\cite{Baru:2022xne}.
}
\label{fig:fpm}
\end{figure*}

The authors of work \cite{Bondar:2011ev}, announced shortly after the Belle publication \cite{Belle:2011aa}, pointed out the fact that the newly discovered $Z_b$'s states resided near the open-bottom $B^{(*)}\bar{B}^*$ thresholds and assigned them both as $I^G(J^P)=1^+(1^+)$ $B\bar{B}^*$ and $B^*\bar{B}^*$ molecules, that is, resonances containing a dominating (or the only) hadron--hadron component in their wave functions. Then, starting from the spin wave functions of the $B^{(*)}$ mesons in the form\footnote{Here $S^{P}_{q_1 \bar{q}_2}$ denotes the wave function of the $q_1\bar{q}_2$ pair with the total spin $S$ and parity $P$.}
\be
B\sim 0_{q\bar{b}}^-,\quad \bar{B}\sim 0_{b\bar{q}}^-,\quad
B^*\sim 1_{q\bar{b}}^-,\quad \bar{B}^*\sim 1_{b\bar{q}}^-,
\ee
then building their combinations with the quantum numbers\footnote{Hereinafter the charge conjugation operation applied to the $B^{(*)}$ mesons is defined as $CB=\bar{B}$ and $CB^*=-\bar{B}^*$, respectively.} $J^P=1^{+-}$,
\be
Z_b\sim B\bar{B}^*-\bar{B}B^*,\quad Z_b'\sim B^*\bar{B}^*,
\ee
and, finally, employing a Fierz transformation to re-shuffle the spins of the heavy and light quarks, one readily arrives at two orthogonal combinations,
\be
\begin{split}
Z_b&=\frac{1}{\sqrt 2}\Bigl[(1^-_{b \bar b}\otimes 0^-_{q \bar q})_{S=1}+
(0^-_{b \bar b}\otimes 1^-_{q \bar q})_{S=1}\Bigr],\\
Z_b'&=\frac{1}{\sqrt{2}}\Bigl[(1^-_{b \bar b}\otimes 0^-_{q \bar
q})_{S=1}-(0^-_{b \bar b}\otimes 1^-_{q \bar q})_{S=1}\Bigr], \label{Zbs}
\end{split}
\ee
where $0_{b\bar{b}}$ and $1_{b\bar{b}}$ can be readily recognized
as the spin wave functions of the bottomonia $h_b$ and $\Upsilon$, respectively. Relations in Eq.~\eqref{Zbs} provide an appealing interpretation of the experimental results by Belle concerning the di-pion transitions from $\Yfive$. Indeed, if the $Z_b$ bottomonium-like hadrons appear as intermediate states in such transitions, they facilitate the latter and enlarge the corresponding probabilities. On the other hand, since both components with $S_{b\bar{b}}=0$ and $S_{b\bar{b}}=1$ are present in the wave functions of both $Z_b$'s, then the di-pion transitions proceed as cascades (see Fig.~\ref{fig:Ups5Sdecays}),
\begin{align}
\Yfive&\to \pi Z_b^{(\prime)}\to \pi [\pi\Upsilon(nS)],\quad n=1,~2,~3,
\label{dipi1}\\
\Yfive&\to \pi Z_b^{(\prime)}\to \pi [\pi h_b(mP)],\quad m=1,~2,
\label{dipi2}
\end{align}
thus implying no heavy quark spin flip and, therefore, no suppression for the probabilities of the decays in Eq.~\eqref{dipi2} in comparison with the decays in Eq.~\eqref{dipi1}. 
In addition, as was noticed in Ref.~\cite{Bondar:2011ev}, the interference patterns between the $Z_b$ and $Z_b'$ contributions 
are distinctly different for the transitions in Eqs.~\eqref{dipi1} and \eqref{dipi2}: 
while in the processes $\Yfive\to \pi\pi\Upsilon$ the interference is destructive in the energy range between the peaks (see Fig.~\ref{fig:5SpipiUps}) and constructive outside of this region, the corresponding pattern is strictly opposite in the decays $\Yfive\to \pi\pi h_b$ (see Fig.~\ref{fig:5Spipihb}). Indeed, as can be seen from Figs.~\ref{fig:5SpipiUps} and \ref{fig:5Spipihb} and in agreement with the natural expectations that follow from the patterns just outlined, the dip between the two peaks is noticeably deeper in the line shapes for the $\Upsilon$-transitions than in those for the $h_b$-transitions. The latter observation implies that the implementation of HQSS in the system at hand is less trivial than could be na{\"i}vely anticipated.
Indeed, on the one hand, the cornerstone relations in Eq.~\eqref{Zbs} rely on exact HQSS and this way ``enable'' a sizeable signal in the $\pi\pi h_b$ final states.
On the other hand, the strict HQSS limit implies {\it inter alia} exact vanishing of all spin-dependent interactions that involve the heavy quark. In particular, the mass splitting between the $B^*$ and $B$ mesons has to vanish and the $B\bar{B}^*$ and $B^*\bar{B}^*$ thresholds must merge. In this case,
the window of the positive interference in the $\pi\pi h_b$ amplitudes between the two open-bottom thresholds shrinks to zero and the remaining destructive interference between the contributions from the two $Z_b$'s washes away the signal in both $\pi\pi h_b$ decay modes. Thus, a visible signal in these modes is only possible as a result of a nontrivial interplay between the effects that respect HQSS and the effects that violate it. This issue is addressed in detail in Ref.~\cite{Baru:2022xne}, where the ratio
\be
r=\frac{m_Z'-m_Z}{\Gamma_Z}
\label{rratio}
\ee
is introduced, with $m_Z^{(\prime)}$ for the masses of the $Z_b$'s and $\Gamma_Z$ for their widths predicted to be the same for both twin states in the strict HQSS limit \cite{Bondar:2011ev}. In Fig.~\ref{fig:fpm}, the quantities $f_\pm(r,x)$ (which mimic the amplitudes of the transitions $\Yfive\to \pi\pi\Upsilon$ and $\Yfive\to \pi\pi h_b$, respectively, --- see Ref.~\cite{Baru:2022xne} for their strict definition and further details) are plotted for several typical values of the ratio $r$ in Eq.~\eqref{rratio} as functions of a dimensionless quantity,
\be
x=\frac{2}{\Gamma_Z}\left(M_5-\bar{m}_Z-\omega\right),
\ee
with $M_5$ for the mass of the $\Upsilon(10860)$, $\bar{m}_Z=\frac12(m_Z+m_{Z'})$, and $\omega$ for the energy of one of the pions. The figure demonstrates the effect of the constructive (the function tagged with ``$+$'') and destructive (the function tagged with ``$-$'') 
interference between the contributions from the $Z_b$ and $Z_b'$ on the resulting line shapes \cite{Baru:2022xne}. In particular,
one can see the signal in the $\pi\pi h_b$ channel to gradually vanish in the strict HQSS limit of $r\to 0$. Meanwhile, it is further argued in Ref.~\cite{Baru:2022xne} that the experimentally observed ratio of the branching fractions for the reactions in Eqs.~\eqref{dipi1} and \eqref{dipi2} is naturally achievable for the physical value $r_{\rm phys}\approx 3$ obtained from the parameters of the $Z_b$'s quoted in RPP \cite{ParticleDataGroup:2024cfk}. The importance of the
HQSS-breaking contributions for a proper description of
the $Z_b$'s line shapes was also stressed in Ref.~\cite{Mehen:2013mva}.

\begin{figure*}[t!]
\centering
\includegraphics*[width=0.6\textwidth]{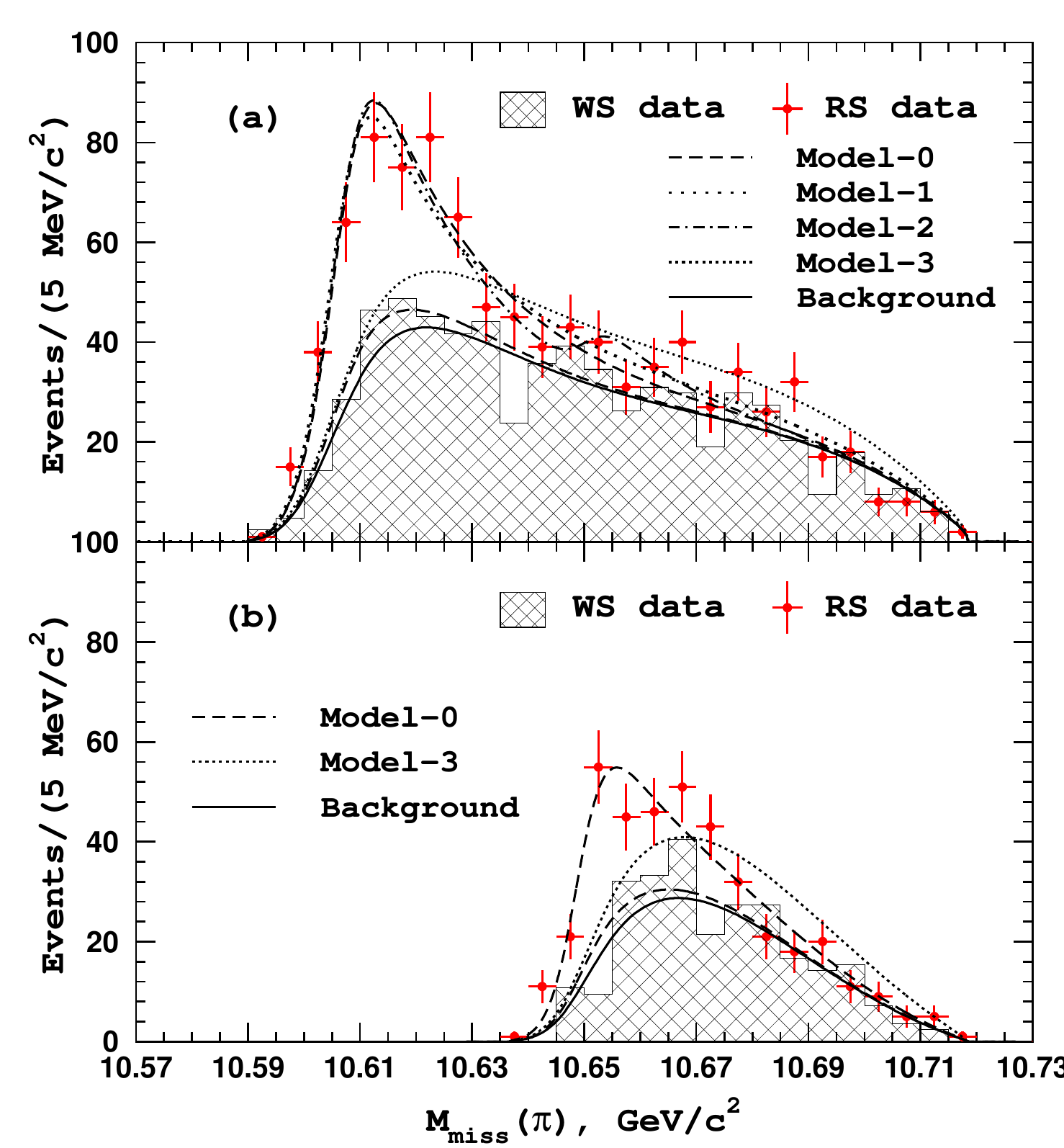}
\caption{Belle Collaboration data for the transitions to the open-bottom final states (a) $B\bar{B}^*$ and (b) $B^*\bar{B}^*$ 
\cite{Belle:2015upu}.}\label{fig:5SBB}
\end{figure*}

The proximity of the nominal masses of the $Z_b(10610)$ and $Z_b(10650)$ resonances to the $B\bar{B}^*$ and $B^*\bar{B}^*$ thresholds, respectively, and their molecular interpretation suggested in Ref.~\cite{Bondar:2011ev} emphasize the relevance of experimental searches for these twin exotic bottomonia in the open-bottom final states. The corresponding results were reported by Belle in 2015 \cite{Belle:2015upu} --- see Fig.~\ref{fig:5SBB}. As anticipated, the line shapes revealed peaking structures around 10.61~\gev~in 
the $B\bar{B}^*$ channel and around 10.65~\gev~in the $B^*\bar{B}^*$ one. In the meantime, unexpectedly, no clear signal excess was observed in the first channel near 10.65~\gev. This observation looks surprising given that the transitions between the $B\bar{B}^*$ and $B^*\bar{B}^*$ pairs in the $1^{+-}$ state are allowed by symmetries and, what is more, they are even enhanced by the pion exchange as compared to the diagonal transitions mediated by pions within the same channels. No satisfactory theoretical explanation to this observation has been suggested so far that could go beyond the unnatural light-quark spin symmetry as a new approximate symmetry of QCD as conjectured in Ref.~\cite{Voloshin:2016cgm}.

\begin{figure*}[t]
\centering
\includegraphics*[width=0.3\textwidth]{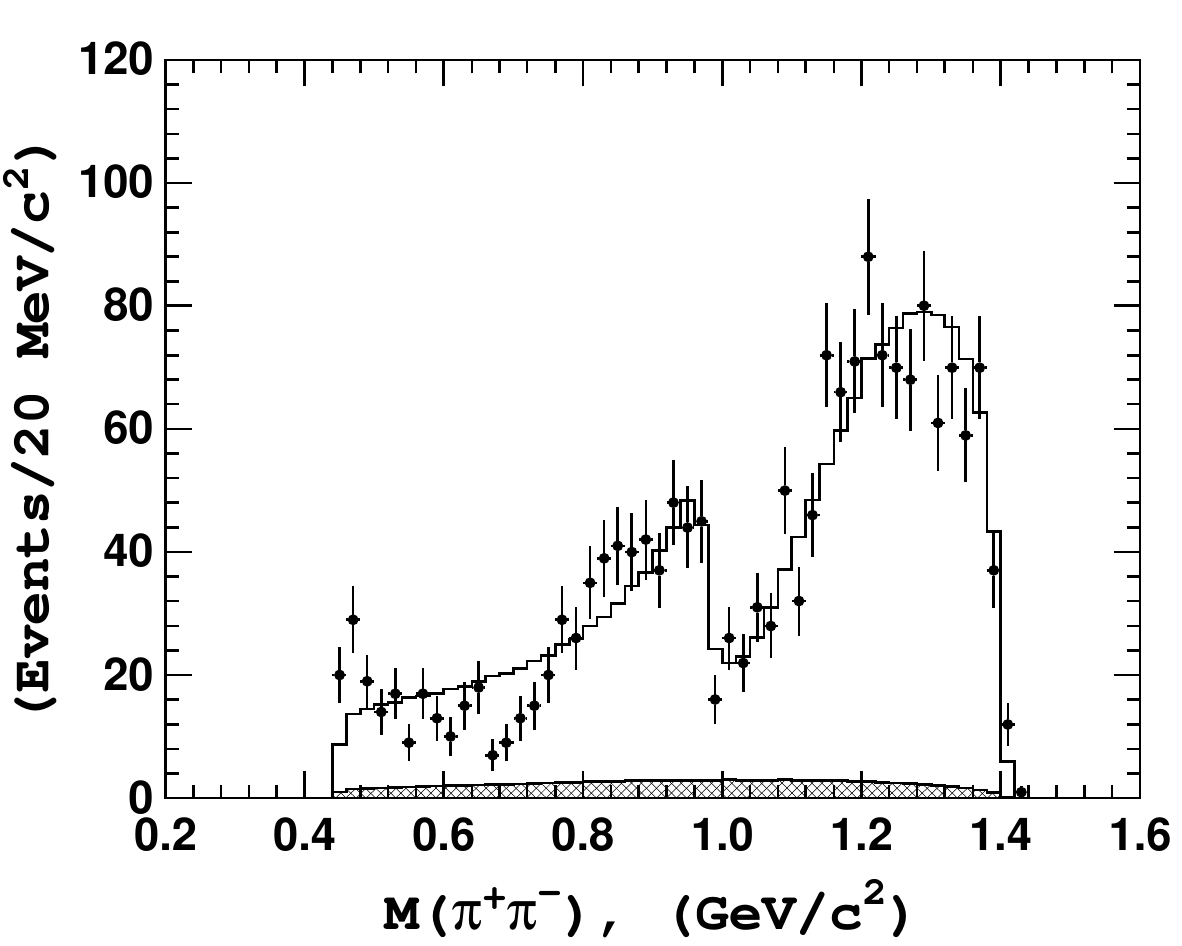}
\includegraphics*[width=0.3\textwidth]{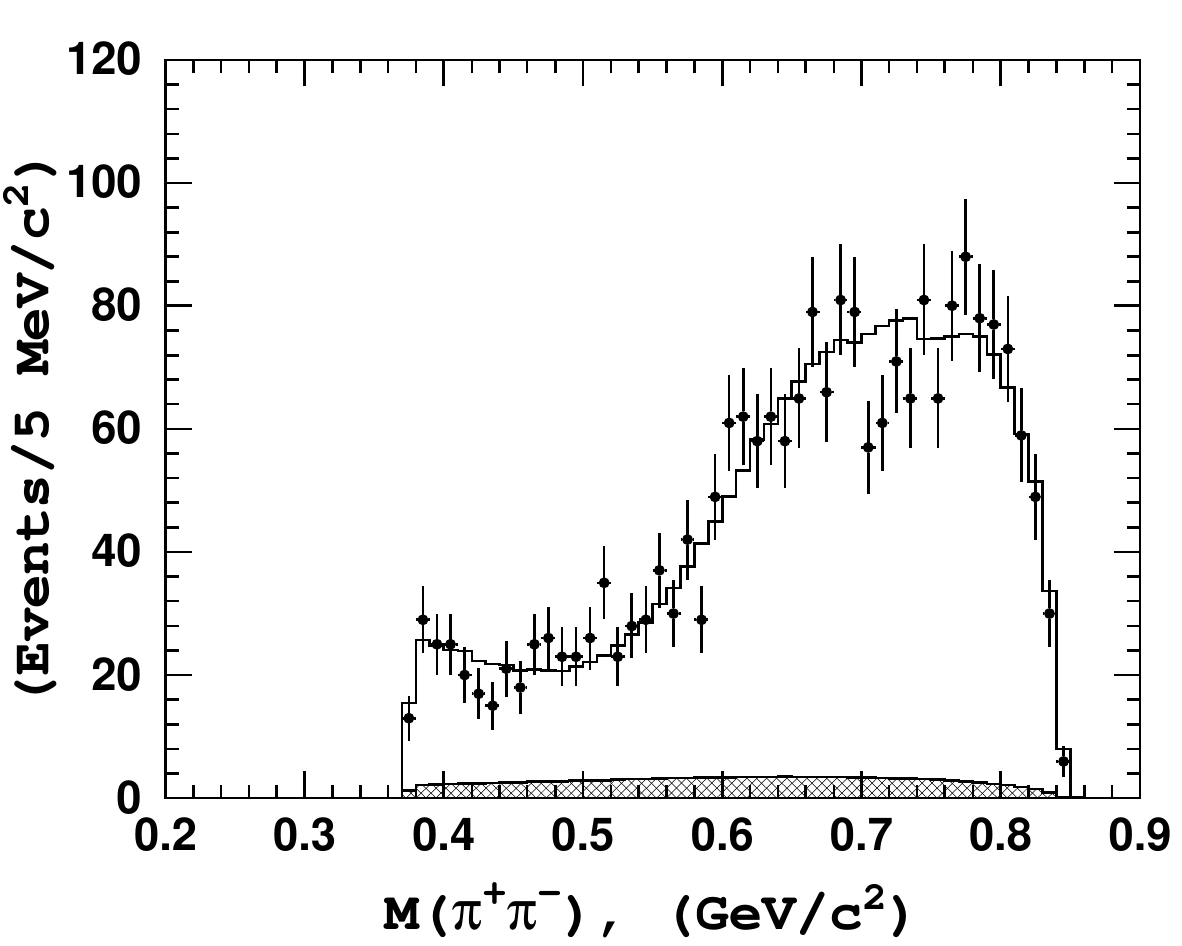}
\includegraphics*[width=0.3\textwidth]{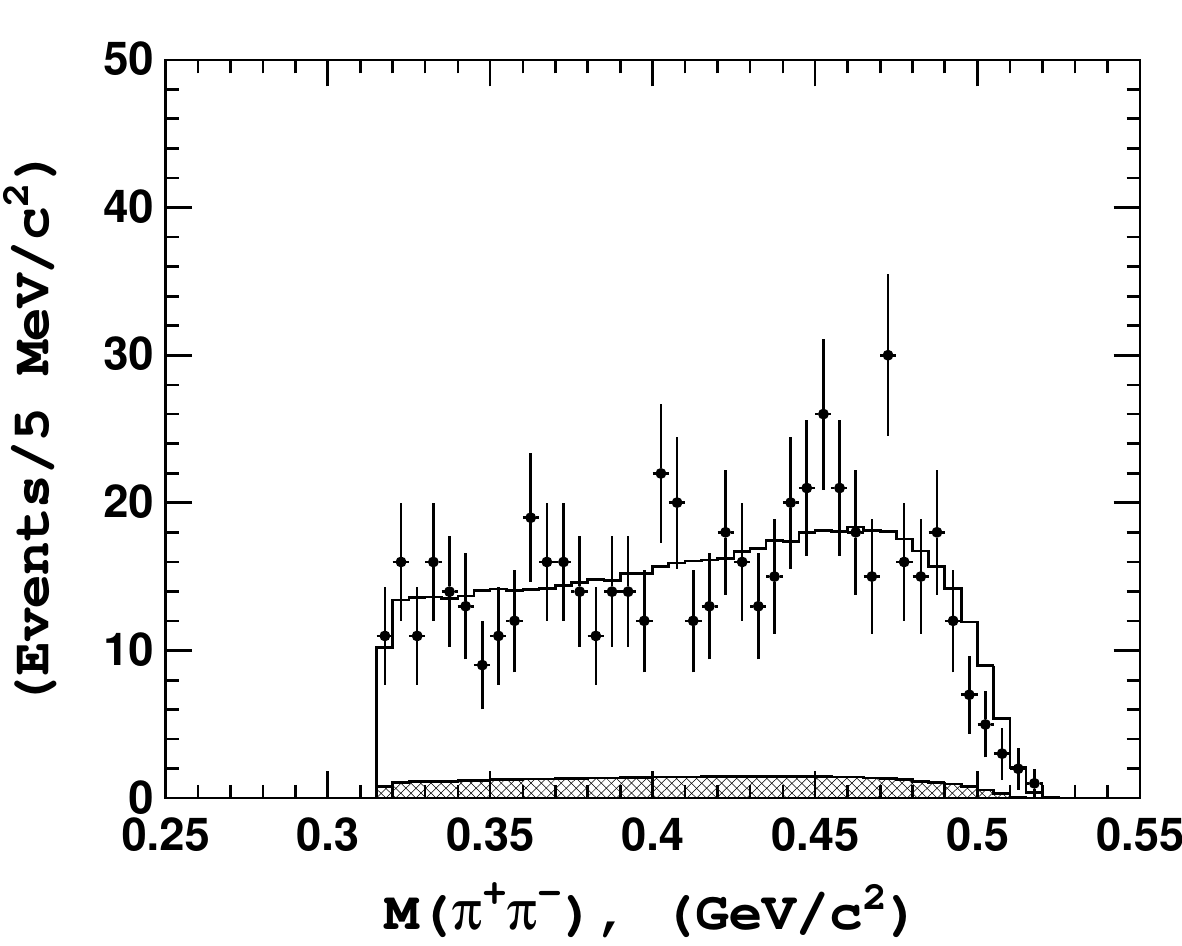}
\caption{Belle Collaboration data for the di-pion mass $M(\pi^+\pi^-)$ in the transitions from the $\Yfive$ state to $\pi\pi\Upsilon(1S)$ (first plot),
$\pi\pi\Upsilon(2S)$ (second plot), and $\pi\pi\Upsilon(3S)$ (third plot)
\cite{Belle:2011aa}.
}\label{fig:5Spipiexp}
\end{figure*}

A combined theoretical analysis of the experimental line shapes for the $Z_b$'s in the four observation channels, $\{B\bar{B}^*$, $B^*\bar{B}^*, \pi h_b(1P),\pi h_b(2P)\}$, was performed in Refs.~\cite{Hanhart:2015cua,Guo:2016bjq}, within a practical parameterization, and in Ref.~\cite{Wang:2018jlv},
within an EFT framework. In Ref.~\cite{Baru:2020ywb}, the latter analysis was further extended to include the $\pi\Upsilon(nS)$ ($n=1,~2,~3$) channels, where the analysis is hindered by a strong dependence of the experimental reconstruction efficiency on the di-pion mass, so the entire two-dimensional Dalitz plots need to be analysed rather than their one-dimensional projections in Figs.~\ref{fig:5SpipiUps} and~\ref{fig:5Spipiexp}. For soft emitted pions,
the approach of effective chiral Lagrangians augmented with the dispersive technique can be employed to describe the experimental line shapes \cite{Chen:2016mjn,Baru:2020ywb}. Meanwhile, already at the lowest order, the $\Upsilon'\to\pi\pi\Upsilon$ transition amplitudes bring several additional unknown parameters fitted to the data. The role played by the $Z_b$'s in the di-pion transitions from low-lying vector bottomonia $\Upsilon(3S)$ and $\Yfour$ is investigated in Refs.~\cite{Chen:2015jgl,Chen:2016mjn} while a production mechanism for the $Z_b$'s through $B$-meson loops from the high-lying bottomonia $\Yfive$ and $\Ysix$ is proposed in Ref.~\cite{Wu:2018xaa}.

The molecular interpretation for the $Z_b$'s entails a straightforward prediction based on HQSS that four additional isovector sibling
states $W_{bJ}$ with the quantum numbers $J^{PC}=J^{++}$ ($J=0,~1,~2$) should exist in the spectrum of bottomonium. Their mutually orthogonal spin wave functions can be written in the form \cite{Bondar:2011ev,Voloshin:2011qa,Mehen:2011yh,Bondar:2016hva,Baru:2017gwo}
\be
\begin{split}
&W_{b0}=\frac{\sqrt{3}}2 (1_{b\bar{b}}\otimes 1_{q\bar{q}})_{S=0}
-\frac12(0_{b\bar{b}}\otimes 0_{q\bar{q}})_{S=0},\\
&W_{b0}'=\frac12(1_{b\bar{b}}\otimes 1_{q\bar{q}})_{S=0}+
\frac{\sqrt{3}}2(0_{b\bar{b}}\otimes 0_{q\bar{q}})_{S=0},\\[1mm]
&W_{b1}=(1_{b\bar{b}}\otimes 1_{q\bar{q}})_{S=1},\\[2mm]
&W_{b2}=(1_{b\bar{b}}\otimes 1_{q\bar{q}})_{S=2}.
\end{split}
\label{WbJs}
\ee
Given the $W_{bJ}$'s opposite $G$-parity ($C$-parity of their electrically neutral components) with respect to the $Z_b$'s, these spin partners can not be produced in pion decays from the $\Yfive$ but rather in the radiative transitions from it --- see the scheme in Fig.~\ref{fig:Ups5Sdecays}. In Ref.~\cite{Mehen:2011yh},
a unified EFT framework for the molecular $Z_b$'s and $W_{bJ}$'s was introduced based on HQSS-symmetric short-range contact interactions and various predictions for the binding energies, decay widths, and line shapes were made. In Ref.~\cite{Guo:2013sya}, consequences of the additional heavy flavor symmetry (HFS) are outlined for various hadronic molecules in the charmonium and bottomonium spectra, including the $Z_b$ states. The possibility of HFS to provide systematically improvable predictions for molecular states with heavy quarks is, however, questioned in Ref.~\cite{Baru:2018qkb}. In Ref.~\cite{Baru:2019xnh}, the EFT-based approach to the $Z_b$'s previously developed in Refs.~\cite{Wang:2018jlv,Baru:2020ywb} is employed to make parameter-free predictions for the line shapes, pole positions, and branching fractions of various open- and hidden-bottom decays of the $W_{bJ}$ states. In particular, the importance of the pion exchange between the $B^{(*)}$ mesons is emphasized in the above publications and it is demonstrated that, while the pionless theory predicts all molecules residing near the $B^{(*)}\bar{B}^{(*)}$ thresholds (see Fig.~\ref{fig:Ups5Sdecays}) as virtual states, a richer structure of the near-threshold poles (including above-threshold resonances) arises in the pionful theory.
For studies of the $Z_b$ bottomonium-like states as hadronic molecules within the framework of QCD sum rules see, for example, Refs.~\cite{Zhang:2011jja,Wang:2013llv,Wang:2013zra,Wang:2013daa,Wang:2014gwa,Ozdem:2017exj,Ozdem:2023rkx}. For a comprehensive review of the molecular model and, in particular, its application to the $Z_b$'s see the review \cite{Guo:2017jvc}.

An alternative scenario for exotic states in the spectrum of bottomonium is provided by the compact tetraquark picture. In this approach, $Z_b$'s appear naturally as heavy replicas of the similar states in the spectrum of charmonium $Z_c(3900)$ and $Z_c(4020)$. Yet, in Ref.~\cite{Wu:2018xdi}, predictions of the compact tetraquark model for the $Z_b$'s are critically reconsidered and the conclusion is drawn on a better consistency of their properties with the molecular assignment. The molecular picture is also supported by the work of Ref.~\cite{Dias:2014pva}, using an extension to the $b$-quark sector of the local hidden gauge approach. In Ref.~\cite{Ali:2011ug}, the effective diquark-antidiquark
Hamiltonian is combined with the meson-loop induced effects to explain the masses and decay properties of the $Z_b$ states.
In this picture, the anomalous properties of the di-pion transitions in the mass range around the $\Yfive$ resonance are explained by the presence in this region of a compact $[bq][\bar{b}\bar{q}]$ diquark-antidiquark state tagged as $Y_b(10890)$ \cite{Ali:2009es,Ali:2010pq}. It still remains to be seen whether the data indeed support the co-existence of two different vector states with very close masses. In the meantime, an alternative option is to assume the above compact tetraquark component as part of the $\Yfive$ resonance wave function in addition to its generic $b\bar{b}$ component. For a comprehensive overview of the compact tetraquark model and its application to exotic hadrons, including the $Z_b$'s, see the review \cite{Esposito:2016noz}. Studies of the $Z_b$ states in the framework of a constituent quark model augmented with the coupled-channel effects can be found in Ref.~\cite{Ortega:2021xst}.

\subsection{$\Xb$ states}
\label{sec:Xb}

\subsubsection{Theoretical expectations}

The $\X$ state was first observed by Belle in 2003 through its decay into $\pi^+\pi^-J/\psi$ in the process $B^+\to K^+\pi^+\pi^-J/\psi$~\cite{Belle:2003nnu}. This discovery underscored the growing complexity of meson spectroscopy, since the $\X$ fails to conform to the conventional $q\bar{q}$ or $qqq$ quark-model assignments~\cite{Brambilla:2019esw}. It is natural to search for a similar state with $J^{PC} = 1^{++}$, hereinafter denoted by $\Xb$,
in the spectrum of bottomonium as well.
Indeed, the fundamental interaction between quarks in QCD mediated by gluons is quark-mass-blind and the quark mass enters only through the corresponding kinetic term in the Lagrangian of theory. Then, na{\"i}vely, proceeding from one heavy-quark sector to the other as prescribed by HFS may look straightforward. However, given the general complexity of theory of strong interactions in the nonperturbative regime,
the possibility to deduce robust theoretical conclusions on the spectrum and properties of hadrons based on HFS may not be a general rule but be limited to particular models for the exotic hadrons. 
Under these circumstances, one must employ various effective approaches. However, their predictions depend non-trivially on the quark mass. Furthermore, relating such predictions across different heavy-quark sectors and estimating the corresponding uncertainty can be a difficult task. For example, as mentioned above for a singly-heavy quark system like a heavy-light meson, the limit of an infinitely heavy quark, $m_Q\to\infty$, is well defined and allows for a systematic expansion of the observables in the inverse powers of $m_Q$. Then the properties of the physical system are expressed in terms of universal, $m_Q$-independent parameters describing the kinetic energies of the light degrees of freedom, spin-dependent interactions, and so on. 
Once these parameters are calculated theoretically, computed numerically via lattice techniques, or extracted from experimental data, they can be used on equal footing in both the charm and bottom sectors of theory. However, the uncertainties of the predictions may differ due to the different masses of the charm and bottom quarks $m_c$ and $m_b$, and consequently, the different precision afforded by the HQET expansion.
Meanwhile, the situation changes dramatically
once one proceeds to doubly-heavy systems like the $\X$ or $\Xb$. Indeed, whereas the reduced mass of a singly-heavy system approaches a finite value in the heavy-quark limit --- determined by the mass of the light subsystem --- the kinetic term in a doubly-heavy system retains an explicit dependence on the heavy-quark mass which regulates the infrared behaviour of the system. Consequently, the contributions associated with the unitary cut in the amplitude no longer exhibit a simple scaling with $m_Q$ and therefore can not be straightforwardly translated from the charmonium sector to the bottomonium sector, or {\it vice versa}. It may seem appealing to employ model-independent and systematically improvable EFT frameworks at the hadronic level (for example, chiral EFT aka $\chi$EFT), supplemented by the pertinent symmetries (chiral symmetry, heavy-flavor symmetry, light-flavor symmetry, and heavy-quark spin symmetry), in order to relate the properties of flavor and spin partner states both within a given heavy-quark sector and across different heavy-quark sectors. However, as demonstrated in Ref.~\cite{Baru:2018qkb}, the $\chi$EFT methods do not allow one to relate different heavy-quark sectors in a controllable way since the $m_Q$-dependence of the contact interactions can not be derived in a model-independent way within this framework. Thus, the generic $\chi$EFT approach requires that data in the complimentary channels with different quantum numbers exist and can be used to fix the parameters of the interaction. As discussed in Sec.~\ref{sec:Zb} above, at present the required experimental data in the spectrum of bottomonium only exist in the isospin-1 channel $1^{+-}$
\cite{Belle:2011aa,Adachi:2011mks,Belle:2012koo,Belle:2015upu},
so no reliable $\chi$EFT prediction for the isospin-0 $\Xb$ state can be made as based on these data. Furthermore, straightforward predictions for near-threshold exotic states deduced solely from the symmetries inherent in the original Lagrangian with undressed parameters may not be accurate enough \cite{Baru:2016iwj,Cincioglu:2016fkm,Ortega:2021zgk}.
Additional complications come from different mass hierarchies inherent in the spectra of the $D$ and $B$ mesons: while the decay $D^*\to D\pi$ is kinematically allowed there is no phase space for a similar decay $B^*\to B\pi$. As a result, the pion exchange, coupled-channel, and threshold effects are different in the $\X$ and the gedanken $\Xb$, so given general fragility of near-threshold molecular states, this difference may strongly affect the predictions. 
Under such circumstances, there are several ways to proceed. 

On the one hand, one can rely on models. 
For instance, in the compact tetraquark picture for the $\Xb$, with the interquark interaction inferred from the quark model, the existence of the $\Xb$ has been predicted in the mass range from 10 to 11 \gev~\cite{Ebert:2005nc,Ebert:2008se,Ali:2009pi,Matheus:2006xi}. In the coupled-channel scheme of Ref.~\cite{Ortega:2021zgk} based on a nonrelativistic constituent quark model augmented with the $^3P_0$ light-quark pair creation mechanism, the dynamically generated $\Xb$ is predicted to be predominantly (with the probability around 94\%) a $B\bar{B}^*$ state with the mass 10.599~\gev. In Ref.~\cite{Zhou:2018hlv}, a suitable extension of the Friedrichs model \cite{Friedrichs} is employed to predict the $\Xb$ as a resonance at 10.615~\gev, which is just above the $B\bar{B}^*$ threshold. 

In the hadronic EFT approach, if a particular $m_Q$-scaling is artificially prescribed to the short-range interactions in the system of two heavy-light mesons, predictions can be done simultaneously for both heavy-quark sectors. In this case, the existence of the $\Xb$ in the spectrum of bottomonium near the $B\bar{B}^*$ threshold comes as a direct consequence of the existence of the $\X$ in the spectrum of charmonium in the vicinity of the $D\bar{D}^*$ threshold.
For example, by assuming an $m_Q$-independent interaction in the two-meson system, one concludes that the $\Xb$ is a molecular state more strongly bound than the $\X$
\cite{Tornqvist:1993ng,Nieves:2011zz,Guo:2013sya,Karliner:2013dqa,Ozpineci:2013zas,Karliner:2015ina,Yamaguchi:2019vea}.
On the contrary, if a $1/m_Q$ scaling of the short-range interaction in the $D\bar{D}^*/B\bar{B}^*$ system is adopted then the binding energy of the resulting bound state scales as $1/m_Q^3$ and the binding energy of the $\Xb$ can be expressed through the binding energy of the $\X$ \cite{AlFiky:2005jd}. Thus, although in general the hadronic EFT supplied with some $m_Q$-scaling of the interaction naturally predicts a molecular $\Xb$ to reside near the $B\bar B^*$ threshold, the model uncertainty of this prediction remains vague. 

An alternative EFT in coordinate space can be derived from QCD exploiting the idea of the Born--Oppenheimer separation of the slow (provided by heavy quarks) and fast (provided by the light quarks and gluons) dynamics. Such the Born--Oppenheimer EFT (BOEFT) is under active development and already demonstrates a strong potential in understanding exotic hadrons --- see Ref.~\cite{Berwein:2015vca} for BOEFT application to hybrids and Ref.~\cite{Berwein:2024ztx} for more recent BOEFT studies
of quarkonium and exotic states with two heavy quarks
as well as for a general pedagogical introduction to the subject. In the given approach, the properties of hadronic states are addressed through systems of coupled Schr{\"o}dinger equations. The non-perturbative low-energy gauge-invariant correlators used as input in such equations can be obtained through parameterizations (see, for example, Refs.~\cite{Brambilla:2024imu,Alasiri:2024nue}) of the quantities calculated using the QCD lattice technique.
The predictions of the BOEFT approach for the $\Xb$ can be found, for example, in Refs.~\cite{Brambilla:2024imu,Brambilla:2026ujo}.

The order-of-magnitude theoretical estimates in Ref.~\cite{Guo:2014sca} give the $\Xb$ production cross section at LHC at a nanobarn level. In Ref.~\cite{Liu:2024ets}, the radiative decays\footnote{See Sec.~\ref{sec:Yb} below for a comprehensive discussion of the properties of the vector bottomonium state $\Yb$.} $\Yb\to\gamma\Xb$ and, in particular, studies of the annihilation reaction $e^+e^-\to\gamma\Xb$ near the energy $\sqrt{s}=10.754$~GeV are argued as a promising way of hunting for the $\Xb$ while the radiative decays of the $\Xb$ itself are studied in Ref.~\cite{Li:2014uia}. 
In Ref.~\cite{Hou:2006it}, the $\pi^+\pi^-\Upsilon$ final states, analogous to the $\pi\pi J/\psi$ channel through which the $\X$ was first discovered, were proposed as promising modes for searching for the $\Xb$ at hadron colliders. However, in these processes the pion pair is produced in an isovector configuration, whereas both the $\Xb$ and the $\Upsilon$ are isoscalars. As a result, the corresponding transitions violate isospin symmetry and can occur only due to the nonvanishing mass splitting between the charged and neutral $B$ mesons, in close analogy with the decay $\X\to\pi\pi J/\psi$, which is driven by the mass difference between the charged and neutral $D$ mesons. However, since $m_{B^0}-m_{B^\pm} \ll m_{D^\pm}-m_{D^0}$, isospin-breaking effects in $\Xb$ decays are expected to be significantly smaller than in $\X$ decays and can therefore be safely neglected. Accordingly, and in contrast to the $\X$ case, the $\Xb$ is expected to decay preferentially into $\pi^+\pi^-\pi^0\Upsilon(1S)$ rather than $\pi^+\pi^-\Upsilon(1S)$~\cite{Bondar:2016hva,Voloshin:2011qa,Guo:2013sya,Ortega:2021zgk,Guo:2014sca,Zhou:2018hlv,Karliner:2015ina}.

\subsubsection{Experimental status}

Using 20.7 fb$^{-1}$ and 16.2 fb$^{-1}$ of $pp$ collision data at C.M. energy $\sqrt{s}$ = 8 TeV, CMS and ATLAS studied the invariant mass distributions of $\pi^+\pi^-\Upsilon(1S)$. Clear signals for both $\Upsilon(2S)$ and $\Upsilon(3S)$ were observed but no evidence for an additional state was found~\cite{CMS:2013ygz,ATLAS:2014mka}. This result by CMS and ATLAS can be naturally interpreted as no signal of the possible isovector states $W_{bJ}$ --- the spin partners of the $Z_b$'s~\cite{Bondar:2016hva,Voloshin:2011qa,Guo:2013sya,Ortega:2021zgk}; see Eq.~\eqref{WbJs} above and the discussion around it.
Upper limits are set at 95\% C.L.\ on the ratio of the inclusive
production cross sections times the branching fractions to $\pi^+\pi^-\Upsilon(1S)$ for the $\Xb$ and $\Upsilon(2S)$.
The upper limits on the ratio range from 0.9\% to 5.4\% for $\Xb$ masses between 10 and 11 \gev~from CMS, and from 0.8\% to 4.0\% for $\Xb$ masses in the regions of 10.05--10.31 \gev~and 10.40--11.00 \gev~from ATLAS.

\begin{figure}[t!]
\centering
\begin{tabular}{cc}
\includegraphics[width=0.5\textwidth]{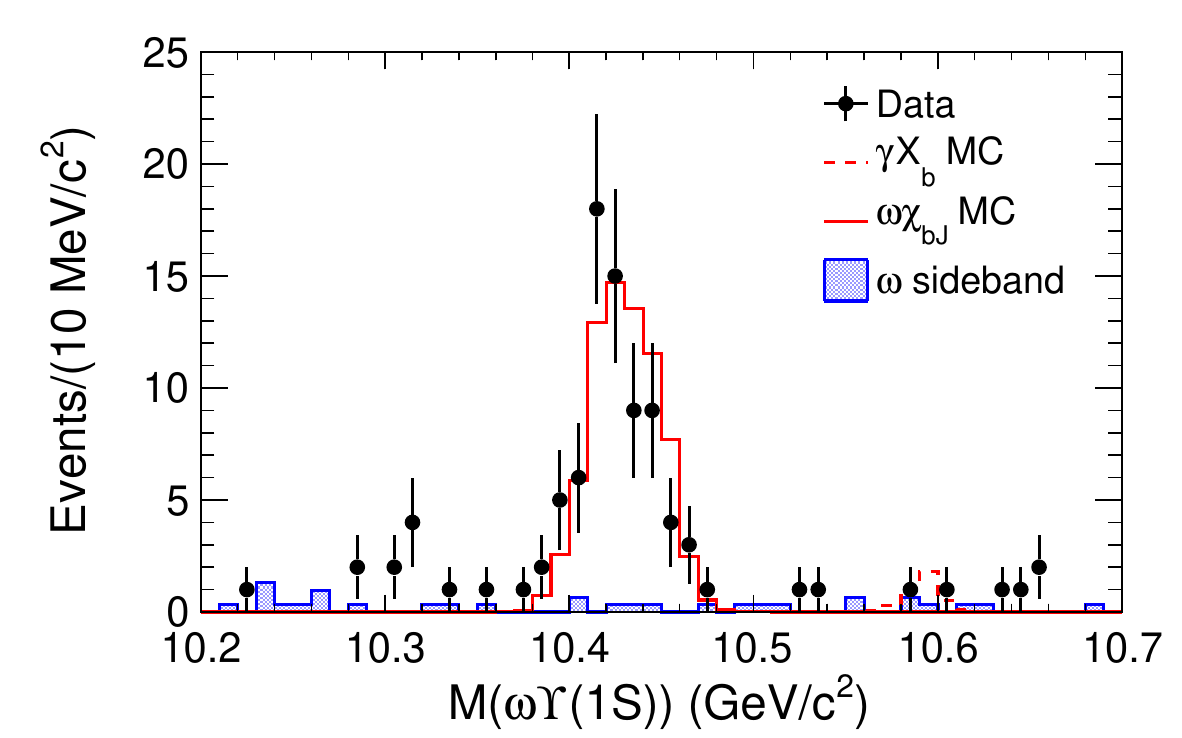} &
\includegraphics[width=0.44\textwidth]{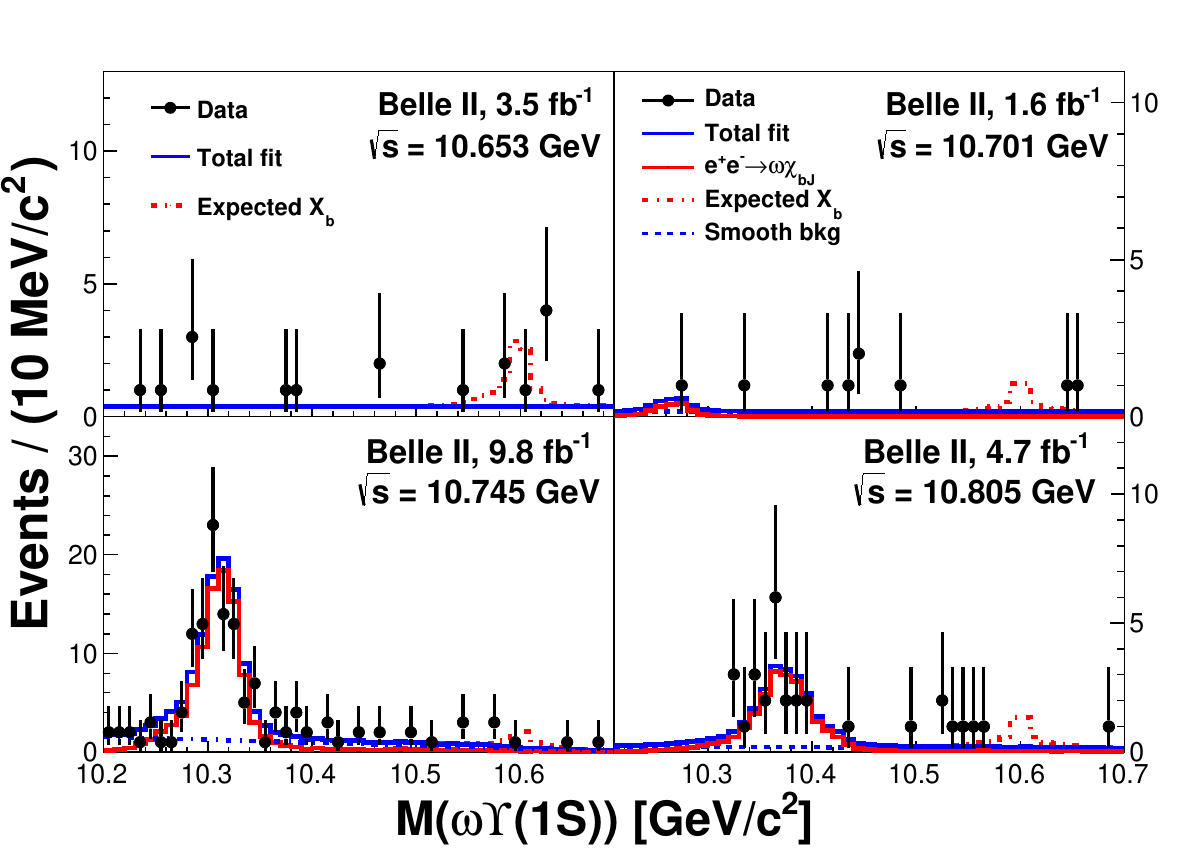}
\end{tabular}
\caption{Distributions of $\omega\Upsilon(1S)$ invariant mass at (left plot) $\sqrt{s}$ = 10.867 GeV at Belle and (right plot) $\sqrt{s}$ = 
10.653, 10.701, 10.745, and 10.805 GeV at Belle II~\cite{Belle:2014sys,Belle-II:2022xdi}. The red dashed
histograms are from simulated events $e^+e^-\to\gamma \Xb(\to\omega\Upsilon(1S))$
with the $\Xb$ mass fixed at 10.6 \gev~and yields fixed at the 90\% C.L.\ upper limit values.}\label{fig3}
\end{figure}

Belle and Belle II searched for $\Xb\to\omega(\to\pi^+\pi^-\pi^0)\Upsilon(1S)$ using a 118 fb$^{-1}$ data sample at $\sqrt{s}$ = 10.867 GeV and a 19.6 fb$^{-1}$ data sample at $\sqrt{s}$ near 10.75 GeV~\cite{Belle:2014sys,Belle-II:2022xdi}.
The invariant mass distributions of $\omega\Upsilon(1S)$ are shown in Fig.~\ref{fig3}.
Prominent reflections of $e^+e^-\to\omega\chi_{bJ}$~\footnote{Hereinafter throughout this review, if not explicitly stated otherwise, $\chi_{cJ}$ and $\chi_{bJ}$ denote $\chi_{cJ}(1P)$ and $\chi_{bJ}(1P)$, respectively, for any $J=0,\;1,\;2$.} signals were observed, as shown by the red solid histograms, but no narrow structure as expected from a $\Xb$ signal was found.
The red dashed histograms are from simulated events $e^+e^-\to\gamma \Xb(\to\omega\Upsilon(1S))$
with the $\Xb$ mass fixed at 10.6 \gev~($B\bar B^*$ mass threshold) and yields fixed at the 90\% C.L.\ upper limit values.
The upper limits at 90\% C.L. on the product of the Born cross section for $e^+e^-\to\gamma \Xb$ and branching fraction for $\Xb\to\omega\Upsilon(1S)$ at $\sqrt{s}$ = 10.653, 10.701, 10.745, 10.805 GeV with $\Xb$ masses between 10.45 and 10.65 \gev~range in (0.14\dash0.55) pb, (0.25\dash0.84) pb, (0.06\dash0.14) pb, and (0.08\dash0.37) pb, and at $\sqrt{s}$ = 10.867 GeV with $\Xb$ masses between 10.55 and 10.65 \gev~range in (0.01\dash0.02) pb.

\begin{figure}[t!]
\centering
\begin{tabular}{cc}
\includegraphics[width=0.45\textwidth]{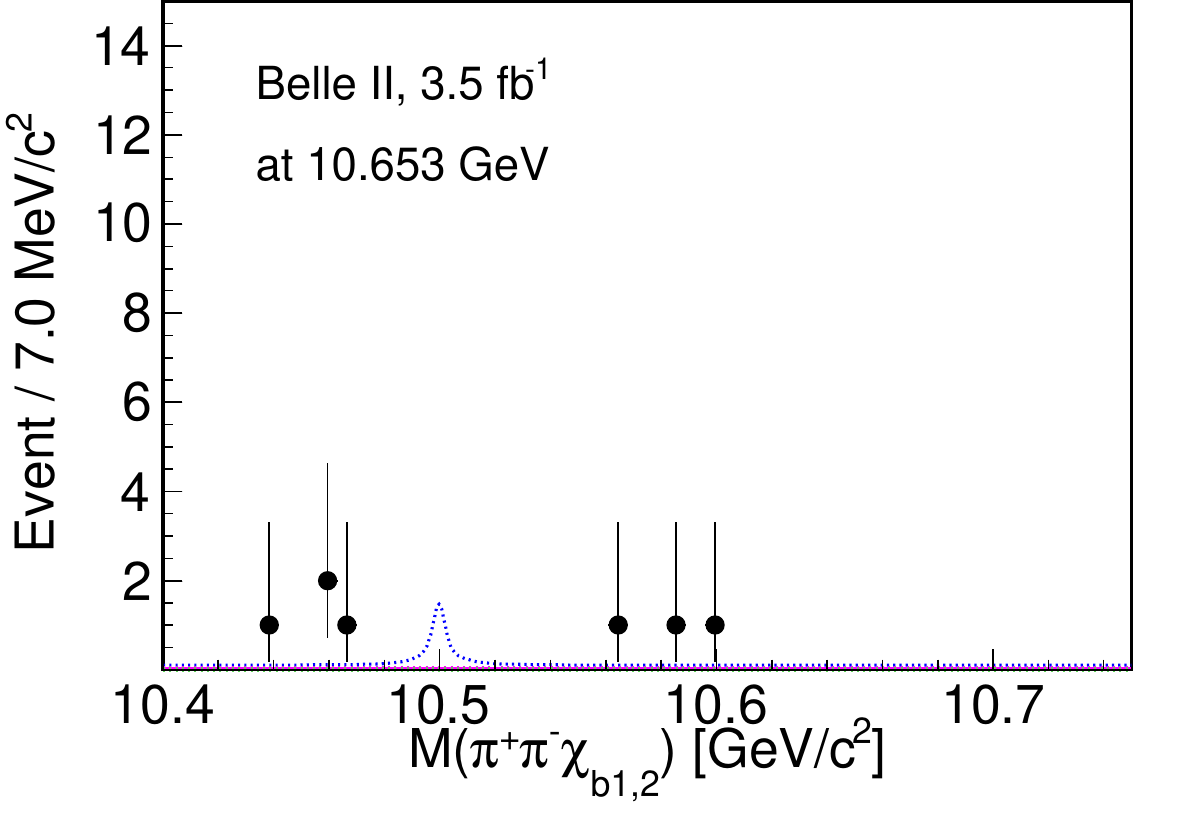} &
\includegraphics[width=0.45\textwidth]{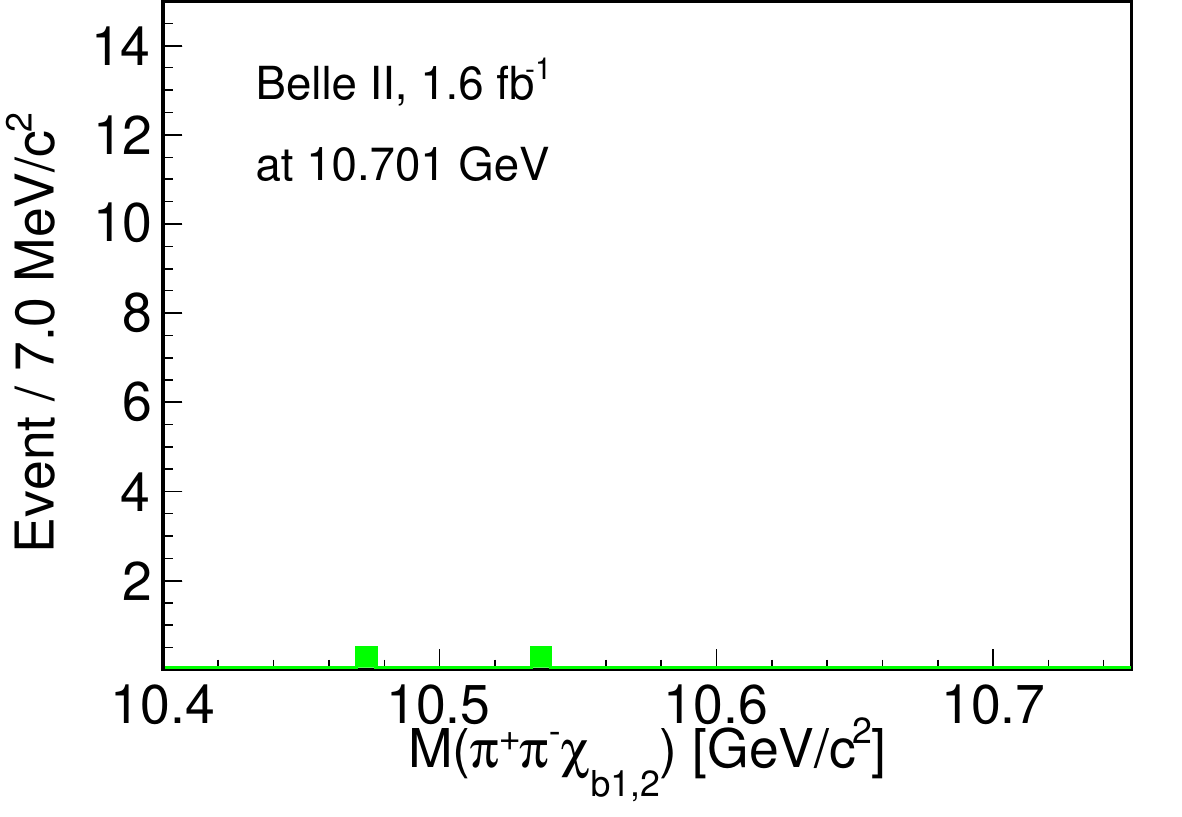}\\
\includegraphics[width=0.45\textwidth]{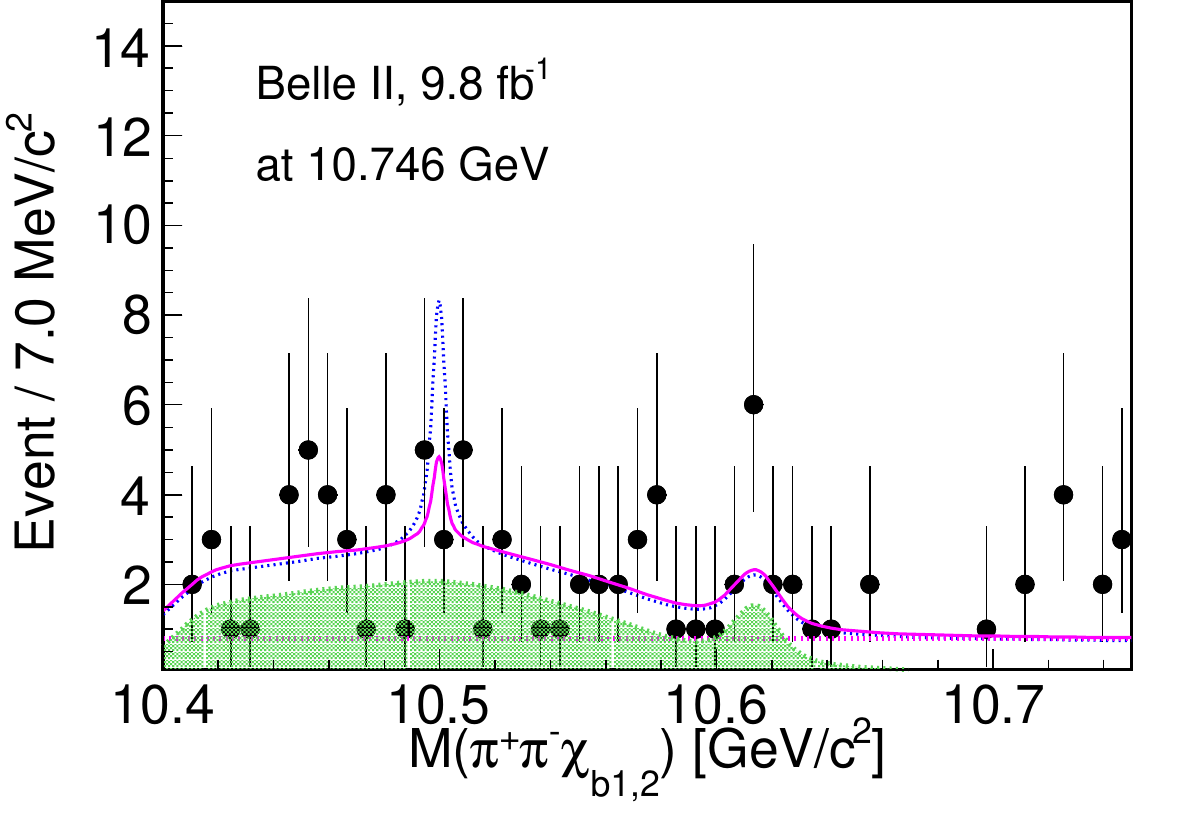} &
\includegraphics[width=0.45\textwidth]{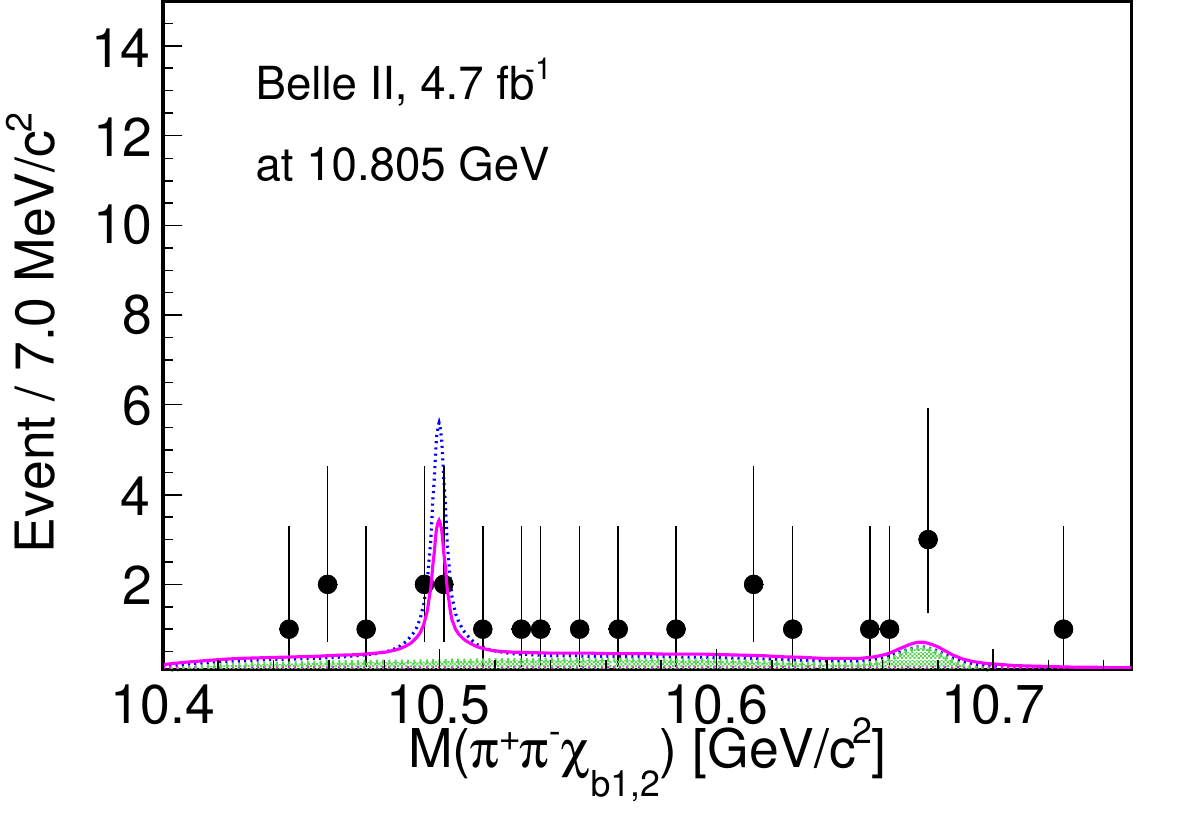}
\end{tabular}
\caption{Invariant mass distributions of $\pi^+\pi^-\chi_{bJ}$ superimposed with the fit results~\cite{Belle-II:2025ubm}. Dots with error bars
are from data. The solid lines present the nominal fit results, and the shaded histograms represent the backgrounds from $e^+e^-\to\omega\chi_{bJ}$ and
$e^+e^-\to\pi^+\pi^-\Upsilon(2S)(\to\gamma\chi_{bJ})$. Blue dashed lines represent the fit results with the contribution of $\Xb$ set to the 90\% C.L.\ upper limits.}\label{fig4}
\end{figure}

The radiative decay width for the $\Upsilon(10753)\to\gamma \Xb$ was predicted to be
order of 10 keV, corresponding to a branching fraction of $10^{-4}$~\cite{Liu:2024ets}.
A search for the $\Xb$ in the process $e^+e^-\to\gamma \Xb$ with $\Xb\to\pi^+\pi^-\chi_{b1}$ at Belle II near $\sqrt{s}$ = 10.75 GeV was strongly suggested~\cite{Liu:2024ets}.
Using 3.5, 1.6, 9.8, and 4.7 fb$^{-1}$ data samples at $\sqrt{s}$ = 10.653, 10.701, 10.746, and 10.804 GeV, Belle II searched for the $\Xb$ with $\Xb\to\pi^+\pi^-\chi_{b1,2}$ in $e^+e^-\to\gamma \Xb$~\cite{Belle-II:2025ubm}.
No evident signal of $\Xb$ was found, as shown in Fig.~\ref{fig4}.
The dominant backgrounds are $e^+e^-\to\omega\chi_{bJ}$ and
$e^+e^-\to\pi^+\pi^-\Upsilon(2S)(\to\gamma\chi_{bJ})$, which are shown in this figure by shaded histograms.
Different hypotheses of the mass of $\Xb$ were evaluated, and the largest upper limit was found when $m(\Xb) = 10.50$ \gev.
The upper limits at 90\% C.L. on the product of the Born cross section for $e^+e^-\to\gamma \Xb$ and branching fraction for $\Xb\to\pi^+\pi^-\chi_{b1,2}$ with $m(\Xb) = 10.50$ \gev~at $\sqrt{s}$ = 10.653, 10.701, 10.746, and 10.804 GeV are 0.14, 0.09, 0.17, and 0.32 pb, respectively.

\subsection{$Y_b$ states (focus on $\Yb$ also known as $\Upsilon(10753)$)}
\label{sec:Yb}

\subsubsection{Experimental observation}\label{10753_Belle2}

Using data samples collected in $e^+e^-$ collisions at the C.M. energy $\sqrt{s}$ ranging from 10.52 to 11.02 GeV, Belle reported the measurement of the Born cross sections for $e^+e^-\to\pi^+\pi^-\Upsilon(nS)$ $(n=1,~2,~3)$~\cite{Belle:2019cbt}. In addition to the $\Yfive$ and $\Upsilon(11020)$, a new resonance at $\sqrt{s}\sim10.75$ GeV was observed for the first time, with a signal significance of 5.2$\sigma$.
The mass and width of this resonance were measured to be $(10752.7\pm5.9^{+0.7}_{-1.1})$~\mev~and $(35.5^{+17.6+3.9}_{-11.3-3.3})$ MeV, respectively. We call this new state $\Yb$.

To confirm this state and to study its properties, the Belle II experiment performed an energy scan, collecting 
data at the following C.M. energies with the corresponding integrated luminosities quoted in parentheses: 10.653 GeV (3.5 fb$^{-1}$), 10.701 GeV (1.6 fb$^{-1}$), 10.746 GeV (9.8 fb$^{-1}$), and 10.804 GeV (4.7 fb$^{-1}$).
Belle II collected data at energy points located in the gaps between those previously collected by Belle.
At Belle, only approximately 1 fb$^{-1}$ data was collected at each energy point.

\begin{figure}[t!]
\centering
\includegraphics[width=0.7\textwidth]{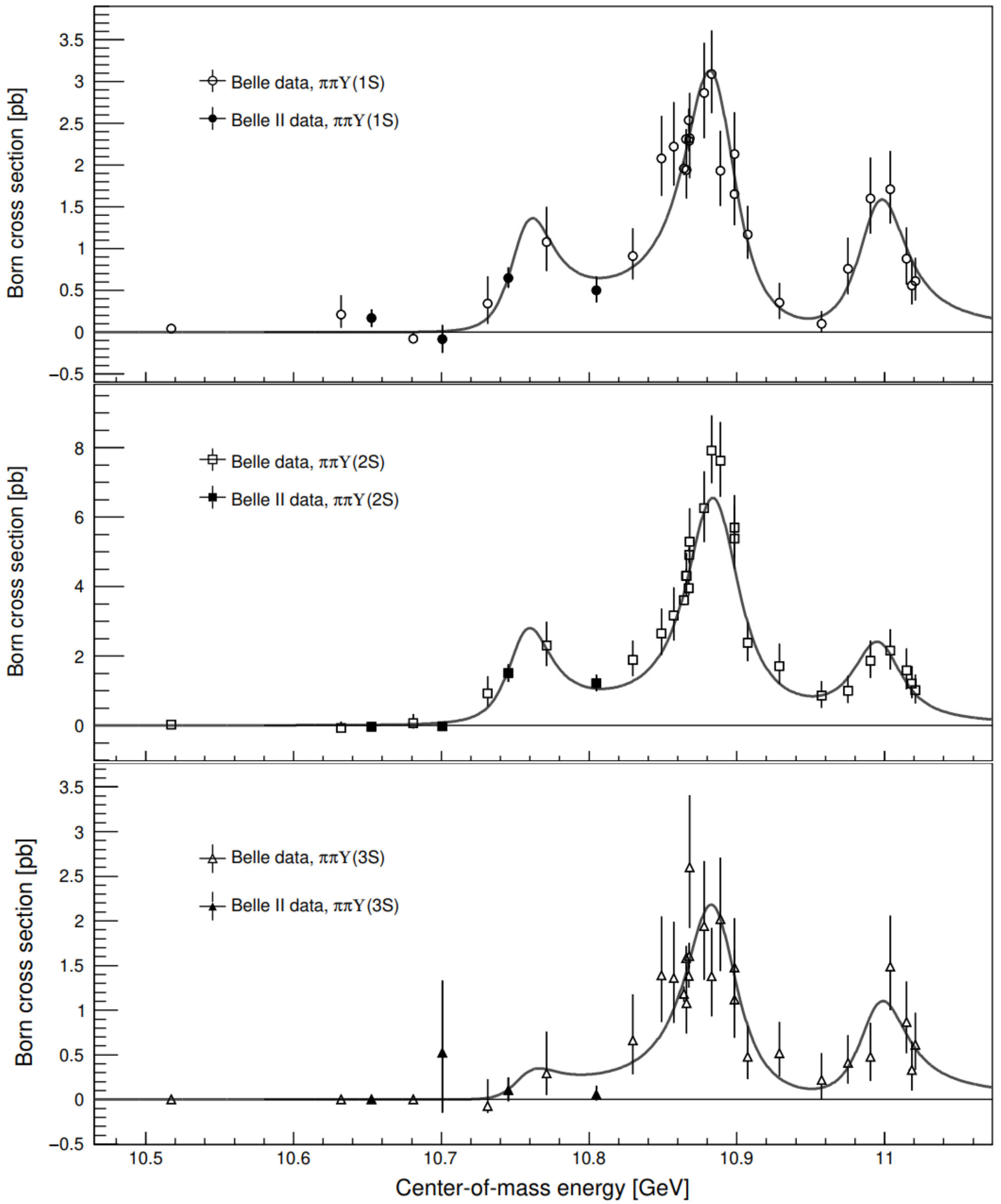}
\caption{Energy dependence of the $e^+e^-\to\pi^+\pi^-\Upsilon(nS)$ Born cross sections ($n = 1,~2,~3$ from top
to bottom)~\cite{Belle-II:2024mjm}. Points with error bars are the data.
The solid curves show the simultaneous fitted results.
}\label{pipiYnS}
\end{figure}

With larger data samples above, Belle II updated the measurement of $e^+e^-\to\pi^+\pi^-\Upsilon(nS)$ $(n=1,~2,~3)$~\cite{Belle-II:2024mjm}.
In comparison with the previous measurement~\cite{Belle:2019cbt}, significantly enhanced signals for $\Yb\to\pi^+\pi^-\Upsilon(nS)$ were observed, with significances of $4.1\sigma$ for $\pi^+\pi^-\Upsilon(1S)$ and $7.5\sigma$ for $\pi^+\pi^-\Upsilon(2S)$, as shown in Fig.~\ref{pipiYnS}.
No evidence for $\Yb$ decaying into $\pi^+\pi^-\Upsilon(3S)$ was found.
The mass and width of the $\Yb$ were measured to be $(10756.6\pm2.7\pm0.9)$~\mev~and $(29.0\pm8.8\pm1.2)$ MeV, respectively, with improved precision.
The intermediate states in the $\pi^+\pi^-\Upsilon(1S,2S)$ transitions were investigated.
No significant $Z^{\pm}_b\to\pi^{\pm}\Upsilon(1S,2S)$ signals were found.
Also, no prominent $f_0(980)\to\pi^+\pi^-$ signal was observed.

\begin{figure}[t!]
\centering
\includegraphics[width=0.45\textwidth]{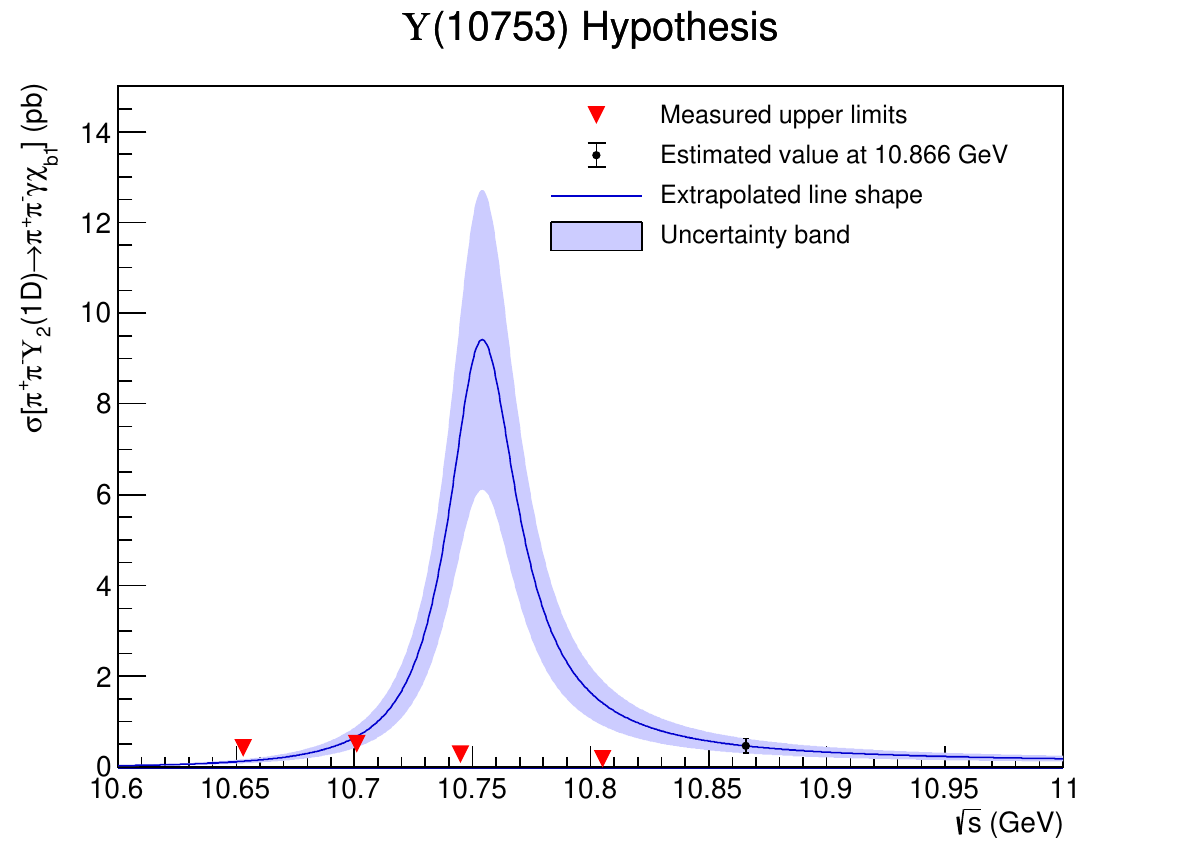}
\put(-180, 120){\bf (a)}
\includegraphics[width=0.45\textwidth]{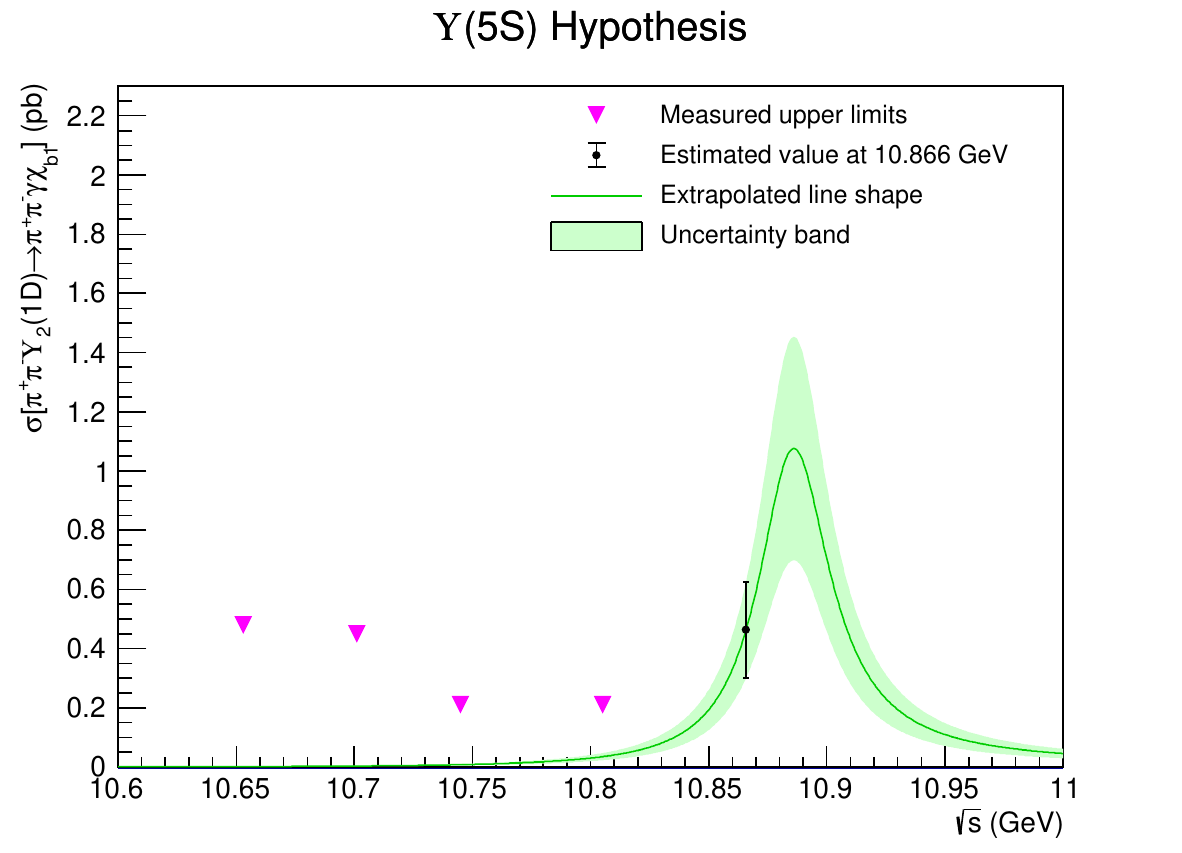}
\put(-180, 120){\bf (b)}
\caption{Measured 90\% C.L. upper limits on the product $\sigma(e^+e^-\to\pi^+\pi^-\Upsilon_2(1D))\,\BR(\Upsilon_2(1D)\to\gamma\chi_{b1})$ as function of C.M. energy~\cite{Belle-II:2026oyq}, evaluated under the (a) $\Yb$ resonance hypothesis and the (b) $\Yfive$ resonance hypothesis. Inverted triangles denote the measured upper limits. The dark dots with error bars indicate the estimated values at 10.866 GeV,
derived from Belle results~\cite{Belle:2011wqq}. The solid curves represent the extrapolated cross sections under the
respective resonance hypotheses, with the filled bands illustrating the uncertainties associated with
these extrapolations.}\label{Y1D}
\end{figure}

Since the $\Yb\to\pi^+\pi^-\Upsilon(nS)$ has been clearly observed~\cite{Belle:2019cbt,Belle-II:2024mjm}, it is natural to search for the $D$-wave $\Upsilon_J(1D)$ state in the same di-pion transition.
Using 19.6 fb$^{-1}$ of data at $\sqrt{s}$ near 10.75 GeV, Belle II presented a study for the process $e^+e^-\to\pi^+\pi^-\Upsilon_{2,3}(1D)$~\cite{Belle-II:2026oyq}.
The $\Upsilon_{2,3}(1D)$ was exclusively reconstructed by $\gamma\chi_{b1,b2}$. 
In the recoil mass distributions of $\pi^+\pi^-$, no significant signal was observed.
The upper limits at the 90\% C.L. were set on the products of the cross sections and branching fractions, 
$\sigma(e^+e^-\to\pi^+\pi^-\Upsilon_2(1D))\,\BR(\Upsilon_2(1D)\to\gamma\chi_{b1})$ and
$\sigma(e^+e^-\to\pi^+\pi^-\Upsilon_3(1D))\,\BR(\Upsilon_3(1D)\to\gamma\chi_{b2})$ at each C.M. energy.
The values of $\sigma(e^+e^-\to\pi^+\pi^-\Upsilon_2(1D))\,\BR(\Upsilon_2(1D)\to\gamma\chi_{b1})$ are shown by inverted triangles in Fig.~\ref{Y1D}.
The extrapolated lines and their bands can be obtained by utilizing the Born cross section for $e^+e^-\to\pi^+\pi^-\Upsilon_J(1D)$ near the peak of the $\Yfive$ resonance~\cite{Belle:2011wqq}, and the masses and widths of $\Yb$ and $\Yfive$~\cite{ParticleDataGroup:2024cfk}.
A direct comparison reveals a pronounced suppression in the coupling of the $\Yb$ resonance to $\Upsilon_J(1D)$ states via di-pion transitions.
The upper limits at 90\% C.L. on the $\sigma(e^+e^-\to\pi^+\pi^-\Upsilon_2(1D))\,\BR(\Upsilon_2(1D)\to\gamma\chi_{b1})$ do not conflict with the $\Yfive$ line shape.

\begin{figure}[t!]
\centering
\begin{tabular}{c}
\includegraphics[width=0.9\textwidth]{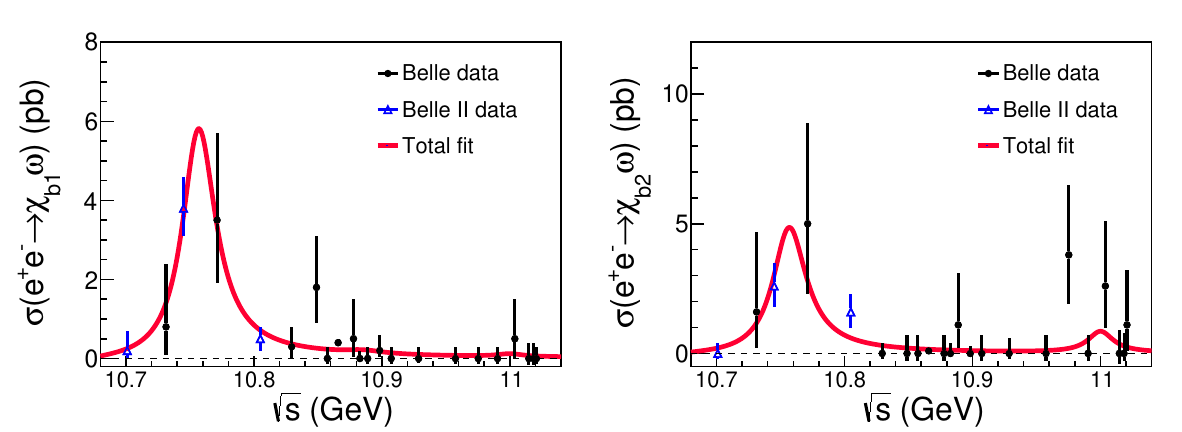}
\put(-375, 120){\large \bf (a)}
\put(-165, 120){\large \bf (b)}\\

\includegraphics[width=0.9\textwidth]{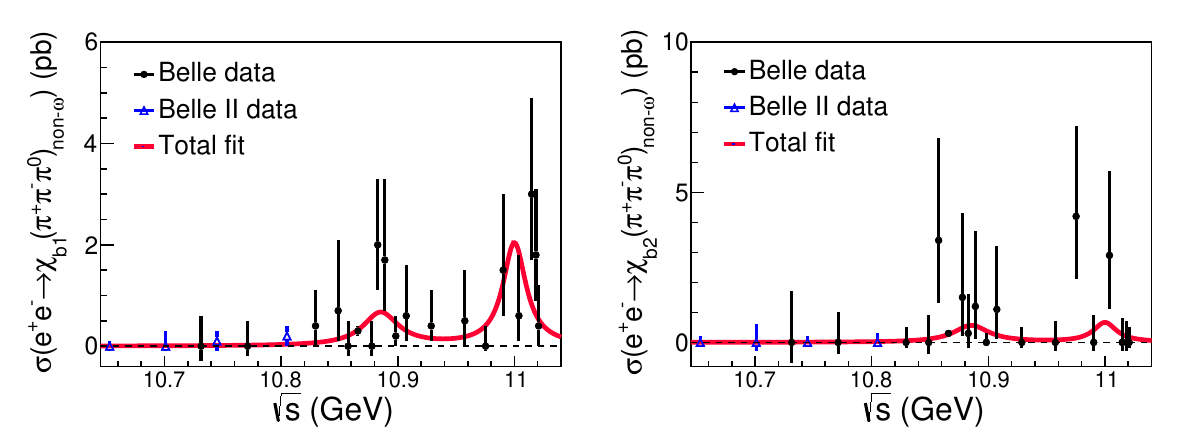}
\put(-245, 125){\large \bf (c)}
\put(-35, 125){\large \bf (d)}
\end{tabular}
\caption{Energy dependence of the Born cross sections for (a) $e^+e^-\to \chi_{b1}\,\omega$, (b) $e^+e^-\to \chi_{b2}\,\omega$, (c) $e^+e^-\to \chi_{b1}\,(\pi^+\pi^-\pi^0)_{\rm non-\omega}$, and (d) $e^+e^-\to \chi_{b2}\,(\pi^+\pi^-\pi^0)_{\rm non-\omega}$~\cite{Belle-II:2025jus}. Filled circles show the measurements at Belle, and triangles show the measurements at Belle II. The error bars show the statistical and additive systematic uncertainties, and the uncertainties in the radiative correction, added in quadrature.
Curves show the fit results.}\label{dependency}
\end{figure}

Belle II utilized the data at $\sqrt{s}$ near 10.75 GeV to study the process $e^+e^-\to\omega\chi_{bJ}~(J = 0,~1,~2)$ and reported the first observation of $\omega\chi_{bJ}$ signals at $\sqrt{s}$ = 10.745 GeV~\cite{Belle-II:2022xdi}.
The energy dependences of the Born cross sections for $e^+e^-\to\omega\chi_{b1,b2}$ are consistent with the shape of the $\Yb$ state, which confirms the existence of the $\Yb$ and reveals a new $\Yb$ decay mode $\Yb\to\omega\chi_{bJ}$.
By combining Belle II data at $\sqrt{s}$ near 10.75 GeV and Belle data at $\sqrt{s}$ from 10.73 to 11.02 GeV, the energy dependences of the Born cross sections for $e^+e^-\to\omega\chi_{b1,b2}$ and $e^+e^-\to(\pi^+\pi^-\pi^0)_{\rm non-\omega}\chi_{b1,b2}$ are shown in Fig.~\ref{dependency}~\cite{Belle-II:2025jus}. The $(\pi^+\pi^-\pi^0)_{\rm non-\omega}$ events were selected by requiring events outside the $\omega$ signal region.
The energy dependences of the $e^+e^-\to\omega\chi_{b1,b2}$ cross sections show a prominent $\Yb$ signal, but no significant signal of the $\Yfive$ or $\Ysix$. The energy dependence of the $e^+e^-\to(\pi^+\pi^-\pi^0)_{\rm non-\omega}\chi_{b1,b2}$ cross sections shows peaks for the $\Yfive$ and $\Ysix$, but no signal for the $\Yb$.
The $\Yfive$ state, and possibly the $\Ysix$ state, decay with a noticeable probability to $(\pi^+\pi^-\pi^0)_{\rm non-\omega}\chi_{bJ}$. This decay pattern is expected if the $(\pi^+\pi^-\pi^0)_{\rm non-\omega}\chi_{bJ}$ final states are produced via intermediate $Z_b(10610)$ and $Z_b(10650)$ states~\cite{Belle:2011aa}, $\Yfive,\Ysix \to Z_b\pi \to \chi_{bJ}\rho\pi$. The decay $Z_b\to\chi_{bJ}\rho$ was predicted in Ref.~\cite{Li:2014cia}.

Using a 19.8 fb$^{-1}$ data sample at $\sqrt{s}$ = 10.746 GeV, Belle II searched for the $\eta_b(1S)$ and $\chi_{b0}$ in the recoil mass distribution of $\omega$~\cite{Belle-II:2023twj}. No evidences were found for $e^+e^-\to\omega\eta_b(1S)$ and $e^+e^-\to\omega\chi_{b0}$ at $\sqrt{s}$ = 10.746 GeV, and the upper limits at 90\% C.L. on the Born cross sections are set to be 2.5 pb and 7.8 pb, respectively.
The obtained upper limit on the Born cross section for $e^+e^-\to\omega\eta_b(1S)$ is close to the Born cross sections of 1\dash3 pb for $e^+e^-\to\pi^+\pi^-\Upsilon(1S,2S,3S)$~\cite{Belle:2019cbt}, which is consistent with the prediction from the $4S$\dash$3D$ mixing model for $\Yb$~\cite{Liu:2023gtx}, but in tension with the tetraquark-model prediction that the $\Yb\to\omega\eta_b(1S)$ decay is strongly enhanced~\cite{Wang:2019veq}.
For a $4S$\dash$3D$ mixed $\Yb$ state, the decay rate to $\omega\chi_{b0}$ is expected to be comparable to the decay rates to $\omega\chi_{b1}$
and $\omega\chi_{b2}$~\cite{Li:2021jjt}; the upper limit is consistent with this expectation.

\begin{figure}[t!]
\centering
\begin{tabular}{ccc}
\includegraphics[width=0.32\textwidth]{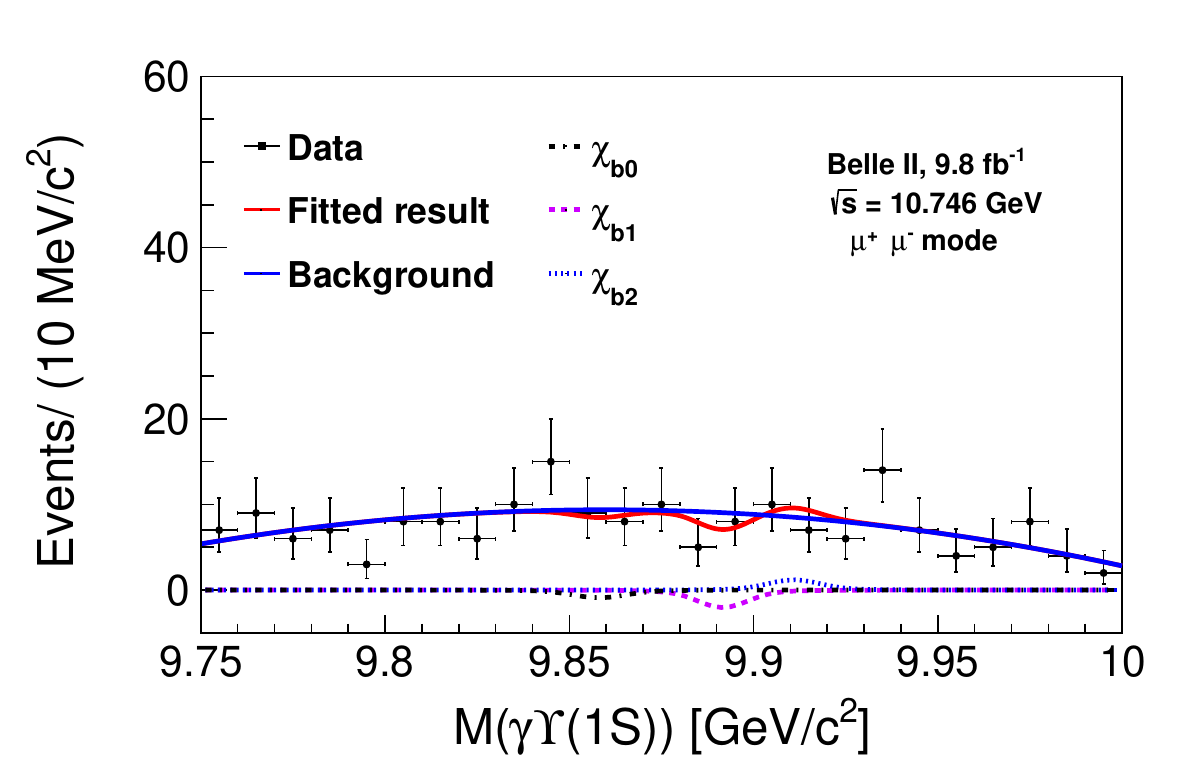}&
\includegraphics[width=0.32\textwidth]{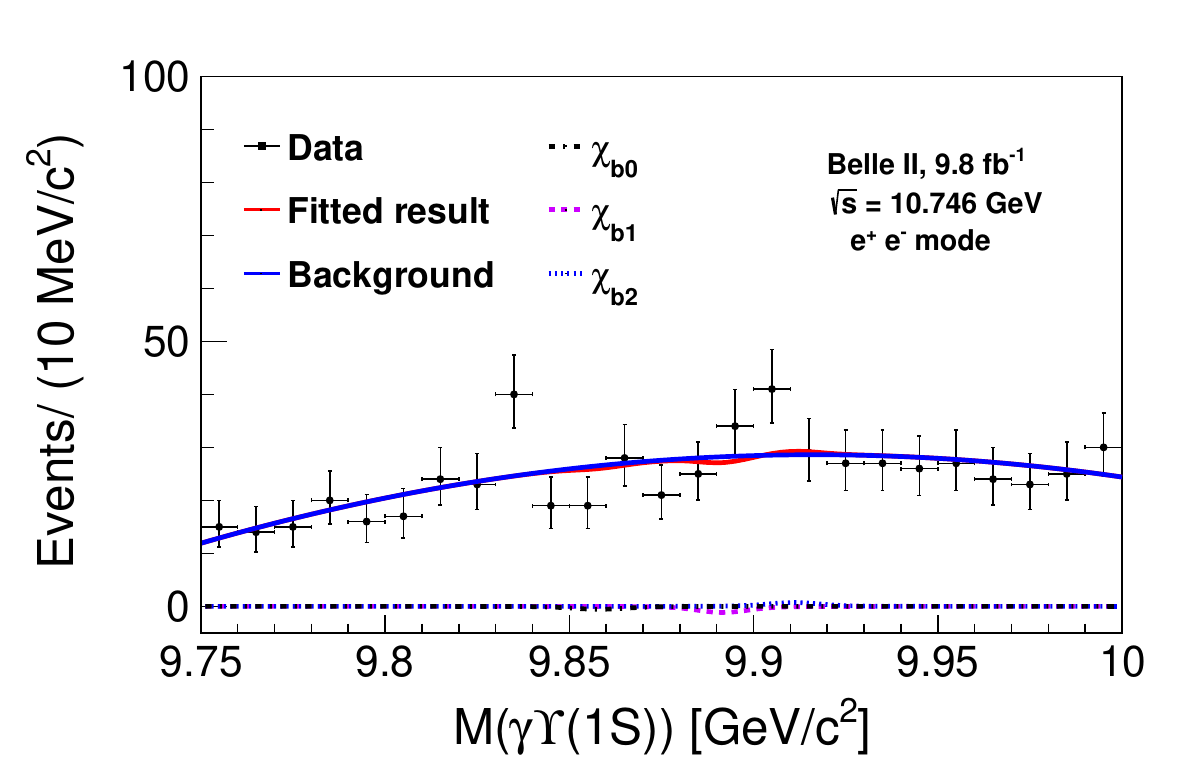}&
\includegraphics[width=0.31\textwidth]{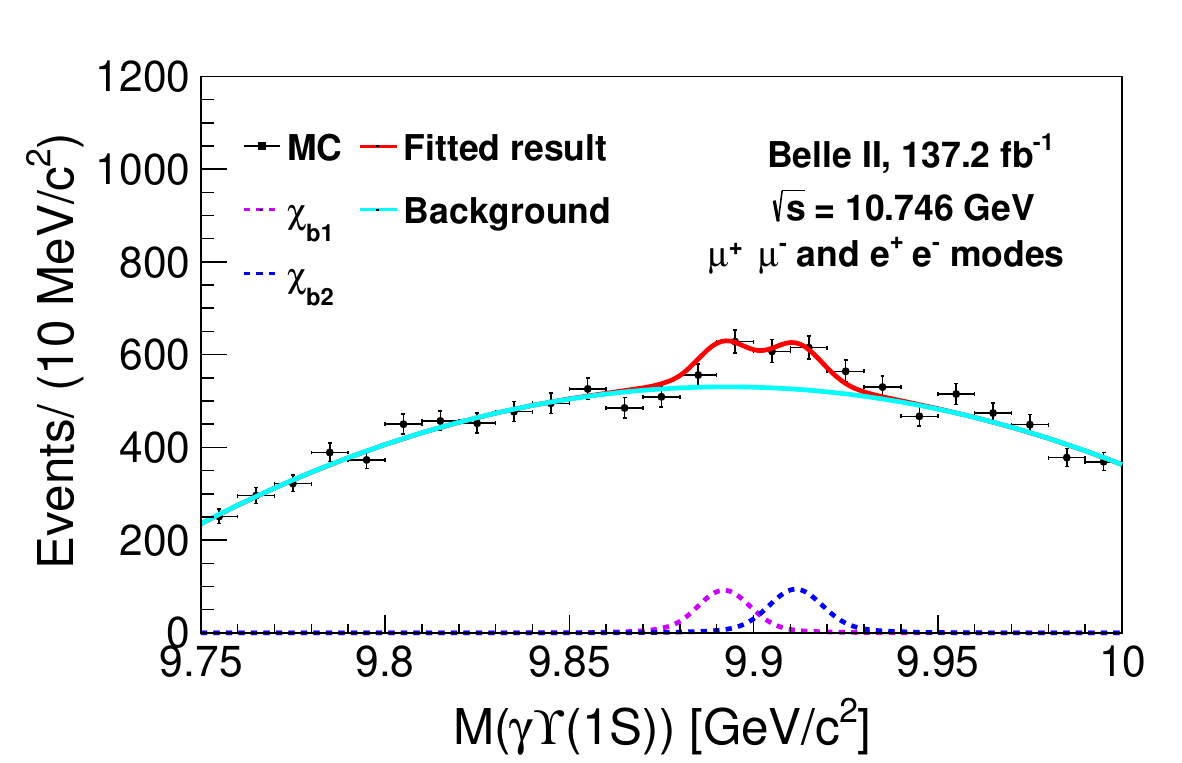}
\end{tabular}
\caption{The invariant mass distributions of $\gamma\Upsilon(1S)$ from a 9.8 fb$^{-1}$ data sample at $\sqrt{s}$ = 10.746 GeV in (left) $\mu^+\mu^-$ and (middle) $e^+e^-$ modes~\cite{Belle-II:2025iil}, and (right) from an assumed 137.2 fb$^{-1}$ data sample at $\sqrt{s}$ = 10.746 GeV in both $\mu^+\mu^-$ and $e^+e^-$ modes with fit results overlaid.
Note that the rightmost plot contains simulated events rather than real data.
}\label{xs_y2s}
\end{figure}

Belle II searched for $e^+e^-\to\gamma\chi_{bJ}$ ($J$ = 0,~1,~2) processes at $\sqrt{s}$ = 10.653, 10.701, 10.746, and 10.804 GeV using 3.5, 1.6, 9.8, and 4.7 fb$^{-1}$ data samples~\cite{Belle-II:2025iil}.
No significant signals were observed at any energy point.
Figures~\ref{xs_y2s}(a) and~\ref{xs_y2s}(b) show the $\gamma\Upsilon(1S)$ invariant mass distributions at $\sqrt{s}$ = 10.746 GeV.
The upper limits at 90\% C.L. on the Born cross sections for $e^+e^-\to\gamma\chi_{bJ}$ were set.
In particular, the upper limit at 90\% C.L. on the Born cross section for $e^+e^-\to\gamma\chi_{b1}$ at $\sqrt{s}$ = 10.746 GeV is 0.26 pb. This value is one order of magnitude smaller than the Born cross sections of $(3.6\pm0.9)$ and $(2.8\pm1.3)$ pb for $e^+e^-\to\omega\chi_{b1}$ and $e^+e^-\to\omega\chi_{b2}$~\cite{Belle-II:2022xdi}.

The sensitivity of observing the $e^+e^-\to\gamma\chi_{b1,b2}$ at $\sqrt{s}$ = 10.746 GeV can be estimated as follows. The upper limits at 90\% C.L. on the Born cross sections for $e^+e^-\to\gamma\chi_{b1}$ and $e^+e^-\to\gamma\chi_{b2}$ at $\sqrt{s}$ = 10.746 GeV are 0.26 and 0.79 pb~\cite{Belle-II:2025iil}.
Here we assume the central values of the Born cross sections for $e^+e^-\to\gamma\chi_{b1}$ and $e^+e^-\to\gamma\chi_{b2}$ at $\sqrt{s}$ = 10.746 GeV are 0.26 and 0.79 pb. We increase the integrated luminosity from 9.8 fb$^{-1}$ in Ref.~\cite{Belle-II:2025iil} by a factor of 14 to be 137.2 fb$^{-1}$.
Under the above assumption, the invariant mass distribution of $\gamma\Upsilon(1S)$ is shown in Fig.~\ref{xs_y2s}(c).
The signal yields for $e^+e^-\to\gamma\chi_{b1}$ and $e^+e^-\to\gamma\chi_{b2}$ are $181\pm38$ and $192\pm39$ with signal significances of 5.0$\sigma$ and 5.2$\sigma$, respectively.
We also perform $10^3$ sets of simulated pseudoexperiments with the above configuration. We find that 95\% of the pseudoexperiments yield the signal significance larger than $3\sigma$.

\begin{figure}[t!]
\centering
\begin{tabular}{cc}
\includegraphics[width=0.5\textwidth]{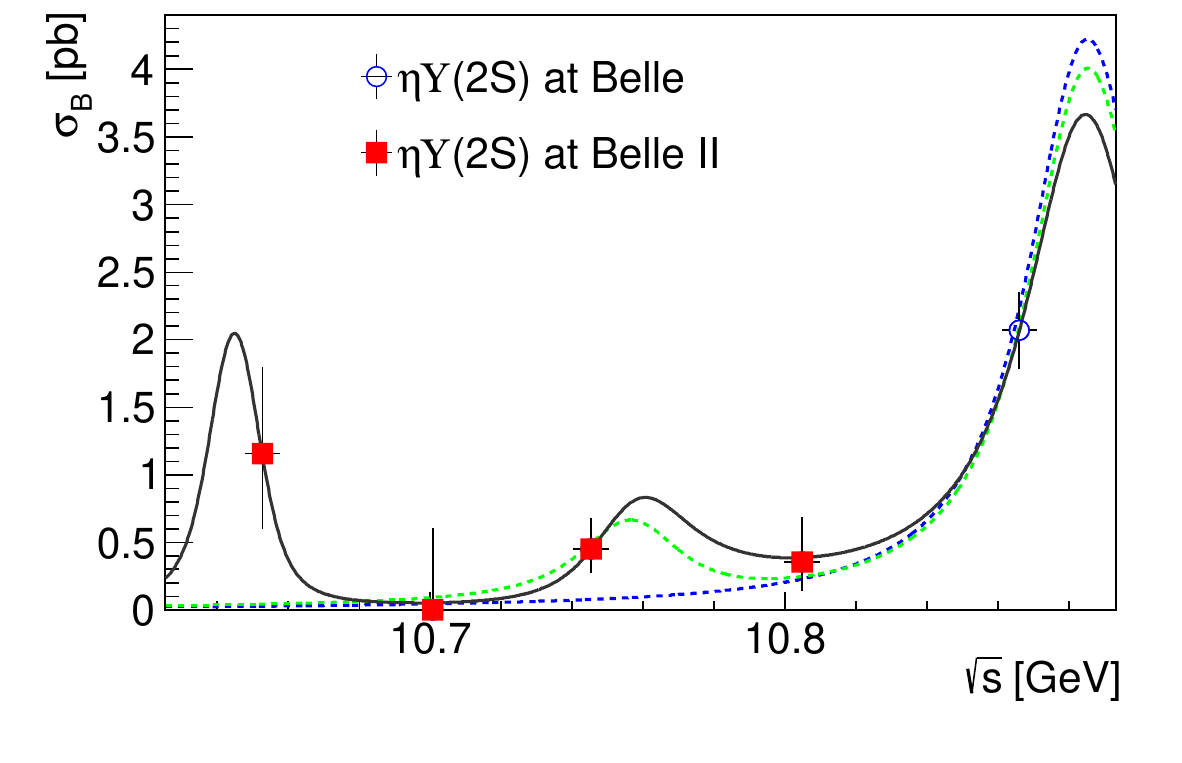}
\end{tabular}
\caption{The Born cross-sections for $e^+e^-\to\eta\Upsilon(2S)$ at Belle~\cite{Belle:2021gws} and Belle II~\cite{Belle-II:2025ubm} with fit results overlaid.
Points with error bars show measured cross sections, the
solid curve is the nominal fit result with $\Yfive$, $\Yb$, and a state near $B^*\bar B^*$
threshold, the green dashed curve is the fit result
with $\Yfive$ and $\Yb$, and the blue dashed curve is the fit
result with only $\Yfive$.
}\label{eta_y2s}
\end{figure}

Belle II performed the study of $e^+e^-\to\eta\Upsilon(1S)$ and $e^+e^-\to\eta\Upsilon(2S)$ using 19.6 fb$^{-1}$ of Belle II data collected at four energy points near the peak of the $\Yb$ resonance~\cite{Belle-II:2025ubm}.
No significant signal was observed for $e^+e^-\to\eta\Upsilon(1S)$ at $\sqrt{s}\sim$ 10.75 GeV.
A signal of $e^+e^-\to\eta\Upsilon(2S)$ at $\sqrt{s}\sim$ 10.75 GeV was observed with a signal significance greater than 6.0$\sigma$.
The energy dependence of the Born cross sections for $e^+e^-\to\eta\Upsilon(2S)$ is shown in Fig.~\ref{eta_y2s}(a).
The central value of the Born cross section at 10.653 GeV was measured to be higher than those at other energy points.
The energy of 10.653 GeV is close to the $B^*\bar B^*$ mass threshold.
In the first 2\dash5 MeV above $B^*\bar B^*$ threshold, the $e^+e^-\to B^*\bar B^*$ cross section increases rapidly~\cite{Belle-II:2024niz}.
The above enhancements close to the $B^*\bar B^*$ threshold in both $e^+e^-\to\eta\Upsilon(2S)$ and $e^+e^-\to B^*\bar B^*$ channels may indicate the presence of a pole close to the threshold.
Three fits to the energy dependence of the Born cross sections for $e^+e^-\to\eta\Upsilon(2S)$ are presented in Fig.~\ref{eta_y2s}(a), including three coherent relativistic BWs representing $\Yfive$, $\Yb$, and a possible state near the $B^*\bar B^*$ threshold (nominal fit); two BWs for $\Yfive$ and $\Yb$; and one BW for $\Yfive$.
Hypotheses that attribute the full
signal ($\Yfive$, $\Yb$, and a state near the $B^*\bar B^*$ threshold) to the $\Yfive$ and/or $\Yb$ states are rejected at the level of at least 3.6$\sigma$.
At $\sqrt{s}$ = 10.653 GeV, the cross section ratio
of $\eta\Upsilon(2S)$ to $\pi^+\pi^-\Upsilon(2S)$ is larger than 23 at the 90\% C.L., indicating a strong violation of HQSS~\cite{Bondar:2016hva}.
After the large cross section near the $B^*\bar B^*$ threshold was reported by Belle~\cite{Belle-II:2025ubm}, a bottomonium-like $Y_b(10650)$ state residing near the $B^*\bar B^*$ mass threshold was predicted in Refs.~\cite{Wang:2025kpm,Chen:2025mgg,Wang:2025jec}.

The summary for experimental results related to $\Yb$ is listed in Table~\ref{summary_10753}. More interpretations from theoretical perspective is detailed in Sec.~\ref{theory_10753}.

\begin{table*}[t]
\renewcommand\arraystretch{1.2}
\setlength{\tabcolsep}{2pt}

\centering
\caption{Summary of experimental results related to $\Yb$.}\label{summary_10753}
\vspace{0.2cm}
\begin{tabular}{lcl}
\hline
\hline
Channel in $e^+e^-$ annihiliation& \hspace*{2mm}&Results \\
\hline
\multirow{2}{*}{$\pi^+\pi^-\Upsilon(nS)$ $(n=1,~2,~3)$~\cite{Belle-II:2024mjm}} && Measure the mass and width of $\Yb$ with higher precision. \\
&& Neither $\pi^{\mp}Z^{\pm}_b$ nor $f_0(980)\Upsilon(nS)$ intermediate states were observed. \\
\hline
$\pi^+\pi^-\Upsilon_{2,3}(1D)$~\cite{Belle-II:2026oyq} && A pronounced suppression compared to $e^+e^-\to\pi^+\pi^-\Upsilon(nS)$. \\\hline
\multirow{2}{*}{$\omega\chi_{b1,b2}$~\cite{Belle-II:2022xdi,Belle-II:2025jus}} && Large discrepancy of $\sigma(e^+e^-\to\omega\chi_{b1,b2})/\sigma(e^+e^-\to\pi^+\pi^-\Upsilon(nS))$ at $\Yb$ and \\
&& $\Yfive$. Compatible cross sections between $e^+e^-\to\omega\chi_{b1}$ and $e^+e^-\to\omega\chi_{b2}$. \\
\hline
$(\pi^+\pi^-\pi^0)_{\rm non-\omega}\chi_{b1,b2}$~\cite{Belle-II:2025jus} && The excess may be due the $\Yfive,\Ysix \to Z_b\pi \to \chi_{bJ}\rho\pi$. \\
\hline
$\omega\eta_b(1S)$~\cite{Belle-II:2023twj} && The upper limit on the cross section is compatible with $\sigma(e^+e^-\to\pi^+\pi^-\Upsilon(nS))$. \\
\hline
$e^+e^-\to\gamma\chi_{b1}$~\cite{Belle-II:2025iil} && The radiative transition is lower than the hadronic transitions at $\sqrt{s}\sim$ 10.75 GeV. \\
\hline
$\eta\Upsilon(2S)$~\cite{Belle-II:2025ubm} && The cross section near the $B^*\bar B^*$ threshold is large. \\
\hline
$ B^*\bar B^*$~\cite{Belle-II:2024niz} && Rapid increase of $\sigma(e^+e^-\to B^*\bar B^*)$ above the $B^*\bar B^*$ threshold. \\
\hline
\hline
\end{tabular}
\end{table*}

\subsubsection{Theoretical interpretations}\label{theory_10753}

The experimental observation of the state $\Yb$ in the spectrum of bottomonium appeared to be an unexpected discovery for theorists. Indeed,
given its unambiguously established quantum numbers $J^{PC}=1^{--}$, a straightforward identification for the $\Yb$ would be a radially excited $n^3S_1$ or $n^3D_1$ generic $b\bar{b}$ state or a certain mixture of those. However, the spin-dependent interactions in heavy quarkonia are strongly suppressed by the heavy quark mass \cite{Eichten:1980mw,Gromes:1984ma} and as such can not provide a noticeable mixing between the $S$- and $D$-wave quarkonia --- see, for example, Ref.~\cite{Radford:2007vd}. In the meantime,
the experimentally established mass of the $\Yb$ significantly deviates from the predictions of the quark models for generic vector bottomonia (see, for example, Ref.~\cite{Godfrey:2015dia} for the predictions of the relativized quark model).
Also, for example, using the results of Ref.~\cite{Belle:2019cbt}, one can estimate
\be
\frac{\sum_{n=1}^3\BR(\Yb\to\pi^+\pi^-\Upsilon(nS))}{\sum_{J=1}^2\BR(\Yb\to\omega\chi_{bJ})}<0.9
\ee
while
\be
\frac{\sum_{n=1}^3\BR(\Yfive\to\pi^+\pi^-\Upsilon(nS))}{\sum_{J=1}^2\BR(\Yfive\to\omega\chi_{bJ})}>28.
\ee
Such different decay patterns for the two states with the same $J^{PC}$ quantum numbers and separated by only 100\,MeV indicate that they must have different internal structures. Thus, the discovery of the $\Yb$ started an active discussion in theoretical community regarding its nature. Then, although in some works, the $\Yb$ is still assigned as a pure or predominantly $D$-wave bottomonium
\cite{Li:2019qsg,Liang:2019geg,Chen:2019uzm,Giron:2020qpb,Zhao:2023hxc,Kaushal:2025kbz}, a $4S$\dash$3D$ mixture is often discussed in the literature as a more reliable option for the $\Yb$ \cite{Li:2021jjt,Bai:2022cfz,Li:2022leg,Liu:2023gtx,Bokade:2025voh,Luo:2025kid} (a more exotic $6S$\dash$4D$ mixing scheme is advocated for this state in Ref.~\cite{Kher:2022gbz}).
Indeed, the fundamental principles of quantum mechanics suggest that states with identical quantum numbers exhibit stronger mixing when their corresponding eigenenergies are closely spaced.
This situation arises naturally for the $(n+1)S$ and $nD$ levels, as radial and orbital excitations contribute comparably to the mass of heavy quarkonium states. However, to get a nonvanishing transition matrix element between such states, one has to proceed beyond the traditional quenched quark model scheme and unquench it by including the effects from the open-flavor loops \cite{Heikkila:1983wd}. Then one can naturally understand, for instance, the puzzling phenomena in the decays of the charmonium $\psi(3770)$, if the latter is treated as a $1D$ state mixed with a $2S$ radial excitation $\psi(3686)$ \cite{Rosner:2001nm}. Further implications of the $4S$\dash$3D$ mixing scheme for categorising the members of the family of vector charmonia can be found, for example, in Ref.~\cite{Wang:2019mhs}. The mixing mechanism described above, mediated by open-bottom loops, is also found to play a significant role for the $(n+1)^3S_1$ and $n^3D_1$ bottomonium states with $n \geqslant 3$, leading to a sizable mixing angle between them; see, for example, Refs.~\cite{Badalian:2008ik,Lu:2016mbb}.
This opens new possibilities for understanding the nature of the newly discovered resonances and incorporating them into the existing framework of vector bottomonia.

An essential role of the open-bottom channels,
that goes beyond the above $S$\dash$D$ mixing effect,
for understanding the nature of the $\Yb$ resonance is emphasized in Refs.~\cite{Bicudo:2020qhp,Bicudo:2022ihz,Bicudo:2022cqi,Ortega:2024rrv,Ni:2025gvx}. In particular, in the unquenched quark model of
Ref.~\cite{Ortega:2024rrv}, the $\Yb$ emerges as a dressed hadronic resonance, in which conventional $b\bar{b}$ configuration and $B^*\bar{B}^*$ molecular configurations contribute with comparable weights.
The authors of Ref.~\cite{Ni:2025gvx} also interpret the $\Yb$ as a dynamical resonance generated by a strong coupling between the $\Yfour$ and the $B^*\bar{B}^*$ open-bottom channel, in agreement with the conclusions of Ref.~\cite{Ortega:2024rrv}.
In Ref.~\cite{Bicudo:2020qhp}, a system of coupled Schr{\"o}dinger equations is derived and solved with the interaction motivated by the
lattice Quantum Chromodynamics (LQCD) string potentials incorporating string breaking effects. The authors find $\Yb$ as a predominantly dynamical meson--meson resonance with around 76\% of the meson--meson content.
Further models for the $\Yb$ state found in the literature include its compact tetraquark \cite{Wang:2019veq,Ali:2019okl,Parkhomenko:2023fkp,Wang:2025sic,Zhao:2025kno} or bottomonium hybrid \cite{TarrusCastella:2019lyq,Chen:2019uzm,TarrusCastella:2021pld} assignment.

Although the existing experimental results are not sufficient to discriminate between all the existing theoretical models, some of them can still be ruled out.
In particular, identification of the $\Yb$ as a $\Upsilon(2D)$ state fails since in this case the branching fraction for the decay $\Yb\to\gamma\chi_{b1}$ can reach 12\%~\cite{Godfrey:2015dia,Wang:2018rjg}, which is almost impossible if other observed decay modes of the $\Yb$ are taken into account. Furthermore, the measured ratio,
\be
\frac{\BR(\Yb\to\omega\chi_{b1})}{\BR(\Yb\to\omega\chi_{b2})}=1.1\pm0.4\pm0.3,
\ee
contradicts the expectation for this ratio to be around 15 for a pure $D$-wave bottomonium state~\cite{Guo:2014qra}. There is also a 1.8$\sigma$ tension with the value 0.2 predicted in Ref.~\cite{Li:2021jjt} for a $S$\dash$D$-mixed state.

In Ref.~\cite{Husken:2022yik}, a combined $K$-matrix analysis of the data in all measured open-bottom channels in the energy range relevant for the $\Yb$ is performed. The authors find
a significant contribution of this resonance in the theoretical fit, but conclude that the data presently available are not able to constrain the parameters of the $\Yb$ and allow for a
broad range of its mass and width, all describing the lineshapes similarly well. 
Searches for the $\Yb$ in the radiative return from $\Yfive$ and around are advocated in Ref.~\cite{Hou:2006it}.
As was mentioned in Sec.~\ref{sec:Xb}, a promising proposal of searching for the exotic state $\Xb$ in the radiative decay $\Yb\to\gamma\Xb$ is put forward in Ref.~\cite{Liu:2024ets} --- the estimated decay width around 10~keV (that translates into $10^{-4}$ for the corresponding branching fraction) may potentially be measurable at Belle II. 

\section{Exotic states in $B$ hadron decays}
\label{sec:Bmeson}

\begin{figure}[t!]
\centering
\begin{tabular}{ccc}
\includegraphics[width=0.31\textwidth]{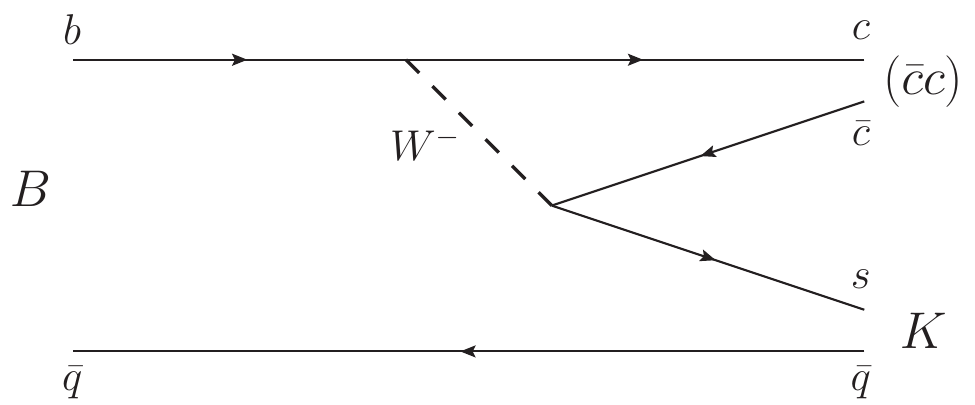}&
\includegraphics[width=0.31\textwidth]{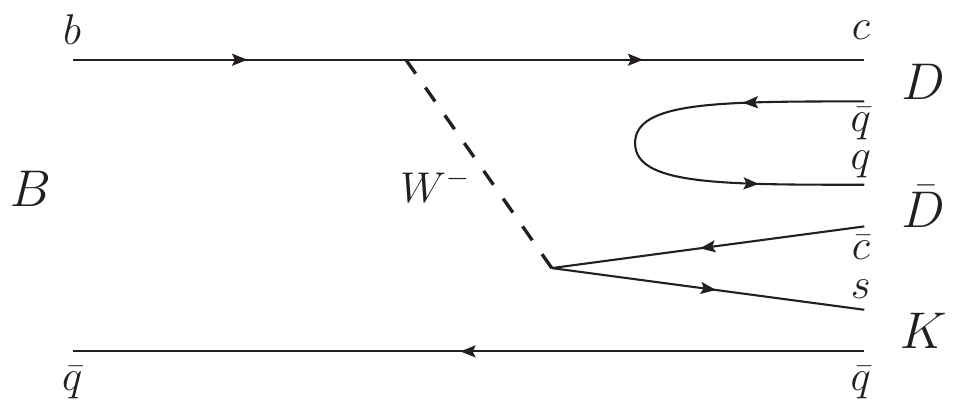}
&\includegraphics[width=0.31\textwidth]{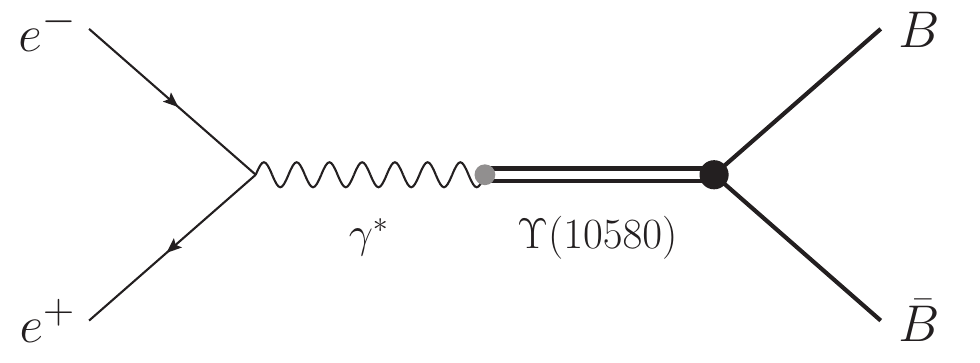}
\put(-455, 65){\large \bf (a)}
\put(-305, 65){\large \bf (b)}
\put(-140, 65){\large \bf (c)}
\end{tabular}
\caption{First plot: Feynman diagram of a weak $B$-meson decay with a charmonium or charmonium-like hadron (denoted as $(\bar{c}c)$) directly produced in the final state and accompanied by a kaon. Second plot: Feynman diagram of a weak $B$-meson decay with a pair of charmed mesons (denoted as $D$ and $\bar{D}$) produced in the final state and accompanied by a kaon. Third plot: Production of $B\bar{B}$ pairs at $B$-factories working at the collisions energy $\sqrt{s}$ close to the nominal mass of the vector $\Upsilon(10580)$ resonance. }\label{fig:BKcharm}
\end{figure}

The hadron decays of the $B$ mesons provide a promising doorway to the production of exotic hadrons with different quantum numbers $J^{PC}$ in the spectrum of charmonium. Indeed, while a direct production of many exotic states in the $e^+e^-$ annihilation may also be possible, either the quantum numbers of the produced resonance are limited to $1^{--}$ or the corresponding production probability per unit of time appears to be small. On the other hand, charmonium-like states with various quantum numbers, including exotic configurations, can be produced via the weak $b$-quark transitions $b\to c\bar{c}\,s(d)$. Figure~\ref{fig:BKcharm}(a) illustrates the corresponding weak $B$-meson decay, in which a charmonium or charmonium-like hadron (denoted as $(\bar{c}c)$) is directly produced in the final state together with a kaon. Figure~\ref{fig:BKcharm}(b) shows a similar weak $B$-meson decay, but with a pair of open-charm mesons (denoted for brevity as $D$ and $\bar{D}$)
produced in the final state, again accompanied by a kaon. Strong interactions between the hadrons in this final state may subsequently lead 
to the formation of exotic configurations with various quantum numbers. In this 
way, any additional dynamical suppression of the production amplitude can be avoided, 
which is why many exotic states were first observed in $B$-meson decays. For example,
the progenitor of the entire family of exotic hadrons --- the $\X$ --- was first observed by Belle in the reaction $B^+\to K^+\X\to K^+\pi^+\pi^- J/\psi$ \cite{Belle:2003nnu}. Later, the $\X$ produced in weak decays of the $B$ mesons was further studied in the same hidden-charm mode and in the open-charm final state $KD\bar{D}^*$ (more precisely, in $KD^0\bar{D}^0\pi^0$ or a sum of the modes $KD^0\bar{D}^0\pi^0$ and $KD^0\bar{D}^0\gamma$) at both existing at that time $B$-factories Belle \cite{Belle:2006olv,Belle:2008fma,Belle:2008oko} and BaBar \cite{BaBar:2007cmo,BaBar:2008qzi}. 
It should also be noted that $B$-meson decays offer clean experimental signatures since the relatively long lifetime of the $B$ meson leads to a displaced decay vertex, which allows for an efficient background suppression. In addition, the well-defined initial state in $B$-meson decays facilitates detailed amplitude analyses, enabling the determination of the quantum numbers of the exotic states and providing crucial information on their internal structure. This is particularly important for disentangling genuine resonant states from kinematic effects near open-charm thresholds.

The process of a copious production of $B\bar{B}$ pairs at $B$-factories, working at the energies $\sqrt{s}$ close to the nominal mass of the vector bottomonium $\Upsilon(10580)$, is visualized in Fig.~\ref{fig:BKcharm}(c). 
The LHC, and in particular the LHCb experiment, also provides an excellent environment for studying exotic charmonium-like states produced in $B$-meson decays.
First, the production cross section for $b\bar b$ pairs at the LHC is large, resulting in an enormous yield of $B$ mesons. 
Since typical production rates and branching fractions for exotic states are small, accumulating high statistics is essential for precise studies. This is facilitated by the fact that many such states decay into final states containing a $J/\psi$, which can be reconstructed with high efficiency and excellent resolution at LHCb through its decay into a muon pair.
To summarize, studies of $B$-meson decays at the $e^+e^-$ (Belle and Belle II experiments) and hadron-hadron (LHCb experiment) colliders are complementary to each other. While the former provides a cleaner environment and fixed C.M. energies, the latter offers substantially larger data samples and access to rare decay modes. Thus, in this section we give an overview of the exotic states with the charmed quark(s) that can be produced in hadron decays of $B$ mesons and studied in the Belle/Belle II and LHCb experiments.

In this section we mainly use the new naming scheme for exotic hadrons suggested by PDG with an exception for the $\chi_{c1}(3872)$ for which we resort to its historical name $\X$ for a better consistency with the previous section.

\subsection{Exotic charmed--strange states}

\subsubsection{Observation of $T^*_{cs0}(2870)^0$ and $T^*_{cs1}(2900)^0$}

In 2020, the LHCb Collaboration reported the observation of two new resonant structures, $T^*_{cs0}(2870)^0$ and $T^*_{cs1}(2900)^0$, in the $D^-K^+$ invariant mass spectrum of the $B^+ \to D^+D^-K^+$ decay~\cite{LHCb:2020bls,LHCb:2020pxc}. Owing to their quark content $ud\bar{s}\bar{c}$, these states 
provide unambiguous evidence of genuine tetraquark configurations. 
The existence of these states was first confirmed in the $D^+K^-$ invariant-mass spectrum of the $B^- \rightarrow D^{*-}D^+K^-$ decay~\cite{LHCb:2024vfz}. Theoretically, the two states may be interpreted as compact quark-model tetraquark states~\cite{Karliner:2020vsi,He:2020jna,Wang:2020prk,Yang:2020atz,Tan:2020cpu,Wang:2020xyc,Zhang:2020oze,Albuquerque:2020ugi,Guo:2021mja,Agaev:2021knl,Yang:2021izl,Agaev:2022eeh,Wei:2022wtr,Liu:2022hbk,Ozdem:2022ydv,Ortega:2023azl,Yang:2023evp}, hadronic molecules~\cite{Hu:2020mxp,Liu:2020nil,Kong:2021ohg,Xiao:2020ltm,Chen:2020aos,Huang:2020ptc,Molina:2010tx,Molina:2020hde,Agaev:2020nrc,Dong:2020rgs,Mutuk:2020igv,Wang:2021lwy,Chen:2021xlu,Qi:2021iyv,Chen:2021tad,Ke:2022ocs,Yue:2022mnf,Agaev:2022eyk,Duan:2023qsg,Wang:2023hpp,Yu:2023avh,Ding:2024dif}, including three-body hadronic bound states \cite{Wu:2020job,Ikeno:2022jbb}, kinematically generated singularities arising from intermediate triangle diagrams \cite{Liu:2020orv}, or threshold effects \cite{Ge:2022dsp,Molina:2022jcd}. In Ref.~\cite{Burns:2020xne}, ways to discriminate among interpretations for the charmed--strange tetraquark states are outlined.
In the compact tetraquark and molecular scenarios, the $T^{*0}_{cs}$ is expected to decay into both $D^+K^-$ and $D^0{\bar K}^0$ final states with comparable rates under the assumption of isospin symmetry. In contrast, in the kinematic-origin hypothesis, the $T^{*0}_{cs}$ structure produced in the $B^{-}\rightarrow D^{-}T^{*0}_{cs}$ decay is predicted to couple more strongly to the $D^+K^-$ final state than to the $D^0\bar{K}^0$ one, this way offering a possible experimental tool to discriminate among these interpretations. In any case, the absence of sibling states in the $D^*\bar K$ channels, as predicted by approximate HQSS, poses a serious obstacle to their identification as predominantly $D\bar K$ molecular states.

\begin{figure}[t]
\centering
\begin{tabular}{ccc}
\includegraphics[width=0.32\columnwidth]{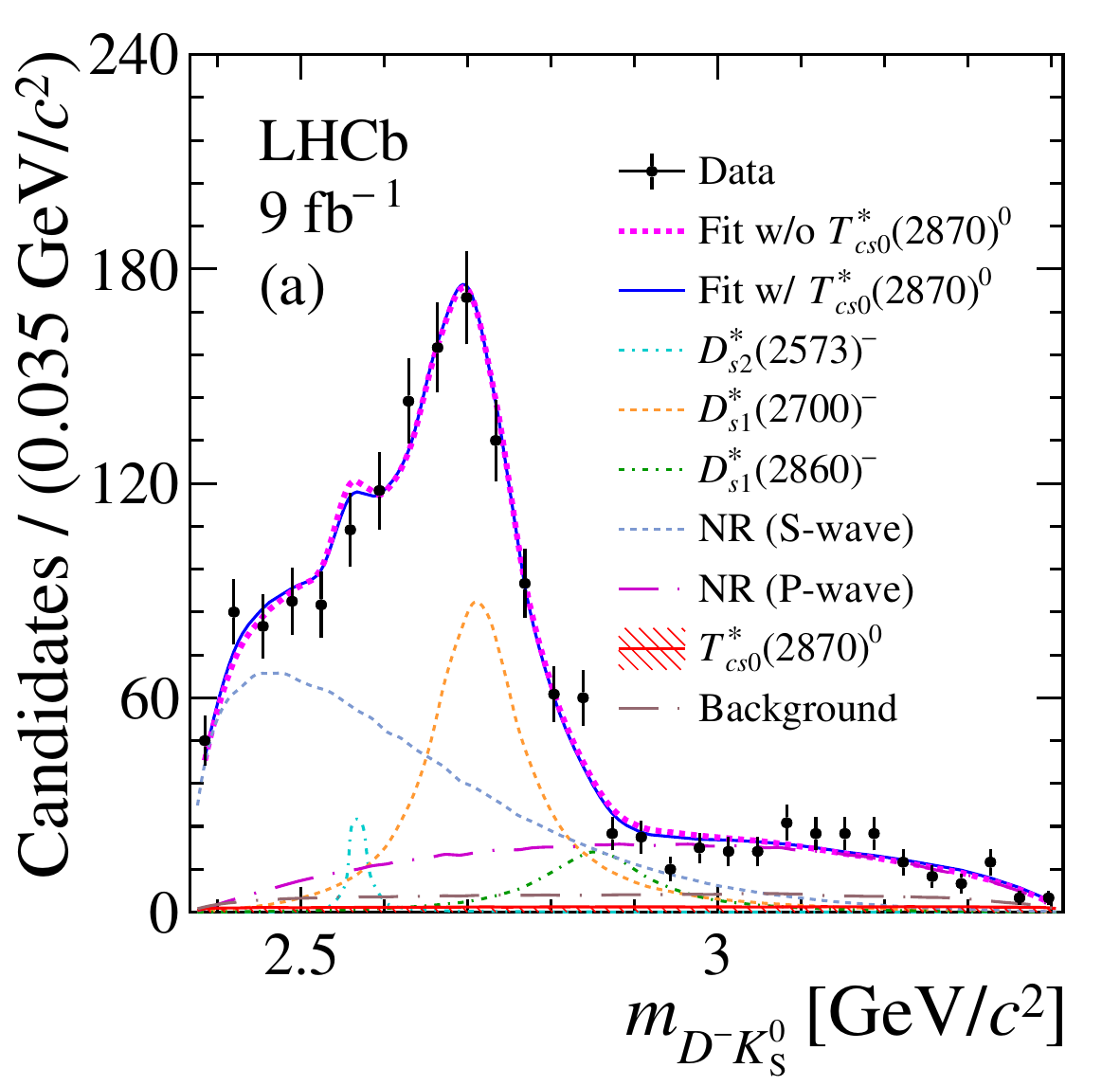}&
\includegraphics[width=0.32\columnwidth]{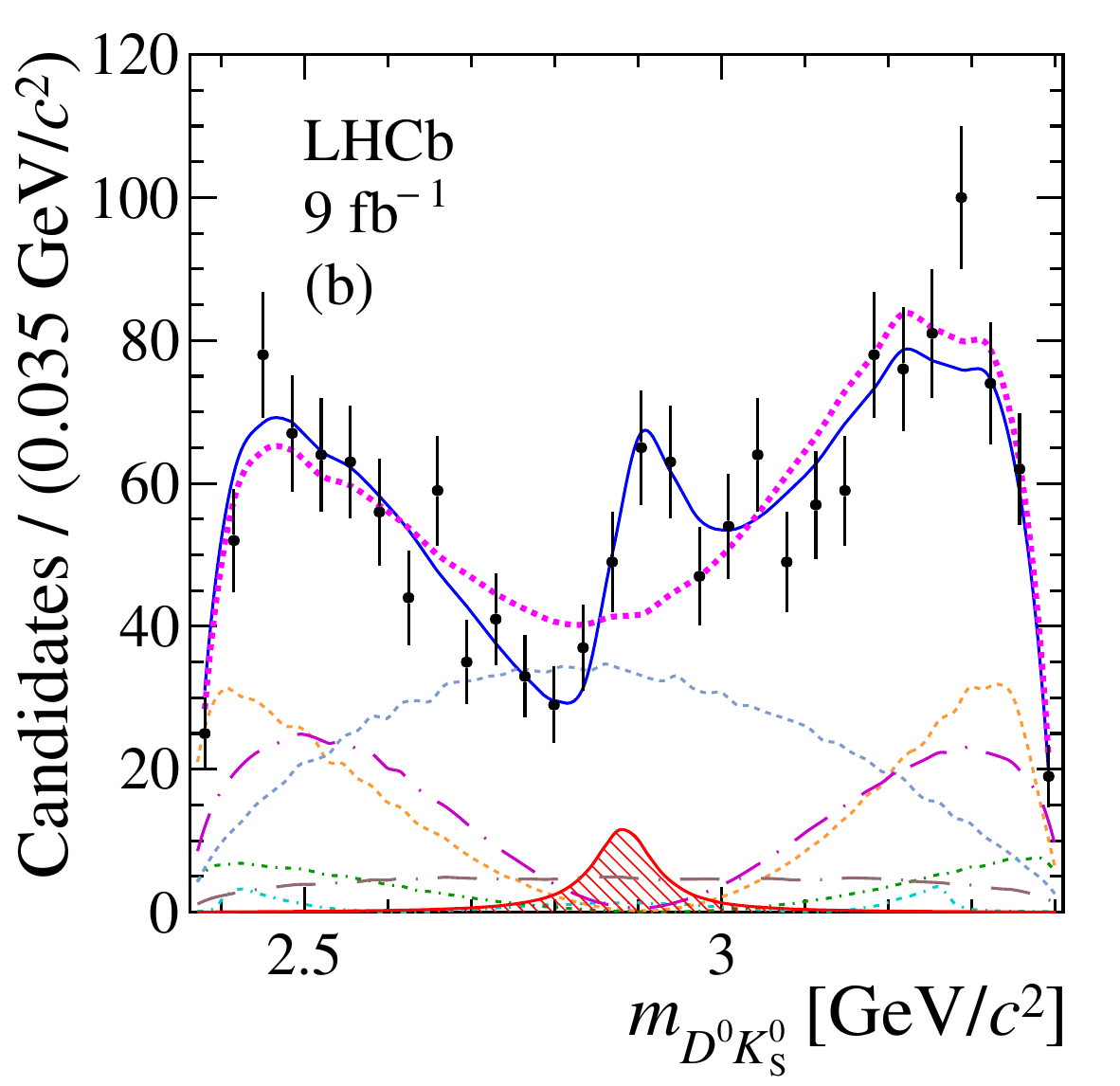}&
\includegraphics[width=0.32\columnwidth]{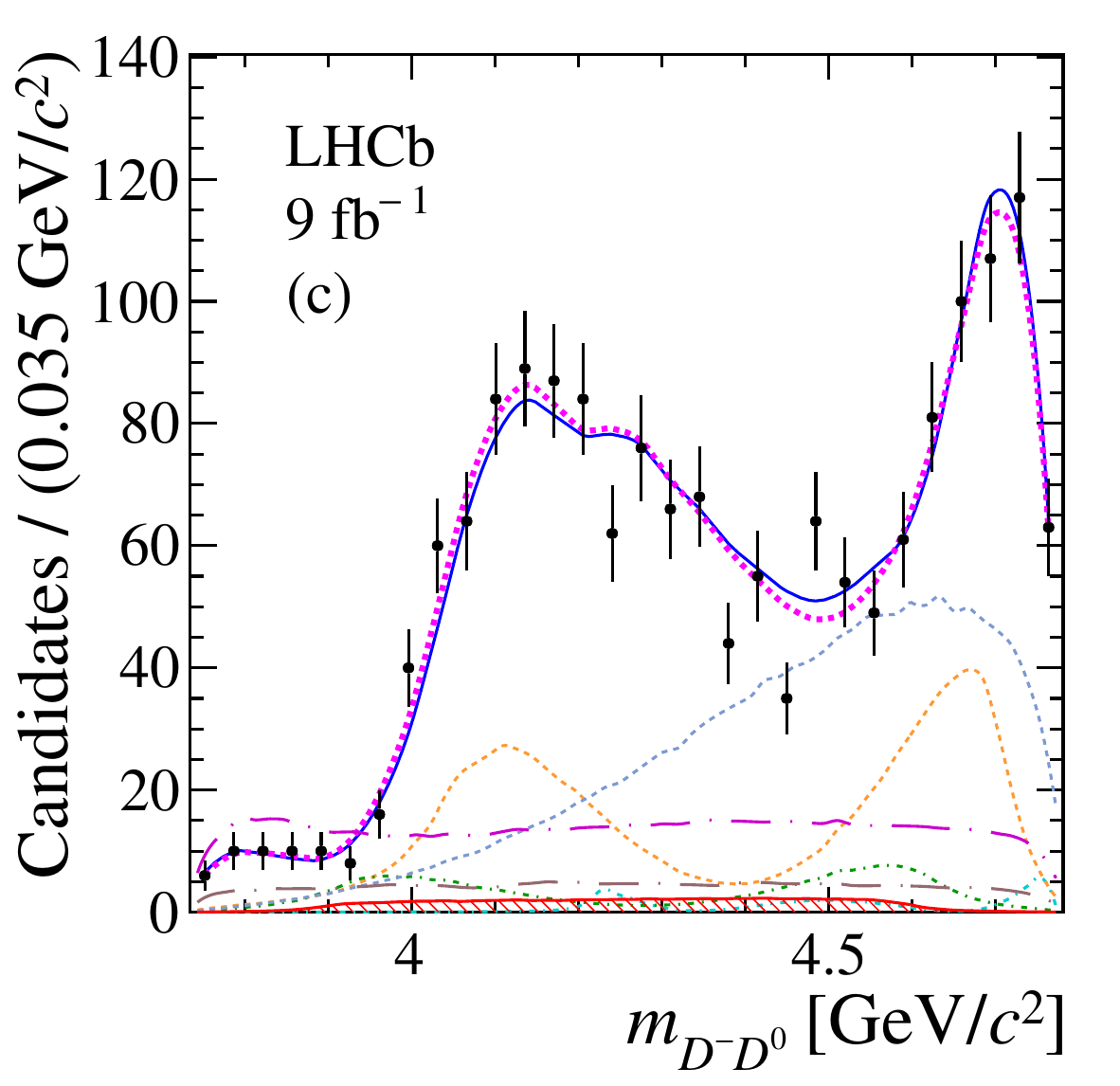}
\end{tabular}
\caption{Mass distributions of (a) $D^- K_\mathrm{S}^0$, (b) $D^0 K_\mathrm{S}^0$, and (c) $D^- D^0$, together with the fit projections with (thick blue) or without (dashed magenta) the $T_{cs0}^{*}(2870)^0$ state.
The subcomponents correspond to the fit including the $T_{cs0}^{*}(2870)^0$ structure~\cite{LHCb:2024xyx}.
}
\label{fig:Tcs2870}
\end{figure}

A recent amplitude analysis performed by the LHCb Collaboration on the $B^{-}\rightarrow D^{-}D^0\bar{K}^0$ decay, where the $\bar{K}^0$ meson is reconstructed via the $K^0_{\rm S}$ mass eigenstate, was designed to search for the $T^{*0}_{cs}$ states~\cite{LHCb:2024xyx}. The analysis uses proton--proton collision data corresponding to an integrated luminosity of $9~\mathrm{fb^{-1}}$ collected at C.M.\ energies of 7, 8, and 13~TeV. The baseline amplitude model, consisting solely of conventional $D_{sJ}^{*-}$ resonances and nonresonant contributions in the $D^-K_{\rm S}^0$ channel, provides an adequate description of the $M(D^-D^0)$ distribution, but fails to reproduce the $M(D^0K^0_{\rm S})$ spectrum, particularly in the region around 2.9~\gev. Modifying the $D_{sJ}^{*-}$ model assumptions or introducing charmonium-like tetraquark candidates decaying to the $D^-D^0$ system does not improve the fit quality.
A better fit is obtained when including an additional resonant contribution in the $D^0K_{\rm S}^0$ channel, modelled with a relativistic BW function with floating mass and width, as shown in Fig.~\ref{fig:Tcs2870}. The preferred solution corresponds to a spin-parity assignment of $J^P=0^{+}$, yielding the fitted parameters
\begin{equation}
 M = (2883 \pm 11 \pm 8)~\mbox{\mev},\qquad \Gamma = (87^{+22}_{-47} \pm 17)~\mathrm{MeV},\nonumber
\end{equation}
which are consistent with those of the $T^{*0}_{cs0}(2870)$ state previously observed in the $D^{+}K^{-}$ invariant-mass distribution of the $B^{-}\rightarrow D^{-}D^{+}K^{-}$ decay.
To test the possible presence of a $J^P = 1^-$ companion state, both $T^{*0}_{cs0}$ and $T^{*0}_{cs1}$ contributions were included in the model. However, no significant improvement in the fit quality is observed compared to the model containing only the $0^+$ component.
To test the isospin symmetry between the 
$T^{*0}_{cs} \rightarrow D^{+}K^{-}$ and 
$T^{*0}_{cs} \rightarrow D^{0}\bar{K}^{0}$ decay channels, both the 
$T^{*0}_{cs0}$ and $T^{*0}_{cs1}$ states are included in the amplitude model of the 
$B^{-} \rightarrow D^{-}D^{0}\bar{K}^{0}$ decay.
The ratio of the partial decay widths is defined as
\begin{equation}
R_\text{I}\!\left(T^{*0}_{cs}\right)
=
\frac{
\Gamma\!\left(B^{-}\rightarrow D^{-}T^{*0}_{cs},\,T^{*0}_{cs}\rightarrow D^{+}K^{-}\right)
}{
\Gamma\!\left(B^{-}\rightarrow D^{-}T^{*0}_{cs},\,T^{*0}_{cs}\rightarrow D^{0}\bar{K}^{0}\right)
}.
\label{eq:RI_definition}
\end{equation}
For the $T^{*0}_{cs0}$ state, the measured value is compatible with the expectation of isospin invariance within the current precision.
In contrast, the $R_\text{I}$ for the $T^{*0}_{cs1}$
and the resulting double ratio,
$
\frac{R_\text{I}\left(T^{*0}_{cs1}\right)}{R_\text{I}\left(T^{*0}_{cs0}\right)},
$
is significantly smaller than unity, as summarized in Table~\ref{tab:Tcs2870_isospin}. 
This deviation from unity suggests substantial isospin violation in the decays of the 
$T^{*0}_{cs1}$ state, consistent with expectations from the kinematic-origin interpretation rather than a compact multiquark or molecular assignment.

\begin{table}[t]
\centering
\caption{
Measurements of the relative rates of decays into the $D^0 \bar{ K}^{0} $ and $D^+ K^-$ final states for $T_{cs0}^*(2870)^0$ and $T_{cs1}^*(2900)^0$ states, $R_\text{I}[T_{cs0}^*(2870)^0]$ and $R_\text{I}[T_{cs1}^*(2900)^0]$, and the double ratio $R_\text{I}\left(T^{*0}_{cs1}\right)$/$R_\text{I}\left(T^{*0}_{cs0}\right)$~\cite{LHCb:2024xyx}.
The first, second, and third uncertainties are statistical, systematic, and due to the external inputs of $B^-\rightarrow D^- D^+ K^-$ and $B^- \rightarrow D^- D^0 K_\mathrm{S}^0$ branching fractions, respectively.} 
\label{tab:Tcs2870_isospin}
\begin{tabular}{lc}
\hline \hline
Observable &Result \\
\hline
$R_\text{I}[T_{cs0}^*(2870)^0]$ &$3.3\phantom{0}\phantom{0}\pm1.1\phantom{0}\phantom{0}\pm1.1\phantom{0}\phantom{0}\pm1.1\phantom{0}\phantom{0}$ \\
$R_\text{I}[T_{cs1}^*(2900)^0]$ &$0.15\phantom{0}\pm0.15\phantom{0}\pm0.05\phantom{0}\pm0.05\phantom{0}$ \\
$R_\text{I}\left(T^{*0}_{cs1}\right)$/$R_\text{I}\left(T^{*0}_{cs0}\right)$ &$0.044\pm0.035\pm0.020\phantom{0000000}$ \\
\hline\hline
\end{tabular}
\end{table}

\subsubsection{Observation of $T^*_{c\bar s0}(2900)^0$ and $T^*_{c\bar s0}(2900)^{++}$}

Motivated by the discoveries of the $cs\bar{u}\bar{d}$ states, theoretical predictions were subsequently made for a doubly-charged partner state with the quark composition $[c\bar{s}u\bar{d}]$ and its isospin counterpart $[c\bar{s}\bar{u}d]$~\cite{He:2020jna,Lu:2020qmp,Burns:2020xne,Agaev:2021knl,Agaev:2021jsz,Azizi:2021aib}.
A dedicated search for such candidates was performed by the LHCb Collaboration using proton–proton collision data corresponding to a total integrated luminosity of $9\,\mathrm{fb}^{-1}$, collected at C.M.\ energies of 7, 8, and 13~TeV. The analysis 
was based on the decay modes $B^0 \to \bar{D}^0 D_s^+\pi^-$ and 
$B^+ \to D^-D_s^+\pi^+$, for which a combined amplitude analysis was carried out to explore the resonant structure across the Dalitz plot~\cite{LHCb:2022sfr}.

\begin{figure}[t!]
\centering
\begin{tabular}{ccc}
\includegraphics[width=0.38\linewidth]{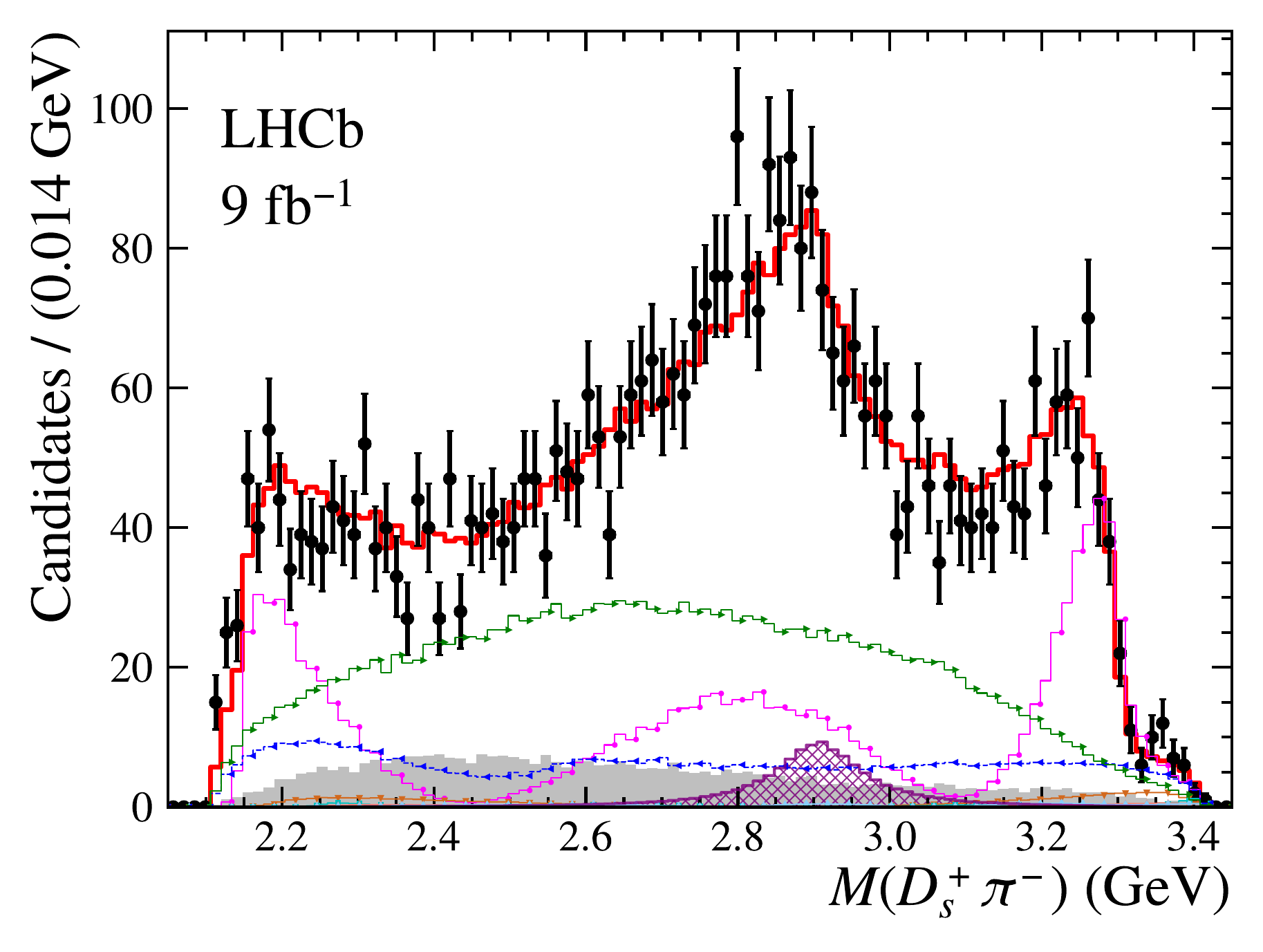}&
\includegraphics[width=0.38\linewidth]{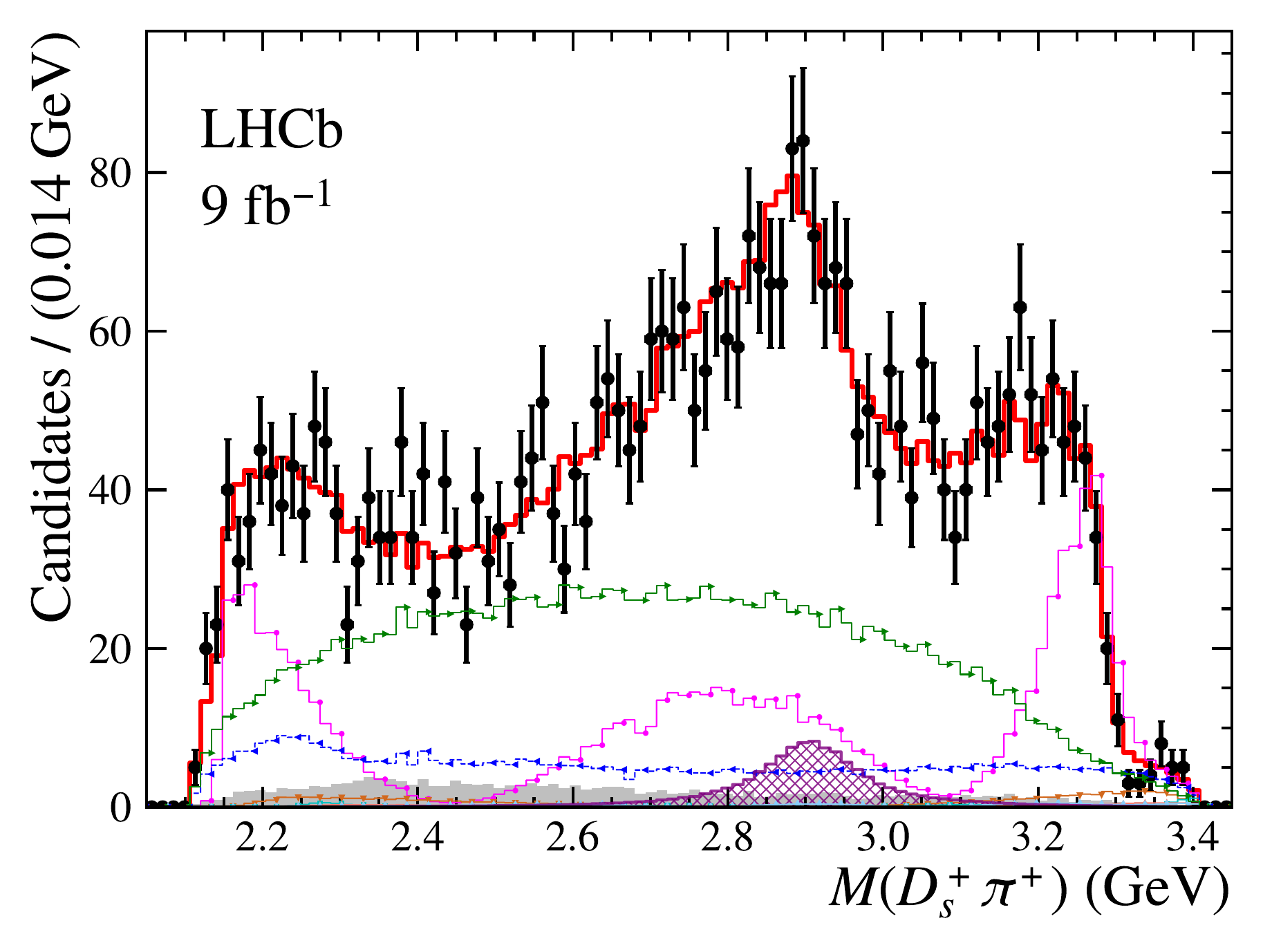}&
\includegraphics[width=0.18\linewidth]{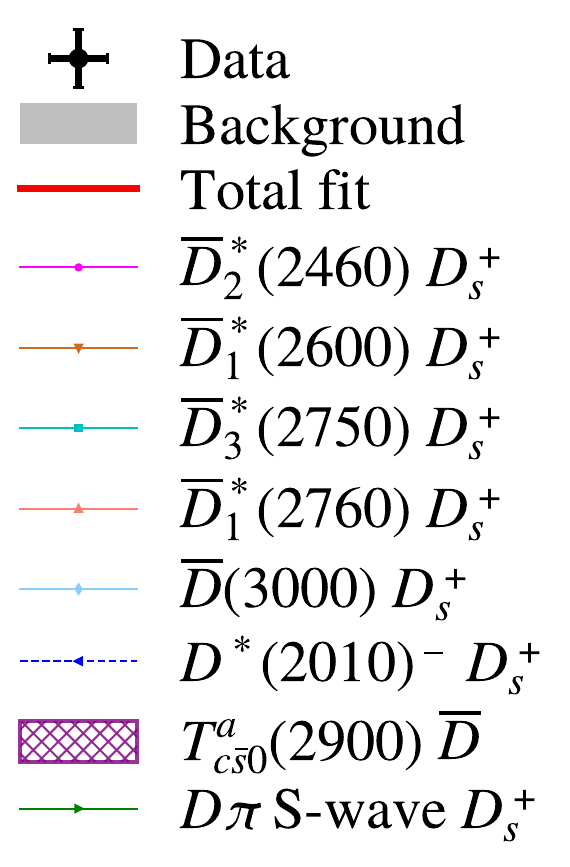}
\put(-320, 110){\large \bf (a)}
\put(-130, 110){\large \bf (b)}
\end{tabular}
\caption{Distributions of (a) $M(D_s^+\pi^-)$ of $B^0 \to \bar{D}^0 D_s^+\pi^-$ decays; and (b) $M(D^+_s\pi^+)$ for the $B^+ \to D^-D_s^+\pi^+$ sample. The data are overlaid with the fit results with the inclusion of the new $0^+$ $D_s^+\pi$ resonant states~\cite{LHCb:2022sfr}.}
 \label{fig:Tcs2900}
\end{figure}

Initial amplitude models included conventional contributions from excited $\bar{D}^\ast$ states decaying to $\bar{D}^0\pi^-$ and $D^-\pi^+$, described by relativistic BW functions, 
together with the $\bar{D}\pi$ $S$-wave parameterized using the quasi-model-independent approach~\cite{LHCb:2016lxy}. While these components provided a good description of most invariant mass projections, except for the structure near 2.9~\gev~in the $D_s^+\pi$ spectrum. Attempts to introduce additional $\bar{D}^\ast$ resonances with freely varying spin hypotheses did not resolve the discrepancy.
An improved description of the data was achieved only after introducing two 
new exotic states, denoted $T_{c\bar{s}0}^\ast(2900)^0$ and 
$T_{c\bar{s}0}^\ast(2900)^{++}$, decaying to $D_s^+\pi^-$ and $D_s^+\pi^+$, respectively. 
Under the assumption of isospin symmetry, their masses, widths, and complex couplings were constrained to be common parameters across the two decay channels. The $M(D^+_s\pi)$ distributions of the fit results are shown in Fig.~\ref{fig:Tcs2900}. The statistical significance of $T_{c\bar{s}0}^\ast(2900)$ contributions exceeds $9\sigma$, establishing these structures as compelling candidates for doubly charged and neutral open-charm tetraquarks.

The mass and width of the new states are measured to be 
$(2.909 \pm 0.010)$~\gev~and 
$(0.134 \pm 0.019)$~GeV, respectively. Their quantum numbers are determined to be $J^P = 0^+$. An alternative amplitude model in which the 
$T_{c\bar{s}0}^\ast(2900)^0$ and 
$T_{c\bar{s}0}^\ast(2900)^{++}$ states are allowed to have independent resonance parameters yields results consistent with the isospin-symmetric hypothesis.
These structures represent a new class of open-charm tetraquark candidates containing a $c$ quark and an $\bar{s}$ antiquark. The measured mass of the 
$T_{c\bar{s}0}^\ast(2900)$ is compatible with that of the previously reported $T^*_{cs0}(2870)^0$, also assigned $J^P=0^+$ and observed in the $D^+K^-$ final state~\cite{LHCb:2020bls,LHCb:2020pxc}. However, their widths and flavor compositions differ, indicating that they may correspond to distinct configurations within the spectrum of open-charm exotic hadrons.

\subsubsection{Properties of $D_{s0}^*(2317)^+$ and $D_{s1}(2460)^+$}

In the standard quark model scheme, open-charm mesons are conventionally assigned the name $D$ and are usually identified by their charge or, more rarely, explicitly by their light-quark content,
\be
D^0\equiv D_u\sim c\bar{u},\quad 
D^+\equiv D_d\sim c\bar{d},\quad
\bar{D}^0\equiv \bar{D}_u\sim u\bar{c},\quad 
D^-\equiv \bar{D}_d\sim d\bar{c}.
\ee
In this family, the ground state and its first excitation are the pseudoscalar and vector mesons $D$ and $D^*$, respectively, where in both cases the orbital angular momentum in the quark--antiquark pair is $l=0$ and the total spin of the pair is $S=0$ (for $D$) or $S=1$ (for $D^*$). 
Next in the spectrum of excitations of the $c\bar{q}$ (with $q=u$, $d$) mesons come positive-parity states $D_J$, with $J=0,~1,~2$, which are sums of the angular momentum $l=1$ and the spin of the quark–antiquark pair $S=0$ or $S=1$. As a result, one has a quartet of $P$-level states: one scalar $D_0^*$, two axial vectors $D_1$, and one tensor $D_2^*$, where we resort to the standard PDG notations \cite{ParticleDataGroup:2024cfk}. Since the charmed quark is much heavier than the $u$ and $d$ quarks, HQET applies to these mesons. In particular, it is easy to see that, in the gedanken limit of an infinitely heavy charmed quark, $m_c\to\infty$, the quartet above naturally splits into two degenerate doublets corresponding to $j_l=1/2$ and $j_l=3/2$ (therefore, often denoted as $D_{1/2}$ and $D_{3/2}$), with $s_q=\frac12$ and $\vej_l=\vel+\ves_q$ for the spin and total momentum of the light degrees of freedom, respectively. Once the charmed quark mass decreases from infinity to its physical value, the masses of the $D_J$ mesons acquire $1/m_c$ corrections that stem from spin-dependent interactions and, therefore, the four $D_J$'s split from each other. Since quark--antiquark mesons composed of different quarks are not eigenstates of the $C$-parity operator, the two resulting physical $D_1$ mesons can be understood as certain dynamical (that is, not constrained by any symmetry) mixtures of the $^1P_1$ and $^3P_1$ $c\bar{q}$ terms.
The corresponding mixing angle can in principle be evaluated in the framework of a particular quark model once its parameters are fixed to the data --- there exists a vast amount of papers on the subject. In particular, one of the questions broadly discussed in the literature is the ordering of the physical heavy-light meson levels since different models predict it differently. 

One of the most distinct predictions of HQET for the $D_J$ mesons is the hierarchy of their total widths. The states belonging to the $D_{1/2}$ doublet can decay to the $D^{(*)}\pi$ final states in the $S$ wave, so they must be broad. On the contrary, the states belonging to the $D_{3/2}$ doublet decay to these final states in the $D$ wave, so their total widths are expected to be relatively small. Indeed, the outlined pattern takes place in nature since the $D_0^*(2300)$ and $D_1(2430)$ mesons possess the widths of the order of 200\dash300~MeV while the widths of the $D_1(2420)$ and $D_2^*(2460)$ mesons are as small as only 30\dash50~MeV \cite{ParticleDataGroup:2024cfk}. 
At first glance, the simple (and somewhat na{\"i}ve --- see below) picture just drawn equally applies to the generic positive-parity $D_{sJ}$ mesons with $J=0,~1,~2$ and the quark content $c\bar{s}$ given that the required mass hierarchy, $m_s\ll m_c$, does hold for them as well. However, the situation with these two charmed--strange mesons appears to be more tricky than that with their nonstrange cousins. 
Indeed, already the first observation of the $\Dszero^+$ made by BaBar in 2003 in the final state $D_s^+\pi^0$ \cite{BaBar:2003oey} attracted a lot of attention to this meson since the quark model had predicted its mass at around 2.48~\gev~\cite{Godfrey:1985xj}, which is about 160~\mev~higher than the experimentally observed value. To resolve the puzzle one needs to take into account the following arguments. 
The phenomenon of spontaneous breaking of chiral symmetry in the vacuum of QCD entails the existence of eight light pseudo-Goldstone bosons in the spectrum of theory: three pions, four kaons, and one $\eta$-meson. All above states are supposed to become massless in the strict chiral limit of $m_u=m_d=m_s=0$. However, while the masses of the $u$ and $d$ quarks are indeed tiny on the typical hadronic scale, the mass of the $s$ quark is not. As a result, in the physical world, the kaons appear to be noticeably heavier than the pions and the $D^{(*)}K$ thresholds lie higher than the $D^{(*)}\pi$ thresholds by approximately $m_K-m_\pi\simeq 350$~\mev. The latter observation implies that the physical $\Dszero$ and $D_{s1}(2460)$ mesons reside not far (within just $\simeq 30$\dash40~\mev) from the $DK$ and $D^*K$ thresholds, respectively. Therefore, a strong influence of the corresponding two-hadron channels on the nature of these two $D_{sJ}$ mesons has to be anticipated and taken into account --- see, for example, Ref.~\cite{Guo:2023wkv}
and references therein for a discussion of the $\Dszero$ and $D_{s1}(2460)$ mesons as hadronic molecules rather than generic $c\bar{s}$ states. In particular, a strong coupling to the $DK$ channel allows one to understand the position of the $\Dszero$ pole on the complex energy plane \cite{vanBeveren:2003kd}. No surprise that the internal structure of the $\Dszero^+$ and $D_{s1}(2460)^+$ mesons has been under active debate since their discovery in 2003~\cite{Godfrey:2003kg, Colangelo:2003vg,Bardeen:2003kt,Fayyazuddin:2003aa,Ishida:2003gu,Azimov:2004xk,Colangelo:2005hv,Close:2005se,Liu:2006jx,Wang:2006zw}. 

The predominantly molecular nature of the $\Dszero^+$ and $D_{s1}(2460)^+$ resonances has received a strong and consistent theoretical support over the last decade~\cite{Albaladejo:2016lbb, Du:2017zvv, Guo:2017jvc}. A key step in this direction was provided in Ref.~\cite{Albaladejo:2016lbb}, where $D^{(*)}$\dash$\phi$ (with $\phi$ for Goldstone bosons) scattering in the $J^P=0^+$ channel and in the $(S,I)=(0,1/2)$ sector was investigated. The analysis revealed compelling evidence for a two-pole structure in the $D_0^\ast(2300)$ region, which was subsequently corroborated by the excellent agreement between parameter-free predictions of low-lying finite-volume energy levels obtained from next-to-leading-order (NLO) unitarized heavy-meson chiral perturbation theory (HMChPT) amplitudes~\cite{Guo:2008gp} and LQCD determinations~\cite{Moir:2016srx}. Additional LQCD support was reported in Ref.~\cite{Albaladejo:2018mhb}, where the same unitarized NLO HMChPT amplitudes were shown to successfully reproduce the $0^+$ and $1^+$ charmed–strange finite-volume spectra obtained in Ref.~\cite{Bali:2017pdv}. The latter lattice study explicitly incorporated the nearby $DK$ and $D^\ast K$ thresholds via four-quark interpolating operators, providing further evidence for the molecular interpretation of the $\Dszero$ and $D_{s1}(2460)$ states. A detailed analysis of the light-flavor SU(3) structure of the $D$\dash$\phi$ interaction clarified the dynamical origin of the two-pole structure~\cite{Albaladejo:2016lbb}, identifying the lower pole as the SU(3) partner of the $\Dszero^+$. This naturally implies a hadronic molecular interpretation for the corresponding non-strange state as well. Analogous patterns were later observed in the $J^P=1^+$ axial channel and in the bottom sector, suggesting that such structures are a generic consequence of the $D^{(*)}$\dash$\phi$ chiral dynamics. These findings offer a natural resolution to a long-standing puzzle in charm spectroscopy, namely the unexpectedly high masses of the non-strange $D_0^\ast(2300)$ and $D_1(2430)$ mesons relative to their strange counterparts. Moreover, the molecular picture provides a coherent explanation for two additional puzzles: the unusually low masses of the $\Dszero$ and $D_{s1}(2460)$ compared to quark-model expectations, and the striking near equality (within about 2 MeV) between the $D_{s1}(2460)$\dash$\Dszero$ mass splitting and the hyperfine splitting between the ground-state vector $D^{*+}$ and pseudoscalar $D^+$ mesons. Further support for this molecular framework was obtained in Refs.~\cite{Du:2017zvv, Du:2019oki}, where the chiral amplitudes constrained in Ref.~\cite{Albaladejo:2016lbb} were shown to be fully consistent with high-quality experimental data on the $B^- \to D^+\pi^-\pi^-$ and $B_s^0 \to \bar{D}^0 K^-\pi^+$ decay channels measured by the LHCb Collaboration~\cite{LHCb:2016lxy, LHCb:2014ioa}. Taken together, these results point toward a paradigm shift in heavy–light meson spectroscopy, challenging the traditional constituent $q\bar q$ interpretation and highlighting the central role of hadronic molecular components and coupled-channel chiral dynamics~\cite{Du:2017zvv}. Overall, these developments strongly support the existence of a two-pole structure in the $D_0^\ast(2300)$ region and an SU(3)-driven pattern in the $0^+$ and $1^+$ heavy–light sectors~\cite{Albaladejo:2016lbb}. They establish a coherent framework in which coupled-channel dynamics and chiral symmetry play a central role in shaping the isoscalar scalar and axial heavy–light meson spectrum~\cite{Du:2017zvv}. Subsequent theoretical studies have further explored the implications of chiral symmetry for scalar charmed mesons~\cite{Du:2019oki, Du:2020pui}, consolidating this emerging molecular-driven perspective. In addition, Ref.~\cite{Asokan:2022usm} discussed whether the two-pole structure observed for the $D_0^\ast(2300)$ can be understood from lattice studies of $D\pi$ scattering at different pion masses~\cite{Moir:2016srx}, where only one pole was reported in the $D_0^\ast$ channel, and Ref.~\cite{Du:2025beb} examined the role played by the $D_0^\ast(2100)$ in $B\to D\pi \ell \nu$ semileptonic decays. In this context, it is also worth mentioning that Ref.~\cite{Flynn:2007ki} showed that the LQCD data for the $D\to K \ell \nu$ semileptonic decay were compatible with the $\Dszero$ state, while the LQCD scalar form factor of the $D\to \pi \ell \nu$ transition was used in that work to predict the existence of an $I=1/2$ $S$-wave resonance at $(2.2 \pm 0.1)$ GeV/$c^2$, compatible with the lower state found in unitarized chiral studies.

The total widths of the $\Dszero$ and $D_{s1}(2460)$ are small (appear at a 1~MeV level or less) since they mainly decay through the isospin-breaking hadronic channels $\Dszero^+\to D_s^+\pi^0$ and $D_{s1}(2460)^+\to D_s^{*+}\pi^0$, respectively. Note also that different assignments for these mesons result in different predictions for their decay patterns --- for instance, for the ratios of their partial radiative and (isospin breaking) one-pion decay widths ${\cal B}(\Dszero \to D_s^*\gamma)/{\cal B}(\Dszero \to D_s\pi )$ and ${\cal B}(D_{s1}(2460)\to D_s\gamma)/{\cal B}(D_{s1}(2460)\to D_s^*\pi)$. For evaluation of these ratios in the molecular model for both $\Dszero$ and $D_{s1}(2460)$ mesons see Refs.~\cite{Lutz:2007sk,Cleven:2014oka}
and Refs.~\cite{Fu:2021wde,Su:2025aiz} for recent updates. In Ref.~\cite{Fu:2025lfo}, the three-body radiative decays $D_{s1}(2460)\to \gamma DK$ are argued to be able to provide additional insight into the nature of the $D_{s1}(2460)$ meson and allow one to assess the admixture of the $D^*K$ molecule in its wave function. The above decays have also been comprehensively studied in the chiral doublet model under the assumption of the $\Dszero$ and $D_{s1}(2460)$ being chiral partners of the pseudoscalar $D_s$ and vector $D_s^*$ mesons, respectively \cite{Bardeen:2003kt}.
A reanalysis of the existing experimental data on the radiative decays of the $D_{s1}(2536)$ --- a would-be $J=1$ quark-model sibling of the $D_{s1}(2460)$ --- is performed in Ref.~\cite{Bondar:2025gsh}. In Ref.~\cite{Wang:2006mf}, the radiative decays $\Dszero\to D_s^*\gamma$ and $D_{s1}(2460)\to D_s\gamma$ are studied by employing the QCD sum rule technique. In Refs.~\cite{Bondar:2025qzg,Bondar:2025jop,Obraztsov:2026jge}, the widths of the radiative decays $D_{s1}(2460)\to D_s^{(*)}\gamma$ and $D_{s1}(2536)\to D_s^{(*)}\gamma$ are calculated, with relativistic corrections taken into account, assuming all involved charm-strange mesons to be conventional quark--antiquark states. In Ref.~\cite{Zhang:2024usz}, the decays $D_{s1}(2460)\to D_s^{(*)}\gamma$ are studied including both $c\bar s$ and two-hadron components. Interestingly, as argued in Refs.~\cite{Tang:2023yls,Roca:2025lij,FKinpreparation}, the shape of the $\pi^+\pi^-$ invariant mass distribution in the di-pion decay $D_{s1}\to D_s\pi\pi$ may shed light on the nature of the decaying meson since a double-hump structure is expected for the $D_{s1}$ as a $D^*K$ molecule versus a single-hump structure for a generic $c\bar{s}$ meson. Electromagnetic properties of the $D_{s1}(2460)$ and $D_{s1}(2536)$ mesons as hadronic molecules are studied in Ref.~\cite{Ozdem:2025olj} by employing the QCD sum rules technique.

\begin{figure}[t!]	
\centering
\includegraphics[width=0.45\textwidth]{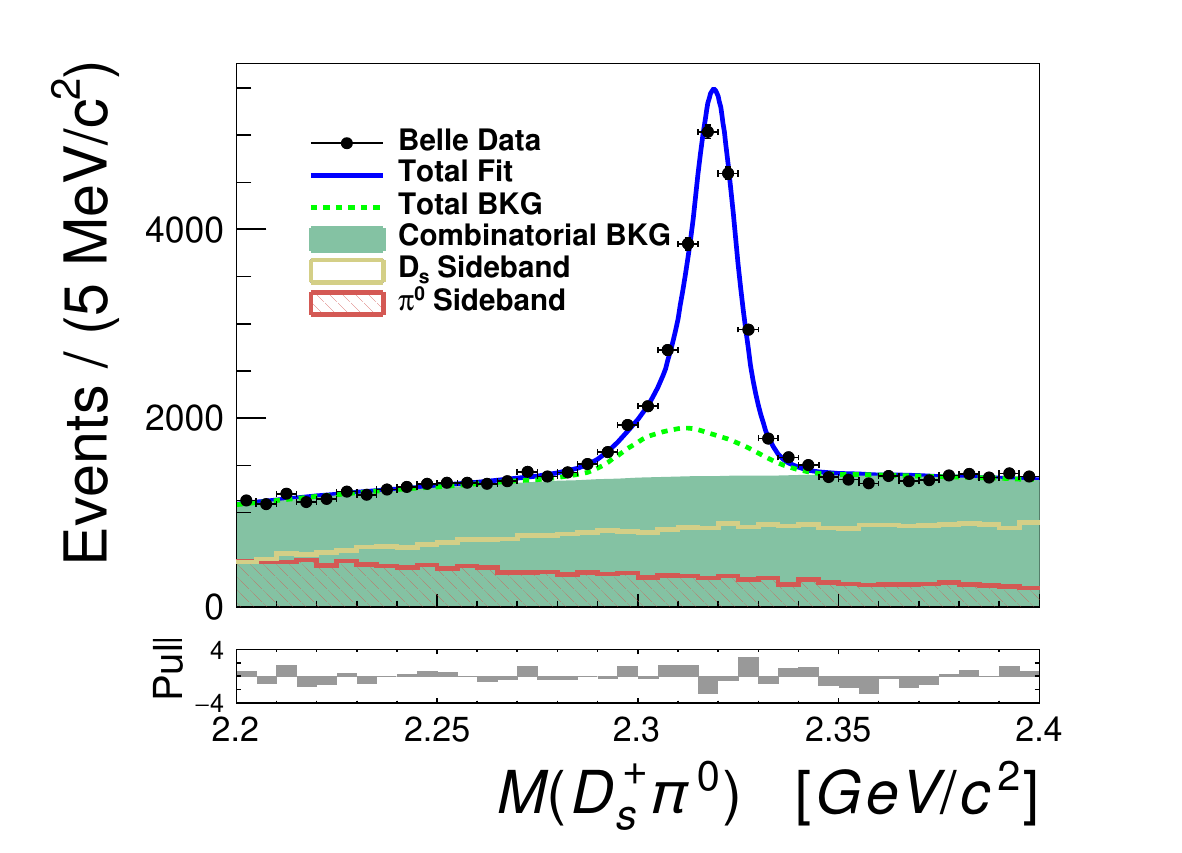} 
\includegraphics[width=0.45\textwidth]{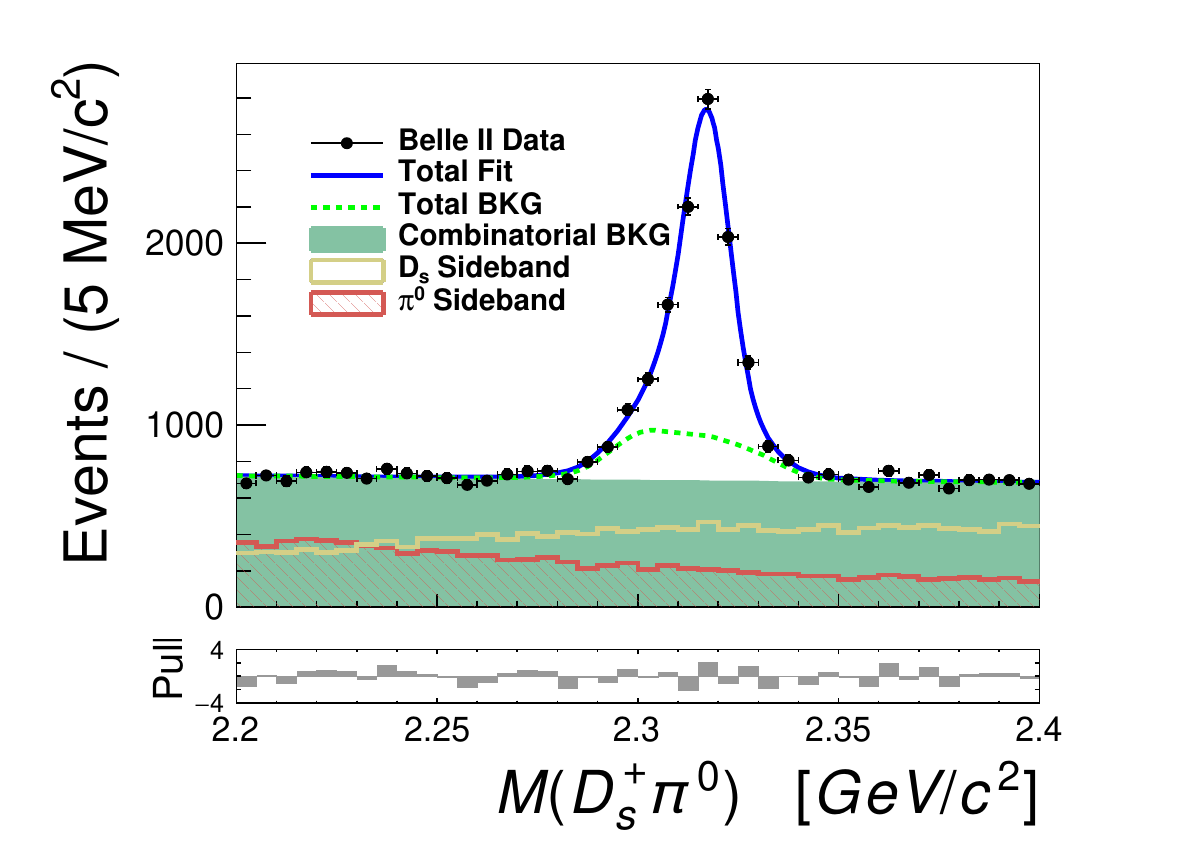} 
\includegraphics[width=0.45\textwidth]{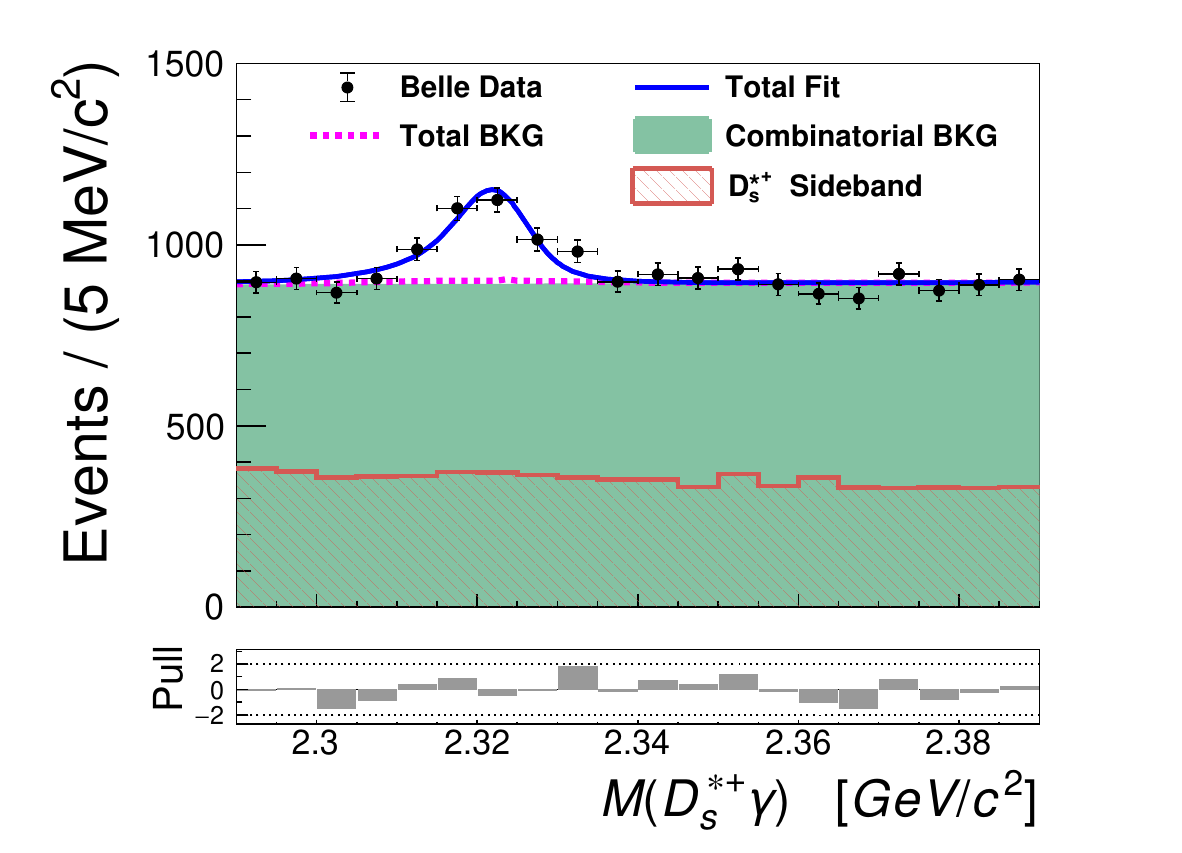} 
\includegraphics[width=0.45\textwidth]{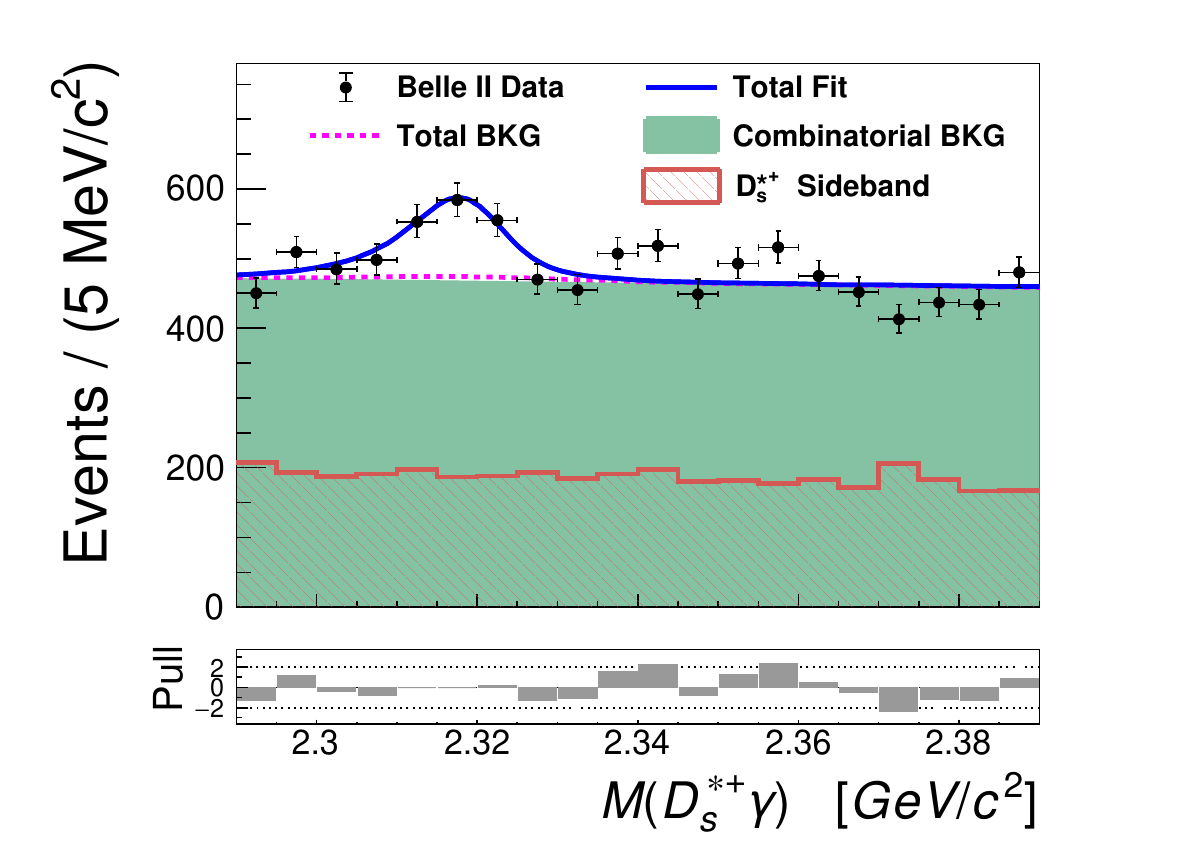} 
\caption{The invariant mass distributions of (top) $D_s^{+}\pi^0$ and (bottom) $D_s^{*+}\gamma$ from (left) Belle and (right) Belle II data samples~\cite{Belle-II:2025dzk}. 
The blue and violet curves present the best fit results and the fitted total
backgrounds, respectively. The filled green areas are the fitted combinatorial backgrounds. 
The histograms in red slashes represent the normalized $D_s^{*+}$ sidebands.}
\label{fig:2317radiative}
\end{figure}

The decay $\Dszero^+\to D_s^{*+}\gamma$ was searched for by CLEO, Belle, and BaBar, but no signals were found~\cite{BaBar:2003oey,CLEO:2003ggt,BaBar:2006eep}. Using large data samples of 980.4 fb$^{-1}$ and 427.9 fb$^{-1}$ at $\sqrt{s}$ around 10.6 GeV, Belle and Belle II updated this branching fraction ratio measurement in the continuum $e^+e^-\to c\bar c$ process~\cite{Belle-II:2025dzk}. Figure~\ref{fig:2317radiative} shows the invariant mass distributions of $D_s^+\pi^0$ and $D_s^{*+}\gamma$ from Belle and Belle II data samples. The signal yields for $\Dszero^+\to D_s^{*+}\gamma$ are $712\pm69$ and $387 \pm38$ events for Belle and Belle II, respectively, with a signal significance larger than 10.1$\sigma$.
The mentioned above ratio of the branching fractions that may play a key role in discriminating between various models for the $D_{s0}^*$ was measured to be~\cite{Belle-II:2025dzk}
\be
{\cal R}_{D_{s0}^*} = \frac{\BR(\Dszero^+\to D_s^{*+}\gamma)}{\BR(\Dszero^+\to D_s^+\pi^0)} = (7.14\pm0.70\pm0.23)\%.
\ee

\begin{figure}[t!]	
\centering
\includegraphics[width=0.7\textwidth]{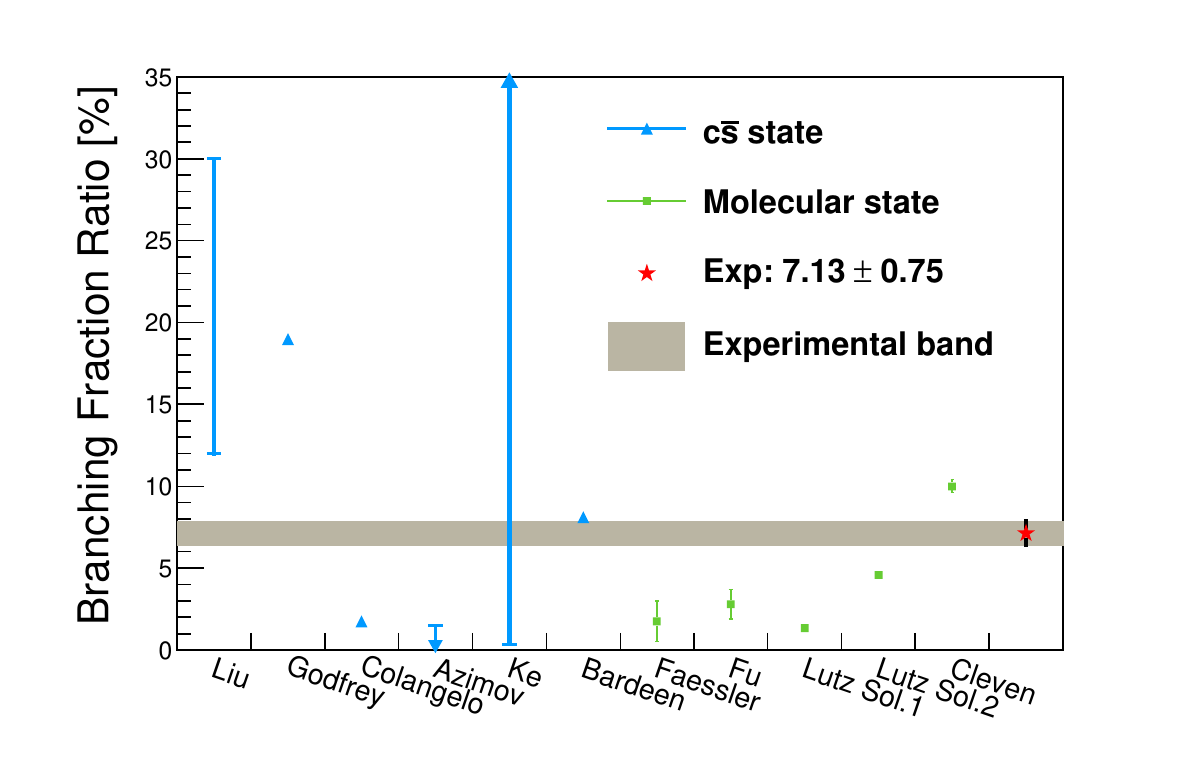}
\caption{Comparison between the measured ${\cal R} = \BR(\Dszero^+\to D_s^{*+}\gamma)/\BR(\Dszero^+\to D_s^+\pi^0)$ in Ref.~\cite{Belle-II:2025dzk} and different theoretical predictions. The cyan points show the $c\bar s$ state predictions from
Liu~\cite{Liu:2006jx}, 
Godfrey~\cite{Godfrey:2003kg}, 
Colangelo~\cite{Colangelo:2003vg}, 
Azimov~\cite{Azimov:2004xk},
Ke~\cite{Ke:2013zs}, and Bardeen~\cite{Bardeen:2003kt}. The green inverted triangles show the molecular state predictions from Faessler~\cite{Faessler:2007gv}, Fu~\cite{Fu:2021wde}, Lutz~\cite{Lutz:2007sk}, and Cleven~\cite{Cleven:2014oka}.
The prediction of Ke~\cite{Ke:2013zs}
is a lower limit. The uncertainty of the experimental measurement is the combination of statistical and systematic
uncertainties.
}
\label{fig:2317comparison}
\end{figure}

The comparison between this branching-fraction ratio and various theoretical predictions is shown in Fig.~\ref{fig:2317comparison}, adapted from a recent Belle paper \cite{Belle-II:2025dzk}. The measured ratio is generally larger than the predictions obtained for a predominantly molecular $\Dszero^+$ state~\cite{Faessler:2007gv,Fu:2021wde,Lutz:2007sk,Su:2025aiz}, while remaining below the expectation for a pure $c\bar s$ state in conventional quark models~\cite{Godfrey:2003kg}. Belle~(II) further argues that the predictions of Refs.~\cite{Ke:2013zs,Bardeen:2003kt}, which are also classified there as quark-model results, are consistent with the measurement under the assumption that the $\Dszero^+$ is a pure $c\bar s$ state. From the theoretical perspective, however, this conclusion should be interpreted with caution. Indeed, within the $c\bar s$ assignment for the $\Dszero^+$, Ref.~\cite{Ke:2013zs} predicts a radiative width $\Gamma(\Dszero\to D_s^{*+}\gamma)\approx (17.1\pm3.9)$~keV,
which exceeds the estimate for the total width of a conventional $c\bar s$ $\Dszero^+$ state, $\Gamma_{\rm tot}(\Dszero)\approx 7$~keV,
obtained in Ref.~\cite{Colangelo:2003vg}. Such a comparison suggests a significant tension between these predictions. Furthermore, the results of Ref.~\cite{Bardeen:2003kt} were obtained within a specific framework combining heavy-quark symmetry with a particular realization of chiral symmetry for heavy mesons. Consequently, they should not be regarded as  predictions of a generic quark model.

A possible interpretation of the Belle~(II) result in Ref.~\cite{Belle-II:2025dzk} is that the $\Dszero^+$ could be an admixture of pure $c\bar s$ and
molecular state, which was suggested in Refs.~\cite{Bicudo:2005de,MartinezTorres:2011pr,Mohler:2013rwa, Albaladejo:2018mhb}. Belle and Belle II are performing the measurement of the ratio of branching fractions for $D_{s1}(2460)^{+}\to D^{*+}_s\gamma$ and $D_{s1}(2460)^{+}\to D^{*+}_s\pi^0$ to help understand the nature of the $D_{s1}(2460)^{+}$.
The invariant mass distribution of $\pi^+\pi^-$ in $D_{s1}(2460)^{+}\to\pi^+\pi^-D^+_s$ is also being investigated to distinguish a pure $D^*K$ molecular state and a compact state model for the $D_{s1}(2460)^{+}$. In $B$ decays, using the recoil mass technique, the absolute branching fractions for $B^0\to \Dszero^+ D^-$ and $B^0\to D_{s1}(2460)^{+} D^-$ can be measured in the recoil mass spectrum of $D^-$. Further, the absolute branching fractions for $\Dszero^+$ and $D_{s1}(2460)^{+}$ decaying into different final states can be determined. With large data samples anticipated at Belle II in the near future, the above absolute branching fractions are expected to be achieved. 
Complementary precision studies of the $D_{sJ}$ mesons performed in the $e^+e^-$ annihilation in the ongoing experiment BESIII may potentially constitute an essential addition to its future physics program \cite{BESIII:2020nme}. The first successful steps in this direction have already been reported recently~\cite{BESIII:2023wsc}.

\begin{figure}[t!]
\centering
\begin{tabular}{cc}
\includegraphics[width=0.45\textwidth]{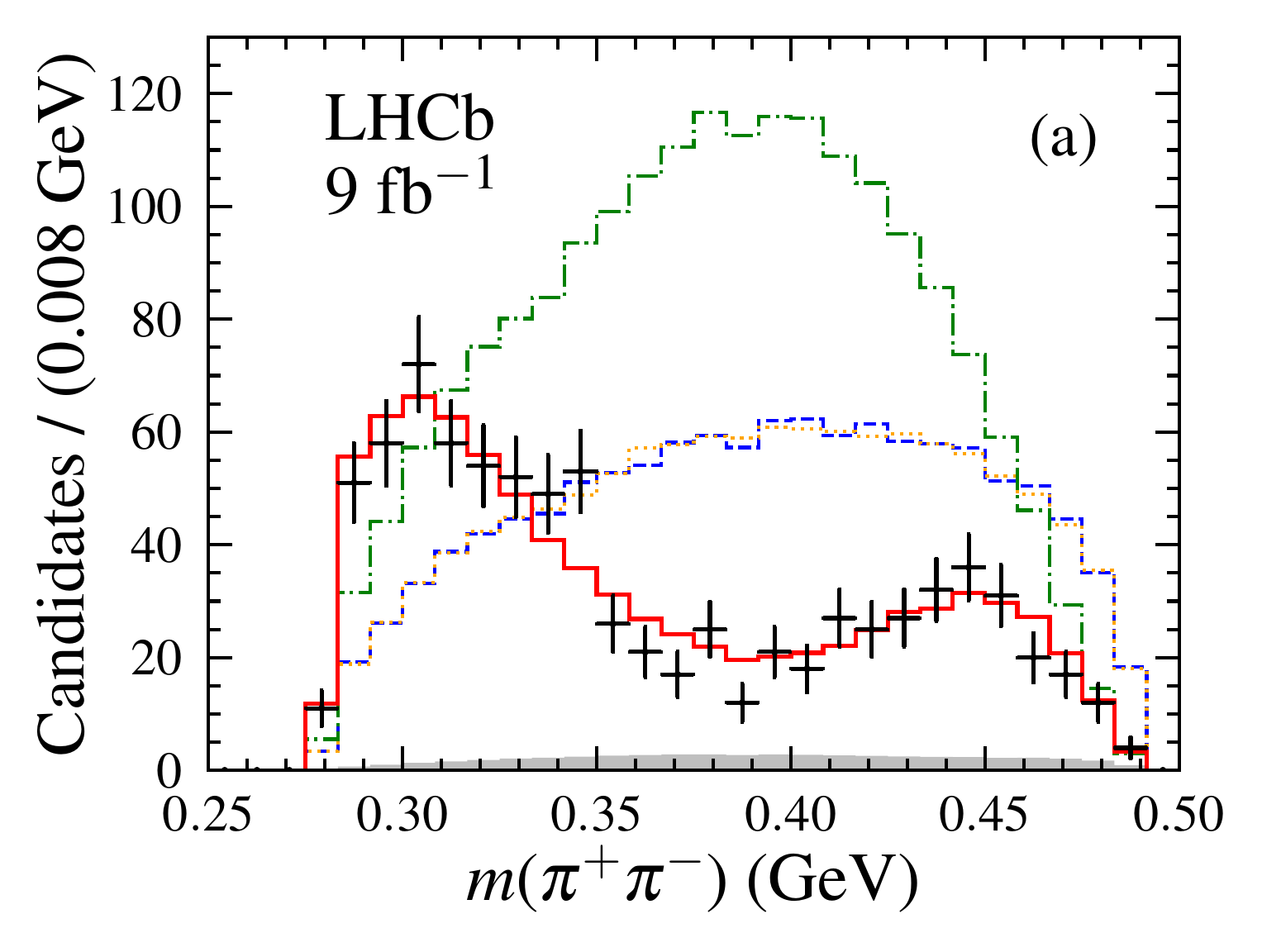}&
\includegraphics[width=0.45\textwidth]{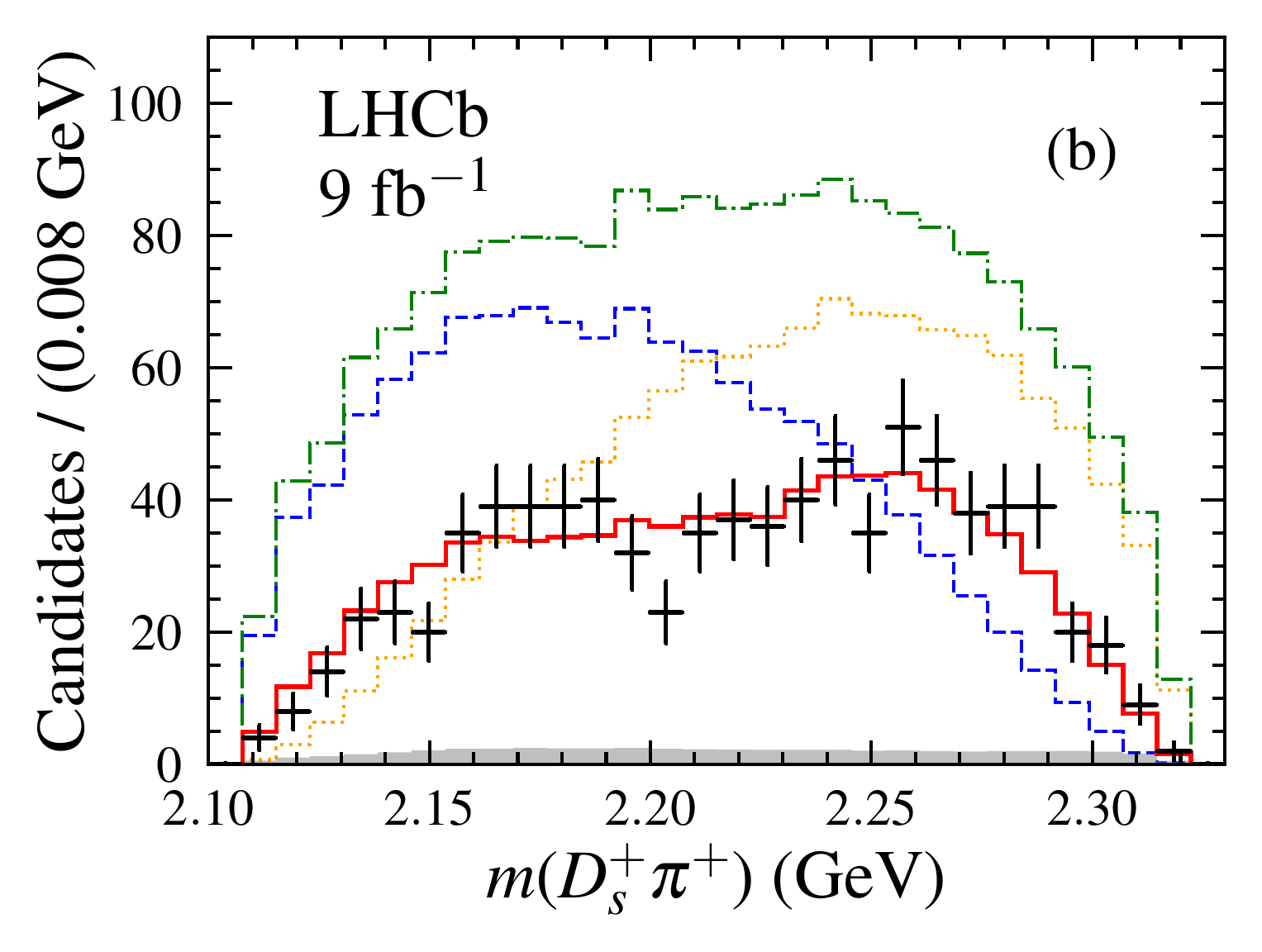}\\
\includegraphics[width=0.45\textwidth]{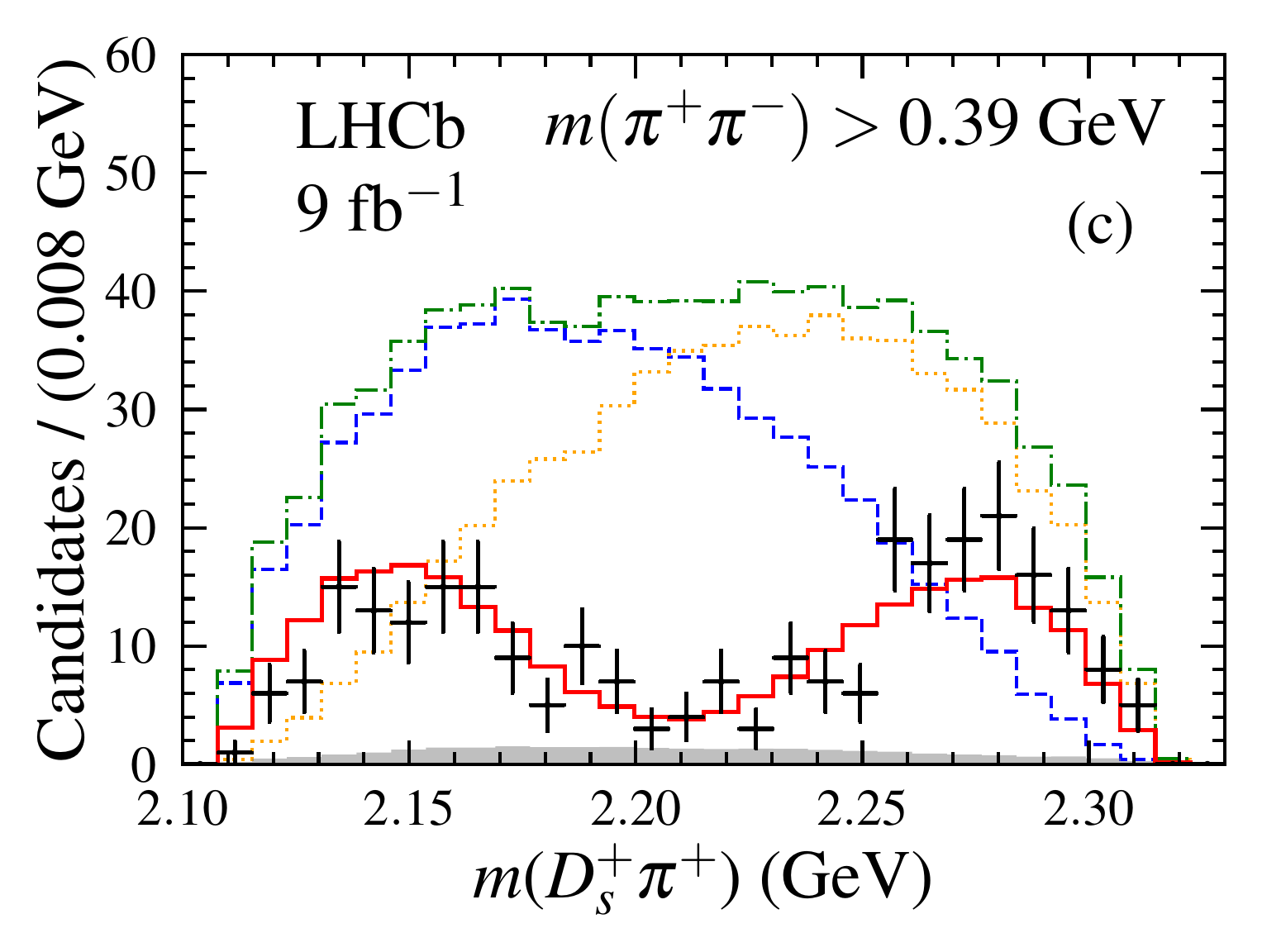}&
\raisebox{15mm}{\includegraphics[width=0.25\linewidth]{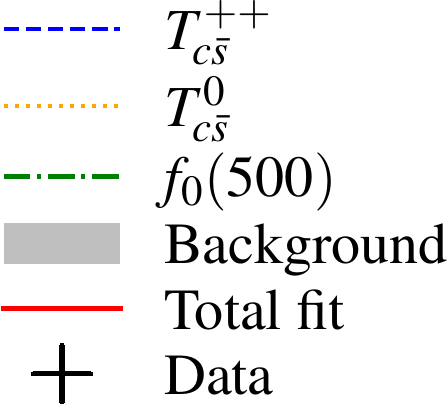}}
\end{tabular}
\caption{Comparison between data (black dots with error bars) and results of the fit with the $f_0(500)+K\text{-matrix}~T_{c\bar{s}}(0^+)$ model (red solid line). The distributions are for the three channels combined in (a)~$m(\pi^{+}\pi^{-})$, (b)~$m(D_{s}^{+}\pi^{+})$, and (c)~$m(D_{s}^{+}\pi^{+})$ requiring $m(\pi^{+}\pi^{-})>0.39$~\gev. Individual components, corresponding to the background contribution estimated from $m(D_{s}^{+}\pi^{+}\pi^{-})$ sideband regions (gray-filled) and the different resonant contributions (colored dashed lines), are also shown as indicated in the legend~\cite{LHCb:2024iuo}.}
\label{fig:AmAn_Ds12460}
\end{figure}

Using $9\,\mathrm{fb^{-1}}$ of $pp$ collision data at C.M.\ energies of $\sqrt{s}=7$, $8$, and $13\,\mathrm{TeV}$, LHCb performed a combined amplitude analysis of the $D_{s1}(2460)^+ \to D_s^+ \pi^+ \pi^-$ transition reconstructed in $B^0 \to D^-D_s^+\pi^+\pi^-$, $B^+ \to D^0D_s^+\pi^+\pi^-$, and $B^0 \to D^{\ast-}D_s^+\pi^+\pi^-$ decays~\cite{LHCb:2024iuo}. Since conventional resonances can only contribute in the $\pi\pi$ system, the analysis first tested models containing only known $\pi\pi$ states. Although these models describe the data statistically well, they require implausibly large contributions from the $f_2(1270)$ and $f_0(980)$ resonances despite the limited phase space, and exhibit unusually strong interference, indicating they may not reflect the true decay dynamics.
To further investigate the decay model, additional exotic contributions decaying to $D_s^+\pi^\pm$, denoted as $T_{c\bar{s}}^{++}$ and $T_{c\bar{s}}^{0}$, were introduced under isospin symmetry constraints. Models including a $J^P = 0^+$ $T_{c\bar{s}}$ resonance provide a fit quality comparable to models employing only $\pi\pi$ resonances, as shown in Fig.~\ref{fig:AmAn_Ds12460}. The $T_{c\bar{s}}$ contribution was parametrized using both relativistic BW and $K$-matrix approaches. These parametrisations yield consistent mass values, whereas the extracted widths show substantial model dependence, as shown in Table~\ref{tab:AmAn_Ds12460}. The masses obtained using the $K$-matrix parametrisation agree with the expectation in Ref.~\cite{Maiani:2024quj}, where the $T_{c\bar{s}}$ states are interpreted as members of an isotriplet. However, both the fitted mass and width deviate from the values reported in Ref.~\cite{Guo:2009ct}. 
The distributions in Fig.~\ref{fig:AmAn_Ds12460} are consistent with the results of Ref.~\cite{FKinpreparation} 
that contains further development of the ideas previously published in 
Ref.~\cite{Tang:2023yls}.
In theoretical work \cite{Roca:2025lij}, the di-pion transition $D_{s1}(2460) \to D_s \pi^+ \pi^-$ was investigated under the assumption of a molecular $D_{s1}(2460)$ built 
mostly from the $D^* K$ and $D_s^* \eta$ components. In the same picture, the $D_{s1}(2536)$ is then argued to contain a dominant $DK^*$ molecular component, which results in specific predictions for the mass distributions in the decays of this state \cite{Dias:2025izv}. Thus, the entire approach and particular conclusions and predictions of Refs.~\cite{Roca:2025lij,Dias:2025izv} can be critically assessed as soon as experimental data on the two-pion decays of $D_{s1}(2536)$ become available at LHCb. On the contrary, only a little molecular admixture (at the level of 2\%) in the wave function of the $D_{s1}(2536)$ is obtained in Ref.~\cite{Yang:2025dcg}, where this meson comes out as a predominantly quark--antiquark state as based on the approach previously suggested in Ref.~\cite{Yang:2021tvc}. Thus, further experimental studies of the state $D_{s1}(2536)$ and, in particular, the decay modes most sensitive to its nature would be allow one to shed more light on its internal structure and draw certain conclusions concerning the quartet of the positive-parity states $D_{sJ}$.

\begin{table}[t!]
\centering
\caption{Summary of fit results for different models described in detail~\cite{LHCb:2024iuo}. 
The two sources of uncertainty are statistical and systematic. For the models containing $T_{c\bar{s}}$ states the quoted fit fraction is the value for each of the isospin partners, and the quoted $T_{c\bar{s}}$ mass and width parameters are the pole mass and width. }
\label{tab:AmAn_Ds12460}

\begin{tabular}{lcccc}
\hline
Model & Resonance & Mass (\mev) & Width (MeV) & Fractions (\%) \\
\hline
\multirow{2}{*}{$f_0(500)+\text{RBW}~T_{c\bar{s}}(0^+)$} & $f_0(500)$ & $464\pm23\pm14$ & $214\pm28\pm8$ & $199\,^{+42}_{-47}\pm39$ \\
 & $T_{c\bar{s}}^{++}/T_{c\bar{s}}^{0}$ & $2312\pm27\pm11$ & $264\pm46\pm21$ & $126\,^{+27}_{-17}\pm20$ \\
\hline
\multirow{2}{*}{$f_0(500)+K\text{-matrix}~T_{c\bar{s}}(0^+)$} & $f_0(500)$ & $474\pm30\pm18$ & $224\pm23\pm16$& $248\,^{+40}_{-54}\pm39$ \\
 & $T_{c\bar{s}}^{++}/T_{c\bar{s}}^{0}$ & $2327\pm13\pm13$ & $96\pm16\pm23$ & $156\,^{+27}_{-38}\pm25$ \\
\hline
\end{tabular}

\end{table}

\subsection{Hidden charm tetraquarks}

\subsubsection{Observation of $T_{c\bar c1}(4430)^+$}

The first evidence for the existence of a charged charmonium-like state $T_{c\bar{c}1}(4430)^+$ (also known as $Z_c(4430)^+$), with a minimal quark content of $c\bar{c}u\bar{d}$, was obtained in the $\psi(2S)\pi^+$ final state of the decay $B \to\psi(2S)K\pi$~\cite{Belle:2007hrb,Belle:2009lvn}. The amplitude analysis has established its spin--parity quantum numbers to be $J^P = 1^+$. Despite extensive theoretical and experimental studies, the underlying nature of the $T_{c\bar{c}1}(4430)^+$ structure remains unresolved. Dynamical interpretations generally treat the $T_{c\bar c1}(4430)^+$ structure as a genuine exotic state such as 
hadrocharmonium \cite{Dubynskiy:2008mq},
compact tetraquark \cite{Liu:2008qx},
or a hadronic molecule ~\cite{Meng:2007fu,Liu:2007bf,Ding:2008mp}. In Refs.~\cite{Pakhlov:2011xj,Pakhlov:2014qva,Uglov:2016nql,Nakamura:2019btl}, it is argued to be an effect of the $\bar{D}^{*0}D^+$\dash$\psi(2S)\pi^+$ re-scatterings. In the molecular picture~\cite{Meng:2007fu,Liu:2007bf,Ding:2008mp}, the $T_{c\bar{c}1}(4430)^+$ is typically interpreted as an $S$-wave $D^\ast D_1$ bound state with possible spin-parity assignments $J^P=0^-,\,1^-,\,2^-$; however, all of these assignments are incompatible with the LHCb measurement. 
The compact tetraquark scenario predicts an octet of related states~\cite{Maiani:2014aja,Deng:2015lca,Wang:2014vha}, none of which have been conclusively observed, although the $T_{c\bar c1}(4430)^+$ is a possible candidate. The kinematical interpretations attribute the observed structure to rescattering effects in $B$-meson decays, which can reproduce the measured properties without introducing a genuine resonance~\cite{Pakhlov:2011xj,Pakhlov:2014qva,Uglov:2016nql,Nakamura:2019btl}. At present, none of the proposed scenarios provides a fully consistent description of the $T_{c\bar c1}(4430)^+$.

\begin{figure}[t!]
\begin{center}
\raisebox{15mm}{\includegraphics[width=0.5\textwidth]{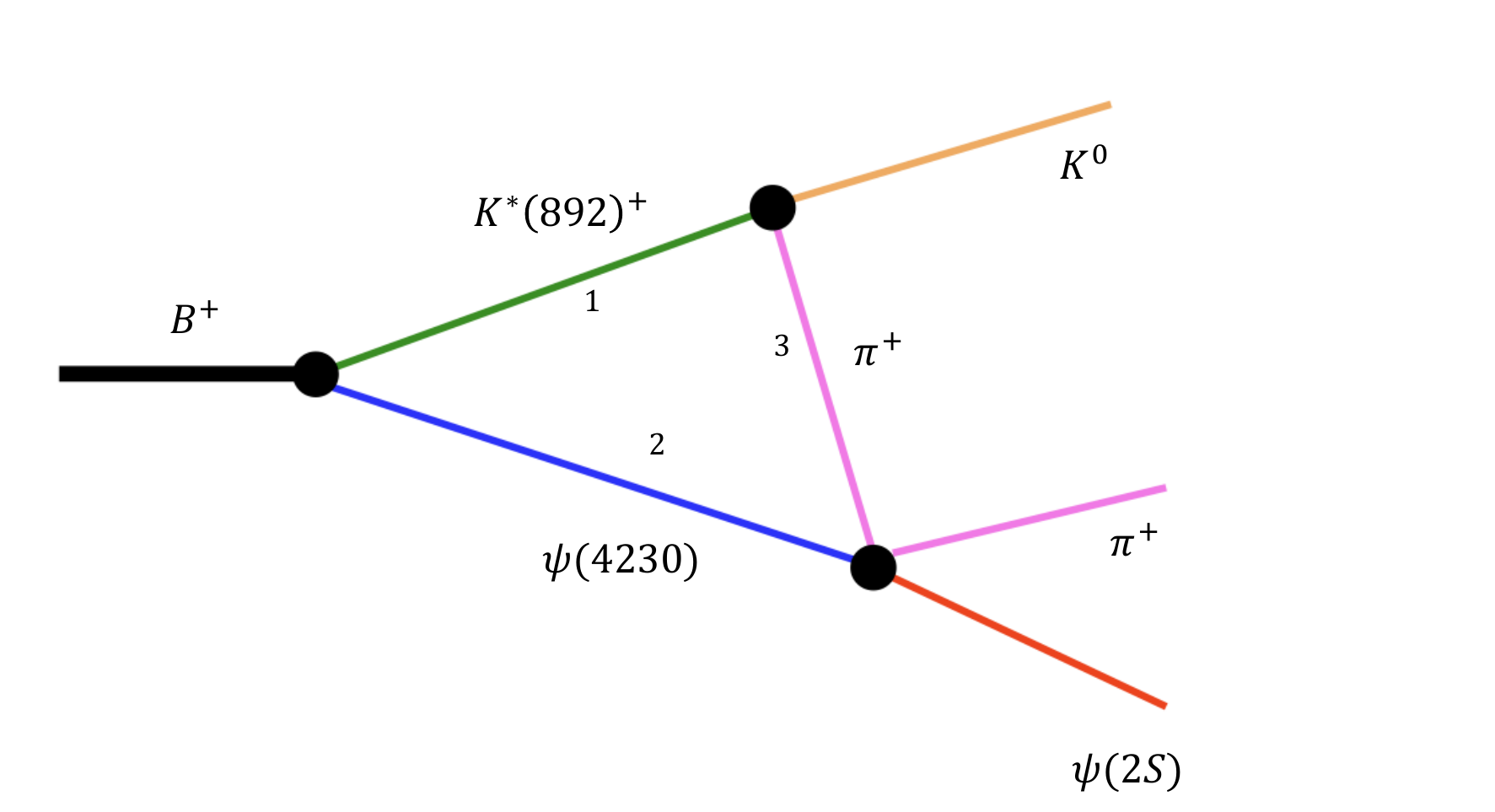}}
\includegraphics[width=0.45\textwidth]{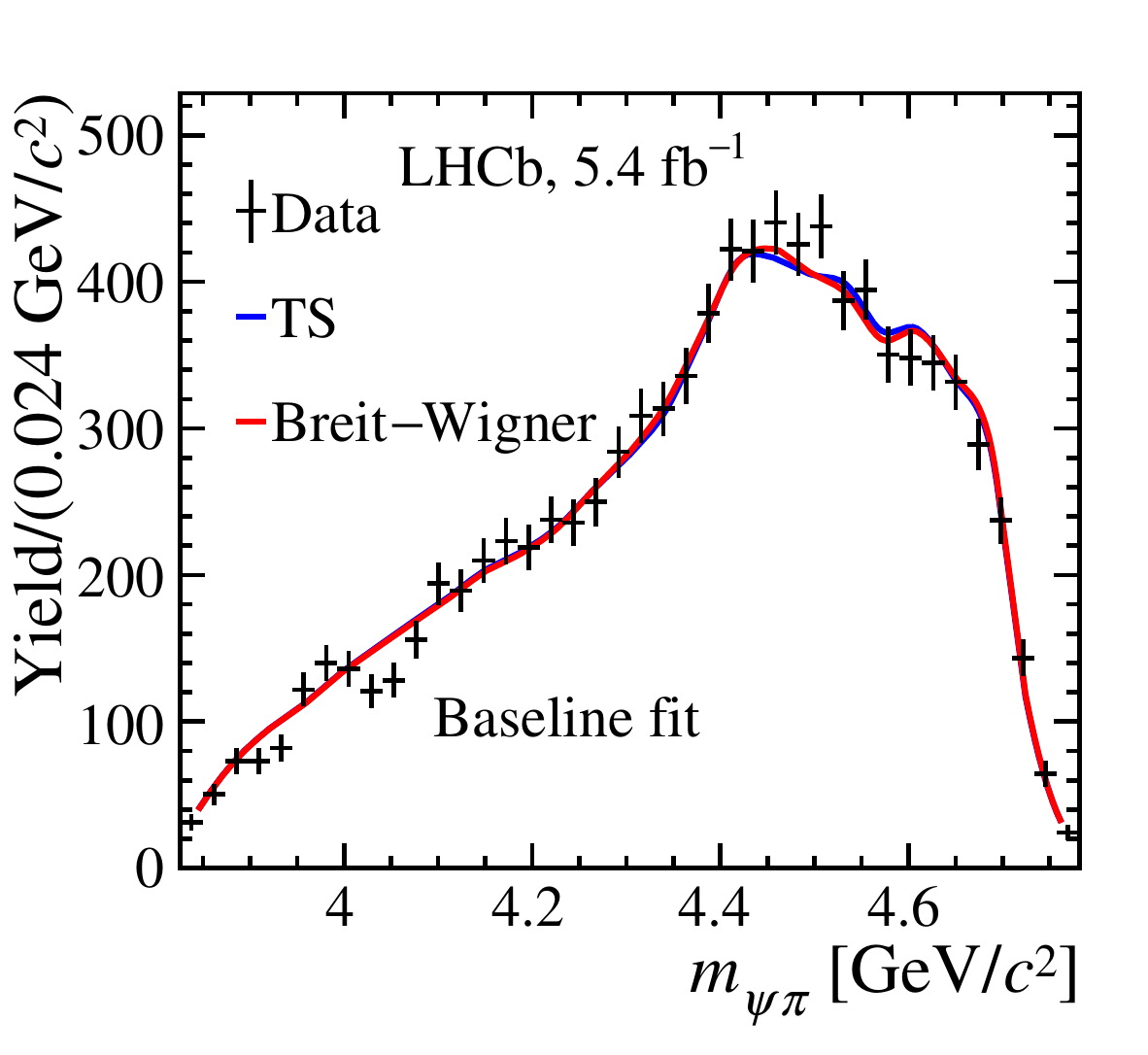}
\end{center}
\caption{ The left plot illustrates the triangle diagram involving the $K^{*}(892)^+$, $\psi(4230)$, and $\pi^+$ intermediate states, which contributes to the $B^+ \to \psi(2S)K^0_S\pi^+$ decay. The right plot shows the invariant-mass distribution of the $\psi(2S)\pi$ system for background-subtracted data, overlaid with the fit projections. The baseline fit includes either a relativistic BW parametrization (red) or a triangle-singularity amplitude (blue)~\cite{LHCb:2025kxf}.}
\label{fig:Tcc4430}
\end{figure}

Recently, the LHCb Collaboration has performed an amplitude analysis of the $B^+ \to \psi(2S)K^0_S\pi^+$ decay using $pp$ collision data collected during the period 2016\dash2018~\cite{LHCb:2025kxf}. The analysis shows that the inclusion of a $T^+_{c\bar c}$ is required to describe the $\psi(2S)\pi^+$ invariant mass spectrum, in particular in the region
$4.2 < M(\psi(2S)\pi^+) < 4.7$~\gev.
Using a relativistic BW parameterization and considering different assumptions for the orbital angular momentum, the preferred solution yields a mass and width
\begin{equation}
M = (4.452 \pm 0.016^{+0.055}_{-0.033})~\mbox{\gev}, \qquad
\Gamma = (0.174 \pm 0.019^{+0.083}_{-0.020})~\mbox{GeV},\nonumber
\end{equation}
with spin-parity quantum numbers $J^P = 1^+$. These parameters are consistent with those of the previously observed $T_{c\bar c1}(4430)^+$ state~\cite{ParticleDataGroup:2024cfk}.
Motivated by a possible molecular interpretation, the coupling of the $T^+_{c\bar c}$ state to the $\bar{D}^{\ast}_1(2600)^0 D^+$ channel is also investigated.
In this framework, the structure is described using a Flatt\'e-like parameterization with two coupled channels~\cite{Flatte:1976xu}, $\psi(2S)\pi^+$ and $\bar{D}^{\ast}_1(2600)^0 D^+$, with coupling strengths $g_1$ and $g_2$, respectively. The fit yields a mass
\begin{equation}
M_{\mathrm{Flatt\acute e}} = (4.452 \pm 0.022^{+0.103}_{-0.005})~\mbox{\gev},\nonumber
\end{equation}
and constrains the relative coupling strength
$R \equiv |g_2/g_1|$ at the 95\% C.L.\ to be $R < 6.8$. 
In addition, a kinematical interpretation based on a triangle singularity mechanism is examined~\cite{Nakamura:2019btl}. In this scenario, the enhancement in the $\psi(2S)\pi^+$ mass spectrum
arises from $\psi(4230)\pi^+ \to \psi(2S)\pi^+$ rescattering in the decay $B^+ \to \psi(4230)K^{\ast}(892)^+$. This model provides a fit quality and phase shift behaviour comparable to that obtained with the BW description, as shown in Fig.~\ref{fig:Tcc4430}. Larger data samples will be required to further discriminate between the two
interpretations.

\subsubsection{Possible consequence for the $B^+\to\psi(2S)K^+\pi^+\pi^-$}

The decay $B^+ \to \psi(2S)K^+\pi^+\pi^-$ provides a rich environment to search for a variety of exotic hadron contributions. Owing to its multi-body topology, this channel is sensitive to charged charmonium-like states $T_{c\bar c}$, previously observed in $B^0 \to T_{c\bar c}^- K^+ \to [\psi(2S)\pi^-]K^+$ decays~\cite{Belle:2009lvn,LHCb:2014zfx,Belle:2013shl}, as well as to strange partners $T_{c\bar c\bar s}^+$ reported in $B^+ \to T_{c\bar c\bar s}^+ \phi \to [J/\psi K^+]\phi$ decays~\cite{LHCb:2021uow}. In addition, contributions from neutral exotic states decaying to $\psi(2S)\pi^+\pi^-$ such as $X^0 \to \psi(2S)\pi^+\pi^-$, or from $T_{c\bar c\bar s}^0 \to \psi(2S)K^+\pi^-$ are also possible. The decay topology further allows for more complex exotic cascade processes, for example $X^0 \to [T_{c\bar c}^- \to \psi(2S)\pi^-]\pi^+$. The LHCb Collaboration has recently reported an amplitude analysis of the decay
$B^+ \to \psi(2S)K^+\pi^+\pi^-$ using the full Run~1 and Run~2 data sets~\cite{LHCb:2024cwp}. The baseline model, comprising 53 interfering amplitudes, provides an adequate description of the data. A rich spectrum of exotic contributions is included in the fit: four $X^0 \to \psi(2S)\pi^+\pi^-$ candidates, three charged $T^{\pm}_{c\bar c} \to \psi(2S)\pi^\pm$ structures, one $T^{+}_{c\bar c\bar s} \to \psi(2S)K^+$ component, and three neutral $T^{0}_{c\bar c\bar s} \to \psi(2S)K^+\pi^-$
candidates. The masses and widths of the exotic states are treated as free parameters in the fit, with the exception of $T^{\ast}_{c\bar c1}(4055)^+$ and $T_{c\bar c\bar s1}(4000)^+$, whose parameters are fixed due to limited experimental sensitivity.

\begin{figure}[t!]
\includegraphics[width=0.329\textwidth,height=!]{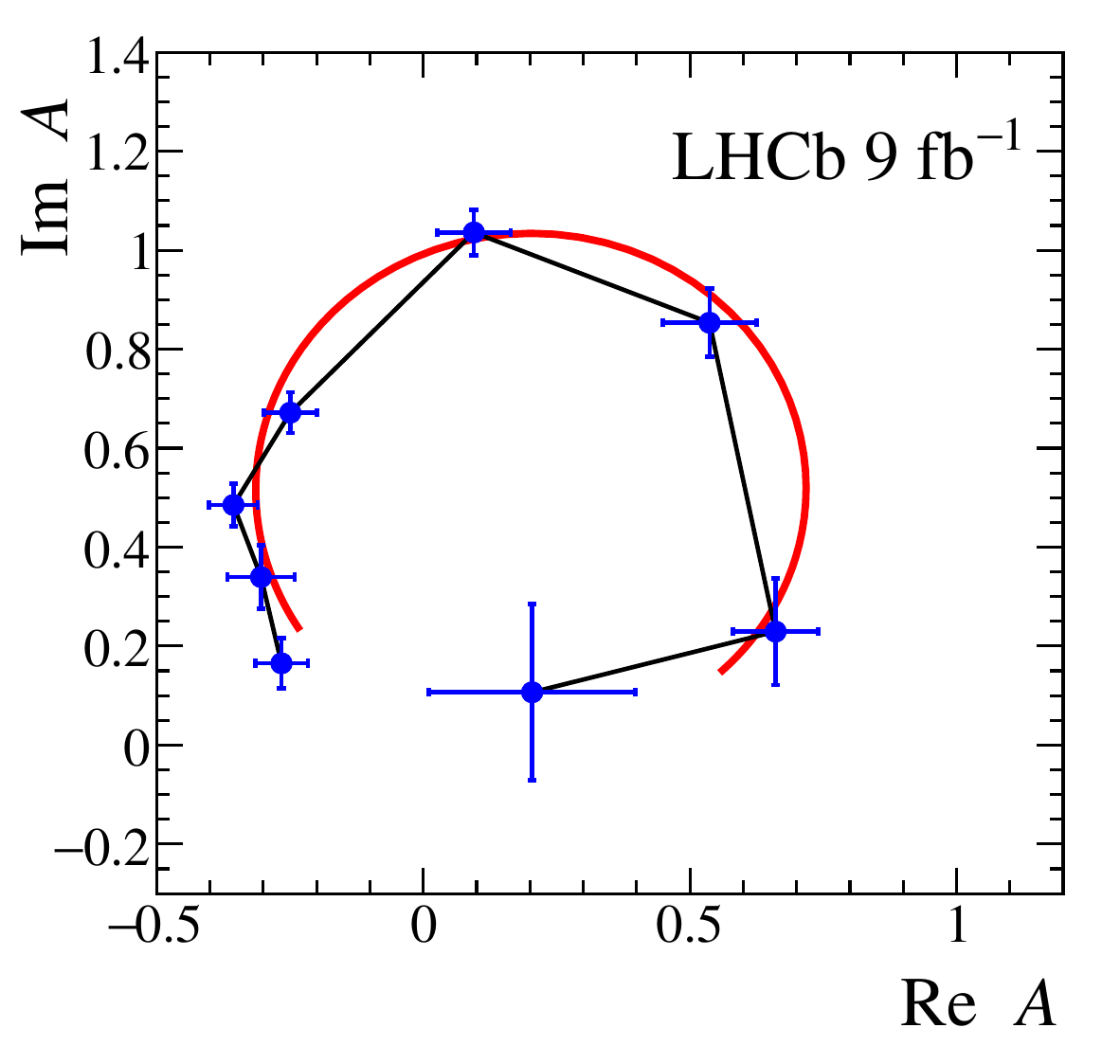}
\includegraphics[width=0.329\textwidth,height=!]{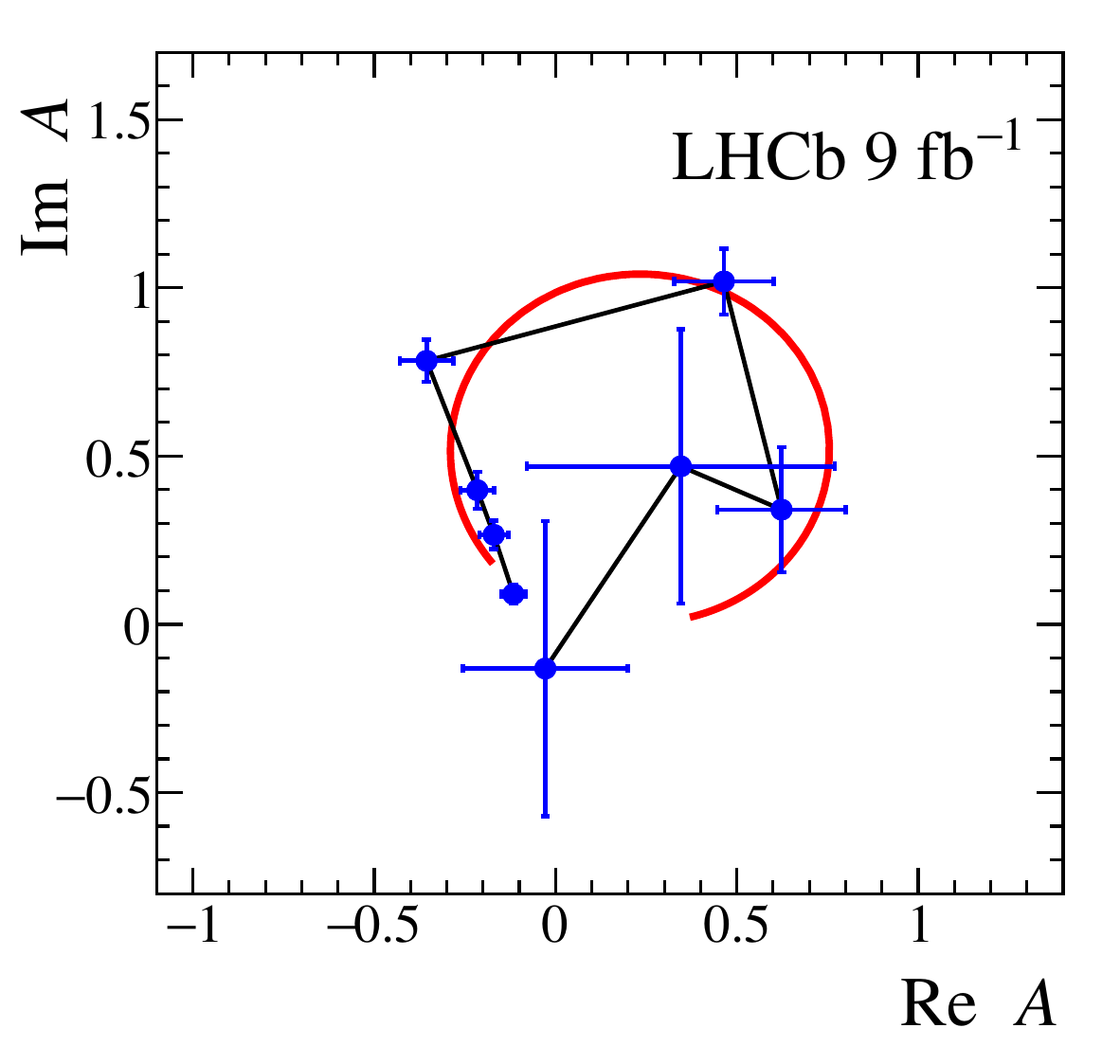}
\includegraphics[width=0.329\textwidth,height=!]{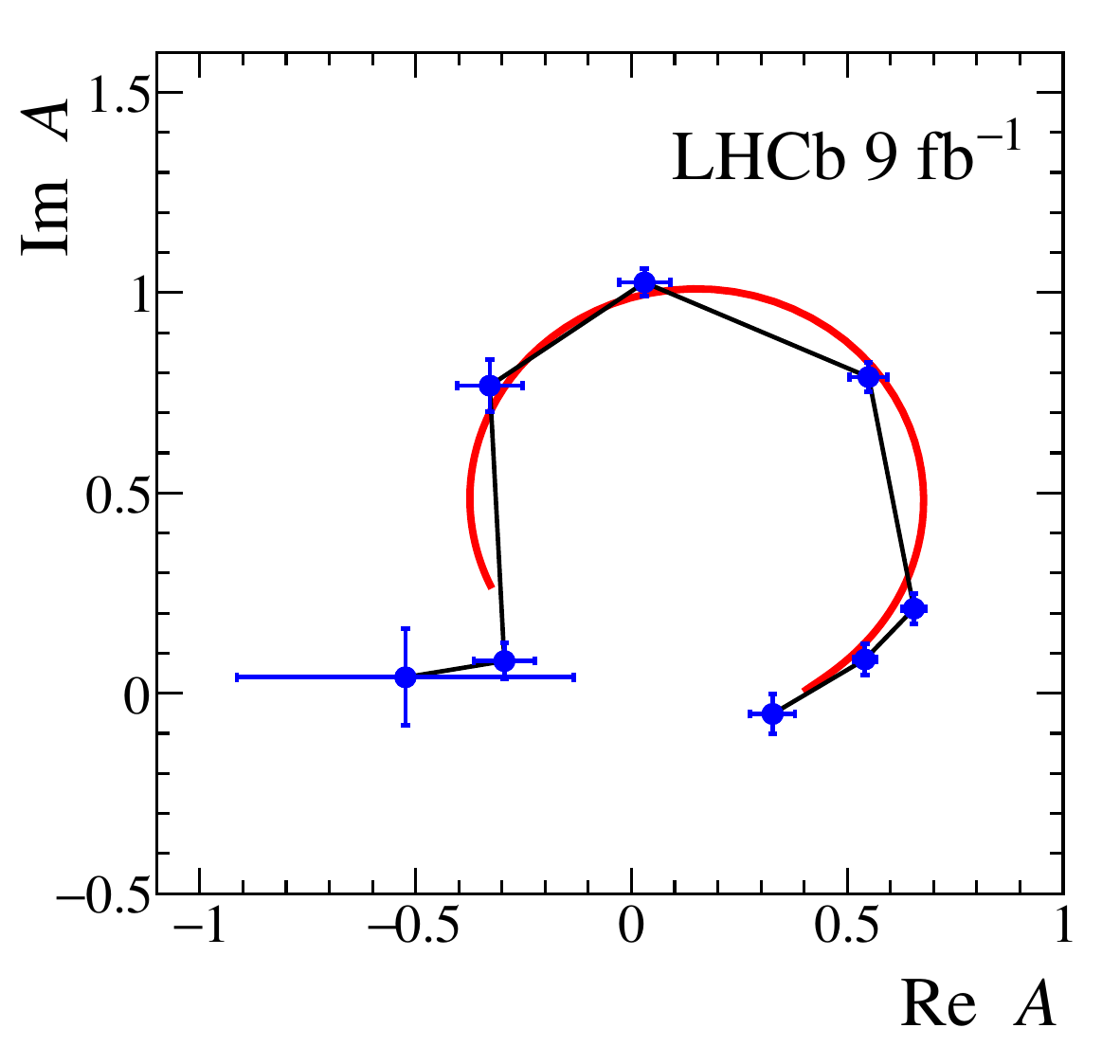}
\caption{
Argand diagrams for the quasi-model-independent partial-wave analysis for the (left) $T^{*}_{c\bar c 0}(4475)^0$, (middle) $T_{c\bar{c}\bar{s}1}(4650)^0$, and (right) $T_{c\bar{c}1}(4200)^+$ resonances~\cite{LHCb:2024cwp}.
The fitted lineshape knots are displayed as connected points with error bars and the invariant mass increases counterclockwise.
The BW lineshape with the mass and width from the nominal fit is superimposed (red line). }
\label{fig:AmAn_psi2s_argand}
\end{figure}

The dominant exotic contribution in the baseline model is $T^{*}_{c\bar c 0}(4475)^0$, which is found to decay predominantly through the $\psi(2S)\rho(770)^0$ channel. Its fitted mass is $(4475 \pm 7 \pm 12)$~\mev, compatible with the $\chi_{c0}(4500)$ structure previously reported in $B^+ \to \chi_{c0}(4500)(\to J/\psi\phi)K^+$~\cite{LHCb:2021uow}. However, its BW width, $(231 \pm 19 \pm 32)~\text{MeV}$, is significantly larger than the reported $\Gamma = (77^{+12}_{-10})~\text{MeV}$ of the $\chi_{c0}(4500)$.
Assuming isospin conservation, one could interpret $T^{\!*}_{c\bar c 0}(4475)^0$ and $\chi_{c0}(4500)$ as different states with $I=1$ ($c\bar c(u\bar u - d\bar d)$) and $I=0$ ($c\bar c s\bar s$), respectively; yet under this hypothesis, the expected mass difference would be much larger than the observed difference. 
The $T_{c\bar{c}1}(4650)^0$ and $T^{*}_{c\bar{c}0}(4710)^0$ states are found to be compatible, within uncertainties, with the $\chi_{c1}(4685)$ and $\chi_{c0}(4700)$ resonances previously reported in $B^+ \to J/\psi\phi\,K^+$ decays (see Table~\ref{tab:AmAn_psi2skpipi}). Given that these final states carry different isospin, identifying the corresponding pairs as the same physical states would require significant isospin-violating effects, for which no compelling mechanism has yet been established.
Besides direct production in $B$ decays, charged hidden-charm exotic states are also produced via cascade decays of higher-mass exotic resonances, such as
$T^{*}_{c\bar{c}} \to T^{\pm}_{c\bar{c}}\pi^\mp$ and 
$T^{0}_{c\bar{c}\bar{s}} \to T^{-}_{c\bar{c}}K^{+}$,
mediated by the intermediate states $T_{c\bar{c}1}(4200)^+$ and $T_{c\bar{c}1}(4430)^+$. The fitted mass and width of the $T_{c\bar{c}1}(4430)^+$ are consistent with those obtained in $B^0 \to \psi(2S) K^- \pi^+$ analyses~\cite{LHCb:2014zfx,Belle:2013shl}, while the $T_{c\bar{c}1}(4200)^+$ shows similar properties to the $T_{c\bar{c}}(4250)^+$ structure reported by Belle in $B^0 \to \chi_{c1}(1P)K^- \pi^+$ decays~\cite{Belle:2008qeq}; its spin--parity is determined to be $J^P=1^+$ for the first time.
In addition, the hidden-charm strange states $T_{c\bar{c}\bar{s}1}(4600)^0$ and $T_{c\bar{c}\bar{s}1}(4900)^0$ are observed for the first time. If interpreted as tetraquark candidates, their minimal quark configuration would be $c\bar{c}s\bar{d}$, and could be radial excitations of the $T_{c\bar{c}\bar{s}1}(4000)^0$ state observed in $B^0 \to T_{c\bar{c}\bar{s}1}(4000)^0(\to J/\psi K^0_S)\phi$ decays~\cite{LHCb:2023hxg}.
The resonant nature of all these exotic candidates is investigated through a 
quasi-model-independent partial-wave analysis. The Argand diagrams for the dominant 
components --- $T^*_{c\bar{c}0}\to \psi(2S)\pi^+\pi^-$, $T_{c\bar{c}\bar{s}1}^{0} \to \psi(2S)K^+\pi^-$, 
and $T_{c\bar{c}1}^{\pm} \to \psi(2S)\pi^\pm$ --- are presented in Fig.~\ref{fig:AmAn_psi2s_argand}. 
In each case, a characteristic counter-clockwise trajectory is observed, exhibiting 
phase motion consistent with the expectation for a resonance. 
Details of other components are available in Ref.~\cite{LHCb:2024cwp}.
Despite high statistical significances of all reported signals, the broad line shapes and the complexity of the amplitude model --- comprising 53 interfering components --- introduce substantial model dependence, which complicates a precise determination of the underlying nature of these structures.

\begin{table}[t]
\caption{Left: Resonance parameters determined from the amplitude analysis~\cite{LHCb:2024cwp}.
Right: If available, the states listed in the PDG~\cite{ParticleDataGroup:2024cfk} with the same $J^P$ quantum numbers and closest mass are listed for comparison. 
}
\resizebox{1.0\linewidth}{!}{
\tiny
\centering
\begin{tabular}{l l r@{}c@{}c@{}c@{}l r@{}c@{}c@{}c@{}l | l r@{}c@{} l r@{}c@{} l}
\hline \hline
Resonance & $J^P$ & \multicolumn{5}{c}{$m_0$ [\mev]} & \multicolumn{5}{c}{$\Gamma_0$ [MeV]} & Res. PDG & \multicolumn{3}{c}{$m_0$ [\mev]} & \multicolumn{3}{c}{$\Gamma_0$ [MeV]} \\ 
\hline

$T^{*}_{c\bar c 0}(4475)^0$ & $0^+$ & $4475 $&$\pm$&$ 7 $&$\pm$&$ 12$ & $231 $&$\pm$&$ 19 $&$\pm$&$ 32$ &$\chi_{c0}(4500) $ & $4474 $ &$\pm$ &$ 4 $ & $77 $ &$^{+}_{-}$ &$ ^{12}_{10}$ \\
$T_{c\bar{c}1}(4650)^0$ & $1^+$ & $4653 $&$\pm$&$ 14 $&$\pm$&$ 27$ & $227 $&$\pm$&$ 26 $&$\pm$&$ 22$ & $\chi_{c1}(4685)$ & $4684$&$^{+}_{-}$&$^{15}_{17} $ & $126 $&$\pm$&$ 40$ \\
$T^{*}_{c\bar c 0}(4710)^0$ & $0^+$ & $ 4710 $&$\pm$&$ 4 $&$\pm$&$ 5$& $64 $&$\pm$&$ 9 $&$\pm$&$ 10$ & $\chi_{c0}(4700)$& $4694$&$^{+}_{-}$&$^{16}_{5} $ & $87$&$^{+}_{-}$&$^{18}_{10}$ \\
\hline 
$T_{c\bar{c}1}(4200)^+$ & $1^+$ & $4257 $&$\pm$&$ 11 $&$\pm$&$ 17$ & $308 $&$\pm$&$ 20 $&$\pm$&$ 32$ & $T_{c\bar{c}1}(4200)^+$ & $4196$&$^{+}_{-} $&$^{35}_{32}$ & $370$&$^{+}_{-}$&$^{100}_{150}$ \\
$T_{c\bar{c}1}(4430)^+$ & $1^+$ & $4468 $&$\pm$&$ 21 $&$\pm$&$ 80$ & $251 $&$\pm$&$ 42 $&$\pm$&$ 82$ & $T_{c\bar{c}1}(4430)^+$ & $4478$&$^{+}_{-}$&$^{15}_{18}$ & $181 $&$\pm$&$ 31$ \\	
\hline	
$T_{c\bar{c}\bar{s}1}(4600)^0$ & $1^+$ & $4578 $&$\pm$&$ 10 $&$\pm$&$ 18$ & $133 $&$\pm$&$ 28 $&$\pm$&$ 69$ & \\
$T_{c\bar{c}\bar{s}1}(4900)^0$ & $1^+$ & $ 4925 $&$\pm$&$ 22 $&$\pm$&$ 47$& $255 $&$\pm$&$ 55 $&$\pm$&$ 127$ & \\
\hline \hline
\end{tabular}
}
\label{tab:AmAn_psi2skpipi}
\end{table}

\subsubsection{Radiative decays of $\X$}

The photon emission vertex is described by Quantum Electrodynamics that is the most studied and well understood field theory entering the Standard Model. For this reason, the radiative decays of hadrons attract special attention of both experimentalists and theorists and are often regarded as a strong investigation tool to probe the nature and properties of various hadronic systems. Since the $\X$ definitely contains a $c\bar{c}$ pair, its radiative decays to hidden- and open-charm final states have been under active studies for many years. 

The LHCb Collaboration has recently reported a measurement of the radiative decays of the $\X$ into the $\gamma\psi(2S)$ and $\gamma J/\psi$ final states, using $pp$ collision data collected during Run~1 and Run~2, corresponding to a total integrated luminosity of $9~\mathrm{fb}^{-1}$~\cite{LHCb:2024tpv}. The analysis is based on the decay chain $B^+ \to \X(\to \gamma\psi)K^+$, where $\psi$ denotes either $J/\psi$ or $\psi(2S)$ reconstructed through their $\mu^+\mu^-$ decay modes.
A signal significance of $4.8\sigma$ and $6.0\sigma$ for the decay $\X\to \gamma \psi(2S)$ is obtained in the Run~1 and Run~2 data sets, respectively, constituting the first observation of this radiative transition. The ratio of the partial radiative widths,
\be
{\cal R}_X=\frac{{\cal B}(\X\to\gamma \psi(2S))}{{\cal B}(\X\to \gamma J/\psi)},
\label{R}
\ee
is measured to be 
\be
{\cal R}_X^{\rm LHCb}=1.67 \pm 0.21 \pm 0.12 \pm 0.04,
\label{RXLHCb}
\ee
for the combined Run~1 and Run~2 data. 

Belle II is also performing the measurement of ${\cal R}_X$. In the previous measurement using Belle data only, the $\X\to \gamma J/\psi$ was observed with a significance of 5.5$\sigma$ including systematic uncertainties. But no significant $\X\to\gamma \psi(2S)$ signal was found~\cite{Belle:2011wdj}.
In this updated work on $\X \to \gamma \psi(2S)$ in $B^{\pm}\to \X K^{\pm}$ and $B^{0}\to \X K^{0}$,
a Boosted Decision Tree (BDT) using the XGBoost algorithm is employed to suppress the dominant $B^{\pm,0}\to\psi(2S)K^{*\pm,0}$ background. Variables such as $M(\pi^0)$, $M(K^{*\pm,0})$, the recoil mass of the kaon, the mass difference between the $B$ and $K$ mesons, the kaon momentum, and the polar angle between $K$ and $\gamma$ are used.
By applying the BDT classifier requirement, the background rejection rate is approximately 80\% with a high signal efficiency of 80\%.
Belle II performed a sensitivity study of observing $\X\to\gamma \psi(2S)$ by using 5,000 sets of simulated pseudoexperiments.
In this study, the signal yield for $\X\to \gamma \psi(2S)$ is determined under an assumption of ${\cal R}_X=1.67$~\cite{LHCb:2024tpv}.
The background Monte Carlo (MC) samples including $e^+e^-\to B^+B^-$ and $e^+e^-\to B^0\bar B^0$ inclusive decay events with the same integrated luminosity as the data are used. Here, the integrated luminosities at Belle and Belle II are 711 fb$^{-1}$ and 363 fb$^{-1}$, respectively.
The mean statistical significance is 3.2$\sigma$ for $X(3872)\to\gamma\psi(2S)$ from 5,000 sets of simulated pseudoexperiments.
With the large data samples collected at Belle II in the near future, it is promising to observe the $\X\to\gamma \psi(2S)$ signal.

\begin{figure}[t!]
\setlength{\unitlength}{1mm}
\centering
\begin{picture}(150,80)
\put(15, 2){
\includegraphics*[width=45mm]{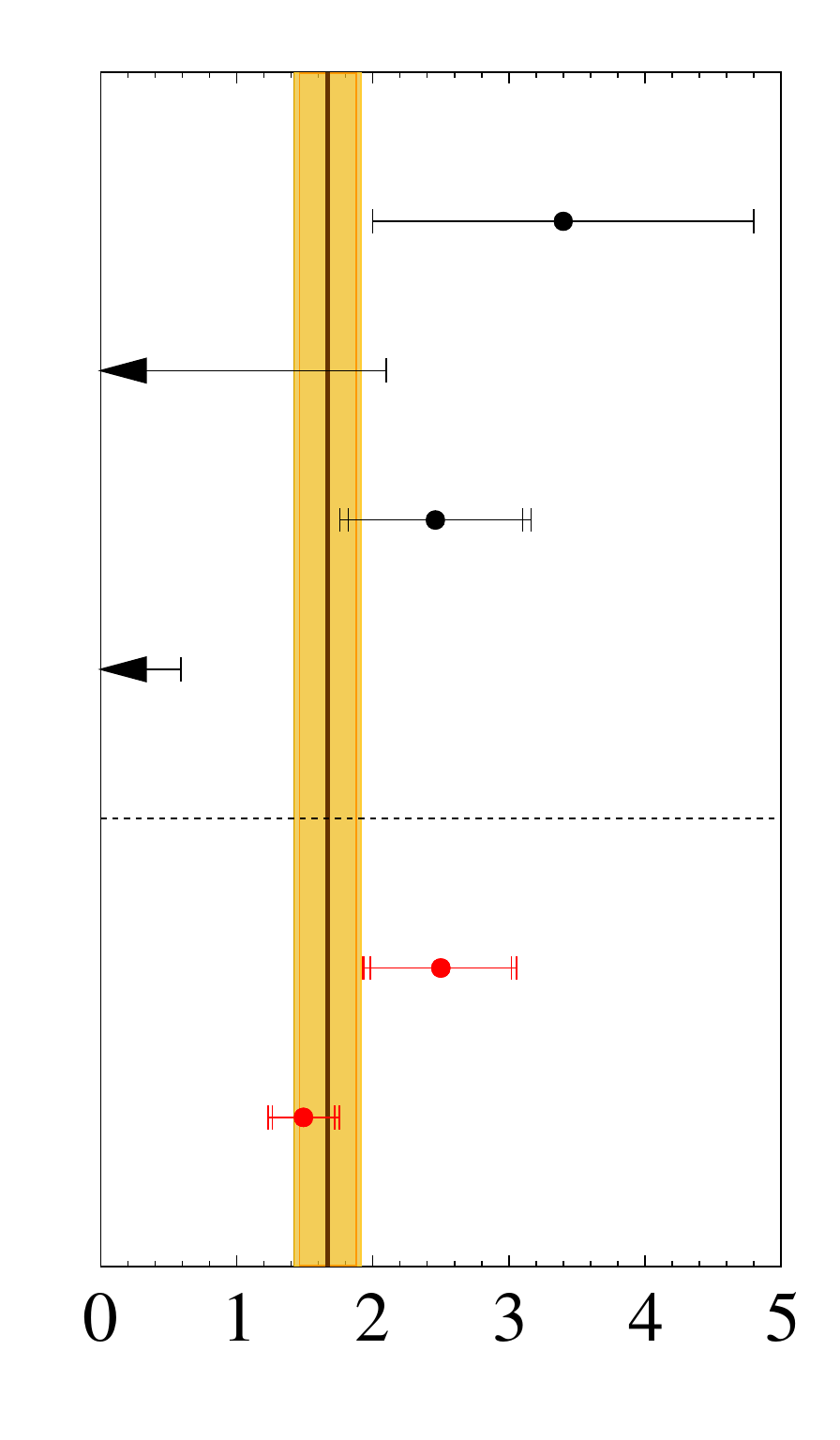} 
}
\put(65,42){\Large\begingroup\renewcommand*{\arraystretch}{1.27}\begin{tabular}{lll} 
 BaBar & 2008 & \cite{BaBar:2008flx} \\
 Belle & 2011 & \cite{Belle:2011wdj} \\
 LHCb/Run\,1 & 2014 & \cite{LHCb:2014jvf} \\ 
 BESIII & 2020 & \cite{BESIII:2020nbj} \\
 & & 
 \\
 LHCb/Run\,1 & 2024 & \\
 LHCb/Run\,2 & 2024 & 
\end{tabular}
\endgroup}

\put(38,2){\large$\mathcal{R}_{X}$}
\end{picture}
\caption{
Summary of the experimental results for the ratio of the partial decay widths in Eq.~\eqref{R}. Adapted from Ref.~\cite{LHCb:2024tpv}.
The results from Ref.~\cite{LHCb:2024tpv}
 for the Run\,1 and Run\,2~data sets
 are shown as red points with error bars.
 The~colored band corresponds to the average of the LHCb results in Ref.~\cite{LHCb:2024tpv}. For the LHCb measurements, the outer error bars represent the total uncertainties, while the inner ones indicate the statistical uncertainties only.
 }
\label{fig:X3872}
\end{figure}

The experimental progress in measuring the ratio ${\cal R}_X$ from Eq.~\eqref{R} over the years is summarized in Table~\ref{tab:rad}, and a comparison of the results is shown in Fig.~\ref{fig:X3872}.
Noticeably, the recent LHCb result in Eq.~\eqref{RXLHCb} lies below the upper limit set by the Belle Collaboration and is consistent with earlier measurements from BaBar and the previous LHCb analysis. Meanwhile, it still demonstrates a significant tension\footnote{In a recent theoretical work \cite{Nakamura:2026xax}, this tension is explained by the contribution of a predicted $2^{-+}$ charmonium candidate, $\eta_{c2}$, residing
slightly above the $D^{*0}\bar{D}^0$ threshold.} with the strong upper limit reported by the BESIII Collaboration\footnote{A recent concise review and prospects concerning the study of the $\X$, including its radiative decays, in the BESIII experiment can be found in Ref.~\cite{Zhou:2025dqk}.}. 
Thus, the most recent LHCb result in Eq.~\eqref{RXLHCb} \cite{LHCb:2024tpv} is conclusive and establishes ${\cal R}_X\simeq 1..2$ that calls for a theoretical interpretation.

\begin{table}[t!]
\caption{Experimental measurements of the ratio ${\cal R}_X$ in Eq.~\eqref{R} for the $\X$ radiative decays.}
\label{tab:rad}
\centering
\begin{tabular}{ccccc}
\toprule
BaBaR \cite{BaBar:2008flx} & Belle \cite{Belle:2011wdj} & LHCb (old) \cite{LHCb:2014jvf} & BESIII \cite{BESIII:2020nbj} & LHCb (new) \cite{LHCb:2024tpv}\\
\midrule
$3.4\pm 1.4$ & $<2.1$ & $2.46\pm 0.64\pm 0.29$ & $<0.59$ & $1.67\pm 0.21\pm 0.12\pm 0.04$\\
\bottomrule
\end{tabular}
\end{table}

It is claimed in an early theoretical work \cite{Swanson:2004pp}
that the ratio in Eq.~\eqref{R} is strongly sensitive to the nature of the $\X$. In particular, ${\cal R}_X\sim 1$ was found to be consistent with the $\X$ as a compact quark state. Indeed, various quark model calculations demonstrate that such ratio around unity can be naturally gained for the $\X$ assigned as a generic charmonium (see, for example, recent works \cite{Bokade:2024tge,Colangelo:2025uhs} and further results and papers quoted therein) or a compact tetraquark \cite{Grinstein:2024rcu}. 
In the meantime, a tiny ratio ${\cal R}_X^{\rm mol}\ll 1$ was argued in Ref.~\cite{Swanson:2004pp} to be inevitable for the $\X$ as a $D\bar{D}^*$ molecule. This claim as well as the conclusion of the experimentally measured ratio ${\cal R}_X$ in Table~\ref{tab:rad} being at odds with the molecular interpretation of the $\X$ is further repeatedly stressed in some recent works \cite{Grinstein:2024rcu,Esposito:2025hlp}. This conclusion is scrutinized in Ref.~\cite{Guo:2014taa} where it is explained that in fact the radiative decays of the molecular $\X$ to charmonia proceed via $D^{(*)}$-meson loops with the photon emitted by a $D^{(*)}$ meson inside the loop rather than through the light-quark annihilation mechanism suggested in Refs.~\cite{Swanson:2004pp,Grinstein:2024rcu,Esposito:2025hlp}. Then, according to the numerical estimates in Ref.~\cite{Guo:2014taa}, even a purely $D\bar{D}^*$ molecule assignment for the $\X$ is consistent with the ratio of the order of unity in Eq.~\eqref{R} under a not restrictive assumption for the couplings of the vector charmonia $J/\psi$ and $\psi(2S)$ to the $D^{(*)}$ mesons being numerically close to each other. Thus, the estimates of the ratio ${\cal R}_X^{\rm mol}\ll 1$ in Refs.~\cite{Swanson:2004pp,Grinstein:2024rcu}, based on the overlap of the molecular $\X$ wave function, do not apply to the studied case and as such can not be regarded as decisive (see also a detailed discussion in Ref.~\cite{Guo:2026ngz}). Furthermore, the authors of Ref.~\cite{Guo:2014taa} conclude on a strong sensitivity of the ratio ${\cal R}_X$ in Eq.~\eqref{R} to the short-range physics implying that it is mainly governed by the compact component of the $\X$ wave function. The size of this component, traditionally measured in terms of the Weinberg field renormalisation factor $Z$ \cite{Weinberg:1962hj}, is estimated in many theoretical works to vary from small values of the order 10\% or less (see, for example, Refs.~\cite{Chen:2013upa,Brambilla:2024imu}) to larger admixtures at the level of 20\% (see, for example, Ref.~\cite{Shen:2024npc}) and up to as large values as $Z\simeq 0.5$ (see, for example, Refs.~\cite{Kalashnikova:2009gt,Xu:2023lll}).
In either case, given a strong sensitivity of ${\cal R}_X$ to this short-range component established in Ref.~\cite{Guo:2014taa}, even a little admixture of a compact state in the $\X$ wave function, irrespective of its particular nature, leads to an improved description of the radiative decays and readily brings the ratio ${\cal R}_X$ to the experimental value
\cite{Dong:2009uf,Guo:2014taa,Cincioglu:2016fkm,Brambilla:2024imu}. 
In particular, the one-node radial wave function of a generic $\chi_{c1}(2P)$ charmonium state naturally exhibits a larger overlap with the radial wave function of the $\psi(2S)$, which also has one node, than with the nodeless radial wave function of the $J/\psi$. Therefore, values of the ratio ${\cal R}_X$ exceeding unity arise naturally if the short-range component of the $\X$ wave function is dominated by a $c\bar{c}$ configuration.
Furthermore, the presence of a short-range component in the $\X$ wave function resolves the issue with a low probability of its prompt production in high-energy hadronic collisions directly through the delocalized molecular component. This relevant worry was put forward in
Refs.~\cite{Bignamini:2009sk,Esposito:2017qef} and then addressed in Refs.~\cite{Artoisenet:2009wk,Albaladejo:2017blx} from the molecular model perspective.
It should also be noted in conclusion that, given the confirmed total spin of the $\X$, $J=1$, its two-photon decays are forbidden by the Landau-Yang theorem.

Further information of various models for the $\X$ and their compliance with the existing experimental data can be found in the dedicated reviews on the $\X$ (see, for example, recent ones
\cite{Kalashnikova:2018vkv,Esposito:2025hlp}) as well as general reviews on hadronic exotics (see, for example, Refs.~\cite{Esposito:2016noz,Guo:2017jvc,Brambilla:2019esw,Sazdjian:2022kaf,Hanhart:2025bun,Liu:2024uxn}).
For a compilation of various theoretical results for the ratio in Eq.~\eqref{R} and the corresponding references see Table~1 in Ref.~\cite{LHCb:2024tpv}. 

\subsubsection{Pionic transitions from $\X$}

\begin{figure}[t!]	
\centering
\includegraphics[width=0.45\textwidth]{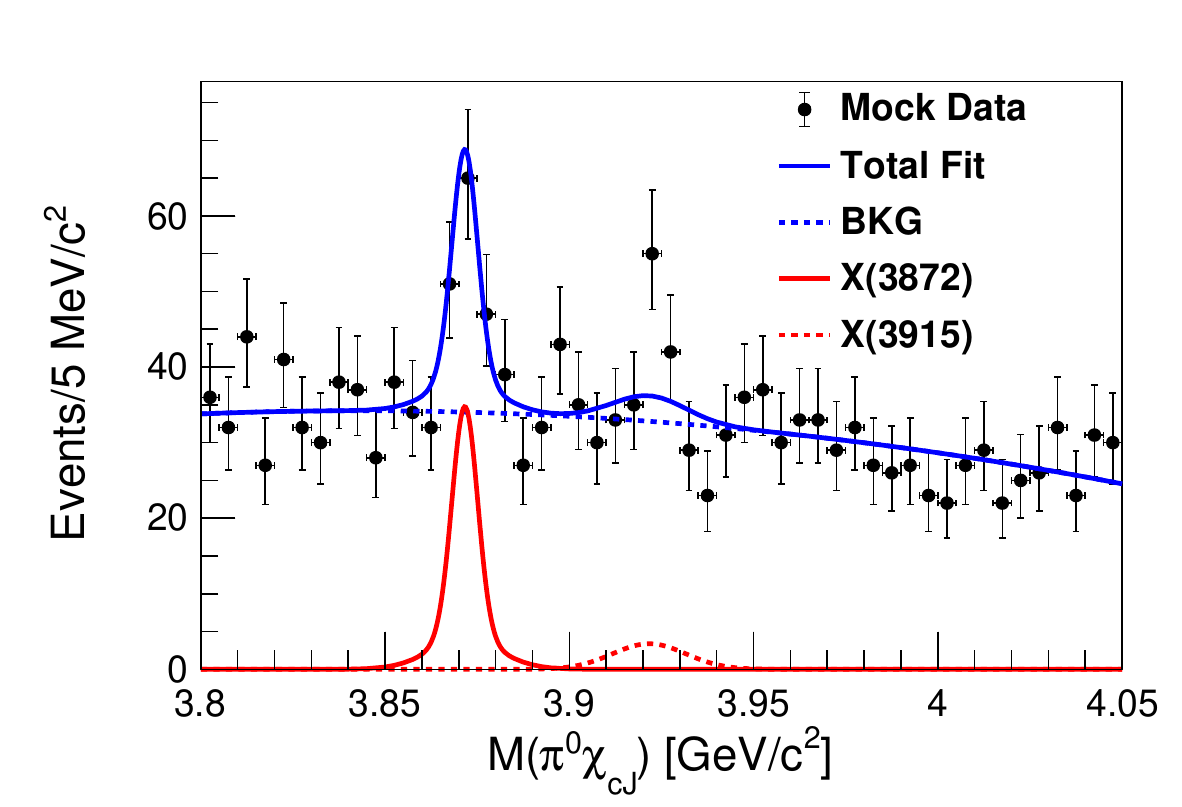} \put(-165,115){{\large \bf (a)}} 
\includegraphics[width=0.45\textwidth]{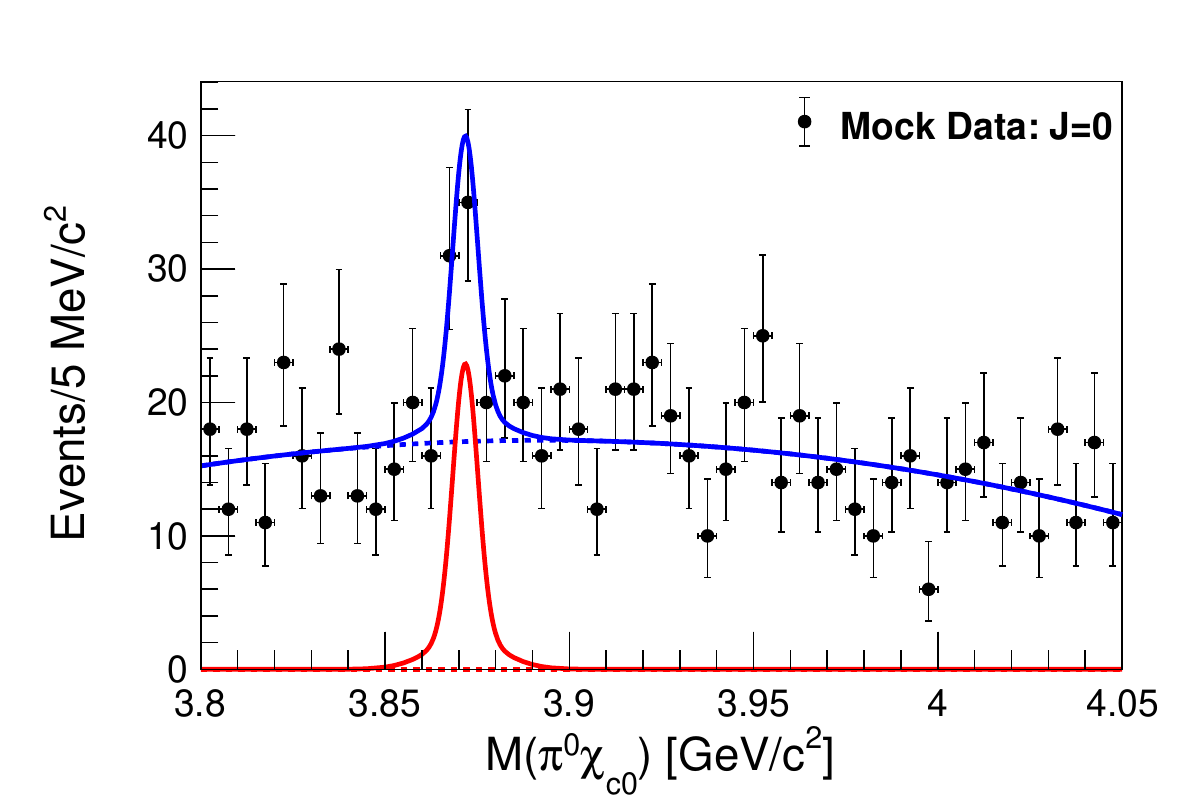} \put(-165,115){{\large \bf (b)}}

\includegraphics[width=0.45\textwidth]{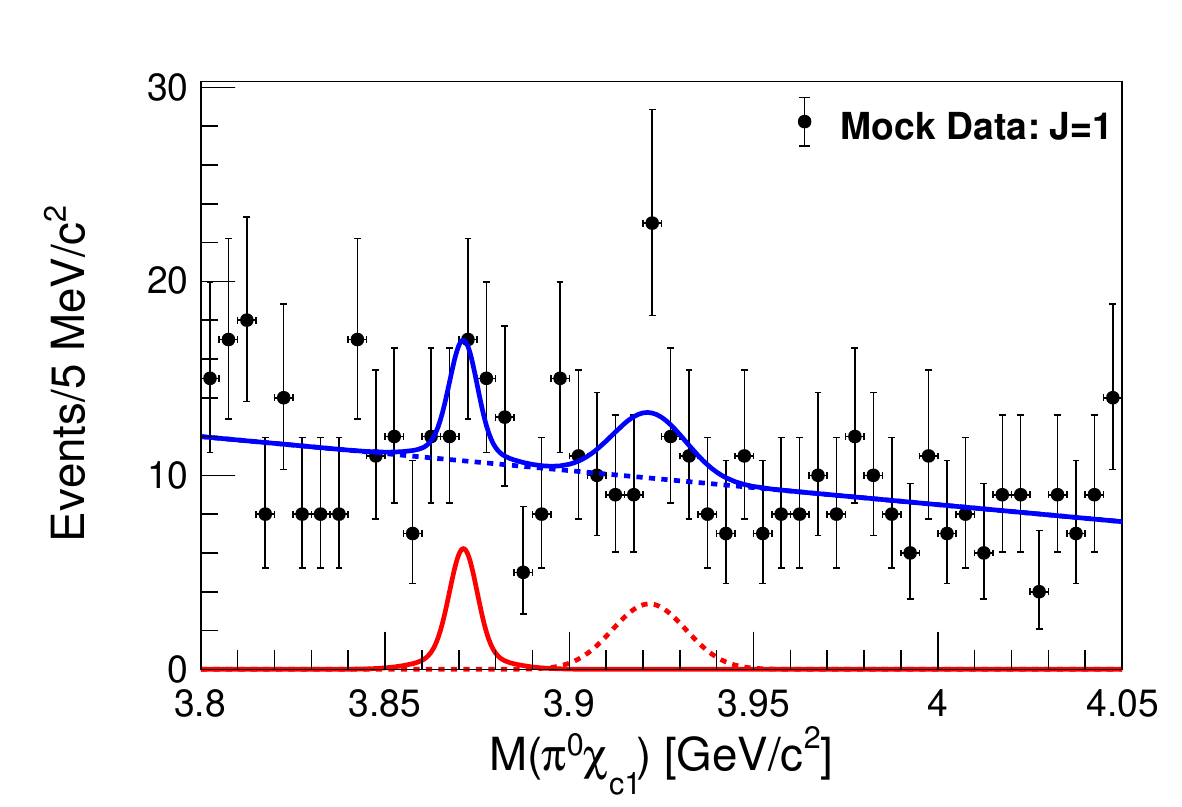} \put(-165,115){{\large \bf (c)}} 
\includegraphics[width=0.45\textwidth]{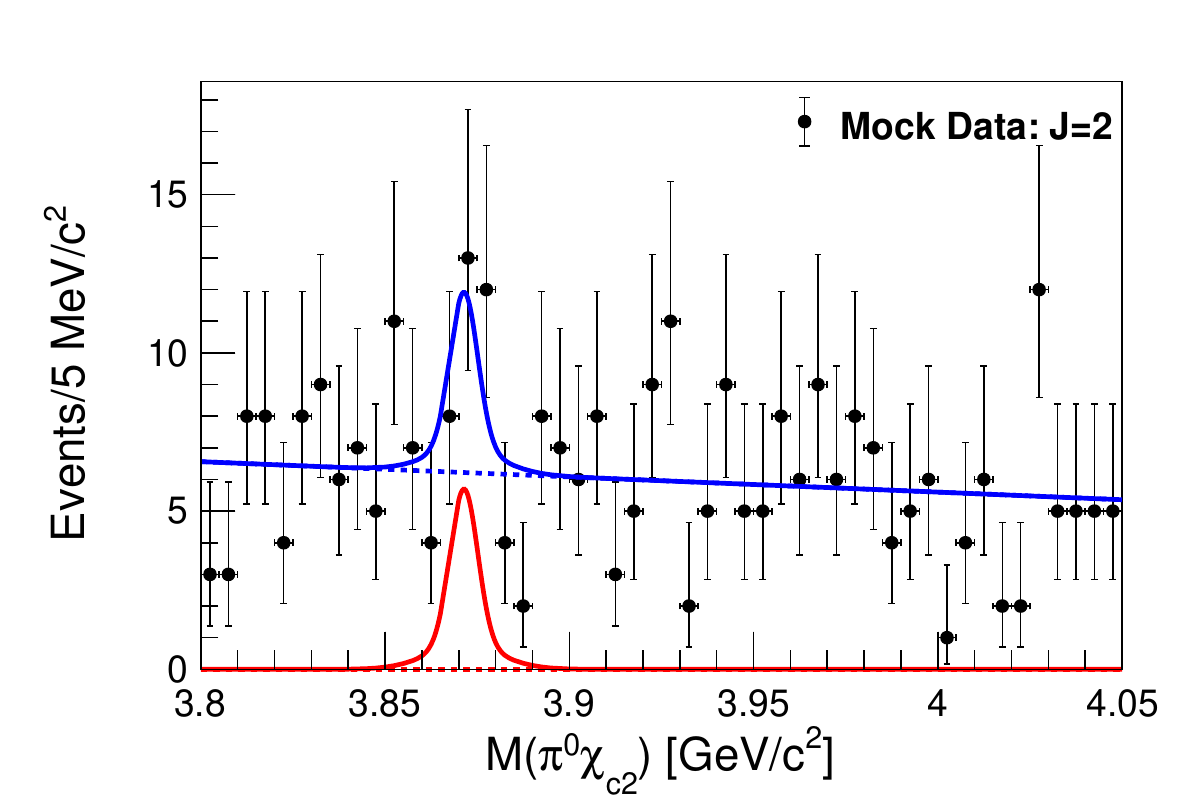} \put(-165,115){{\large \bf (d)}} 
\caption{The invariant mass spectra of $\pi^0\chi_{cJ}$ in $B^+\to \pi^0\chi_{cJ}K^+$ decays for (a) all $J$ summed up, and (b) $J=0$, (c) $1$, and (d) $2$, respectively, in combined Belle and Belle~II mock data.
The dots with error bars represent mock data samples, blue lines show the total fit results, blue dashed lines show the total fitted backgrounds, and red solid and dashed lines are fitted $\X$ and $X(3915)$ signals, respectively.
Note that all the plots are from simulated events, not from the real data.}
\label{fig:pi0chicj}
\end{figure}

The pionic transitions from the $\X$ to $\chi_{cJ}$ $(J = 0,~1,~2)$ are expected to provide a sensitive probe of different theoretical interpretations for the $X(3872)$. Although these decay rates are expected to be very small for a conventional $c\bar{c}$ state hypothesis~\cite{Dubynskiy:2007tj}, they can be sizeable 
if the $\X$ is a tetraquark or hadronic molecule~\cite{Dubynskiy:2007tj,Fleming:2008yn}.
The ratios $\Gamma_0:\Gamma_1:\Gamma_2$ of the decay widths ($\Gamma_0=\Gamma(\X\to\pi^0\chi_{c0})$, $\Gamma_1=\Gamma(\X\to\pi^0\chi_{c1})$, and $\Gamma_2=\Gamma(\X\to\pi^0\chi_{c2})$)
are predicted by different theoretical models~\cite{Dubynskiy:2007tj,Fleming:2008yn,Dong:2009yp,Zhou:2019swr,Wu:2021udi}, which offer a complementary probe to discriminate among different scenarios for $\X$. 
Previously, only the transition to $J=1$ state was observed by BESIII~\cite{BESIII:2019esk}, and no significant signals were seen by BESIII and Belle for the other two modes~\cite{Belle:2019jeu,BESIII:2022kow}. Recently the $\X\to\pi^0\chi_{cJ}$ and $X(3915)\to\pi^0\chi_{c1}$ decays were searched for in $B^+\to \pi^0\chi_{cJ}K^+$ decays using a total data sample of 1.2 ${\rm ab}^{-1}$ in Belle and Belle~II collected at the $\Yfour$ resonance.
The $\chi_{c0}$ candidates are constructed using $\pi^+\pi^-$, $K^+K^-$, and $K^+K^-K^+K^-$ events, and the $\chi_{c1,c2}$ candidates are constructed using $\gamma J/\psi(\to\ell^+\ell^-)$ ones.
Figure~\ref{fig:pi0chicj} shows the invariant mass distributions of $M(\pi^0\chi_{cJ})$ for all $J$ summed up and $J=0,1,2$, respectively, in mock data sample.
The mock data consists of generic MC and signal MC samples. The generic MC samples, including $e^+e^-\to B^+B^-$ and $e^+e^-\to B^0\bar B^0$ inclusive decay events, are scaled to the integrated luminosity of the data. The $X(3872)$ and $X(3915)$ signal yields are generated by assuming the ratios of branching fractions $\mathcal{B}(\X\to\pi^0\chi_{c1})/\mathcal{B}(\X\to\pi^+\pi^-J/\psi)=3.6$,~0.88,~1.1 for $J=0$, 1, 2~\cite{BESIII:2019esk,BESIII:2022kow}, respectively, and $\mathcal{B}(B^+\to X(3915)K^+)\times\mathcal{B}(X(3915)\to\pi^0\chi_{c1})=\mathcal{B}(B^+\to \X K^+)\times\mathcal{B}(\X\to\pi^+\pi^-J/\psi)$, with the previous Belle measurement of $\mathcal{B}(B^+\to \X K^+)\times\mathcal{B}(\X\to\pi^+\pi^-J/\psi)=8.63\times10^{-6}$~\cite{Belle:2011vlx} as input. A simultaneous fit to the mass spectra is performed to obtain the signal yields.
We perform $10^{3}$ sets of simulated pseudoexperiments with the above configuration. 
The fitted results in Fig.~\ref{fig:pi0chicj} represent the average signal yields from $10^{3}$ sets of simulated pseudoexperiments.
In Fig.~\ref{fig:pi0chicj}, clear $\X\to\pi^0\chi_{cJ}$ and $\X\to\pi^0\chi_{c0}$ signals are observed, with the significances larger than 6.5$\sigma$ and 6.3$\sigma$, respectively, with systematic uncertainties included.
No significant signals with systematic significance larger than $3\sigma$ are found from other modes. Thus, if the above assumed ratios are correct, with the currently available Belle and Belle II data, it is promising to observe the $\X\to\pi^0\chi_{cJ}$ decay. 

Further studies of the di-pion transitions from the $\X$ may also provide valuable information on the nature of the latter and reveal its possible partners. In particular, it is argued in Ref.~\cite{Ji:2025hjw} that a comparative study of the $\pi\pi J/\psi$ subsystem produced in the decays $B\to K \pi\pi J/\psi$ between the neutral and the charged $B$'s may show a signal of the isovector partner of the $\X$, advocated to exist in Ref.~\cite{Zhang:2024fxy}. Furthermore, the predictions of the molecular and compact tetraquark models for the isospin partner of the $\X$ considerably differ from each other thus providing an additional option to probe the nature of the $\X$ \cite{Ji:2025hjw}.

\subsubsection{The absolute branching fraction of $\X\to\pi^+\pi^-J/\psi$}

In $B$-factory experiments, the absolute branching fraction of $\X\to\pi^+\pi^-J/\psi$ can be measured.
The $\Yfour$ almost exclusively decays into a $B\bar B$ pair. By exclusively reconstructing one $B$ and identifying the $K^+$ in the decay of the other $B$ meson, one can obtain the branching fraction for $B^+\to K^+\X$. The product of the branching fraction of $B^+\to K^+\X$ and the branching fraction of $\X\to\pi^+\pi^-J/\psi$ has been determined before using the exclusive reconstruction method. Therefore, one can further obtain the absolute branching fraction of $\X\to\pi^+\pi^-J/\psi$. A precise determination of the latter is essential for elucidating the nature of the $\X$. In particular, if the $\X$ is a tetraquark state, this branching fraction should be approximately 50\%~\cite{Maiani:2004vq} while it
is less than 10\% for a molecule 
~\cite{Braaten:2005ai,Barnes:2003vb}.

\begin{figure}[t!]
\centering
\includegraphics[width=0.6\textwidth]{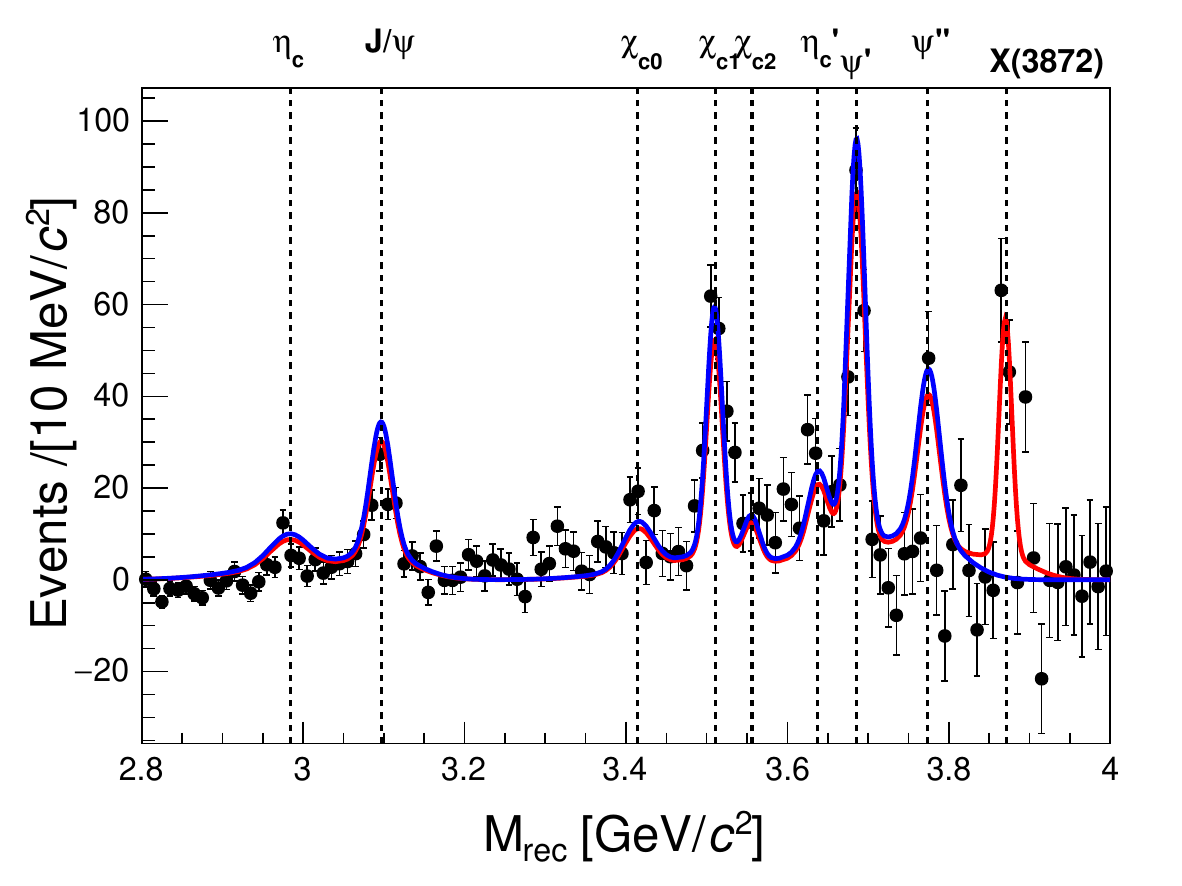}
\caption{
The background-subtracted recoil mass distribution between 2.8 and 4 GeV/$c^2$ from 711 fb$^{-1}$ mock MC samples at Belle. The fit function (red) peaks for nine particles, indicated by the dashed lines. The fit function where the $\X$ yield is forced to zero is drawn in blue. The $\X$ signal is scaled according to the PDG value of $\mathcal{B}(B^+ \to K^+ \X)$~\cite{ParticleDataGroup:2024cfk}.
Note that the plotted distribution contains simulated events rather than real data.
}
\label{fig:sig_bkgRej}
\end{figure}

In 2006 and 2017, using $232\times10^6$ and $772\times10^6$ $B\bar B$ pairs, BaBar and Belle searched for the $\X$ in $B$ decays using a recoil mass technique~\cite{BaBar:2005pcw,Belle:2017psv}. No clear signals were observed. The
90\% C.L. upper limits on the $\BR(B^+\to K^+\X)$ were set to be 3.2$\times10^{-4}$ and 2.6$\times$10$^{-4}$ at BaBar and Belle, respectively.
BaBar updated this measurement in 2020~\cite{BaBar:2019hzd}.
In this analysis, if more than one $B$ candidates were found in an event, all candidates were retained to increase the efficiency and then two dedicated neural networks were applied to suppress backgrounds.
As a result, BaBar reported the measurement 
${\mathcal B}(B^+ \to K^+ \X) = (2.1 \pm 0.6 \pm 0.3) \times 10^{-4}$ with a significance of $3\sigma$. This entails $\BR(\X\to\pi^+\pi^-J/\psi)=(4.1\pm1.3)\%$, supporting the hypothesis of a molecular nature for this resonance.
Although $B^+ \to K^+ \X$ has been measured inclusively, the low significance ($3\sigma$) and 32\% relative uncertainty are still insufficient to firmly discriminate between competing theoretical models and limit the precision of absolute branching fraction measurements for $\X$ decays. Using the large data sets, Belle and Belle II propose a new inclusive study of $B^+ \to K^+ \X$ with an improved $B$-tagging method and a DeepSets-based neural network~\cite{Zaheer:2017wmg} for background suppression. 
At a fixed background rejection level of 80\%, the new method achieves a signal efficiency of approximately 94\%, a considerable improvement over the corresponding efficiency of 56\% in the BaBar measurement~\cite{BaBar:2019hzd}. After a selection criterion is applied to the output of the DeepSets network, the resulting recoil mass distribution after subtracting the background from 711 fb$^{-1}$ mock data at Belle is shown in Fig.~\ref{fig:sig_bkgRej}.
The mock data consists of generic MC and signal MC samples. The generic MC samples include $e^+e^-\to B^+B^-$, $e^+e^-\to B^0\bar B^0$, and $e^+e^-\to q\bar q$ ($q=u,~d,~s,~c$) events. The signal samples are generated according to the PDG value of $\mathcal{B}(B^+ \to K^+ \X)$~\cite{ParticleDataGroup:2024cfk}.
After subtracting the fitted background with a second-order exponential function, the spectrum is fitted with double-Gaussian functions to describe the nine visible charmonium(-like) states: $\eta_c$, $J/\psi$, $\chi_{c0}$, $\chi_{c1}$, $\chi_{c2}$, $\eta_c(2S)$, $\psi^{\prime}$, $\psi(3770)$, and $\X$. The parameters of these signal shapes are fixed to those obtained from independent signal MC fits. The statistical significance of the $\X$ signal exceeds $5\sigma$, as illustrated in Fig.~\ref{fig:sig_bkgRej}.
Belle and Belle II anticipate that this analysis will improve the precision of ${\mathcal B}(B^+ \to K^+ \X)$ and thereby contribute to understanding the nature of the $\X$.

\subsubsection{Observation of $T_{c\bar c\bar s1}(4000)$}

\begin{figure}[t!]
\centering
\includegraphics[width=1.0\textwidth]{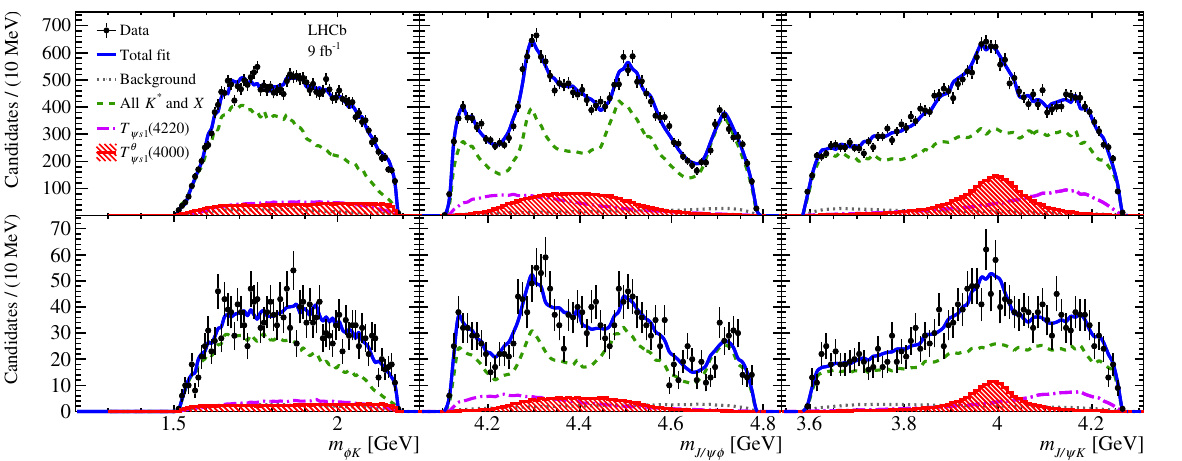}
\caption{
Distributions of (left) $m_{\phi K}$, (middle) $m_{J/\psi \phi}$, and (right) $m_{J/\psi K}$, overlaid with the corresponding projections of the default fit model. 
The upper and lower rows correspond to the $B^{+}\to J/\psi \phi K^{+}$ and $B^{0}\to J/\psi \phi K_S^{0}$ decays, respectively~\cite{LHCb:2023hxg}.
}
\label{fig:Tpsis4000}
\end{figure}

Since 2021, several tetraquark candidates with open charm and strange quark content $(c\bar{c}u\bar{s})$ have been reported. 
The charged state $Z_{cs}(3985)^{+}$ was first observed by the BESIII Collaboration in the
$D_s^{+}D^{*0}$ and $D_s^{*+}D^{0}$ invariant-mass spectra~\cite{BESIII:2020qkh}. Subsequently, the LHCb experiment reported two charged structures, $T^{\theta}_{\psi s1}(4000)^{+}$ and
$T_{\psi s1}(4220)^{+}$, in the $B^{+}\to J/\psi \phi K^{+}$ decay~\cite{LHCb:2021uow}\footnote{Here we employ both naming conventions $Z_{cs}$ and $T_{c\bar c\bar s1}/T^{\theta}_{\psi s1}$ simultaneously to better comply with the notations used in the cited works.}. Although the masses of the $Z_{cs}(3985)^{+}$ and $T^{\theta}_{\psi s1}(4000)^{+}$ are
very close to each other, their measured widths differ significantly, indicating that the nature underlying these structures may not be identical. 
In the meantime, it is claimed in theoretical work \cite{Ortega:2021enc} that both line shapes from BESIII and LHCb can be reconciled within a particular coupled-channel scheme as governed by the same pole of the amplitude. The existence of a near-threshold state in this energy region is put in question in Ref.~\cite{Ikeno:2020mra}, where the interaction in the system is found to be not strong enough to produce a bound state or resonance but sufficient to produce a large accumulation of strength at the threshold. Yet, the most popular theoretical interpretations for the $Z_{cs}$ include its compact tetraquark assignment \cite{Maiani:2021tri} or a hadronic molecule picture \cite{Meng:2020ihj,Yang:2020nrt,Wang:2020htx,Baru:2021ddn,Du:2022jjv}, both predicting the existence of a partner state in the same mass range. In the compact tetraquark model, the observed state $Z_{cs}(3985)$ and its predicted partner $Z_{cs}(4003)$ mirror the situation observed for the $\X$ and $Z_c(3900)$ in the non-strange sector \cite{Maiani:2021tri}. In the molecular model, the existence of the $Z_{cs}(3985)$ entails the existence of its spin partner $Z_{cs}'$ near the $\bar{D}_s^*D^*$ threshold \cite{Meng:2020ihj,Yang:2020nrt,Wang:2020htx,Baru:2021ddn}. Meanwhile, an alternative scenario, also consistent with the existing BESIII data but requiring a certain fine tuning of the parameters of the interaction, is additionally discussed in Ref.~\cite{Baru:2021ddn} where no partner state $Z_{cs}'$ appears. Therefore, further experimental studies are essential to clarify the properties of the $Z_{cs}$ family.
One important approach is a search for the corresponding isospin partners of these charged states. 
Recently, the BESIII Collaboration reported evidence
for a neutral partner, $Z_{cs}(3985)^{0}$, in the $D_s^{+}D^{*-}$ and $D_s^{*+}D^{-}$ invariant-mass spectra~\cite{BESIII:2022qzr}.
The decay $B^{0}\to J/\psi \phi K_S^{0}$ provides an excellent environment to
search for neutral $T^0_{\psi s}$ states. An amplitude analysis of
$B^{0}\to J/\psi \phi K_S^{0}$ decays is performed using the full Run~1 and
Run~2 proton--proton collision data collected by the LHCb detector~\cite{LHCb:2023hxg}. Owing to
the limited size of the $B^{0}$ sample and the complex resonant structures across the Dalitz plot, a simultaneous fit is carried out to the
$B^{0}\to J/\psi \phi K_S^{0}$ and $B^{+}\to J/\psi \phi K^{+}$ samples, using the same model developed in Ref.~\cite{LHCb:2021uow}. In the nominal fit, the masses, widths, and
helicity couplings of all components are constrained by isospin symmetry, with the exception of the $T^{\theta}_{\psi s1}(4000)$ contribution. Figure~\ref{fig:Tpsis4000} shows the fit projections of the nominal model.
Evidence for the neutral state $T^{\theta}_{\psi s1}(4000)^{0}$ in $m(J\psi K)$ spectrum is found with a
statistical significance of about $4\,\sigma$. Its mass and
BW width are measured to be
\be
M\!\left(T^{\theta}_{\psi s1}(4000)^{0}\right) = (3991\,^{+12}_{-10}\,
^{+9}_{-17})~\mbox{\mev}, \qquad
\Gamma\!\left(T^{\theta}_{\psi s1}(4000)^{0}\right) = (105\,^{+29}_{-25}\,
^{+17}_{-2})~\mathrm{MeV}.
\ee
The mass difference between the neutral and charged states is measured to be
\begin{equation}
\Delta M = M\!\left(T^{\theta}_{\psi s1}(4000)^{0}\right)
- M\!\left(T^{\theta}_{\psi s1}(4000)^{+}\right)
= (-12\,^{+11}_{-10}\,^{+6}_{-4})\,~\mbox{\mev},\nonumber
\end{equation}
which is consistent with isospin symmetry. Then, if isospin symmetry is imposed on the $T^{\theta}_{\psi s1}(4000)$ states, the significance of the
$T^{\theta}_{\psi s1}(4000)^{0}$ signal increases to $5.4\,\sigma$.

\subsubsection{Observation of $X(3960)$}

The state presently known as $\chi_{c0}(3915)$ (originally named $X(3915)$) was first observed in the $\omega J/\psi$ final
state by the Belle Collaboration~\cite{Belle:2004lle,BaBar:2010wfc,Belle:2009and,BaBar:2012nxg}.
Its properties are difficult to accommodate within conventional charmonium assignments, such as $\chi_{c0}(2P)$ or 
$\chi_{c0}(3P)$, since
decays of a conventional $\chi_{c0}$ state into $\omega J/\psi$ are expected to be
OZI suppressed.
This tension is further reinforced by the long-standing absence of observed decays
into open-charm final states such as $D^{(*)}\bar{D}^{(*)}$, together challenging a simple $c\bar{c}$ or $c\bar{c}q\bar{q}$ ($q=u,~d$) interpretation~\cite{Barnes:2005pb,Radford:2007vd,Li:2009zu,Wang:2014voa,Guo:2012tv}. In Ref.~\cite{Lebed:2016yvr}, the $\chi_{c0}(3915)$ is proposed as the lightest $c\bar{c}s\bar{s}$ state. The QCD sum-rule calculations favour its interpretation as a tetraquark, either with 
a $c\bar{c}q\bar{q}$ ($q=u,~d$) or a $c\bar{c}s\bar{s}$ quark configuration~\cite{Chen:2017dpy}. 
It was conjectured in Ref.~\cite{Zhou:2015uva} that the $\chi_{c0}(3915)$ could be the helicity-0 realisation of a $J^{PC}=2^{++}$ tensor charmonium-like state --- a suitable candidate $\chi_{c2}(3930)$ was indeed previously reported by both Belle \cite{Belle:2005rte} and BaBar Collaborations~\cite{BaBar:2010jfn}. 
However, the proposed scenario called for the helicity-0 dominance in the two-photon decay, which was at odds with the well established property of conventional tensor mesons \cite{Li:1990sx} but could potentially take place for an exotic state. This hypothesis was scrutinized in Ref.~\cite{Baru:2017fgv}, where the corresponding tensor $D^*\bar{D}^*$ molecular state was treated as a HQSS partner of the $\X$, and helicity-2 dominance in the two-photon decay was found to take place for it as well. In 2020, the $\chi_{c0}(3930)$ state was observed by the LHCb Collaboration in the
$D^{+}D^{-}$ invariant-mass spectrum~\cite{LHCb:2020bls,LHCb:2020pxc}. Its measured mass,
width, and spin--parity quantum numbers are compatible with those of the $\chi_{c0}(3915)$ resonance
previously observed in the $\omega J/\psi$ channel, suggesting that the two structures
could correspond to the same underlying state. Some LQCD studies predict a narrow
$0^{++}$ resonance just below the $D_s^{+}D_s^{-}$ threshold, with a strong coupling to
$D_s^{+}D_s^{-}$ and a much weaker coupling to $D\bar{D}$, which may be related to the
narrow $\chi_{c0}(3915)/\chi_{c0}(3930)$ observed experimentally~\cite{Prelovsek:2020eiw}. In parallel, phenomenological
studies have also proposed a $D_s^{+}D_s^{-}$ molecular interpretation~\cite{Hidalgo-Duque:2012rqv,Li:2015iga,Liu:2021xje,Meng:2020cbk,Dong:2021juy,Ji:2022uie,Bayar:2022dqa,Ji:2022vdj,Shi:2024llv}.
Meanwhile, the results of other LQCD calculations demonstrate no evidence for bound-state or resonance singularities in the energy region between the $\chi_{c0}(1P)$ state and the $D_s\bar{D}_s$ threshold
\cite{Wilson:2023hzu,Wilson:2023anv,Wilson:2026bhu}, so further investigations are necessary to clarify the situation.

\begin{figure}[t!]
\centering
\hspace{-0.8cm}
\includegraphics[width=0.33\textwidth]{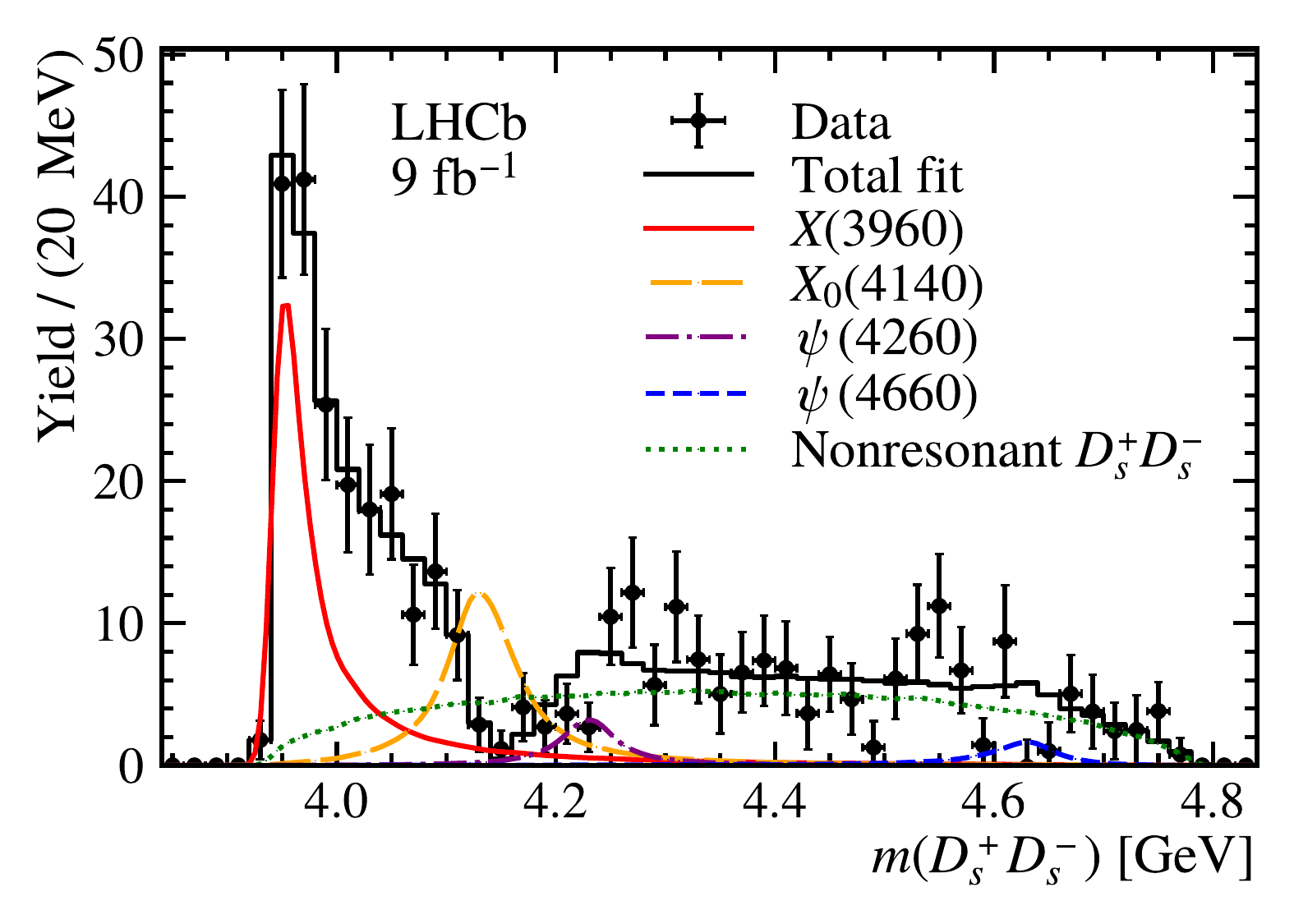} 
\includegraphics[width=0.33\textwidth]{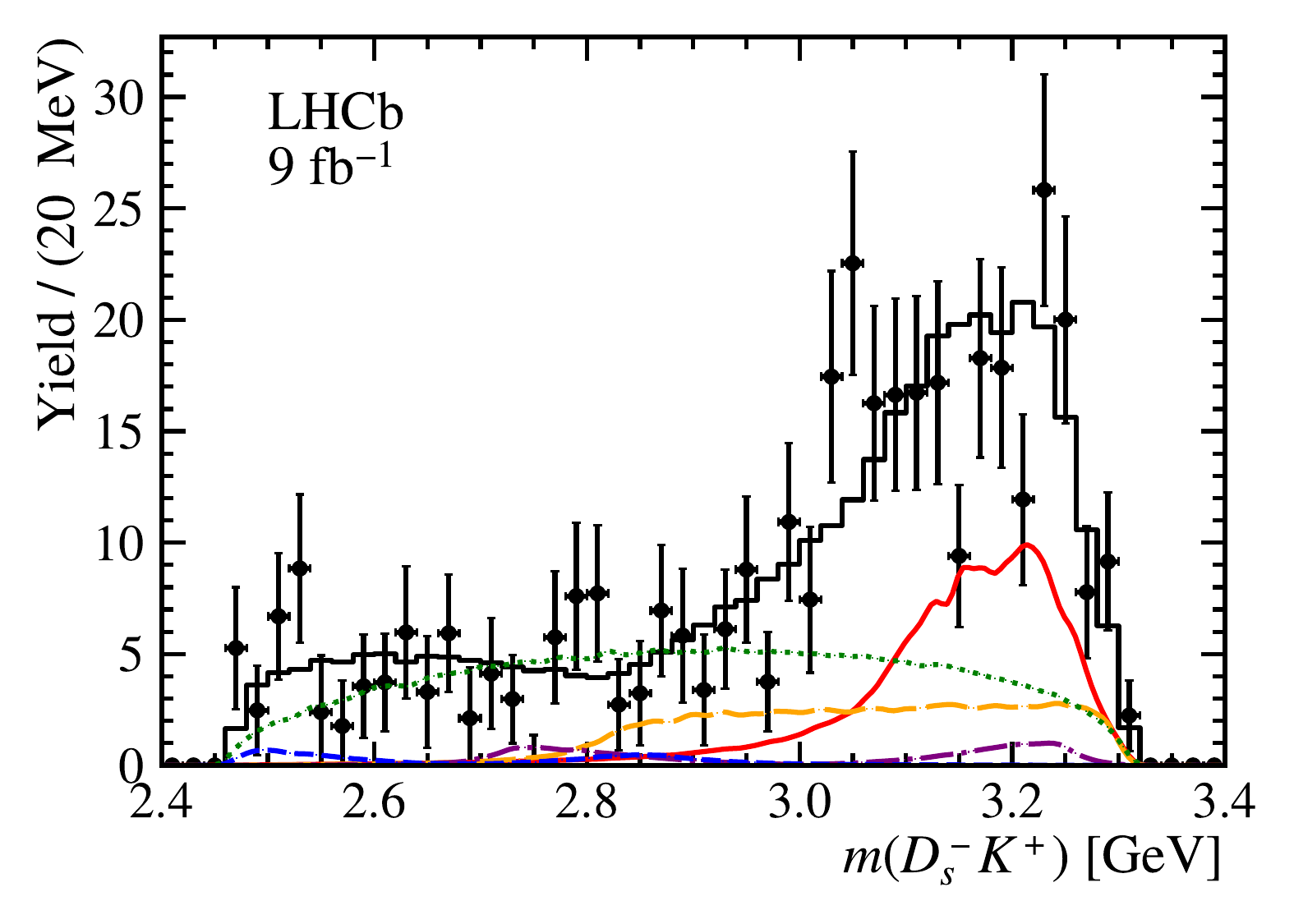}
\includegraphics[width=0.33\textwidth]{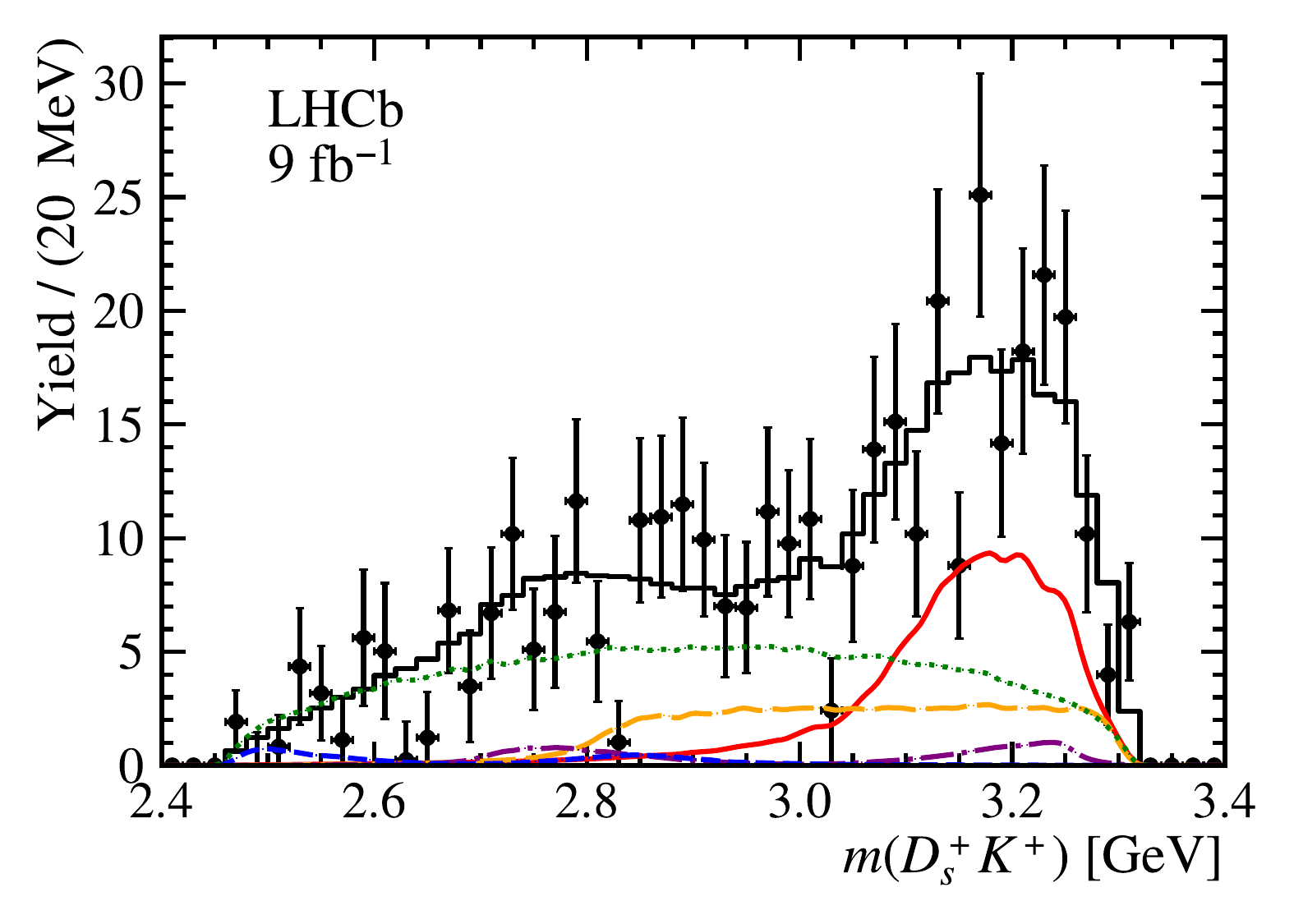}
\caption{Background-subtracted invariant-mass distributions (left) $m(D^+_sD^-_s)$, (middle) $m(D^+_s K^+)$, and (right) $m(D^-_s K^+)$ for the $B^+\to D^+_sD^-_sK^+ $ signal. 
The projections of the fit with the amplitude model are also shown~\cite{LHCb:2022aki}.}
\label{fig:X3960}
\end{figure}

Taken together, the above results point at the presence of a resonant structure in the vicinity
of the $D_s^{+}D_s^{-}$ threshold in the corresponding invariant-mass spectrum. 
LHCb has performed a full amplitude analysis of the
$B^{+}\!\to D_s^{+} D_s^{-} K^{+}$ decay using the complete Run~1 and Run~2 
datasets~\cite{LHCb:2022aki}. The invariant-mass projections in $M(D_s^{+}D_s^{-})$, 
$M(D_s^{+} K^{+})$, and $M(D_s^{-} K^{+})$ are shown in Fig.~\ref{fig:X3960}. 
A pronounced structure near the $D_s^{+}D_s^{-}$ threshold, modeled with a one-channel Flatté-like lineshape including only the $D_s^{+}D_s^{-}$ mode, is identified as the $X(3960)$ state with a significance exceeding $12\,\sigma$ and quantum numbers determined to be $J^{PC}=0^{++}$. The measured mass and width of the $X(3960)$, listed in Table~\Ref{tab:X3960X04140}, are consistent with those of the $\chi_{c0}(3930)$ within $3\,\sigma$~\cite{LHCb:2020bls,LHCb:2020pxc}.
Assuming that $X(3960)$ and $\chi_{c0}(3930)$ correspond to the same state, an additional $D^{+}D^{-}$ coupled channel is introduced in the Flatté parameterization. 
The fit results remain compatible with the one-channel model, 
and the partial width for $X(3960)\to D_s^{+}D_s^{-}$ is found to be significantly larger than that for $X(3960)\to D^{+}D^{-}$. Since producing an $s\bar s$ pair from the vacuum is suppressed relative to $u\bar u$ or $d\bar d$, 
this behavior suggests that the $X(3960)$ has a substantial 
$c\bar c s \bar s$ component, indicating a possible exotic hadronic structure.
Further experimental studies are essential to clarify the nature of the 
$D_s^{+}D_s^{-}$ threshold enhancement. In particular, measuring the relative 
branching fractions to the $D_{(s)}\bar{D}_{(s)}$ and $\omega J/\psi$ channels in a 
different production environment, such as two-photon fusion processes at Belle~II, 
would provide valuable complementary information. In this direction, a study of the $B^- \to K^- \omega J/\psi$ decay was performed in Ref.~\cite{Abreu:2023rye}, where distinct features in the $\omega J/\psi$ mass spectrum were predicted under the assumption that the $X(3960)$ is a molecular state predominantly composed of $D\bar D$ and $D_s\bar D_s$ components. 

In the $M(D_s^{+}D_s^{-})$ spectrum, a dip is observed near 
4140~\mev. This feature can be described either by introducing a $J^{PC}=0^{++}$ state, $X_{0}(4140)$, which interferes destructively with the nonresonant $0^{++}$ component and the $X(3960)$ amplitude, or by invoking coupled-channel effects associated with the $J/\psi\phi \leftrightarrow D_s^{+}D_s^{-}$ rescattering. 
The spin parity $J^{PC}=0^{++}$ of $X_{0}(4140)$ is incompatible with the $1^{++}$ quantum number of the $X(4140)$ resonance (also known as $\chi_{c1}(4140)$) observed in the $J/\psi\phi$ final state~\cite{ParticleDataGroup:2024cfk}. 
Given that the significance of $X_{0}(4140)$ is only $3.5\,\sigma$, additional measurements will be crucial for establishing its existence and understanding the underlying 
phenomena.
Alternative theoretical hypotheses for the measured states should also be scrutinized. For example, if the $\chi_{c0}(3930)$ is assigned as a $D_s\bar D_s$ molecule, 
then the $X(4140)$ can be understood as its spin-partner $D_s^*\bar D_s^*$ molecular state \cite{Branz:2009yt}, 
and the events currently prescribed to the $X(3960)$ could potentially be identified with the tail from the $\chi_{c0}(3930)$ signal.

\begin{table}[t!]
 \centering
 \caption{Properties of $X(3960)$ and $X_0(4140)$ measured in Ref.~\cite{LHCb:2022aki}. }
 \label{tab:X3960X04140}
 \begin{tabular}{lccc}\hline
 Component & $J^{PC}$ & $M_0$~(\mev) & $\Gamma_0$~(MeV) \\ \hline
 $X(3960)$ & $0^{++}$ & $3956\pm5\pm10$ & $43\pm13\pm8$ \\
 $X_0(4140)$ & $0^{++}$ & $4133\pm6\pm6$ & $67\pm17\pm7$ \\
 \hline
 \end{tabular}
\end{table}

\subsubsection{Observation of $\chi_{c1}(4010)$}

\begin{figure}[t!]
\centering
\includegraphics[width=0.9\linewidth]{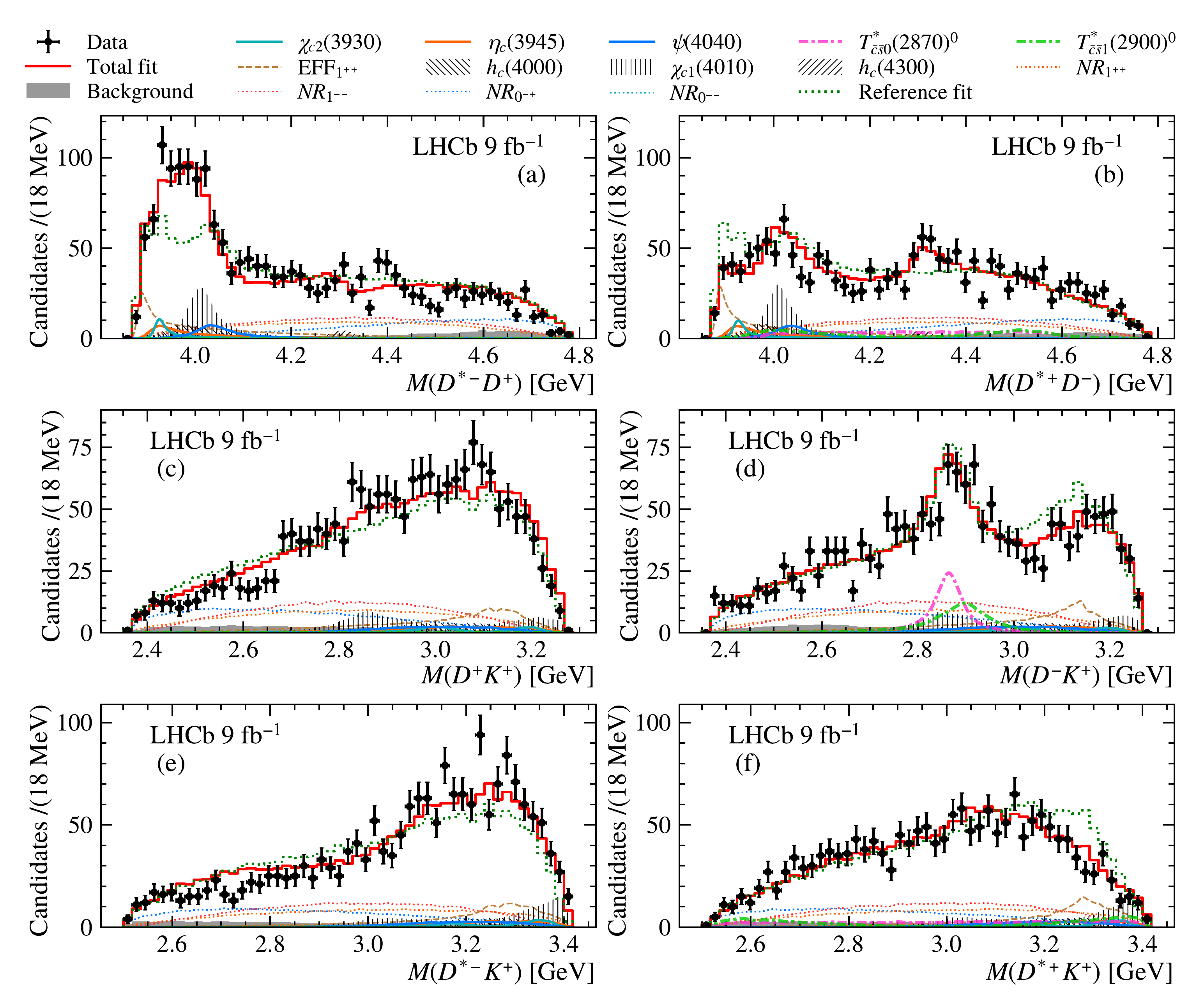}
\caption{Distributions of two-body invariant masses: (a)~$M(D^{*-} D^{+})$, (c)~$M(D^{+} K^{+})$, and (e)~$M(D^{*-} K^{+})$ in the $B^{+}\rightarrow D^{*-}D^{+}K^{+}$ sample; (b)~$M(D^{*+} D^{-})$, (d)~$M(D^{-} K^{+})$, and (e)~$M(D^{*+} K^{+})$ in the $B^{+}\to D^{*+}D^{-}K^{+}$ sample. 
The fit results (red-solid lines) are overlaid on the data distributions. 
Contributions from different components are also shown in different line styles as indicated in the legend. 
The result of fitting the data using a model without the $h_{c}(4000)$, $\chi_{c1}(4010)$, and $h_{c}(4300)$ components is shown with green-dotted lines for comparison~\cite{LHCb:2024vfz}.}
\label{fig:chic14010}
\end{figure}

A simultaneous amplitude analysis of the decays 
$B^{+}\!\to D^{*\pm}D^{\mp}K^{+}$ is performed for the first time using the full Run~1 
and Run~2 $pp$ collision data collected by the LHCb experiment~\cite{LHCb:2024vfz}. The analysis 
exploits the $C$-parity relations between contributions of charmonium(-like) 
resonances appearing in the two charge-conjugate channels. As shown in 
Fig.~\ref{fig:chic14010}, four charmonium(-like) states decaying into $D^{*\pm}D^{\mp}$ are 
observed: $\eta_{c}(3945)$, $h_{c}(4000)$, $\chi_{c1}(4010)$, and $h_{c}(4300)$, 
with their $J^{PC}$ quantum numbers determined to be $0^{-+}$, $1^{+-}$, 
$1^{++}$, and $1^{+-}$, respectively. The $\eta_{c}(3945)$ is compatible with the 
previously reported $X(3940)$ state~\cite{Belle:2005lik,Belle:2007woe}, while the remaining three structures are 
observed for the first time. 
The observed $\chi_{c1}(4010)$ has quantum numbers $J^{PC}=1^{++}$ and is thus naturally associated with the $\chi_{c1}$ charmonium sector. However, its mass does not fit naturally into the established pattern defined by the $\chi_{c1}(3930)$ and $\chi_{c1}(4140)$ states. Being located near the open-charm $D\bar D^\ast$ threshold, the state is expected to be strongly affected by coupled-channel dynamics.
Recent studies, including a reanalysis of the LQCD energy levels reported in Ref.~\cite{Prelovsek:2013cra} and a separate LQCD calculation in Ref.~\cite{Li:2024pfg}, suggest a rich interplay between the $\chi_{c1}(2P)$ charmonium state and the $D\bar D^\ast$ channel. In the former case, it is found that a strong coupling to $D\bar D^\ast$ may generate an additional $1^{++}$ pole above threshold, with a mass around 4.0~GeV, which could be associated with the observed $\chi_{c1}(4010)$~\cite{Shi:2024llv}. In the latter study, mixing between the $\chi_{c1}(2P)$ state and noninteracting $D\bar D^\ast$ scattering levels is shown to shift finite-volume energy levels, pushing one level below threshold while raising another above. A subsequent $K$-matrix analysis of these spectra yields both a $D\bar D^\ast$ bound state, naturally identified with the $X(3872)$, and a resonance pole slightly above 4.0~GeV with a width of approximately 40--60~MeV, compatible with the LHCb observation of the $\chi_{c1}(4010)$. Nevertheless, the nature of the $\chi_{c1}(4010)$ is still not firmly established and requires further theoretical and experimental clarification.


In addition, the two resonant structures $T^{*\,0}_{\bar{c}\bar{s}0}(2870)$ and 
$T^{*\,0}_{\bar{c}\bar{s}1}(2900)$, previously observed in 
$B^{+}\!\to D^{+}D^{-}K^{+}$ decays~\cite{LHCb:2020bls,LHCb:2020pxc}, are incorporated into the 
$B^{+}\!\to D^{*+}D^{-}K^{+}$ amplitude model to describe the enhancement seen in 
Fig.~\ref{fig:chic14010}(e), thereby confirming their presence in this new decay channel. The measured ratio of branching fractions between
$B^{+}\!\to D^{*+}T^{*\,0}_{\bar{c}\bar{s}0}(2870)$ and
$B^{+}\!\to D^{*+}T^{*\,0}_{\bar{c}\bar{s}1}(2900)$ is determined to be
$1.17 \pm 0.31 \pm 0.48$.
This value is larger than the corresponding ratio
$0.18 \pm 0.05$ observed in
$B^{+}\!\to D^{+}T^{*\,0}_{\bar{c}\bar{s}0,1}$ decays, and may provide
further insight into the production mechanisms of the
$T^{*\,0}_{\bar{c}\bar{s}0,1}$ states, if their existence is firmly established.

\subsection{Pentaquarks}
\label{sec:penta}

\subsubsection{Theory overview}

An important milestone in the study of exotic baryons beyond the conventional three-quark configuration was the discovery by the LHCb Collaboration in 2015 of the charmonium pentaquark candidates $P_{c\bar c}(4450)$ and $P_{c\bar c}(4380)$ in the $J/\psi p$ invariant mass spectrum~\cite{LHCb:2015yax}. With increased statistics from Run-2 data, LHCb subsequently reported in 2019~\cite{LHCb:2019kea}, in the decay process $\Lambda_b \to J/\psi p K^-$, the observation of a lighter and narrow state $P_{c\bar c}(4312)$ and resolved the original $P_{c\bar c}(4450)$ structure into two distinct narrow states, $P_{c\bar c}(4440)$ and $P_{c\bar c}(4457)$. Remarkably, the existence of such hidden-charm pentaquark states with a predominantly molecular nature had been predicted nearly a decade earlier~\cite{Wu:2010jy, Wu:2010vk}, in qualitative agreement with the subsequently observed spectrum (see also Ref.~\cite{Xiao:2013yca}). Since their discovery, the nature of the three narrow $P_c$ states has been the subject of intense theoretical debate. In particular, several coupled-channel approaches consistent with HQSS --- some of them including one-pion exchange, --- in which the $P_c$ states are interpreted as $\Sigma_c^{(*)}\bar D^{(*)}$ real or virtual bound state hadronic molecules, successfully reproduce the observed structures, allow for spin–parity assignments, and predict the possible existence of a narrow $P_{c\bar c}(4380)$ state~\cite{Roca:2015dva,Du:2019pij,Yamaguchi:2016ote,Liu:2019tjn,Xiao:2019aya, PavonValderrama:2019nbk, Fernandez-Ramirez:2019koa, Du:2021fmf}. The spectrum and decay patterns of the $P_c$ states —-- including those of their spin partners --- within the BOEFT approach have been recently analyzed in Ref.~\cite{Brambilla:2025xma}. Some studies suggest that structures such as \(P_{c\bar{c}}(4312)\) or \(P_{c \bar{c}}(4457)\) could be explained by “triangle diagrams”, where intermediate particles are exchanged in such a way that they create an artificial peak in data without the need for a real pentaquark~\cite{Nakamura:2021qvy}. Consistency limitations of some of these states being merely a kinematic effect, rather than corresponding to a real resonance, were discussed since the earliest observations by LHCb~\cite{Guo:2015umn, Guo:2016bkl}. On the other hand, first LQCD studies of hidden-charm pentaquarks have started to emerge, providing first-principles insights into the $\Sigma_c\bar D^{(*)}$ systems and their possible bound or resonant nature~\cite{Xing:2022ijm} and the minor role played by the $J/\psi N$ and $\eta_c N$ channels~\cite{Skerbis:2018lew, Lyu:2024ttm}. For the pentaquark studies employing the QCD sum rules technique see, for example, Refs.~\cite{Ozdem:2022kei,Ozdem:2024rch,Ozdem:2025jda}.

\begin{figure}[t!]
\centering
\includegraphics[width=0.49\linewidth]{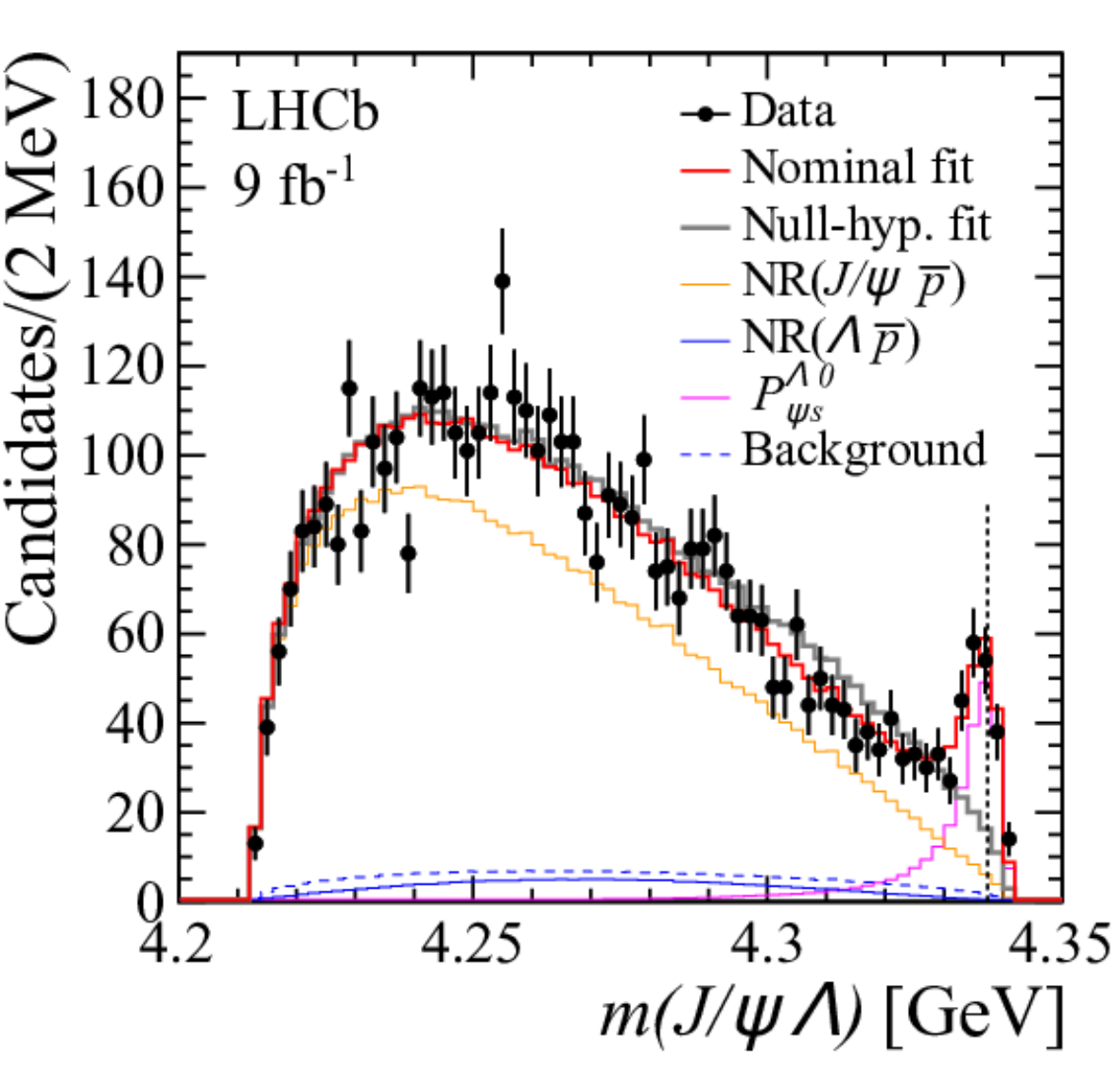}
\includegraphics[width=0.49\linewidth]{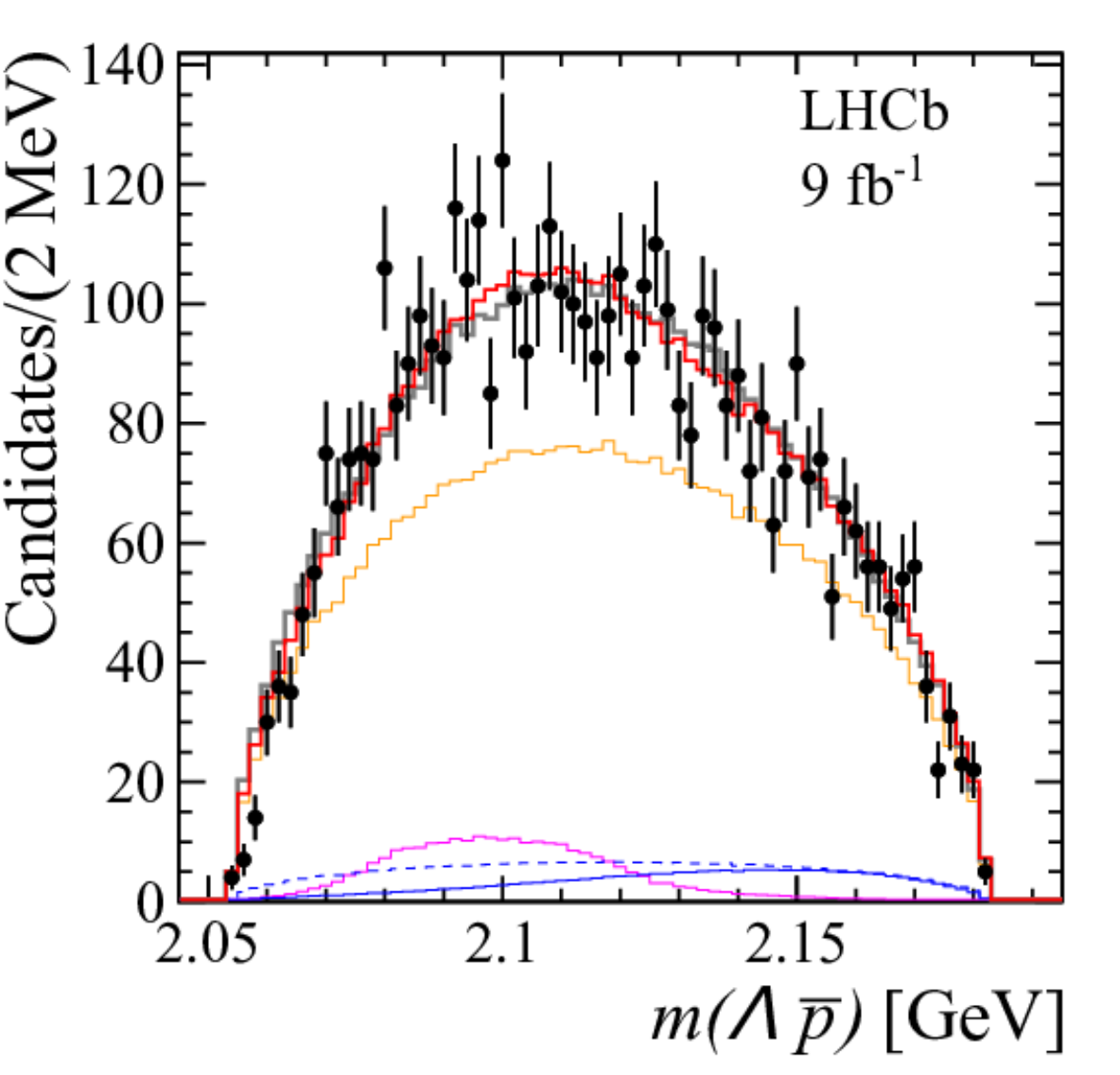} \\
\caption{\label{fig:Pc4338} Distributions of $J/\psi \Lambda$ and $\Lambda \bar{p}$ invariant masses in $B^{-} \to J/\psi \Lambda \bar{p}$ decay. 
Fit results to data using the nominal model are superimposed. The null-hypothesis model fit results are also shown in grey. The $\Xi_c^+D^-$ baryon-meson threshold at 4.337~\gev~is indicated with a vertical dashed line in the $m(J/\psi\Lambda)$ invariant mass distribution~\cite{LHCb:2022ogu}.} 
\end{figure}

Compact pentaquark interpretations based on constituent quark models and diquark correlations have been extensively explored following the LHCb discoveries. In these approaches, the $P_c$ states are described as tightly bound five-quark configurations, often organized in diquark–diquark–antiquark or diquark–triquark structures, and their spectra can be reproduced by effective chromomagnetic Hamiltonians or quark potential models --- see, for example, Refs.~\cite{Lebed:2015tna,Maiani:2015vwa,Santopinto:2016pkp,Ali:2017jda,Ortega:2016syt,Carlson:2003pn}. However, such models typically predict a rich spectrum of compact pentaquark states, many of which have not been observed experimentally. While compact quark-model descriptions provide useful qualitative insights, the proximity of the observed $P_c$ states to $\Sigma_c^{(*)}\bar D^{(*)}$ thresholds, together with coupled-channel and LQCD results, strongly suggests that long-range hadronic dynamics plays a dominant role, making a predominantly molecular interpretation highly plausible. There are also studies of hidden-charm pentaquarks using the QCD sum rule method (see, for example, Refs.~\cite{Azizi:2016dhy,Xiang:2017byz}), although the associated theoretical uncertainties are larger than those in EFT-based approaches.
Reviews containing various theoretical interpretations of the dynamics and structure of these exotic states can be found in Refs.~\cite{Chen:2016qju, Lebed:2016hpi, Guo:2017jvc, Liu:2019zoy}.

All theoretical frameworks mentioned above can naturally accommodate hidden-charm strange resonant structures with the quark content $c\bar c s q q$, and a vast literature on this subject exists. Here we cite only a few representative works such as Refs.~\cite{Wu:2010jy,Wu:2010vk,Santopinto:2016pkp,Chen:2015sxa,Chen:2016ryt,Xiao:2019gjd,Wang:2019nvm,Chen:2020uif,Peng:2020hql,Azizi:2021utt,Wang:2022mxy,Feijoo:2022rxf,Giachino:2022pws,Chen:2020kco,Wang:2022neq,Burns:2022uha,Meng:2022wgl,Chen:2022wkh,Ortega:2022uyu,Yang:2023dzb}. As will be discussed in detail below, two states, $P_{c\bar c s}(4338)^0$ and $P_{c\bar c s}(4459)^0$, have been reported by the LHCb Collaboration in the $J/\psi\,\Lambda$ invariant-mass distributions of different bottom-hadron decays. Within the molecular picture, these exotic baryons arise from coupled-channel dynamics of the $\bar D^{(*)}\Xi_c^{(\prime, *)}$ systems, described by effective Lagrangians respecting HQSS and SU(3) light-flavor symmetry. A distinctive feature of this sector is that, in the heavy-quark limit, the light degrees of freedom couple to spin zero in the $\Xi_c$ baryon as in the $\Lambda_c$, while they couple to spin one in the $\Xi_c^\prime$ as in the non-strange $\Sigma_c$. As a consequence, the $\bar D\Xi_c$\dash$\bar D_s\Lambda_c$ and $\bar D^*\Xi_c$\dash$\bar D_s^*\Lambda_c$ channels, connected by strong transition potentials, cooperate to generate a larger attraction than the corresponding $\bar D\Sigma_c$\dash$\bar D\Lambda_c$ and $\bar D^*\Sigma_c$\dash$\bar D^*\Lambda_c$ channels involved in the formation of the strangenessless $P_c$ states, since in the latter case the transition potentials between these channels are suppressed~\cite{Xiao:2019gjd, Feijoo:2022rxf}. 
 
\subsubsection{Observation of $P_{c\bar cs}(4338)^0$} 

Over the past decade, the LHCb Collaboration has reported the observation and evidence of several pentaquark candidates. More recently, evidence for a new pentaquark state $P_{c\bar{c}}(4338)$ has been observed in the $B_s^0 \to J/\psi p \bar{p}$ decay with a mass of $(4337\,^{+7}_{-4}\,^{+2}_{-2})$~\mev~and width of $(29\,^{+26}_{-12}\,^{+14}_{-14})$ MeV~\cite{LHCb:2019rmd,LHCb:2021chn}.
In addition, evidence for a strange pentaquark candidate $P_{c\bar{c}s}(4459)^0$ with quark content $c\bar{c}uds$ has been found in the $J/\psi\Lambda$ system produced in $\Xi_b^- \to J/\psi\Lambda K^-$ decays \cite{LHCb:2020jpq}. Experimentally, pentaquark candidates are often observed near the thresholds for the production of conventional baryon--meson pairs. For instance, the previously reported $P_{c\bar{c}}^{+}$ states are located close to the $\Sigma_c^{+}\bar{D}^{0}$ and $\Sigma_c^{+} \bar{D}^{*0}$ thresholds~\cite{LHCb:2015yax,LHCb:2019kea}, while the $P_{c\bar{c} s}^{0}$ state appears near the $\Xi_c^{0} \bar{D}^{*0}$ threshold~\cite{LHCb:2020jpq}.
The decay $B^{-} \to J/\psi \Lambda \bar{p}$ provides a unique laboratory to search
simultaneously for the $\bar{P}_{c\bar{c}}^{-}$ and $P_{c\bar{c} s}^{0}$ pentaquark candidates in the
$J/\psi \bar{p}$ and $J/\psi \Lambda$ systems, respectively. Owing to its large available
phase space, this decay allows sensitivity to pentaquark states located near various
baryon--meson thresholds, such as $\Lambda_c^{+} \bar{D}^{0}$ for $P_{c\bar{c}}^{+}$ states and
$\Lambda_c^{+} D_s^{-}$ or $\Xi_c^{+} D^{-}$ for strange pentaquark candidates.

Using the full Run~1 and Run~2 proton--proton collision data sets, the LHCb Collaboration performed a full amplitude analysis of the $B^{-} \to J/\psi \Lambda \bar{p}$ decay~\cite{LHCb:2022ogu}, reconstructing approximately $4400$ high-purity signal events.
In the nominal fit to data, a new narrow structure is observed in the $J/\psi \Lambda$ invariant-mass spectrum with a significance exceeding $15\,\sigma$, as shown in Fig.~\ref{fig:Pc4338}. The observed state, denoted as
$P_{c\bar{c} s}(4338)^{0}$, has a mass and width measured to be
$M = (4338.2 \pm 0.7 \pm 0.4)$~\mev~and
$\Gamma = (7.0 \pm 1.2 \pm 1.3)~\mathrm{MeV}$, respectively, with quantum numbers $J^{P} = 1/2^{-}$ preferred.
The $P_{c\bar{c} s}(4338)^{0}$ state contains at least five valence quarks, $\bar{c} c s u d$, and is located near the $\Xi_c^{+} D^{-}$ baryon--meson threshold, a feature that is highly relevant for interpretations of its internal structure.

\begin{figure}[t!]	
\centering
\includegraphics[width=0.6\textwidth]{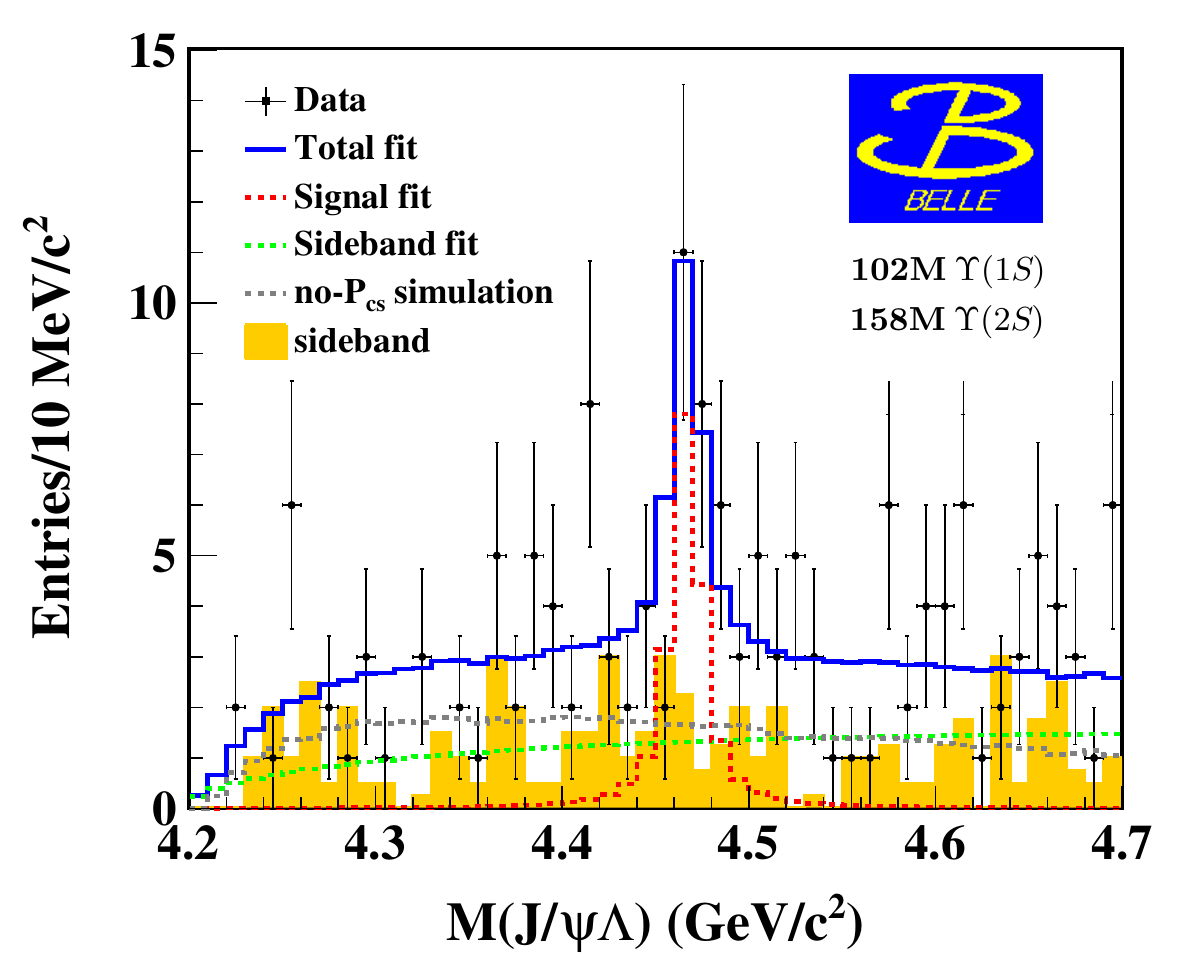} 
\caption{The distribution of the invariant mass of $J/\psi\Lambda$ in the
combined $\Upsilon(1S,2S)$ data sample~\cite{Belle:2025pey} and the fit results with the mass and
width of $P_{c\bar cs}(4459)^0$ constrained according to the LHCb measurement~\cite{LHCb:2020jpq}. The points with error bars are the data, and the
yellow histogram is the background estimated from the two dimensional $J/\psi$ and $\Lambda$ sideband regions. The solid curve shows the best fit
results. The red dashed curve shows the signal. The brown dashed
curve shows the no-$P^0_{c\bar cs}$ component. The green dashed curve
shows the fit to the background estimated from the sidebands.}
\label{fig:pentaquark_Belle}
\end{figure}

\subsubsection{Evidence of the $P_{c\bar cs}(4459)^0$ in $\Upsilon(1S,2S)$ decays}

In 2021, LHCb reported an evidence (3.1$\sigma$) of a pentaquark candidate state with a $udsc\bar c$ quark component, named as the $P_{c\bar cs}(4459)^0$, with
a mass of $(4458.8\pm 2.9^{+4.7}_{-1.1})$~\mev~and a width of
$(17.3\pm 6.5^{+8.0}_{-5.7})$ MeV, in the $J/\psi\Lambda$ substructure of the decay
$\Xi^-_b\to J/\psi\Lambda K^-$~\cite{LHCb:2020jpq}.
The ample gluons in the hadronic decays of $\Upsilon(1S,2S)$ provide an entry
to many potential exotic states, including the 
multi-quark ones~\cite{CLEO:2005mdr,Chen:2024aom,Chen:2024eaq}. Therefore, Belle used 102 million $\Upsilon(1S)$ events and 158 million $\Upsilon(2S)$ events to search for the $P_{c\bar cs}(4459)^0\to J/\psi\Lambda$~\cite{Belle:2025pey} in $\Upsilon(1S,2S)$ inclusive decays.
Clear signal events of $J/\psi\Lambda$ were found, with the number of those being $84\pm11$ and $140\pm17$ for $\Upsilon(1S)\to J/\psi\Lambda+{\rm anything}$ and for $\Upsilon(2S)\to J/\psi\Lambda+{\rm anything}$, respectively. The corresponding branching fractions for $\Upsilon(1S)\to J/\psi\Lambda+{\rm anything}$ and for $\Upsilon(2S)\to J/\psi\Lambda+{\rm anything}$ are $(36.9\pm5.3\pm2.4)\times10^{-6}$ and $(22.3\pm5.7\pm3.1)\times10^{-6}$, respectively, where the charged conjugated mode has been included.

The invarinat mass distribution of $J/\psi\Lambda$ from combined $\Upsilon(1S)$ and $\Upsilon(2S)$ data samples is shown in Fig.~\ref{fig:pentaquark_Belle}.
Since the excess is close to the mass of $P_{c\bar cs}(4459)^0$, a Gaussian constraint using prior knowledge of $P_{c\bar cs}(4459)^0$ mass and width from the LHCb measurement~\cite{LHCb:2020jpq} was included in the binned maximum-likelihood fit.
The total fitted result is shown by the blue curve in Fig.~\ref{fig:pentaquark_Belle}.
The signal yield for $P_{c\bar cs}(4459)^0\to J/\psi\Lambda$ is $21\pm5$ with a significance
of $3.3\sigma$ using a pseudoexperiment technique, including systematic uncertainties.
The authors also performed a fit without the mass and width
constraints. 
The fit yields a mass of $(4471.7\pm4.8\pm
0.6)$~\mev~and a width of $(22\pm13\pm3)$ MeV. 
The mass and width are consistent with those of the $P_{c\bar cs}(4459)^0$ as reported by LHCb,
with differences of 1.8$\sigma$ for the mass and 0.3$\sigma$ for the width.
The local significance is calculated to be 2.8$\sigma$ using a pseudoexperiment technique.
The branching fractions for $\Upsilon(1S)\to P_{c\bar cs}(4459)^0+{\rm anything}$ and $\Upsilon(2S)\to P_{c\bar cs}(4459)^0+{\rm anything}$ are $(3.5\pm2.0\pm0.2)\times 10^{-6}$ and $(2.9\pm1.7\pm0.4)\times 10^{-6}$, respectively.

\subsubsection{Search for pentaquark in $\Lambda_c^+D_s^-$}

\begin{figure}[t!]
\centering
\includegraphics[width=0.45\textwidth]{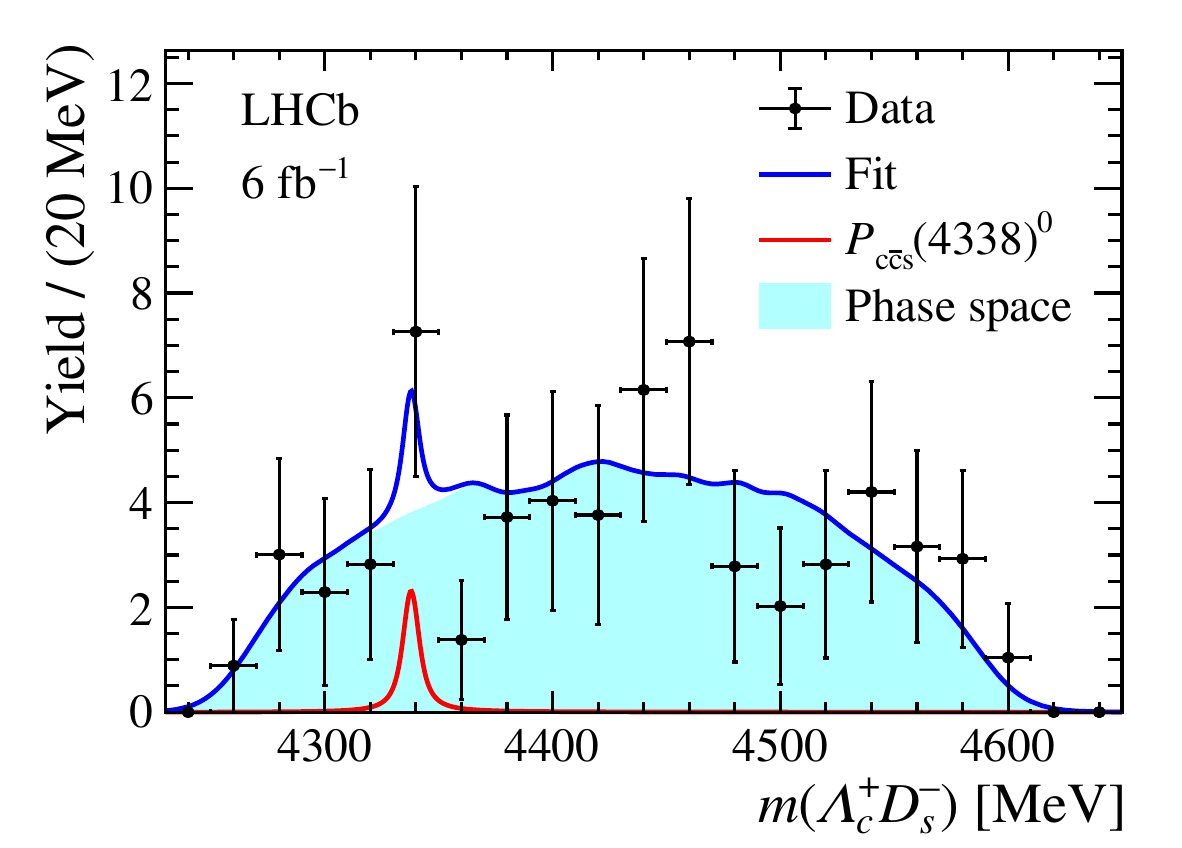}
\includegraphics[width=0.45\textwidth]{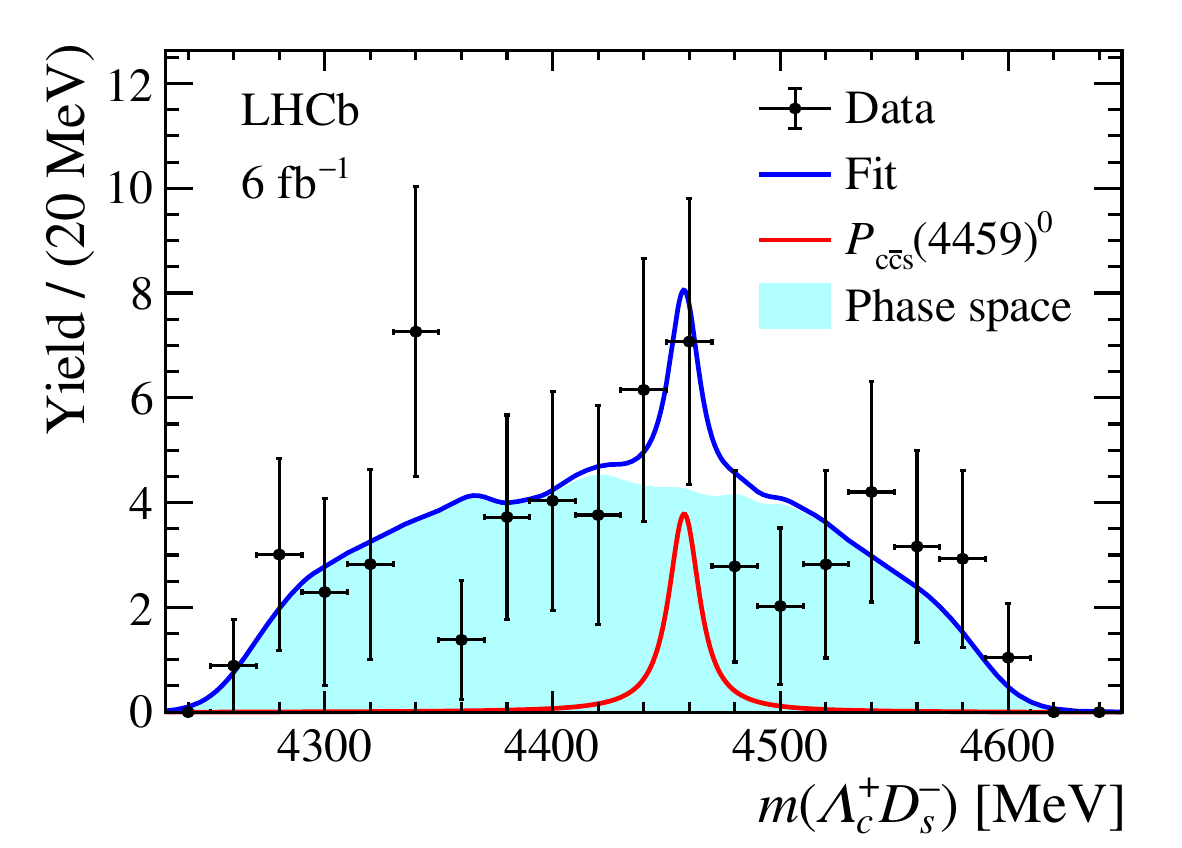}
\caption{Distributions of \mbox{$m(\Lambda_c^+D^-_s)$} for background-subtracted $\Lambda_b^0 \to \Lambda_c^+ D_s^- K^+ K^-$ decays with the fit results also shown. The red lines show the contributions from the (left) $P_{ccs}(4338)^0$ and (right) $P_{ccs}(4459)^0$ states, while the cyan filled curves represent the phase-space distributions~\cite{LHCb:2025lwm}.}
 \label{fig:Lb2LcDsKK}
\end{figure}

The decay $\Lambda_b^0 \to \Lambda_c^+ D_s^- K^+ K^-$ provides an opportunity to investigate possible pentaquark contributions in the $\Lambda_c^+ D_s^-$ system. The accessible phase space of the $\Lambda_c^+ D_s^-$ invariant mass, ranging from 4235 to 4632~\mev, covers the masses of the previously reported pentaquark candidates $P_{ccs}(4338)^0$ and $P_{ccs}(4459)^0$~\cite{LHCb:2022ogu,LHCb:2020jpq}. 
Using the full Run~2 data set, the LHCb experiment has observed the $\Lambda_b^0 \to \Lambda_c^+ D_s^- K^+ K^-$ decay for the first time~\cite{LHCb:2025lwm}. The branching fraction of this decay relative to $\Lambda_b^0 \to \Lambda_c^+ D_s^-$ is measured to be
\be
\frac{\mathcal{B}(\Lambda_b^0 \to \Lambda_c^+ D_s^- K^+ K^-)}{\mathcal{B}(\Lambda_b^0 \to \Lambda_c^+ D_s^-)}
= 0.0141 \pm 0.0019 \pm 0.0012.
\ee
A search for a possible $P_{ccs}^0 \to \Lambda_c^+ D_s^-$ contribution is performed by studying the $\Lambda_c^+ D_s^-$ invariant-mass distribution, with background statistically subtracted using the \textit{sPlot} technique~\cite{Pivk:2004ty}. Fit projections including contributions from the $P_{ccs}(4338)^0$ and $P_{ccs}(4459)^0$ states are shown in Fig.~\ref{fig:Lb2LcDsKK}. In both cases, the fitted significance is below $2\sigma$, and no significant evidence for pentaquark contributions in the $\Lambda_c^+ D_s^-$ system is observed.

\section{Future prospects}
\label{sec:future}

\subsection{The Belle II experiment}

\begin{figure}[t!]
\centering
\includegraphics[width=0.75\textwidth]{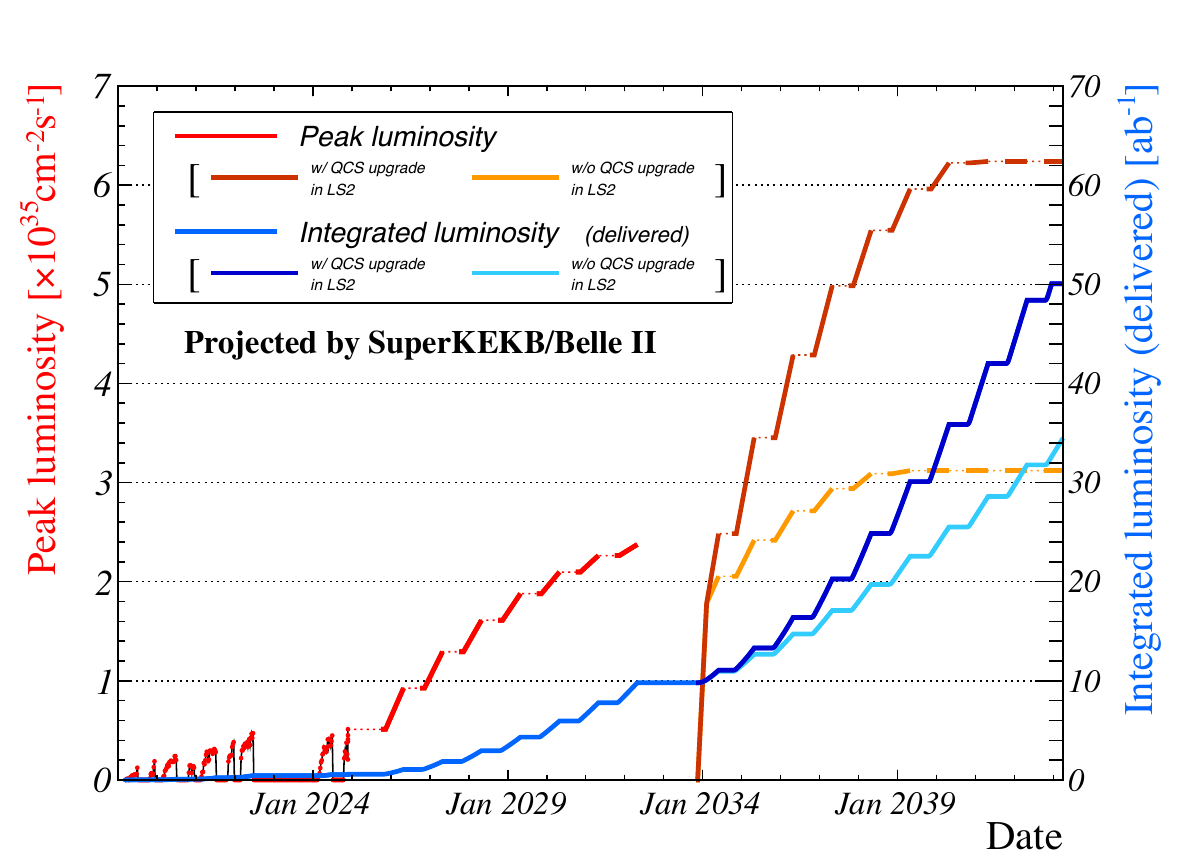}
\caption{Plan of instantaneous and integrated luminosity at SuperKEKB~\cite{luminosity}.}\label{LuminosityProjection_2024Dec}
\end{figure}

Belle II is the only currently operating $B$-factory experiment in the world that has been designed for precise measurements of weak interaction parameters, study exotic hadrons, and search for new phenomena beyond the Standard Model of particle physics. 
It started to record physics data in 2019.
In June 2022, SuperKEKB achieved a new instantaneous luminosity world record of $4.7\times 10^{34}$ cm$^{-2}$s$^{-1}$. 
From 2019 to 2022, Belle II accumulated a 428 fb$^{-1}$ data sample (run 1)~\cite{Belle-II:2024vuc} and, from 2024 to 2025, it accumulated a 165 fb$^{-1}$ data sample (run 2).
The luminosity projection plot for the coming years is shown in Fig.~\ref{LuminosityProjection_2024Dec}~\cite{luminosity}.
The integrated luminosity at Belle ll is expected to achieve 4 ab$^{-1}$ and 10 ab$^{-1}$ in 2029 and 2034, respectively.
With and without the Quadrupole-Corrector-Sextupole (QCS) upgrade,
the instantaneous luminosity can achieve $6.2\times 10^{35}$ cm$^{-2}$s$^{-1}$ and $3.1\times 10^{35}$ cm$^{-2}$s$^{-1}$, and the integrated luminosity can achieve 50 ab$^{-1}$ and 34 ab$^{-1}$ by 2043.

There are four ways to explore the bottomonium-like states at Belle II: (i) direct $e^+e^-$ annihilations, (ii) initial state radiation (ISR) processes~\cite{Baier:1968wvg}, (iii) hadronic transitions, and (iv) radiative transitions. In $e^+e^-$ collisions, the C.M. energy can be set at some specific energy points to search for possible new states and measure the masses and widths of known resonances. The $\Yb$ was discovered in the $\pi^+\pi^-\Upsilon(nS)$ final states using direct $e^+e^-$ collision data and the ISR process from 10.52 to 11.02 GeV at Belle~\cite{Belle:2019cbt}.
The hadronic and radiative transitions are also effective ways to search for new bottomonium-like states.
Noticeably, the $Z_b(10610)$ and $Z_b(10650)$ were observed in the pion transition from $\Yfive$ and $\Ysix$ at Belle~\cite{Belle:2011aa}.
The bottomonium ground state $\eta_b(1S)$ was observed in the radiative decay of $\Upsilon(3S)$~\cite{BaBar:2008dae}.
At Belle II, an energy scan from the $B\bar B$ threshold up to the highest possible energy of 11.24 GeV, with approximately 10 fb$^{-1}$
per point and 10 MeV step, is of great interest in the search for new excited $\Upsilon$ and $Y_b$ states~\cite{Belle-II:2018jsg}. If new bottomonium(-like) states are observed, more data will be collected at the corresponding energy points to perform a detailed study of radiative and hadronic transitions to search
for $X_b$ and other $Z_b$ states.

Below, we discuss future prospects for the $Z_b/W_{bJ}$, $X_b$, and $Y_b$ states in more detail.

\begin{itemize}
\item $Z_b/W_{bJ}$ states:
\begin{itemize}
\item Search for more $Z_b$ states via open flavor ($B\bar B$, $B\bar B^*$, $B^*\bar B^*$) and hidden flavor ($\pi\Upsilon(nS)$, $\pi \eta_b(1S)$, $\pi h_b(1P)$, $\pi\chi_{bJ}(1P)$) final states in single-pion emission decays of higher bottomonium(-like) states.
\item Search for new $Z_{bs}$ states in $e^+e^-\to K Z_{bs}(\to B^{(*)}\bar B^{(*)}_s, K \Upsilon(nS))$.
\item Given the $W_{bJ}$'s negative $G$-parity, the isovector $W_{bJ}$ states can be searched for in the $\rho\Upsilon(nS)$, $\pi\eta_b(nS)$, and $\pi\chi_{bJ}(1P)$ final states reached via the $e^+e^-$ radiative transitions.
\end{itemize}
\end{itemize}

\begin{itemize}
\item $Y_b$ states:
\begin{itemize}
\item Search for more $\Yb$ hadronic decay modes using final states of $K^+K^-\Upsilon(1S)$, $\eta h_b(1P)$, $\eta \Upsilon_J(1D)$, $\eta^{\prime} \Upsilon(1S)$, and $\phi\eta_b(1S)$.
\item Collect at least 137.2 fb$^{-1}$ data sample at $\sqrt{s}$ = 10.746 GeV to report the observation of $e^+e^-\to\gamma\chi_{b1,b2}(1P)$ according to the study of simulated pseudoexperiments discussed in Sec.~\ref{10753_Belle2}. 
Search for more $\Yb$ radiative decays to $\gamma\chi_{bJ}(2P,3P)$ and $\gamma\eta_b(nS)$.
\item Collect at least 17.5 fb$^{-1}$ data sample at $\sqrt{s}$ = 10.653 GeV to report the observation of the $Y_b(10653)\to\eta\Upsilon(2S)$ according to the study of simulated pseudoexperiments in Sec.~\ref{10753_Belle2}. Study the processes $e^+e^-\to\pi^+\pi^-\Upsilon(nS)$, $\eta h_b(1P)$, and $\pi^+\pi^-\pi^0\chi_{bJ}$ at $\sqrt{s}$ near 10.653 GeV to confirm the existence of the $Y_b(10653)$ and provide more decay modes of this state.
\item Measure the energy dependence of the cross sections for exclusive open-flavor ($B\bar B$, $B\bar B^*$, $B^*\bar B^*$, $B^{(*)}\bar B^{(*)}\pi$, $B_s\bar B_s$, and so on) and hidden flavor ($\pi^+\pi^-\Upsilon(nS)$, $\pi^+\pi^-h_b(mP)$, $\eta\Upsilon(nS)$, $\eta\Upsilon_J(1D)$, $\eta h_b(mP)$, $\omega\eta_{b}(nS)$, $\omega\chi_{bJ}(1P)$, and so on) processes to search for new $Y_b$ states and precisely determine the parameters of these new as well as already known $Y_b$ states using Belle II high-statistics scan data.
\end{itemize}
\end{itemize}

\begin{itemize}
\item $X_b$ states:
\begin{itemize}
\item Considering the breaking isospin symmetry may be suppressed, the $X_b$ should be searched for in the final states such as $\pi^+\pi^-\pi^0\Upsilon(1S)$, $\pi^+\pi^-\chi_{bJ}(1P)$, and $\gamma\Upsilon(1S)$.
\item Besides the $e^+e^-$ radiative decay, the $X_b$ can also be searched for in the reactions $e^+e^-\to\omega X_b$ and $e^+e^-\to \rho X_b$, which implies an increase of the SuperKEKB $e^+e^-$ collision energy to at least 11.4 GeV. It may be challenging for Belle II.
\end{itemize}
\end{itemize}

\subsection{The LHCb experiment}

\begin{figure}[t!]
\centering
\includegraphics[width=0.7\textwidth]{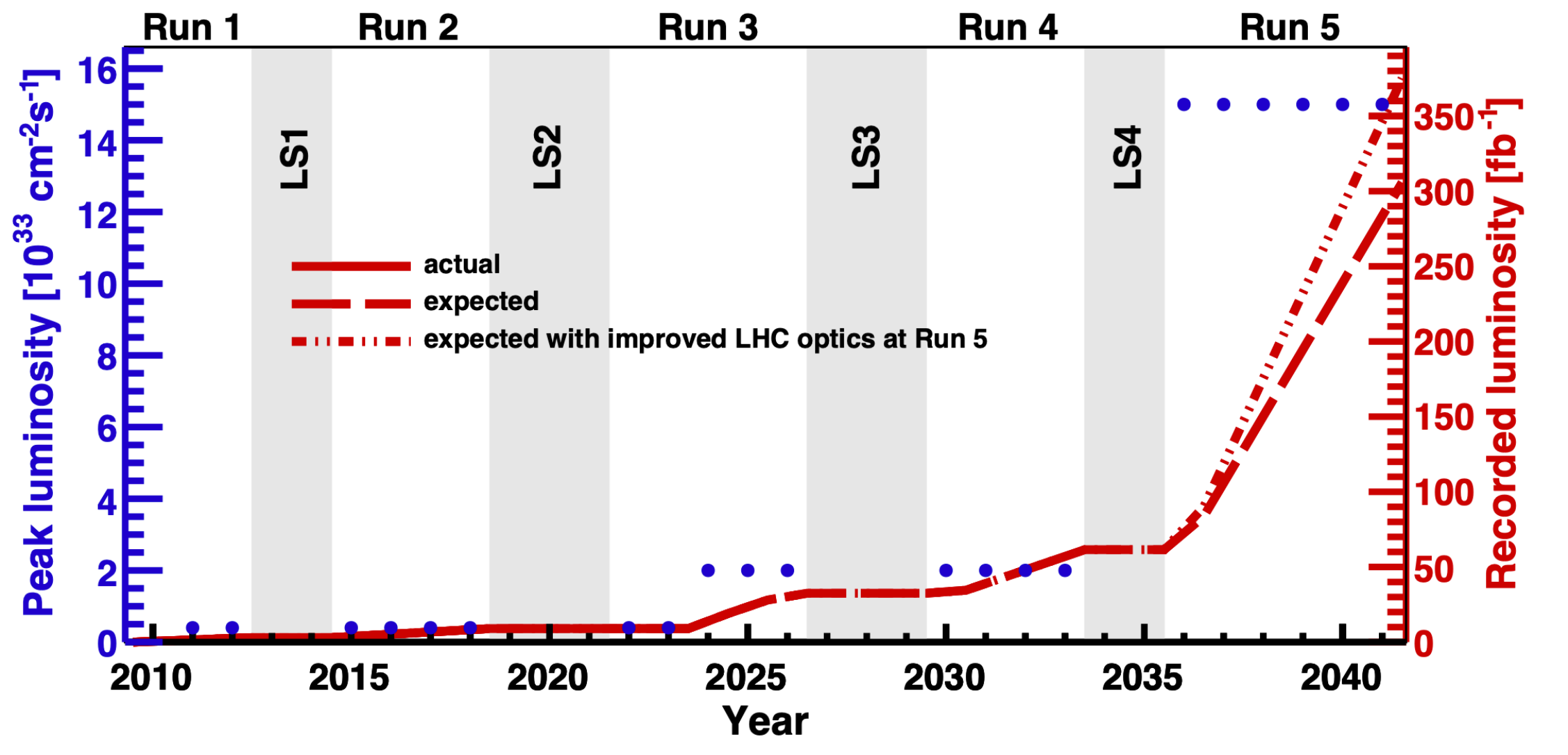}
\caption{Instantaneous (peak) and integrated (recorded) luminosity profiles for the original LHCb (Runs 1 and 2), Upgrade I (Runs 3 and 4) and Upgrade II (Run 5) detectors, in the
context of the LHC and HL-LHC schedule.}
\label{fig:lhcb_lumi_plan}
\end{figure}

The achievements of the LHCb physics program in Run~1\dash2 have already revealed a rich landscape of exotic hadrons, yet the current understanding remains incomplete.
Following Upgrade~I, the LHCb detector operates at an instantaneous luminosity
approximately five times higher than in Run~2 and now employs a fully software-based trigger in place of the former hardware system~\cite{LHCb:2023hlw}. This enables real-time event reconstruction and roughly doubles the trigger efficiency for fully hadronic final states, providing a foundation for both the discovery of new exotic states and major progress in long-standing
questions related to previously observed structures and decay modes.
By the end of 2026, Run~3 has a target integrated luminosity of approximately
$23\,\text{fb}^{-1}$, a goal that is already close to being met with more than
$22\,\text{fb}^{-1}$ recorded so far. An intermediate Upgrade~Ib, scheduled between
Runs~3 and 4, will focus on consolidating detector performance without a further
increase in instantaneous luminosity. By the end of Run~4, the total integrated
luminosity is expected to reach around $50\,\text{fb}^{-1}$, corresponding to an
order-of-magnitude increase in heavy-flavor yields compared to Run~1\dash2
(see Fig.~\ref{fig:lhcb_lumi_plan} and Table~\ref{tab:production_comp}). A major Upgrade~II is foreseen for the
High-Luminosity LHC era~\cite{LHCb:2021glh}, enabling operation at approximately seven times the
current luminosity and aiming for a total integrated luminosity of about
$300\,\text{fb}^{-1}$ over the lifetime of the program. This dataset will provide
a significant step forward for precision measurements and detailed studies of exotic
hadrons.

\begin{table}[t]
 \centering
 \caption{Approximate number of $b$-hadrons produced and expected at the $B$-factories \cite{Belle-II:2018jsg,BaBar:2014omp,Bernlochner:2021vlv} and at the LHCb
 experiment~\cite{Albrecht:2019djb}, including some of the latest developments~\cite{HL-LHC-TDR:2020}. The LHCb numbers take into account an average
 geometrical acceptance of about 15\%. Note that the overall $B$ reconstruction efficiencies at LHCb are usually significantly lower than those at the $B-$factories.
 The two values of integrated luminosities and the C.M. energies shown
 for Belle and Belle II correspond to data taking at the
 $\Yfour$ and $\Yfive$ resonances, respectively. The $B$-factory experiments also
 recorded data sets at lower C.M.\ energies (below the open beauty threshold) that are not included in this table.}
 \label{tab:production_comp}
 \begin{tabular*}{\textwidth}{@{\extracolsep{\fill}} l cc cccc}
 \hline\hline
 \multirow{2}{*}{Experiment} & \multirow{2}{*}{Belle} & \multirow{2}{*}{Belle II} & \multicolumn{4}{c}{LHCb} \\

 & & & Run 1 & Run 2 & Runs 3\dash4 & Runs 5\dash6 \\
 \hline
 Completion date & 2010 & 2041 & 2012 & 2018 & 2031 & 2041 \\
 C.M. energy & 10.58/10.87~GeV & 10.58/10.87~GeV & 7/8~TeV & 13~TeV & 14~TeV & 14~TeV \\
 $b\bar{b}$ cross section [nb] & 1.05/0.34 & 1.05/0.34 & (3.0/3.4)$\times 10^5$ & $5.6\times10^5$ & $6.0\times10^5$ & $6.0\times 10^5$ \\
 Integrated & & & & 6 & & \\
 luminocity [fb$^{-1}$] & 711/121 & $(40/4) \times 10^3$ & 3 & 6 & 40 & 300 \\
 \hline
 $B^0$ mesons [10$^9$] & 0.77 & 40 & 100 & 350 & 2,500 & 19,000 \\
 $B^+$ mesons [10$^9$] & 0.77 & 40 & 100 & 350 & 2,500 & 19,000 \\
 $B^0_s$ mesons [10$^9$] & 0.01 & 0.5 & 24 & 84 & 610 & 4,600 \\
 $\Lambda_b^0$ baryons [10$^9$] & - & - & 51 & 180 & 1,300 & 9,800 \\
 $B^+_c$ mesons [10$^9$] & - & - & 0.8 & 4.4 & 19 & 150 \\
 \hline\hline
 \end{tabular*}
\end{table}

\begin{figure}[t]
\centering
\includegraphics[width=0.5\textwidth]{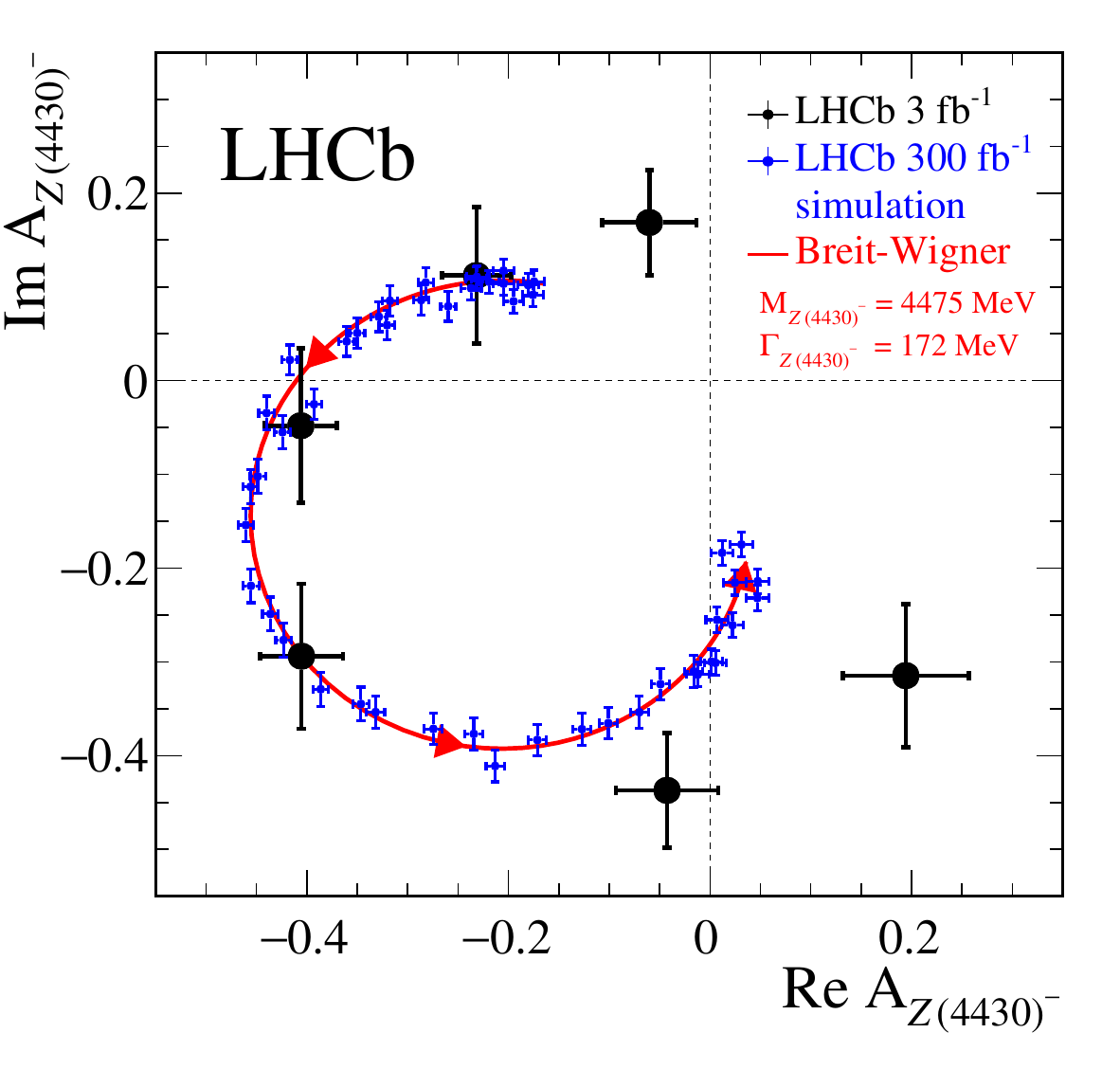}
\caption {
Argand diagram of the $T_{c\bar{c}1}(4430)$ (aslo denoted as $Z(4430)^-$) amplitude in bins of $m^2_{\psi(2S) \pi^-}$ from a fit to the $B^0\to \psi(2S) K^+ \pi^- $ decays. The black points are the results based on Run 1 data~\cite{LHCb:2014zfx} while the blue points correspond to an extrapolation to an integrated luminosity of $300\,\text{fb}^{-1}$ expected at the LHCb Upgrade II.
The red curve is the prediction from the BW formula with a resonance mass of 4475~\mev~and width of 172 MeV~\cite{LHCb:2018roe}.}
 \label{fig:upgrade_argand_Zc4430}
\end{figure}

\paragraph{\textbf{Precise measurement of the exotic candidates}} For the well-established exotic state $\X$, larger data samples will enable studies of additional decay modes and a more precise determination of its branching fractions. In radiative decays, the photon reconstruction efficiency at LHCb is intrinsically low; however, the statistics foreseen in Upgrade~II are expected to compensate for this limitation, allowing significantly improved measurements of radiative decay rates
and, consequently, tighter constraints on the internal structure of the state.
Another approach to probe the nature of the $\X$, not discussed in detail in this review, is the study of its prompt-production cross section at the LHC in comparison with conventional charmonium states. The System for Measuring Overlap with Gas 2~\cite{2707819} further extends these studies to additional colliding systems (for example\ $p$--He, $p$--Ne, $p$--Ar), providing complementary information on production mechanisms.
Compared with $\X$, many other exotic candidates --- such as the $T_{c\bar{c}1}(4430)$ --- remain less well explored. In some cases, measurements from different experiments are inconsistent; in others, the uncertainties are too large to discriminate between theoretical models, or the reported features can not be accommodated by existing models. Clarifying their properties, or even establishing their existence, requires analyses based on larger data samples. 
Amplitude analyses with larger data samples are essential to precisely determine their resonance parameters, thereby providing crucial input to clarify their underlying nature (see for example Fig.~\ref{fig:upgrade_argand_Zc4430}).

\paragraph{\textbf{Search for new exotic states}} Multiquark candidates are often found near thresholds of hadron--hadron systems, motivating dedicated searches in regions such as the $\Lambda_c \bar{D}_s$ and $D\bar{D}$ thresholds. These processes typically involve multi-body final states and low reconstruction efficiency, making large data samples essential. With the increased instantaneous luminosity and upgraded detector performance in Run~3 and Run~4, LHCb is expected to collect samples of order millions for $b$-hadron decays to final states containing $J/\psi$ or $\psi(2S)\!\to\!\mu^+\mu^-$, and tens of thousands of events for decays to double open-charm final states (for example, $D\bar{D}$, $\Lambda_c^+\bar{D}$). 
The Run 3 and Run 4 data will also enable searches for doubly-charmed exotic states and fully-charmed tetraquark candidates, including investigations of structures in $D^{(*)}D^{(*)}$ and $J/\psi J/\psi$ systems.
Some expected data samples and key decay modes are listed in Table~\ref{tab:expected_yields}.
These statistics will broaden the experimental reach for studies of threshold structures and offer a possibility to search for tetraquark and pentaquark candidates in these modes.
Related configurations are expected in the bottom sector. LHCb is pursuing searches for open-bottom $(bqqq)$, hidden-bottom $(b\bar{b}qq)$, and doubly-bottom $(bb\bar{q}\bar{q})$ candidates, as well as their pentaquark analogues, in prompt production. This program will continue throughout Run 3, Run 4, and into the HL-LHC era. Building on the established searches for tetraquark and pentaquark candidates, it is natural to extend the exploration to dibaryons, which represent the next relevant class of multiquark systems. Such searches are included in the current and future LHCb physics program.

\begin{table}[t]
\centering
\label{tab:yields}
\caption{Expected data samples at LHCb upgrade II and Belle II for key decay modes for the spectroscopy of heavy flavored hadrons. The expected yields at Belle II are estimated by assuming similar efficiencies as those at Belle \cite{LHCb:2018roe}.}
\begin{tabular}{l|ccc|c}
\hline
 & \multicolumn{3}{c|}{LHCb} & Belle II\\
Decay mode & 23$\,\text{fb}^{-1}$ & 50$\,\text{fb}^{-1}$ & 300$\,\text{fb}^{-1}$ & 50$\,\text{ab}^{-1}$\\
\hline
$B^+\to \X(\to J/\psi \pi^+\pi^-)K^+$ &14k &30k&180k&11k\\
$B^+\to \X(\to J/\psi(2S) \gamma)K^+$ &500 &1k& 7k & 4k \\
$B^0\to \psi(2S)K^-\pi^+$ &340k &700k&4M& 140k\\
$B_c^+\to D_s^+ D^0 \bar{D}^0 $ & 10 &20& 100 & ---\\
$\Lambda_b^0\to J/\psi p K^-$ &340k &700k&4M& ---\\
$\Xi_b^-\to J/\psi\Lambda K^-$ &4k &10k&55k& ---\\
\hline
\end{tabular}
\label{tab:expected_yields}
\end{table}

\subsection{Discussion on theoretical prospects}

The discovery and rapid proliferation of exotic hadronic states require both updates to traditional theoretical methods and the development of new frameworks appropriate to the present situation.
In this subsection, we briefly review the most developed and successful theoretical approaches to ordinary and exotic hadrons and outline prospects for their future progress.

\paragraph{\textbf{Quark-model approaches}}

Quark models provide one of the most straightforward and intuitive frameworks for the study of ordinary and exotic hadrons. In this approach, hadrons are treated as bound states of quarks and gluons interacting via phenomenological potentials inspired by QCD, typically incorporating confinement and short-range one-gluon exchange. As anticipated in the early years of the quark model, extending the framework from conventional quark–antiquark and three-quark hadrons to multiquark systems naturally leads to the possibility of tetraquarks, pentaquarks, hybrid mesons, glueballs, and other exotic configurations.
Importantly, although such higher configurations qualify as exotic hadrons, their nature is restricted to compact (confinement-governed) bound states, and therefore the unphysical Riemann sheets of the complex energy plane are not accessed. For multiquark configurations, specific internal clustering is often assumed, with the diquark–antidiquark picture of compact tetraquarks being the most widely used. Although the dynamical origin of such clustering is not always well understood, the resulting tetraquark models are relatively simple and can provide reasonable predictions for mass spectra, spin–parity assignments, decay patterns, and related observables. Another attractive feature of quark models is their ability to identify patterns across flavor sectors and to highlight states that do not fit naturally into the conventional quarkonium spectrum.

Despite their undeniable phenomenological success, quark models suffer from intrinsic limitations that often hinder their applicability to the study of exotic hadrons. On the one hand, quark models are not derived directly from QCD and neglect important dynamical effects such as coupled-channel interactions, threshold behavior, final-state interactions, and related nonperturbative phenomena.
Moreover, the frequent appearance of exotic states in close proximity to strong thresholds appears essentially accidental within the quark-model framework. Consequently, quark-model predictions for near-threshold exotic states can be strongly affected by effects lying beyond the scope of the model, rendering its ultimate predictive power uncertain.

Current efforts aim to improve the quark model by unquenching it, that is, by incorporating unitarization and coupled-channel dynamics, as well as mixing between compact multiquark configurations and hadronic molecular components. Such developments seek to bridge the gap between the intuitive constituent-quark picture and more rigorous QCD-based approaches, providing a more realistic description of exotic hadrons while retaining the simplicity and conceptual transparency inherent in the traditional quark model. 

Given the vast and extensive literature on quark models, we mention here only a few representative references that provide a general introduction to the quark model and its application to exotic hadrons. In particular, an overview of the quark model is presented in the dedicated chapter of RPP \cite{ParticleDataGroup:2024cfk}. In Ref.~\cite{Richard:2016eis}, the constituent quark model description of multiquark states is discussed in considerable detail. The review in Ref.~\cite{Esposito:2016noz} also addresses the multiquark sector of hadron spectroscopy and offers a discussion aimed at providing a coherent interpretation of the $X$ and $Z$ resonances within the compact tetraquark paradigm.

\paragraph{\textbf{Effective field theories}}

EFTs provide a systematic and largely model-independent framework for the study of ordinary and exotic hadrons whose dynamics is governed by well-separated energy scales. In particular, EFTs are ideally suited for the description of near-threshold states, where conventional quark-model approaches often encounter significant limitations. In this context, many exotic hadrons are naturally interpreted as hadronic molecules, that is, weakly bound states of hadrons interacting via residual strong forces between color-neutral objects. Here, binding should be understood in a broad sense, implying a large probability to find the studied resonance in a given hadronic channel (approaching unity for a pure molecular state). This definition does not constrain the position of the corresponding pole in the complex energy plane. Accordingly, an exotic state may manifest itself as a bound state, a virtual state (a pole just below threshold on the physical or nearby unphysical Riemann sheet, respectively), or a resonance (a pole near threshold on an unphysical sheet, possibly above threshold). Hadronic EFTs exploit symmetries inherent in the system, such as HQSS, flavor symmetry, and chiral symmetry, to constrain the interactions among heavy mesons and between heavy and light degrees of freedom. This allows for robust predictions of spin partners, decay patterns, and symmetry-breaking effects that are largely insensitive to the poorly known short-distance dynamics. A key advantage of the EFT framework is its ability to treat bound states, virtual states, and resonances on equal footing, while consistently implementing unitarity and analyticity. This is particularly important for exotic states located very close to open-flavor thresholds, where scattering dynamics and coupled-channel effects dominate. Moreover, EFTs provide a transparent connection between experimental observables and underlying interaction parameters, enabling quantitative tests of the molecular interpretation. Interaction parameters and couplings to different hadronic channels extracted in this model-independent way can subsequently be interpreted within more microscopic models of exotic hadrons and their dynamics, thereby providing a bridge between phenomenology and underlying QCD-based descriptions.

The main difficulties and limitations of hadronic EFTs are associated with the convergence of the EFT expansion --- of which the chiral expansion is a well-known example --- and with the scarcity of experimental data needed to constrain the low-energy constants. Moreover, by construction, EFTs can not address the microscopic origin of short-range interactions, which must instead be encoded in contact terms or matched to more fundamental approaches. Future developments should therefore focus on combining EFTs with LQCD and dispersive methods (see below) to further constrain interaction strengths and extend the applicability of EFT frameworks to more complex coupled-channel and multichannel systems. Such synergies are expected to significantly improve the reliability of theoretical predictions for exotic hadrons and clarify the role of molecular components in their internal structure.

An alternative EFT formulated in coordinate space can be derived from QCD by exploiting the Born–Oppenheimer separation between slow (heavy-quark) and fast (light-quark and gluon) degrees of freedom. This BOEFT is currently under active development and has already demonstrated strong potential for understanding exotic hadrons; see, for example, Ref.~\cite{Berwein:2024ztx} for a recent comprehensive study that also provides a pedagogical introduction to the framework. In this approach, hadronic properties are described through systems of coupled Schrödinger equations, whose potentials and couplings are determined by nonperturbative, gauge-invariant correlators. These correlators can be obtained from LQCD calculations and subsequently parameterized for use in the BOEFT framework~\cite{Brambilla:2024imu, Alasiri:2024nue}. Indeed, since the earliest lattice studies, interquark potentials have been among the key observables. In particular, the linearly rising confining potential between two static fundamental sources was already observed in the pioneering lattice work~\cite{Creutz:1980zw}. Subsequent lattice calculations have provided detailed information on static spin-independent potentials (see, for example, Ref.~\cite{Bali:2000gf}), spin-dependent potentials~\cite{Koma:2006fw}, and hybrid potentials associated with excited gluonic degrees of freedom~\cite{Juge:1997nc} and later studies. In this way, the BOEFT framework is firmly anchored in first-principles LQCD results and has the potential to deliver model-independent predictions for exotic hadrons. In the meantime, a systematic account for non-local interactions is somewhat obscure in BOEFT and provides an important direction for future theoretical development.

For a general introduction to EFTs and a comprehensive discussion of the foundations of EFTs for QCD, the reader is referred to the recent monograph~\cite{Meissner:2022cbi} and the references therein. A thorough review of the molecular interpretation of exotic hadrons and its application to specific exotic candidates can be found in Ref.~\cite{Guo:2017jvc}.

\paragraph{\textbf{Functional methods}}
Functional methods provide a continuum, nonperturbative framework for studying hadron properties directly from QCD. These approaches are based on QCD Green’s functions and include, in particular, Dyson–Schwinger equations, Bethe–Salpeter equations, and functional renormalisation group techniques. As such, they offer an alternative to LQCD that does not rely on space–time discretisation and allows for studies at physical quark masses. In the context of exotic hadrons, functional approaches are well suited to investigate the interplay between confinement, dynamical chiral symmetry breaking, and multiquark dynamics. Bound states emerge as poles in appropriate Green’s functions, while multiquark states can be accessed through coupled systems of Bethe–Salpeter equations or via effective interactions generated by gluon exchange. These methods provide insight into the internal structure of exotic states, including the relative importance of compact multiquark components and meson–meson correlations. A key strength of functional approaches is their ability to describe both quark-level dynamics and hadronic bound states within a unified framework. Furthermore, analytic continuation techniques can, in principle, be employed to address resonance properties and time-like observables, which remain challenging for other nonperturbative approaches.

These methods rely on truncations of the infinite hierarchy of coupled equations, and the reliability of the resulting predictions depends critically on the quality of the truncation schemes employed. While systematic strategies for improvement exist, achieving quantitative precision for multiquark systems remains challenging, particularly for states near open-flavor thresholds and in the presence of coupled-channel dynamics. Ongoing efforts therefore focus on refining truncation schemes, incorporating unquenching effects, and forging closer connections with LQCD and EFTs. These developments are expected to enhance the predictive power of functional approaches and clarify their role within a comprehensive theoretical framework for exotic hadrons. For a recent review and a convenient starting point for further study of functional methods, the reader is referred to Ref.~\cite{Eichmann:2025wgs} and the references therein.

\paragraph{\textbf{QCD sum rules}}
QCD sum rules constitute a widely used analytical framework for investigating the properties of exotic hadrons by relating hadronic observables to QCD correlation functions evaluated at the quark and gluon level. The method is based on the operator product expansion (OPE), which systematically separates short-distance perturbative contributions from long-distance nonperturbative effects encoded in vacuum condensates. By matching the OPE representation of the correlator to a phenomenological parametrization of the hadronic spectrum, one can extract estimates for masses, decay constants, and coupling strengths of exotic states. In the context of exotic spectroscopy, QCD sum rules have been extensively applied to tetraquark, pentaquark, and hadronic molecular currents. One of the main advantages of this approach is its direct connection to QCD degrees of freedom, combined with a relative technical simplicity compared to fully numerical methods such as LQCD. Moreover, sum rules can be formulated for a wide range of quantum numbers, making them a flexible tool for exploring possible exotic configurations and guiding experimental searches.

Nevertheless, the application of QCD sum rules to exotic hadrons faces significant challenges. First, the construction of interpolating currents is not unique, and different choices can lead to substantially different predictions. Second, the separation between the ground state and the continuum introduces a considerable model dependence. In addition, the convergence of the OPE can be slow for multiquark systems, where higher-dimensional condensates may play an essential role. As a consequence, theoretical uncertainties are often sizable and difficult to quantify in a systematic way.

Current efforts therefore focus on improving the reliability of QCD sum-rule analyses by refining interpolating currents, extending calculations to higher orders in the OPE, and incorporating constraints from complementary theoretical approaches and experimental data. While QCD sum rules are unlikely to provide high-precision predictions on their own, they remain a valuable complementary tool for surveying the spectrum of exotic hadrons and identifying promising candidates for further experimental and theoretical studies. A comprehensive overview of the method, its applications to exotic hadrons, and relevant references can be found in the recent review~\cite{Wang:2025sic}.

\paragraph{\textbf{Amplitude-based and dispersive methods}} 

Amplitude-based and dispersive approaches provide a largely model-independent framework for the analysis of experimental data on exotic hadrons, ensuring consistency with fundamental principles such as unitarity, analyticity, and crossing symmetry. These methods focus on hadronic scattering and decay amplitudes, enabling a rigorous extraction of resonance parameters, pole positions, and decay couplings directly from experimental observables. Dispersion relations relate the real and imaginary parts of amplitudes and impose strong constraints on their energy dependence, which is particularly important for states located near thresholds or in processes with strong final-state interactions.

Combined with partial-wave analysis and unitarity constraints, dispersive techniques enable the disentanglement of overlapping resonances and the identification of genuine exotic states, as opposed to kinematic enhancements or cusp effects. Amplitude-based analyses have been crucial in the study of near-threshold charmonium-like states, where conventional BW parameterizations are often inadequate. By exploiting analyticity and unitarity, dispersive methods allow for a reliable determination of pole positions in the complex energy plane and provide robust information on the underlying nature of the observed structures.

Although these approaches are computationally and analytically demanding, they offer a rigorous bridge between theory and experiment, complementing LQCD, EFTs, and quark-model predictions. Current efforts focus on extending dispersive frameworks to coupled-channel systems and multi-body decays, with the goal of achieving a more quantitative and unified description of exotic hadrons and their interactions.

Recent progress in the application of partial-wave dynamics combined with chiral EFTs and dispersion relations is reviewed in Ref.~\cite{Yao:2020bxx}. The contributions of the Joint Physics Analysis Center, one of the leading groups employing amplitude-based and dispersive techniques in hadron spectroscopy, are summarized in Ref.~\cite{JPAC:2021rxu}, which also provides a pedagogical introduction to the technical aspects of these methods.

\paragraph{\textbf{LQCD}}

LQCD provides a first-principles, nonperturbative numerical framework for studying hadron spectroscopy directly from the QCD Lagrangian. By formulating theory on a finite Euclidean space–time lattice, numerical simulations enable the computation of correlation functions, energy spectra, and matrix elements with systematically improvable control over theoretical uncertainties. As such, LQCD plays a central role in theoretical investigation of strong interactions. A comprehensive pedagogical introduction to lattice techniques can be found in Ref.~\cite{Gattringer:2010zz}.

In recent years, substantial progress has been achieved in lattice studies of multiquark systems and near-threshold states. The use of large operator bases, including both compact multiquark and two-hadron interpolating fields, has enabled the extraction of finite-volume energy levels relevant for exotic candidates. Through the Lüscher finite-volume formalism, these energy levels can be related to infinite-volume scattering amplitudes and resonance parameters, allowing for quantitative investigations of bound states, virtual states, and resonances. Current lattice techniques are already capable of providing promising predictions for exotic hadrons containing heavy quarks, including the $\X$~\cite{Prelovsek:2013cra, Padmanath:2015era}, the $T_{cc}(3875)^+$~\cite{Cheung:2017tnt,Junnarkar:2018twb,Padmanath:2022cvl, Chen:2022vpo,Lyu:2023xro, Collins:2024sfi,Shrimal:2025ues,Prelovsek:2025vbr,Whyte:2024ihh}, its possible isospin-1 partner
\cite{Meng:2024kkp,PitangaLachini:2025pxr}, its predicted bottom counterparts $T_{bc}$ \cite{Alexandrou:2023cqg,Padmanath:2023rdu,Radhakrishnan:2024ihu} and $T_{bb}$~\cite{Francis:2016hui,Junnarkar:2018twb,Hudspith:2020tdf,Aoki:2023nzp,Alexandrou:2024iwi,Tripathy:2025vao, Vujmilovic:2025czt}, the $Z_b$ states~\cite{Prelovsek:2019ywc, Sadl:2021bme}, and other multiquark configurations.

Despite these advances, LQCD studies of exotic hadrons face substantial challenges. Near-threshold dynamics and coupled-channel effects require high statistical precision, large lattice volumes, and simulations close to the physical pion mass. Moreover, the extraction of resonance properties involves nontrivial analytic continuation and model assumptions, which can limit the precision of the results, especially in the presence of additional complications such as left-hand cuts arising from $t$- and $u$-channel exchanges~\cite{Du:2023hlu}. The computational cost associated with multi-hadron operators and disconnected diagrams remains a major obstacle, and a comprehensive coupled-channel study of exotic states in the bottomonium sector is still beyond current lattice capabilities.

Ongoing developments aim to overcome these limitations through improved algorithms, increased computational resources, and extensions of finite-volume formalisms to more complex coupled-channel systems. In parallel, closer connections between LQCD and EFTs are being established, enabling controlled extrapolations and more transparent interpretations of lattice results. Meanwhile, it should also be noted that already the present lattice simulations of exotic hadrons performed away from the physical point bring additional insights into the underlying physics that is not available in the real experiment, so they are quite informative for phenomenological and theoretical studies.
The progress achieved by lattice collaborations in recent years has been remarkable; see, for example, Ref.~\cite{Ryan:2020iog} or recent reviews \cite{Bicudo:2022cqi,Francis:2024fwf}. As these efforts continue, LQCD is expected to provide increasingly precise and reliable insights into the spectrum and structure of exotic hadrons, serving as a cornerstone for a comprehensive understanding of strong-interaction dynamics in the near future.

\section{Summary}
\label{sec:summary}

At present, hadronic physics is a vibrant and rapidly developing area of strong-interaction physics, yielding a wealth of new and exciting results on exotic hadrons that do not fit within the simple quark-model framework. It therefore brings together extensive experimental and theoretical efforts aimed at gaining deeper insight into the nature and properties of these enigmatic hadronic states.

This review summarizes recent advances in the study of exotic hadrons associated with the $b$-quark over the past several years. In particular, we revisit the current understanding of the well-established $\X$, $\Yb$, and the twin 
$Z_b$ exotic states, and highlight recent experimental progress in the search for additional members of the 
$XYZ$ family, as well as for other tetraquark and pentaquark states. Special emphasis is placed on the capabilities of the Belle/Belle II and LHCb experiments, both of which are uniquely suited to performing comprehensive and complementary studies of exotic hadrons containing the 
$b$ quark or produced in weak decays of 
$B$ mesons.

So far the Belle experiment has discovered the only two charged and undoubtedly exotic bottomonium-like states $Z_b(10610)^{\pm}$ and $Z_b(10650)^{\pm}$~\cite{Belle:2011aa}. 
The most recent neutral bottomonium-like state, $\Yb$, was previously observed by Belle in 2019~\cite{Belle:2019cbt}.
Using a larger dataset collected near $\sqrt{s} = 10.75\;\text{GeV}$, Belle~II has conducted a dedicated study to investigate the nature of the $\Yb$~\cite{Belle-II:2024mjm,Belle-II:2026oyq,Belle-II:2022xdi,
Belle-II:2025jus,
Belle-II:2023twj,
Belle-II:2025iil,
Belle-II:2025ubm,
Belle-II:2024niz}.
Belle and Belle~II have searched for the $X_b$ states through their decays into $\pi^{+}\pi^{-}\pi^{0}\Upsilon(1S)$ and $\pi^{+}\pi^{-}\chi_{b1,b2}$, but no significant signals have been observed~\cite{Belle:2014sys,Belle-II:2022xdi,Belle-II:2025ubm}.
By mid-2026, Belle~II is expected to accumulate approximately $1\;\text{ab}^{-1}$ data at $\sqrt{s}$ = 10.58 GeV~\cite{luminosity}.
Subsequently, the experiment plans to collect additional data at $\sqrt{s} = 10.65\;\text{GeV}$ and in the regions around the $\Yfive$ and $\Ysix$ resonances, in order to further explore the properties of the $X_b$, $Y_b$, and $Z_b$ states and discover new members of these families.

In the past three years, the LHCb experiment has reported a number of results on tetraquark states,
including the observation of new hidden-charm and strange tetraquark candidates
and improved measurements of the properties of previously known exotic hadrons~\cite{LHCb:2022sfr,LHCb:2024xyx,LHCb:2024iuo,LHCb:2024vfz,LHCb:2022aki,LHCb:2023hxg,LHCb:2024tpv,LHCb:2024cwp,LHCb:2025kxf}.
Despite this progress, the internal structure of exotic hadrons remains an open question.
Even for the $\X$, which is the most extensively studied case,
experimental investigations have progressed from measurements of its mass and production rates
to determinations of its quantum numbers and studies of its line shape,
yet no definitive conclusion on its nature has been reached.
For many other exotic states, current experimental constraints are limited,
allowing significant opportunities for future measurements.
Recent progress on pentaquark states has been more limited,
with the $P_{c\bar cs}(4338)^0$ being the only newly reported candidate~\cite{LHCb:2022ogu}.
At the same time, LHCb has observed several new $b$-baryon decay modes with open-charm final states,
such as $\Lambda_b^0 \to \Lambda_c^+ \bar{D}^{(*)0} K^-$~\cite{LHCb:2023eeb} and
$\Lambda_b^0 \to \Sigma_c^{(*)++} D^{(*)-} K^-$~\cite{LHCb:2024fel},
which provide additional channels for pentaquark studies.
With the increased statistics from Run~3,
LHCb can further investigate known pentaquark candidates
and continue searches for new exotic hadrons,
providing experimental input for discriminating between theoretical interpretations.

From a theoretical perspective, we highlight the rapid progress in developing methods to interpret the steadily increasing volume of experimental data. These advances span a wide range of approaches, from the quark model --- which continues to provide an intuitive classification scheme for both conventional and exotic hadrons --- to model-independent, systematically improvable EFT frameworks formulated in both momentum and coordinate space. They also include functional methods and QCD sum rules, which remain powerful tools for exploring hadronic structure, as well as amplitude-based and dispersive techniques that provide complementary, model-independent strategies for data analysis. In parallel, the lattice QCD approach has made significant progress in incorporating multi-hadron dynamics and is expected to deliver increasingly precise first-principles insights into the nature of exotic hadrons.

Finally, we address future directions of the field and emphasize that further advances in understanding the nature and properties of exotic hadrons will depend on sustained and coordinated experimental and theoretical efforts. In particular, progress will rely on comprehensive data analyses, as well as systematic predictions and dedicated searches for various partners of the already established states, including their spin and light- and heavy-flavor partners.

\section*{Acknowledgements}

The authors would like to thank (in alphabetic order) 
Pedro Bicudo, Alexander Bondar,
Nora Brambilla, 
Feng-Kun Guo,
Christoph Hanhart,
Padmanath Madanagopalan, 
Ulf Mei{\ss}ner, Eulogio Oset, 
Sasa Prelovsek for reading the manuscript and valuable comments and suggestions.
This work is supported by the National Key R\&D Program of China under contract Nos. 2024YFA1610501, 2024YFA1610502, 2024YFA1610503, and 2024YFA1610504; National Natural Science Foundation of China under contract
Nos. 12135005 and 12475076; Fundamental Research Funds of China for the Central Universities under
contract Nos. 2242025RCB0014 and RF1028623046; the Spanish MICIU/AEI/10.13039/ 501100011033 under grants PID2023-147458NB-C21 and CEX2023-001292-S and by Generalitat Valenciana under Grant CIPROM 2023/59; Deutsche Forschungsgemeinschaft (DFG) under contract No. 525056915; CAS President’s International Fellowship Initiative (PIFI) under contract No.~2024PVA0004\_Y1.





\bibliographystyle{elsarticle-num}

\bibliography{ref}

\newpage

\end{document}